\documentclass[paper,notoc]{JHEP3}

\usepackage{epsfig}
\usepackage{graphicx}
\usepackage{feynmp}

\usepackage{amsmath}
\usepackage{amsfonts}
\usepackage{amssymb}
\usepackage{graphicx}

\def\beq{\begin{equation}}
\def\eeq{\end{equation}}
\newcommand{\bea}{\begin{eqnarray}}
\newcommand{\eea}{\end{eqnarray}}
\newcommand{\nn}{\nonumber}

\def\Eqn#1{Eq.~(\ref{#1})}
\def\Ref#1{Ref.~\cite{#1}}
\def\eqns#1#2{Eqs.~(\ref{#1}) and~(\ref{#2})}

\def\bsp#1\esp{\begin{split}#1\end{split}}

\newcommand{\eps}{\epsilon}
\newcommand{\ep}{\epsilon}

\newcommand{\ord}{\begin{cal}O\end{cal}}

\newcommand{\cC}{{\cal C}}

\def\P{{\cal P} }

\def\N{{\cal N} }
\def\C{{\cal C} }

\newcommand{\eug}{\gamma_E}

\newcommand{\rd}{\mathrm{d}}

\newcommand{\re}{\textrm{Re}}
\newcommand{\res}{\textrm{Res}}
\newcommand{\umijk}{u_{123}^{(-)}}
\newcommand{\upijk}{u_{123}^{(+)}}
\newcommand{\gcal}{\begin{cal}G\end{cal}}
\newcommand{\hcal}{\begin{cal}H\end{cal}}
\newcommand{\rcal}{\begin{cal}R\end{cal}}
\renewcommand{\log}{\ln}

\newcommand{\mbint}{\int_{-i\infty}^{+i\infty}}

\def\bit#1\eit{\begin{itemize}#1\end{itemize}}
\def\ben#1\een{\begin{enumerate}#1\end{enumerate}}

\newenvironment{sloppyequation}[0]{\sloppy\begin{flushleft}\hspace*{0.75cm}\(}{\)\end{flushleft}\fussy}
\newenvironment{sloppytext}[0]{\sloppy\begin{flushleft}}{\end{flushleft}\fussy}

\newcommand{\beqsloppy}{\begin{sloppyequation}}
\newcommand{\eeqsloppy}{\end{sloppyequation}}
\newcommand{\btxtsloppy}{\begin{sloppytext}}
\newcommand{\etxtsloppy}{\end{sloppytext}}

\title{The Two-Loop Hexagon Wilson Loop in $\N = 4$ SYM}

\author{Vittorio Del Duca\\
PH Department, TH Unit, CERN CH-1211, Geneva 23, Switzerland\\
INFN, Laboratori Nazionali Frascati, 00044 Frascati (Roma), Italy\\
       E-mail: \email{vittorio.del.duca@cern.ch}}

\author{Claude Duhr\\
Institute for Particle Physics Phenomenology,
University of Durham\\ Durham, DH1 3LE, U.K.\\
E-mail: \email{claude.duhr@durham.ac.uk}}

\author{Vladimir A. Smirnov\\
Nuclear Physics Institute of Moscow State University\\
Moscow 119992, Russia\\
E-mail: \email{smirnov@theory.sinp.msu.ru}}

\abstract{In the planar $\begin{cal}N\end{cal}=4$ supersymmetric Yang-Mills theory,
the conformal symmetry constrains multi-loop $n$-edged Wilson loops to be
given in terms of the one-loop $n$-edged Wilson loop, augmented, for $n\ge 6$,
by a function of conformally invariant cross ratios. That function is termed the remainder function.
In a recent paper, we have displayed the first analytic computation of the two-loop six-edged Wilson loop,
and thus of the corresponding remainder function.
Although the calculation was performed in the quasi-multi-Regge kinematics of a pair along the ladder,
the Regge exactness of the six-edged Wilson loop in those kinematics entails that the result is the same as
in general kinematics. We show in detail how the most difficult of the integrals is computed, which
contribute to the six-edged Wilson loop. Finally, the remainder function is given as a
function of uniform transcendental weight four in terms of Goncharov polylogarithms.
We consider also some asymptotic values of the remainder function, and the value
when all the cross ratios are equal.}

\preprint{IPPP/10/21, DCPT/10/42\\ CERN-PH-TH/2010-059}

\begin{document}

\tableofcontents

\section{Introduction}
\label{sec:intro}

In the planar $\begin{cal}N\end{cal}=4$ supersymmetric Yang-Mills (SYM) theory,
Anastasiou, Bern, Dixon and Kosower~\cite{Anastasiou:2003kj} made an ansatz for the iterative
structure of the colour-stripped two-loop scattering amplitude with an arbitrary number $n$ of external
legs in a maximally-helicity violating (MHV) configuration. Writing at any loop order $L$, the amplitude
$M_n^{(L)}$ as the tree-level amplitude, $M_n^{(0)}$, which depends on the helicity configuration,
times a scalar function, $m_n^{(L)}$,
the proposed iteration formula for the two-loop MHV amplitude $m_n^{(2)}(\eps)$ was
\beq
m_n^{(2)}(\eps) = \frac{1}{2} \left[m_n^{(1)}(\eps)\right]^2
+ f^{(2)}(\eps)\, m_n^{(1)}(2\eps) + C^{(2)} + \ord(\eps)\, .\label{eq:ite2bds}
\eeq
Thus the two-loop amplitude is determined in terms of
the one-loop MHV amplitude $m_n^{(1)}(\eps)$ evaluated through to $\ord(\eps^2)$
in the dimensional-regularisation parameter $\eps=(4-d)/2$,
the constant $C^{(2)} = -\zeta_2^2/2$, and the function $f^{(2)}(\eps) = -\zeta_2 - \zeta_3\eps - \zeta_4\eps^2$,
with $\zeta_i = \zeta(i)$ and $\zeta(z)$ the Riemann zeta function. In Ref.~\cite{Anastasiou:2003kj},
the iteration formula (\ref{eq:ite2bds}) was shown to be correct for the two-loop four-point amplitude, which previously
had been evaluated analytically~\cite{Bern:1997nh}. \Eqn{eq:ite2bds} was proven to be correct also for the
two-loop five-point  amplitude through numerical calculations~\cite{Bern:2006vw,Cachazo:2008vp}\footnote{The 
one-loop five-point amplitude to O($\eps^2$) has been computed analytically in the multi-Regge
kinematics~\cite{DelDuca:2009ac,DelDuca:2009ae}. The great computational complexity introduced in the amplitude
by the higher orders in $\epsilon$ makes it desirable to devise regularisation schemes for which such higher-order terms
are not present. A step in this direction has been made in Refs.~\cite{Alday:2009zm,Henn:2010bk} where a mass regulator 
has been introduced, which does not require, in the square of the amplitude, higher-order terms in the regulator.}.

Subsequently, Bern, Dixon and one of the present authors extended the ansatz by proposing an all-loop resummation
formula~\cite{Bern:2005iz} for the colour-stripped $n$-point MHV amplitude, which implies a tower
of  iteration formulae, allowing one to determine the $n$-point amplitude at a given number of
loops in  terms of amplitudes with fewer loops, evaluated to higher orders of $\eps$.
In Ref.~\cite{Bern:2005iz}, the ansatz was shown to be correct for the three-loop four-point amplitude,
by evaluating analytically
$m_4^{(3)}(\eps)$ through to finite terms, as well as $m_4^{(2)}(\eps)$ through to $\ord(\eps^2)$ and
$m_4^{(1)}(\eps)$ through to $\ord(\eps^4)$.

However, \Eqn{eq:ite2bds} fails for the two-loop six-point amplitude: in Ref.~\cite{Bern:2008ap},
it was shown that the finite pieces of the parity-even part of $m_6^{(2)}(\eps)$
are incorrectly determined by \Eqn{eq:ite2bds}\footnote{There were hints of a failure from
the strong-coupling limit of an amplitude with a large number of legs~\cite{Alday:2007he},
from the two-loop six-edged Wilson loop~\cite{Drummond:2007bm}, from
the six-point amplitude analysed in the multi-Regge kinematics in a Minkowski
region~\cite{Bartels:2008ce,Bartels:2008sc,Schabinger:2009bb}.}, although
the parity-odd part of $m_6^{(2)}(\eps)$ does fulfill \Eqn{eq:ite2bds}~\cite{Cachazo:2008hp}.
In particular, it was shown numerically that the two-loop remainder function,
defined as the difference between the two-loop amplitude and the ansatz for it,
\beq
R_n^{(2)} = m_n^{(2)}(\eps) - \frac{1}{2} \left[m_n^{(1)}(\eps)\right]^2
- f^{(2)}(\eps)\, m_n^{(1)}(2\eps) - C^{(2)} + \ord(\eps)\, ,\label{eq:discr}
\eeq
is different from zero for $n = 6$,
where $R_n^{(2)}$ is a function of the kinematical parameters of the $n$-point amplitude,
but a constant with respect to $\eps$. The analytic computation of $R_6^{(2)}$ has been performed
recently in Ref.~\cite{DelDuca:2009au}.

In the strong-coupling limit, Alday and Maldacena~\cite{Alday:2007hr} showed that
planar scattering amplitudes exponentiate like in the ansatz, and
suggested that in the weak-coupling regime the vacuum expectation value of the $n$-edged Wilson
loop could be related to the $n$-point MHV amplitude in $\begin{cal}N\end{cal}=4$ SYM.
At weak coupling, the agreement between the light-like Wilson loop and the (parity-even
part of the) MHV amplitude has been verified for the one-loop four-edged~\cite{Drummond:2007aua}
and $n$-edged~\cite{Brandhuber:2007yx} Wilson loops, and for the two-loop
four-edged~\cite{Drummond:2007cf}, five-edged~\cite{Drummond:2007au} and
six-edged~\cite{Drummond:2007bm,Drummond:2008aq} Wilson loops.

Furthermore, it was shown that the $L$-loop light-like Wilson loop exhibits a conformal symmetry,
and that the solution of the Ward identity for a special conformal boost is given by the ansatz, augmented,
for $n\ge 6$, by a function $R_{n,WL}^{(L)}$ of conformally invariant cross ratios~\cite{Drummond:2007au}.
Because of the duality between Wilson loops and amplitudes at one and two loops, $R_{n,WL}^{(2)}$ 
can be identified as the remainder function of \Eqn{eq:discr}.

In Refs.~\cite{Drummond:2008aq,Anastasiou:2009kna}, the two-loop $n$-edged Wilson loop
has been given in terms of Feynman-parameter-like integrals. Furthermore, in Ref.~\cite{Anastasiou:2009kna}
a numerical algorithm has been set up, which is valid for the two-loop $n$-edged Wilson loop and
by which the two-loop seven-edged and eight-edged Wilson loops have been
computed\footnote{In fact, in a
particular kinematic setup for which only $2n$-edged regular polygons are allowed~\cite{Alday:2009yn},
and for which the simplest non-trivial remainder function is the one of the two-loop eight-edged
Wilson loop, the remainder function has been computed numerically through the algorithm of
Ref.~\cite{Anastasiou:2009kna} for Wilson loops with up to 30 edges~\cite{Brandhuber:2009da}.},
although the corresponding MHV amplitudes are not known\footnote{The parity-even part of
the two-loop $n$-point MHV amplitude has been given in terms of scalar Feynman integrals, yet to be
evaluated~\cite{Vergu:2009tu}.}. Thus, also the remainder
functions $R_{7,WL}^{(2)}$ and $R_{8,WL}^{(2)}$ of the Wilson loops are
known numerically, and the numerical evidence~\cite{Anastasiou:2009kna} confirms that they
are functions of conformally invariant cross ratios only. However, their analytic form is in general unknown.

In Ref.~\cite{DelDuca:2009au}, we gave a brief account of the first analytic computation at weak coupling of the
two-loop six-edged Wilson loop in general kinematics. The computation was performed in the Euclidean region in
$D=4-2\eps$ dimensions, where the result is real, and it was found in agreement with the numerical evaluation
of Ref.~\cite{Anastasiou:2009kna}. For $n=6$, $R_{6,WL}^{(2)}$ is a function of the three conformally
invariant cross ratios, $u_1,u_2,u_3$. However, it is sufficient to compute the two-loop six-edged Wilson loop
in any kinematical limit which does not modify the analytic dependence of $R_{6,WL}^{(2)}$ on
$u_1,u_2,u_3$~\cite{DelDuca:2008jg}.
Among such limits are the ones which feature an exact Regge factorisation of the Wilson loop~\cite{Drummond:2007aua}.
In Ref.~\cite{DelDuca:2009au}, we showed that the exact Regge factorisation is exhibited by the $L$-loop Wilson loops
$w_n^{(L)}$ in the quasi-multi-Regge kinematics (QMRK) of a cluster of $(n-4)$ particles along a Regge-limit 
ladder\footnote{Note that the corresponding $L$-loop amplitudes do not exhibit exact Regge factorisation because the one-loop 
amplitude to higher orders in $\epsilon$ is not Regge exact~\cite{DelDuca:2009ac,DelDuca:2009ae}.},
thus in particular by $w_6^{(L)}$ in the QMRK of a pair along the ladder~\cite{Fadin:1989kf,DelDuca:1995ki},
by $w_7^{(L)}$ in the QMRK of three-of-a-kind along the ladder~\cite{Del Duca:1999ha}, by $w_8^{(L)}$
in the QMRK of four-of-a-kind along the ladder~\cite{claude}.
Then, we illustrated briefly how the Feynman-parameter-like integrals of the two-loop
six-edged Wilson loop have been computed in the QMRK of a pair along the ladder, and commented on the type
of functions which appear in the final result. Because of the exact Regge factorisation, the ensuing remainder function
is valid in general kinematics.
It can be expressed as a linear combination of Goncharov polylogarithms of uniform transcendental weight four.
In Ref.~\cite{DelDuca:2009au}, the remainder function was presented in an electronic form at {\tt www.arxiv.org}
where a text file containing the {\tt Mathematica} expression was provided.
Furthermore, the remainder function for $u_1=u_2=u_3=u$ was computed, and compared to the numerical values
quoted in Ref.~\cite{Anastasiou:2009kna} and to the analytic expression in the strong coupling proposed in
Ref.~\cite{Alday:2009dv}.

In this paper, we provide a detailed account of the most difficult integral we had to evaluate in the analytic computation of the
two-loop six-edged Wilson loop, as well as the explicit expression of the remainder function $R_{6,WL}^{(2)}$ as a function
of the cross ratios $u_1,u_2,u_3$, and for $u_1=u_2=u_3=u$. Furthermore,
we compute the asymptotic values of $R_{6,WL}^{(2)}$ for large or small values of the cross ratios.
Finally, we briefly comment on the exact Regge factorisation of the Wilson loop in the QMRK of a pair along the ladder in backward scattering,
and on the possibility of computing the remainder function in that kinematic setup.

In Sec.~\ref{sec:wl}, we write the two-loop Wilson loop in terms of the one-loop Wilson loop plus a
remainder function $R_{n,WL}^{(2)}$. 
Then we write the six-edged Wilson loop in terms of Feynman-parameter-like
integrals~\cite{Anastasiou:2009kna}, and derive Mellin-Barnes representations for all of them.
Finally, we exploit the Regge exactness of the Wilson loop, and extract the leading
behaviour of the integrals in the QMRK of a pair along the ladder.
In that fashion, the Mellin-Barnes integrals are reduced to one threefold integral plus several twofold and onefold integrals.
In Sec.~\ref{sec:hardcomp} and App.~\ref{app:residues}, we describe the evaluation of the diagram
which generates the threefold integral. The full expression of the remainder function $R_{6,WL}^{(2)}$ is rather
lengthy and is given in App.~\ref{app:R62} as a function of uniform transcendental weight four in terms of
Goncharov polylogarithms.
In Sec.~\ref{sec:Rasympt}, we consider some asymptotic values of the remainder function when
the conformal cross ratios are either large or small and
in Sec.~\ref{sec:Ruuu}, we evaluate the remainder function when all the cross ratios are equal, $u_1=u_2=u_3=u$ and consider some special values of it.
Our conclusions are given in Sec.~\ref{sec:Conclusion}. Definitions of harmonic sums
and Goncharov polylogarithms are recalled in App.~\ref{app:Ssums} and \ref{app:Goncharov}. The multi-Regge and
collinear limits of the remainder function are discussed in App.~\ref{sec:limits}. App.~\ref{app:special_values_Li}, \ref{app:special_values_GPL} and \ref{app:GPL_HPL} collect relations between Goncharov multiple polylogarithms and (harmonic) polylogarithms for several special values of the arguments.

\section{The two-loop Wilson loop}
\label{sec:wl}

\subsection{Definitions}

The Wilson loop is defined through the path-ordered exponential,
\beq
W[\C_n] = {\rm Tr}\ \P\ {\rm exp} \left[ ig \oint {\rm d}\tau \dot{x}^\mu(\tau) A_\mu(x(\tau)) \right]\,,
\label{eq:wloop}
\eeq
computed on a closed contour $\C_n$. In what follows, the closed contour
is a light-like $n$-edged polygonal contour~\cite{Alday:2007hr}.
The contour is such that labelling the $n$ vertices of the polygon as $x_1,\ldots,x_n$, the distance
between any two contiguous vertices, {\em i.e.}, the length of the edge in between, is given by the
momentum of a particle in the corresponding colour-ordered scattering amplitude,
\beq
p_i = x_i - x_{i+1}\, ,\label{eq:dist}
\eeq
with $i=1,\ldots,n$. Because the $n$ momenta add up to zero, $\sum_{i=1}^n p_i = 0$, the $n$-edged contour
closes, provided we make the identification $x_1 = x_{n+1}$.

In the weak-coupling limit, the Wilson loop can be computed as an expansion in the coupling.
The expansion of \Eqn{eq:wloop} is done through
the non-abelian exponentiation theorem~\cite{Gatheral:1983cz,Frenkel:1984pz},
which gives the vacuum expectation value of the Wilson loop as an exponential,
\beq
\langle W[\C_n] \rangle = 1 + \sum_{L=1}^\infty a^L W_n^{(L)} = {\rm exp} \sum_{L=1}^\infty a^L w_n^{(L)}\, ,\label{eq:fgt}
\eeq
where the coupling is defined as
\beq
a = \frac{g^2 N}{8\pi^2}\,.
\eeq
For the first two loop orders, one obtains
\beq
w_n^{(1)} = W_n^{(1)}\,, \qquad
w_n^{(2)} = W_n^{(2)} - \frac{1}{2} \left( W_n^{(1)}\right)^2\, .\label{eq:twowl}
\eeq
The one-loop coefficient $w_n^{(1)}$ was evaluated in Refs.~\cite{Drummond:2007aua,Brandhuber:2007yx},
where it was given in terms of the one-loop $n$-point MHV amplitude,
\beq
w_n^{(1)} = \frac{\Gamma(1-2\eps)}{\Gamma^2(1-\eps)} m_n^{(1)} =
m_n^{(1)} - n \frac{\zeta_2}{2} + \ord(\eps)
\, ,\label{eq:wlamp}
\eeq
where the amplitude is a sum of one-loop {\em two-mass-easy} box functions~\cite{Bern:1994zx},
\beq
m_n^{(1)} = \sum_{p,q} F^{\rm 2m e}(p,q,P,Q)\,, \label{eq:2me}
\eeq
where $p$ and $q$ are two external momenta corresponding to two opposite massless legs,
while the two remaining legs $P$ and $Q$ are massive. The two-loop coefficient $w_n^{(2)}$
has been computed analytically for $n=4$~\cite{Drummond:2007cf}, $n=5$~\cite{Drummond:2007au} and
$n=6$~\cite{DelDuca:2009au}, and
numerically for $n=6$~\cite{Drummond:2008aq} and $n=7, 8$~\cite{Anastasiou:2009kna}.

In Ref.~\cite{Drummond:2007au} it was established that
the Wilson loop fulfils a special conformal Ward identity, whose solution is the BDS ansatz plus, for $n\ge 6$,
an arbitrary function of the conformally invariant cross ratios, defined in \Eqn{eq:crossratios}.
Thus, the two-loop coefficient $w_n^{(2)}$ can be written as
\beq
w_n^{(2)}(\eps) = f^{(2)}_{WL}(\eps)\, w_n^{(1)}(2\eps) + C_{WL}^{(2)} + R_{n,WL}^{(2)}
+ \ord(\eps)\, ,\label{eq:wl2wi}
\eeq
where the constant is the same as in \Eqn{eq:ite2bds}, $C_{WL}^{(2)} = C^{(2)} = -\zeta_2^2/2$,
and the function $f^{(2)}_{WL}(\eps)$ is~\cite{Drummond:2007cf,Anastasiou:2009kna,Korchemskaya:1992je},
\beq
f^{(2)}_{WL}(\eps) = -\zeta_2 + 7\zeta_3\eps - 5\zeta_4\eps^2\, .\label{eq:fdue}
\eeq
With the two-loop coefficient $w_n^{(2)}$ given by \eqns{eq:wl2wi}{eq:fdue} and the two-loop MHV amplitude given by
\eqns{eq:ite2bds}{eq:discr}, the duality between Wilson loops and amplitudes is expressed by the equality of their
remainder functions~\cite{Anastasiou:2009kna},
\beq
R_{n,WL}^{(2)} = R_n^{(2)}\,. \label{eq:remaind}
\eeq

Defining the conformally invariant cross ratios as,
\beq
u_{ij} = \frac{x_{ij+1}^2 x_{i+1j}^2}{x_{ij}^2x_{i+1j+1}^2}\, ,\label{eq:crossratios}
\eeq
for $n=6$, they are~\cite{Drummond:2007au},
\beq
u_{36} = u_1 = \frac{x_{13}^2 x_{46}^2}{x_{36}^2 x_{41}^2}\,, \qquad
u_{14} = u_2 = \frac{x_{15}^2 x_{24}^2}{x_{14}^2 x_{25}^2}\,, \qquad
u_{25} = u_3 = \frac{x_{26}^2 x_{35}^2}{x_{25}^2 x_{36}^2}\,, \label{eq:3cross}
\eeq
where $x_{ij}^2 = (x_i-x_j)^2$, and using \Eqn{eq:dist} one sees that $x_{i,i+2}^2 = s_{i,i+1}$ and $x_{i,i+3}^2 = s_{i,i+1,i+2}$,
where the labels are understood to be modulo 6.

\subsection{The two-loop six-edged Wilson loop}
\label{sec:2loopwl}

The diagrams that enter the computation of the two-loop six-edged Wilson loop have been spelled out in
Ref.~\cite{Anastasiou:2009kna}\footnote{Note that there is a factor 1/8 missing in Eqs.~(4.4) and (B.1) of
Ref.~\cite{Anastasiou:2009kna}.}. In terms of those diagrams, we write the two-loop six-edged Wilson loop as,
\bea
w_6^{(2)} &=& \cC \left[
f_H(p_1,p_2,p_3; 0, p_4+p_5+p_6, 0) + f_H(p_1,p_2,p_4; p_3, p_5+p_6, 0)
\right. \nn\\ && + f_H(p_1,p_2,p_5; p_3+p_4,p_6, 0)
+ (1/3)  f_H(p_1,p_3,p_5; p_4,p_6,p_2) \nn\\
&& + f_C(p_1,p_2,p_3; 0, p_4+p_5+p_6, 0) + f_C(p_1,p_2,p_4; p_3, p_5+p_6, 0)
\nn\\ && + f_C(p_1,p_2,p_5; p_3+p_4,p_6, 0) +
f_C(p_1,p_2,p_6; p_3+p_4+p_5, 0,0) \nn\\
&& + f_C(p_1,p_3,p_4; 0, p_5+p_6, p_2) + f_C(p_1,p_3,p_5; p_4, p_6, p_2) \nn\\
&& + f_C(p_1,p_3,p_6; p_4+p_5, 0, p_2) + f_C(p_1,p_4,p_5; 0, p_6, p_2+p_3)
\nn\\ && + f_C(p_1,p_4,p_6; p_5, 0, p_2+p_3) +
 f_C(p_1,p_5,p_6; 0, 0, p_2+p_3+p_4) \nn\\
 && + f_X(p_1,p_2; p_3+p_4+p_5+p_6, 0) + f_Y(p_1,p_2; p_3+p_4+p_5+p_6, 0)
\nn\\ && + f_Y(p_2,p_1; 0, p_3+p_4+p_5+p_6)
+ f_X(p_1,p_3; p_4+p_5+p_6, p_2) \nn\\ &&
+ f_Y(p_1,p_3; p_4+p_5+p_6, p_2) + f_Y(p_3,p_1; p_2, p_4+p_5+p_6) \nn\\
&& + (1/2) f_X(p_1,p_4; p_5+p_6, p_2+p_3) +  f_Y(p_1,p_4; p_5+p_6, p_2+p_3)
\nn\\ && + (-1/2) f_P(p_1,p_3; p_4+p_5+p_6, p_2)\, f_P(p_2,p_4; p_1+p_5+p_6, p_3) \nn\\
&& + (-1/2) f_P(p_1,p_3; p_4+p_5+p_6, p_2)\, f_P(p_2,p_5; p_1+p_6, p_3+p_4) \nn\\
&& + (-1/4) f_P(p_1,p_4; p_5+p_6, p_2+p_3)\, f_P(p_2,p_5; p_1+p_6, p_3+p_4) \nn\\
&&  \left. +\ {\rm cyclic\ permutations\ of}\ (p_1,p_2,p_3, p_4, p_5,p_6) \right]\,,
\label{eq:w62l}
\eea
where, in the terminology of  Ref.~\cite{Anastasiou:2009kna}, $f_H$ stands for a hard diagram, $f_C$ for a curtain
diagram, $f_X$ for a cross diagram, $f_Y$ for a Y diagram plus half a self-energy diagram,
$f_P$ for a factorised cross diagram. Furthermore,
\beq
\cC = 2 a^2 \mu^{4\eps} \left[ \Gamma(1+\eps) e^{\gamma\eps}\right]^2\,,
\eeq
and the scale $\mu^2$ is given in terms of the Wilson loop scale, $\mu_{WL}^2 = \pi e^\gamma \mu^2$.

The six-edged Wilson loop is a function of the six external momenta $p_i$, $1\le i\le 6$. Imposing momentum conservation and on-shellness reduces the number of independent multi-particle invariants to nine\footnote{Note that in four dimensions we could also impose the Gram determinant constraint. However do not impose this constraint and treat the nine invariants as independent variables.}.
As the basic kinematic invariants, we choose
$s_{12},s_{23},s_{34},s_{45},s_{56},s_{61},s_{123},s_{342},s_{345}$,
all the other kinematic invariants being related to them by the following relations,
\bea
s_{13} &=& -s_{12} + s_{123} - s_{23}, \;\;\;s_{14} = -s_{123} + s_{23} - s_{234} + s_{56},
\nn \\
  s_{15}& =& -s_{16} + s_{234} - s_{56}, \;\;\;s_{24} = -s_{23} + s_{234} - s_{34},
\nn \\
  s_{25}& =& s_{16} - s_{234} + s_{34} - s_{345}, \;\;\;s_{26} = -s_{12} - s_{16} + s_{345},
\nn \\
  s_{35}& =& -s_{34} + s_{345} - s_{45},\;\;\; s_{36} = s_{12} - s_{123} - s_{345} + s_{45},
\nn \\
  s_{46} &=& s_{123} - s_{45} - s_{56}\,. \label{eq:sinv}
\eea

We use the parametric representations of the Wilson loop diagrams given in Ref.~\cite{Anastasiou:2009kna} and we derive appropriate Mellin--Barnes (MB) representations for all of them.
In multi-loop calculations it is sometimes difficult
to find an optimal choice for the MB representation. However, in our case the
MB representations are introduced in a straightforward way using the basic formula
\begin{equation}
{1\over (A+B)^\lambda} =
{1\over \Gamma(\lambda)}\,\int_{-i\infty}^{+i\infty}{\rd z\over 2\pi i}\,
\Gamma(-z)\,\Gamma(\lambda+z)\,{A^{z}\over B^{\lambda+z}}\, .
\label{MB}
\end{equation}
where the contour is chosen such as to separate the poles in $\Gamma(\ldots-z)$ from
the poles in $\Gamma(\ldots+z)$.
Note that in our case $\lambda$ is in general an integer plus an off-set corresponding
to the dimensional regulator $\eps$. In order to resolve the singularity structures in
$\epsilon$, we apply the strategy based on the MB representation and given in
Refs.~\cite{Smirnov:1999gc,Tausk:1999vh,Smirnov:2004ym,Smirnov:2006ry}. To this effect,
we apply the codes {\tt MB}~\cite{Czakon:2005rk} and {\tt MBresolve}~\cite{Smirnov:2009up}
and obtain a set of MB integrals which can be safely expanded in $\eps$ under the integration sign. After applying these codes, all the integration contours are straight vertical lines. Then we proceed and simplify the computation by exploiting the Regge exactness of the Wilson loop~\cite{DelDuca:2009au} and extract the leading quasi-multi-Regge behaviour by applying {\tt MBasymptotics} \cite{CzakonMBA}. Finally, we apply {\tt barnesroutines} \cite{Kosower} to perform
integrations that can be done by corollaries of Barnes lemmas.

To illustrate this procedure, let us consider the hard diagram $f_H (p_1, p_3, p_5; p_4, p_6, p_2)$
of (\ref{eq:w62l}). A parametric representation of this diagram was given in Eqs.~(B.1)--(B.5) of
Ref.~\cite{Anastasiou:2009kna}. We consider separately the nine terms originating from
the decomposition of the numerator (B.1). In particular, for the first part
of the first line of (B.1),
$1/2 s_{13} s_{15} \alpha_1 \alpha_2 (1 - \tau_1)$,
we have the following parametric integral:
\bea
F=\frac{\Gamma (2-2 \ep) (s_{12}-s_{123}+s_{23})
   (s_{16}-s_{234}+s_{56})}{2 \Gamma (1-\ep)^2}
&& \nn \\ &&  \hspace*{-70mm}
\times \int_0^1\ldots\int_0^1 \left(\prod_{i=1}^3 {\rm d}\alpha_i \right)
\left(\prod_{i=1}^3 {\rm d}\tau_i \right)\delta\left(\sum_{i=1}^3\alpha_i -1  \right)
(1-\tau_1) \alpha_1^{1-\ep} \alpha_2^{1-\ep} \alpha_3^{-\ep}
\nn \\ &&  \hspace*{-70mm}
\times
\left[- s_{12}\alpha_1 \alpha_2 (1-\tau_1) (1-\tau_2)
- s_{123}\alpha_1 \alpha_2 (1-\tau_1) \tau_2
- s_{23}\alpha_1 \alpha_2 \tau_1 \tau_2
- s_{16}\alpha_1 \alpha_3 \tau_1 \tau_3
\right.
\nn \\ &&  \hspace*{-70mm}
- s_{234}\alpha_1 \alpha_3 \tau_1 (1-\tau_3)
- s_{56}\alpha_1 \alpha_3 (1-\tau_1) (1-\tau_3)
- s_{34}\alpha_2 \alpha_3  (1-\tau_2) (1-\tau_3)
\nn \\ &&  \hspace*{-70mm}
\left.
- s_{345}\alpha_2 \alpha_3 (1-\tau_2) \tau_3
- s_{45}\alpha_2 \alpha_3 \tau_2 \tau_3
\right]^{2 \ep-2}\,.
\label{H41}
\eea
Separating different terms of the function in the square brackets using \Eqn{MB},
we obtain the following eightfold MB representation:
\bea
&&F=\frac{(s_{12}-s_{123}+s_{23})(s_{16}-s_{234}+s_{56})}
{2\Gamma (1-\ep)^2 \Gamma(\ep+1)}
 \mbint\ldots\mbint \left(\prod_{i=1}^8 {{\rm d} z_i\over 2\pi i} \Gamma(-z_i)\right)
\nn \\ &&
\times
(-s_{12})^{z_1}
(-s_{123})^{z_7}
 (-s_{16})^{z_6}
   (-s_{23})^{z_2} (-s_{234})^{z_8} (-s_{34})^{z_3}
   (-s_{45})^{z_4}
   \nn \\ &&
\times\label{H41-MB}
(-s_{56})^{z_5}
   (-s_{345})^{2\ep-z_1-z_2-z_3-z_4-z_5-z_6-z_7-z_8-2}
\\ &&
\times
\Gamma(\ep-z_1-z_2-z_7-1)\Gamma (2\ep-z_2-z_4-z_5-z_6-z_7-z_8-1)
\nn \\ &&
\times
\Gamma (2\ep-z_1-z_2-z_3-z_5-z_7-z_8-1) \Gamma(-\ep+z_1+z_2+z_5+z_6+z_7+z_8+2)
\nn \\ &&
\times
 \Gamma(\ep-z_5-z_6-z_8)
\Gamma (-2\ep+z_1+z_2+z_3+z_4+z_5+z_6+z_7+z_8+2)
\nn \\ &&
\times
\frac{\Gamma (z_1+z_5+z_7+2) \Gamma(z_2+z_4+z_7+1) \Gamma(z_2+z_6+z_8+1)
\Gamma(z_3+z_5+z_8+1)}{\Gamma(2 \ep-z_1-z_2-z_7) \Gamma(2\ep-z_5-z_6-z_8)
\Gamma(z_1+z_2+z_5+z_6+z_7+z_8+3)}
\nn
\,.
\eea
Then we apply the codes {\tt MB}~\cite{Czakon:2005rk} and
{\tt MBresolve}~\cite{Smirnov:2009up} to resolve the singularity structure in
$\ep$.

It might seem that we have made the situation more complicated because, instead of
the fivefold integral in \Eqn{H41} (one of the six integrations is performed using
the $\delta$ function), we have now the eightfold integral (\ref{H41-MB}).
However, Eq.~(\ref{H41-MB}) as well as the MB representations of the other contributions
to Eq.~(\ref{eq:w62l}) is much more convenient for taking various limits.
In fact, the cornerstone of our approach is to expand \Eqn{eq:wl2wi} in some limit such
that for $n = 6$ the computation of the remainder function
is considerably simplified.  Explicitly, we rewrite \Eqn{eq:wl2wi} as
\beq
 R_{6,WL}^{(2)}  = \left.\left[w_6^{(2)}(\eps)
 -f^{(2)}_{WL}(\eps)\, w_6^{(1)}(2\eps)\right]\right|_{\eps= 0}
 - C_{WL}^{(2)}\,,
\label{eq:R6}
\eeq
and look for a limit in which ({\it i}) the cross ratios the remainder function depends upon
take non-trivial values and ({\it ii}) the two-loop hexagon Wilson loop $w_6^{(2)}(\eps)$
is as simple as possible.
The simplest variant of such a limit
is the quasi-multi-Regge limit (QMRK) of a pair along the ladder~\cite{Fadin:1989kf,DelDuca:1995ki}.
In those kinematics, the outgoing gluons are strongly ordered in rapidity, except for a central pair of gluons along the ladder,
while their transverse momenta are all of the same size. In the physical region,
defining $1$ and $2$ as the incoming gluons, with momenta $p_2=(p_2^+/2,0,0,p_2^+/2)$ and $p_1=(p_1^-/2,0,0,-p_1^-/2)$,
and $3, 4, 5, 6$ as the outgoing gluons, the ordering can be chosen as
\begin{equation}
y_3 \gg y_4 \simeq y_5 \gg y_6;\qquad |p_{3\perp}| \simeq |p_{4\perp}|
\simeq |p_{5\perp}| \simeq|p_{6\perp}|\,
,\label{qmrk6ptc}
\end{equation}
where the particle momentum $p$ is parametrised in terms of the rapidity $y$ and the azimuthal angle $\phi$,
$p=(|p_{\perp}|\cosh y, |p_{\perp}|\cos\phi, |p_{\perp}|\sin\phi,|p_{\perp}|\sinh y)$.
We shall work in the Euclidean region, where the Wilson loop is real.
There the Mandelstam invariants are taken as all negative, and in the QMRK of a pair along the ladder
they are ordered as follows,
\begin{equation}
-s_{12} \gg -s_{34}, -s_{56}, -s_{345}, -s_{123} \gg -s_{23}, -s_{45}, -s_{61}, -s_{234}\, .\label{eq:mrk2lLip}
\end{equation}
Introducing a parameter $\lambda \ll 1$, the hierarchy above is equivalent to the rescaling
\beq
\{ s_{34}, s_{56}, s_{123}, s_{345} \} = \ord(\lambda)\,, \qquad
\{ s_{23}, s_{45}, s_{61}, s_{234} \} = \ord(\lambda^2)\,.  \label{eq:qmrla}
\eeq
It is easy to see that in this limit the three cross ratios (\ref{eq:3cross}) do not take trivial limiting
values~\cite{DelDuca:2008jg},
\beq\bsp
u_1&\,\rightarrow u_1^{QMRK}  = \frac{s_{45}}{(p_4^++p_5^+)(p_4^-+p_5^-)} = \ord(1)\, , \\
u_2&\,\rightarrow u_2^{QMRK}= \frac{|p_{3\perp}|^2  p_5^+p_6^-}{(|p_{3\perp}+p_{4\perp}|^2 + p_5^+p_4^-)
(p_4^++p_5^+)p_6^- } = \ord(1)\, ,\\
u_3&\,\rightarrow u_3^{QMRK} = \frac{|p_{6\perp}|^2  p_3^+p_4^- }{p_3^+ (p_4^-+p_5^-)
(|p_{3\perp}+p_{4\perp}|^2 + p_5^+p_4^-) } = \ord(1)\, .\label{thrinvarqmrkc}
\esp\eeq

Taking this limit on \Eqn{eq:R6}, its right-hand side simplifies. However, the Regge exactness
of the Wilson loop allows us to take, one after the other, not only this limit but also the five limits
obtained from the first one by cyclic permutations of the external
momenta $p_1,\ldots,p_6$~\cite{DelDuca:2009au}. For example, the second limit in this series is,
\beq
\{ s_{45}, s_{61}, s_{234}, s_{123} \} = \ord(\lambda)\,, \qquad
\{ s_{34}, s_{56}, s_{12}, s_{345} \} = \ord(\lambda^2)\,.  \label{eq:qmrlaa}
\eeq
While taking these consecutive limits, we keep in each case the leading power asymptotics
(including all the logarithms), a step which is fully automatized by the code
{\tt MBasymptotics} \cite{CzakonMBA}. We also apply the code {\tt barnesroutines}
\cite{Kosower} whenever possible
to perform integrations that can be done by corollaries of Barnes lemmas.

Finally, we arrive at a set of multiple MB integrals of a much simpler type than the original ones.
After applying our procedure, all integrals are at most threefold and all of them are
explicitly dependent on the cross ratios only\footnote{Note that the
coefficients of the integrals do not only depend on the cross ratios, but on
logarithms of Mandelstam invariants as well. This is to be expected since the contribution to
$w_6^{(2)}$ depends on such quantities.}. We checked numerically that the sum of the
MB integrals in the QMRK equals the sum of all the original parametric
integrals, the latter being evaluated numerically
using {\tt FIESTA}~\cite{Smirnov:2008py, Smirnov:2009pb}.
In particular, for the diagram $f_H (p_1, p_3, p_5; p_4, p_6, p_2)$, the eightfold integral of \Eqn{H41-MB}
reduces to a combination of one threefold integral, 51 twofold integrals and 22 onefold integrals
and a term without any integration left. Note that, after taking the six consecutive limits
described above, this diagram is the only one that involves a threefold integral, all other
contributions to \Eqn{eq:w62l} involving at most twofold integrals. The threefold contribution to
$f_H (p_1, p_3, p_5; p_4, p_6, p_2)$ reads,
\beq\bsp\label{eq:3foldH4}
 -{1\over4}&\,\mbint\mbint\mbint{\rd z_1\over 2\pi i}\,{\rd z_2\over 2\pi i}\,{\rd z_3\over 2\pi i}\,(z_1\,z_2+z_2\,z_3+z_3\,z_1)\, u_1^{z_1}\, u_2^{z_2}\, u_3^{z_3}\\
 &\times \Gamma \left(-z_1\right)^2\, \Gamma
   \left(-z_2\right)^2\, \Gamma \left(-z_3\right)^2\, \Gamma
   \left(z_1+z_2\right)\, \Gamma \left(z_2+z_3\right)\, \Gamma
   \left(z_3+z_1\right)\,,
  \esp \eeq
 where the contours are straight vertical lines such that,
 \beq\label{eq:contours}
 \re(z_1) = -{1\over3},\qquad  \re(z_2) = -{1\over4},\qquad  \re(z_3) = -{1\over5}\,.
 \eeq
The explicit evaluation of this integral is reviewed in the next section, whereas the full analytic expression for the remainder function is given in App.~\ref{app:R62} and is also available in electronic form at {\tt www.arXiv.org}.


\section{Evaluation of the hard diagram}
\label{sec:hardcomp}

In this section we review the computation of the MB integrals we derived in the previous section. Apart from the threefold integral contributing to $f_H (p_1, p_3, p_5; p_4, p_6, p_2)$, \Eqn{eq:3foldH4}, all the integrals are at most twofold and can be computed by closing the integration contours at infinity and summing up residues using the standard techniques. Therefore, in this paper we only concentrate on the case of the hard diagram and present in detail the analytic computation of the integral in \Eqn{eq:3foldH4}.

We rewrite \Eqn{eq:3foldH4} in the form,
\beq\label{eq:3foldH43}
-{1\over8}\,\left(F(u_1,u_2,u_3) + F(u_2,u_3,u_1) + F(u_3,u_1,u_2)\right)\,,
\eeq
where we define,
\beq\bsp
F(u_1,u_2,u_3) =&\, \mbint \mbint \mbint{\rd z_1\over 2\pi i}\,{\rd z_2\over 2\pi i}\,{\rd z_3\over 2\pi i}\, z_3\, u_1^{z_1}\, u_2^{z_2}\, u_3^{z_3}\\
\times&\, \Gamma \left(-z_1\right)^2\, \Gamma
   \left(-z_2\right)^2\, \Gamma
   \left(-z_3\right)^2\, \Gamma \left(z_1+z_2+1\right)\,  \Gamma \left(z_1+z_3\right)\, \Gamma
   \left(z_2+z_3\right)\,.
   \esp\eeq
Note that this function is symmetric in its first two arguments,
\beq
F(u_1,u_2,u_3) = F(u_2,u_1, u_3)\,,
\eeq
so that the expression in \Eqn{eq:3foldH43} is totally symmetric in the three cross ratios. We start with the change of variable $z_3 = z'_3+1$. This also shifts the corresponding contour, $\re(z_3) = -1/5\to\re(z'_3)=-6/5$. Shifting the contour back to $\re(z'_3)=-1/5$, we arrive at the expression,
\beq\label{eq:f123}
F(u_1,u_2,u_3) = \tilde F(u_1,u_2,u_3) + R_{-1}(u_1,u_2,u_3) + R_{-1-z_1}(u_1,u_2,u_3) + R_{-1-z_2}(u_1,u_2,u_3)\,,
\eeq
where the threefold integral $\tilde F$ is given by
\beq\bsp
\tilde F&(u_1,u_2,u_3) = \mbint \mbint \mbint{\rd z_1\over 2\pi i}\,{\rd z_2\over 2\pi i}\,{\rd z_3\over 2\pi i}\,
\frac{1}{1+z_3}\,u_1^{z_1}\, u_2^{z_2}\, u_3^{z_3+1}\\
&\times \Gamma \left(-z_1\right)^2\,
   \Gamma \left(-z_2\right)^2\,  \Gamma
   \left(-z_3\right)^2\, \Gamma \left(z_1+z_2+1\right) \Gamma \left(z_1+z_3+1\right) \Gamma
   \left(z_2+z_3+1\right)\,,
   \esp\eeq
 and we made the relabelling $z'_3\to z_3$. The integration contours are given by \Eqn{eq:contours}.
The functions $R_j(u_1,u_2,u_3)$ arise from taking the residues of the poles in $z'_3 = j$ that we crossed when shifting the contour from $\re(z'_3)=-6/5$ to $\re(z'_3)=-1/5$,
\beq\bsp\label{eq:addres}
R_{-1}(u_1,u_2,u_3)& = -\mbint \mbint{\rd z_1\over 2\pi i}\,{\rd z_2\over 2\pi i}\,
u_1^{z_1}\, u_2^{z_2}\,\Gamma \left(-z_1\right)^2\, \Gamma
   \left(z_1\right)\, \Gamma \left(-z_2\right)^2\, \Gamma \left(z_2\right)\\
   &\times\Gamma \left(z_1+z_2+1\right)\,,\\
   R_{-1-z_1}(u_1,u_2,u_3)& = \mbint \mbint{\rd z_1\over 2\pi i}\,{\rd z_2\over 2\pi i}\,
   u_1^{z_1}\, u_2^{z_2}\, u_3^{-z_1}\, \frac{1}{z_1}\, \Gamma \left(-z_1\right)^2\, \Gamma
   \left(z_1+1\right)^2\\
   &\times \Gamma \left(-z_2\right)^2\, \Gamma
   \left(z_2-z_1\right)\, \Gamma \left(z_1+z_2+1\right)\,,\\
   R_{-1-z_2}(u_1,u_2,u_3)& = \mbint \mbint{\rd z_1\over 2\pi i}\,{\rd z_2\over 2\pi i}\,
   u_1^{z_1}\, u_2^{z_2}\, u_3^{-z_2}\, \frac{1}{z_2}\, \Gamma \left(-z_2\right)^2\,  \Gamma
   \left(z_2+1\right)^2\\
   &\times\Gamma \left(-z_1\right)^2\,\Gamma
   \left(z_1-z_2\right)\, \Gamma \left(z_1+z_2+1\right)\,.
   \esp\eeq
Note that we have the relation,
\beq
R_{-1-z_2}(u_1,u_2,u_3) = R_{-1-z_1}(u_2,u_1,u_3)\,,
\eeq
where we assumed that the contours on both sides are chosen according to \Eqn{eq:contours}.
The computation of $\tilde F(u_1,u_2,u_3)$ is detailed in this section, whereas the computation of the residues $R_{j}(u_1,u_2,u_3)$ is discussed in App.~\ref{app:residues}.
We start by writing $\tilde F(u_1,u_2,u_3)$ as the integral of the derivative,
\beq
\tilde F(u_1,u_2,u_3) = \tilde F(u_1,u_2, 0) + \int_0^{u_3}\rd u\,{\partial\over\partial u}\,\tilde F(u_1,u_2,u)\,.
\eeq
The value for $u_3 = 0$ can be easily obtained by expanding around small values of $u_3$ using {\tt MBasymptotics}. We find,
\beq
\tilde F(u_1,u_2, 0 ) = 0\,.
\eeq
Next, we follow the procedure used in Ref.~\cite{DelDuca:2009ac} and we replace the MB integrations over $z_1$, $z_2$ and $z_3$ by Euler integrations using the formula~(see, \emph{e.g.}, Ref.~\cite{Smirnov:2006ry}),
\beq\bsp\label{eq:MBtoEuler}
\mbint&{\rd z\over 2\pi i}\,\Gamma(-z_1)\,\Gamma(c-z_1)\,\Gamma(b+z_1)\,\Gamma(c+z_1)\,X^{z_1}\\
&=\Gamma(a)\,\Gamma(b+c)\,\int_0^1\rd v\,v^{b-1}\,(1-v)^{a+c-1}\,(1-(1-X)v)^{-a}\,.
\esp\eeq
This leaves us with a fourfold Euler integral,
\beq\bsp\label{eq:3foldEuler}
\tilde F&(u_1,u_2,u_3) = \int_0^{1}\rd v_1\int_0^{1}\rd v_2\int_0^{1}\rd v_3 \int_0^{u_3}\rd u\,
\left(1-(1-u_1)\, v_1\right)^{-1}\\
&\times\left(1-v_2 \left(1-\frac{u_2
   \left(1-v_1\right)}{1-\left(1-u_1\right) v_1}\right)\right)^{-1}
   \left(1-v_3 \left(1-u\, v_1\, v_2\right)\right)^{-1}\,.
   \esp\eeq
Some comments are in order: Firstly, \Eqn{eq:MBtoEuler} is only valid if the contour separates the poles in $\Gamma(\ldots -z_i)$ from the poles in $\Gamma(\ldots+z_i)$. It is easy to observe that our contours, \Eqn{eq:contours}, fulfill this requirement. Secondly, we
tacitly exchanged the order of the integrations in deriving \Eqn{eq:3foldEuler}. We checked numerically that this operation is allowed in the present case.

The integrals over $u$ and $v_3$ in \Eqn{eq:3foldEuler} can be done very easily, resulting in the following twofold integral,
\beq\bsp\label{eq:F2Euler}
\tilde F&(u_1,u_2,u_3) = -\int_0^{1}\rd v_1\int_0^{1}\rd v_2\, v_1^{-1}\, v_2^{-1}\,
\left({\pi ^2\over6}- \text{Li}_2\left(1-v_1 v_2 u_3\right)\right)\\
&\times
 \left(-u_2  v_2+v_1 \left(u_1(v_2-1)+(u_2-1)v_2+1\right)+v_2-1\right)^{-1}\,.
   \esp\eeq
The remaining twofold integral can be computed in terms of Goncharov multiple polylogarithms~\cite{Goncharov:1998, Goncharov:2001}, defined recursively by,
\beq\label{eq:Gonchdef}
G(a,\vec w;z)=\int_0^z{\rd t\over t-a}\,G(\vec w;t)\,,\qquad G(a;z) = \ln\left(1-{z\over a}\right)\,.
\eeq
If all indices are zero we define
\beq
G(\vec 0_n;z)=\int_1^z{\rd t\over t} \,G(\vec 0_{n-1};t)={1\over n!}\,\ln^n z.
\eeq
In particular cases the Goncharov polylogarithms can be expressed in terms of ordinary logarithms and polylogarithms, \emph{e.g.},
\beq
G(\vec a_n;z)=\frac{1}{n!}\,\ln^n\left(1-\frac{z}{a}\right),\quad G(\vec 0_{n-1},a;z) = -\textrm{Li}_n\left(\frac{z}{a}\right).
\eeq
We define the weight of the function $G(\vec w;z)$  as the number of elements in the vector $\vec w$.
The Goncharov polylogarithms form a shuffle algebra, \emph{i.e.}, a product of two $G$ functions of weight $w_1$ and $w_2$ can be expressed as a linear combination of functions of weight $w=w_1+w_2$,
\beq
G(\vec w_1;z)\,G(\vec w_2;z)=\sum_{\vec w=\vec w_1\uplus \vec w_2}\,G(\vec w;z),
\eeq
where $\vec w_1\uplus \vec w_2$ denotes all the mergings of the vectors $\vec w_1$ and $\vec w_2$, \emph{i.e.}, all possible concatenations of $\vec w_1$ and $\vec w_2$ in which relative orderings of $\vec w_1$ and $\vec w_2$ are preserved. Furthermore, if the rightmost element of the weight vector $\vec w$ is non zero, the polylogarithms are invariant under a rescaling of the arguments,
\beq
G(k\,\vec w; k\,z) = G(\vec w;z)\,.
\eeq
A more detailed review of Goncharov polylogarithms and of their properties and special values is presented in App.~\ref{app:Goncharov}, \ref{app:special_values_Li}, \ref{app:special_values_GPL} and \ref{app:GPL_HPL}.

Using the definitions we just introduced, we can rewrite the numerator of the integrand in the form,
\beq\bsp
{\pi ^2\over6}&- \text{Li}_2\left(1-v_1 v_2 u_3\right) = G(1,0; v_1v_2u_3)\\
&= G(1; v_1v_2u_3)\,G(0; v_1v_2u_3) - G(0,1; v_1v_2u_3)\\
&= G\left({1\over v_1u_3};v_2\right)\,\Big[G(0; v_1)+G(0; v_2) + G(0;u_3)\Big] - G\left(0,{1\over v_1u_3}; v_2\right)\\
&=\Big[G(0; v_1) + G(0;u_3)\Big]G\left({1\over v_1u_3};v_2\right) + G\left({1\over v_1u_3},0; v_2\right)\,.
\esp\eeq
After partial fractioning the integrand in $v_2$, we arrive at,
\beq\bsp\label{eq:intv2}
\tilde F(u_1,u_2,u_3) =&\, -\int_0^{1}\rd v_1\int_0^{1}\rd v_2\, v_1^{-1}\,  \left(u_1 v_1-v_1+1\right)^{-1} \\
\times&\,\left\{\Big[G(0; v_1) + G(0;u_3)\Big]G\left({1\over v_1u_3};v_2\right) + G\left({1\over v_1u_3},0; v_2\right)\right\}\\
\times&\,
\left[
\left(v_2 -\frac{u_1
   v_1-v_1+1}{u_1 v_1+u_2 v_1-u_2-v_1+1}\right)^{-1} - v_2^{-1}\right]\,.
   \esp\eeq
We see that the integral over $v_2$ can be reduced to a sum of four terms, each of them consisting of a Goncharov polylogarithm divided by a linear function of $v_2$. This form matches precisely the recursive definition~(\ref{eq:Gonchdef}) of the polylogarithms, and we can easily perform the integral in $v_2$ in terms of those special functions, \emph{e.g.},
\beq\bsp\label{eq:intv2G}
\int_0^{1}&\rd v_2\,  G\left({1\over v_1u_3},0; v_2\right)
\left(v_2 -\frac{u_1
   v_1-v_1+1}{u_1 v_1+u_2 v_1-u_2-v_1+1}\right)^{-1} \\
  =&\, G\left(\frac{u_1
   v_1-v_1+1}{u_1 v_1+u_2 v_1-u_2-v_1+1},{1\over v_1u_3},0; 1\right)\,.
   \esp\eeq
All other terms in \Eqn{eq:intv2} can be integrated in the same way.

We now turn to the remaining integral in $v_1$. From \Eqn{eq:intv2G} it is clear that the integration in $v_2$ has produced an integrand which depends on Goncharov polylogarithms whose weight vectors are rather complicated functions of $v_1$. In order to perform the integration over $v_1$ in the same way as we did for $v_2$, we need all polylogarithms to be of the form $G(\vec w;v_1)$, where $\vec w$ is independent of $v_1$. In App.~\ref{app:Goncharov} we describe an algorithm that allows us to rewrite all the terms in the required form, \emph{e.g.},
\btxtsloppy
\parbox{130mm}{\raggedright\(\displaystyle
G\left(\frac{u_1 v_1-v_1+1}{u_1 v_1+u_2 v_1-u_2-v_1+1}, {1\over v_1u_3},0;1\right)=\)}
\raggedleft\refstepcounter{equation}(\theequation)\label{eq:GRED1}\\
\raggedright\(\displaystyle
-G\left(1,0;u_2\right) G\left(\upijk;v_1\right)+\frac{1}{6} \pi ^2 G\left(\upijk;v_1\right)+G\left(0;u_2\right) G\left(0,\frac{u_2-1}{u_1+u_2-1};v_1\right)-G\left(0;u_2\right) G\left(0,\umijk;v_1\right)-G\left(0;u_2\right) G\left(0,\upijk;v_1\right)+\frac{1}{6} G\left(\umijk;v_1\right) \left(\pi ^2-6 G\left(1,0;u_2\right)\right)+G\left(\frac{u_2-1}{u_1+u_2-1};v_1\right) \left(G\left(1,0;u_2\right)-\frac{\pi ^2}{6}\right)-G\left(0;u_2\right) G\left(\frac{u_2-1}{u_1+u_2-1},\frac{1}{1-u_1};v_1\right)+G\left(0;u_2\right) G\left(\frac{u_2-1}{u_1+u_2-1},\frac{u_2-1}{u_1+u_2-1};v_1\right)+G\left(0;u_2\right) G\left(\umijk,\frac{1}{1-u_1};v_1\right)-G\left(0;u_2\right) G\left(\umijk,\frac{u_2-1}{u_1+u_2-1};v_1\right)+G\left(0;u_2\right) G\left(\upijk,\frac{1}{1-u_1};v_1\right)-G\left(0;u_2\right) G\left(\upijk,\frac{u_2-1}{u_1+u_2-1};v_1\right)-2 G\left(0,0,\frac{1}{u_3};v_1\right)-G\left(0,1,\frac{1}{u_3};v_1\right)+G\left(0,\frac{u_2-1}{u_1+u_2-1},1;v_1\right)-G\left(0,\frac{u_2-1}{u_1+u_2-1},\frac{1}{1-u_1};v_1\right)-G\left(0,\umijk,1;v_1\right)+G\left(0,\umijk,\frac{1}{1-u_1};v_1\right)+G\left(0,\umijk,\frac{1}{u_3};v_1\right)-G\left(0,\upijk,1;v_1\right)+G\left(0,\upijk,\frac{1}{1-u_1};v_1\right)+G\left(0,\upijk,\frac{1}{u_3};v_1\right)-G\left(1,0,\frac{1}{u_3};v_1\right)-G\left(\frac{u_2-1}{u_1+u_2-1},\frac{1}{1-u_1},1;v_1\right)+G\left(\frac{u_2-1}{u_1+u_2-1},\frac{1}{1-u_1},\frac{1}{1-u_1};v_1\right)+G\left(\frac{u_2-1}{u_1+u_2-1},\frac{u_2-1}{u_1+u_2-1},1;v_1\right)-G\left(\frac{u_2-1}{u_1+u_2-1},\frac{u_2-1}{u_1+u_2-1},\frac{1}{1-u_1};v_1\right)+G\left(\umijk,0,\frac{1}{u_3};v_1\right)+G\left(\umijk,\frac{1}{1-u_1},1;v_1\right)-G\left(\umijk,\frac{1}{1-u_1},\frac{1}{1-u_1};v_1\right)-G\left(\umijk,\frac{u_2-1}{u_1+u_2-1},1;v_1\right)+G\left(\umijk,\frac{u_2-1}{u_1+u_2-1},\frac{1}{1-u_1};v_1\right)+G\left(\upijk,0,\frac{1}{u_3};v_1\right)+G\left(\upijk,\frac{1}{1-u_1},1;v_1\right)-G\left(\upijk,\frac{1}{1-u_1},\frac{1}{1-u_1};v_1\right)-G\left(\upijk,\frac{u_2-1}{u_1+u_2-1},1;v_1\right)+G\left(\upijk,\frac{u_2-1}{u_1+u_2-1},\frac{1}{1-u_1};v_1\right)
\,,
\)\etxtsloppy
where we defined,
\beq\bsp\label{eq:sqrt}
u_{jkl}^{(\pm)}=&\,\frac{1-u_j-u_k+u_l\pm\sqrt{\left(u_j+u_k-u_l-1\right)^2-4 \left(1-u_j\right)\left(1-u_k\right) u_l}}{2 \left(1-u_j\right) u_l}\, .
   \esp\eeq
   A comment is in order about the square roots in Eq.~(\ref{eq:sqrt}): It turns out that the square roots become complex for certain values of the cross ratios inside the unit cube, but they always come in pairs such that the sum of the two contributions is real. To emphasize this property, we introduce the following notation,
\beq\bsp\label{eq:gcaldef}
\begin{cal}G\end{cal}(\ldots, u_{ijk},\ldots; z) &\, = G\left(\ldots,u_{ijk}^{(+)},\ldots;z\right) +G\left(\ldots,u_{ijk}^{(-)},\ldots;z\right)\, .
\esp\eeq
Note that this definition follows the same spirit as the definition of Clausen's function, defined as the real or imaginary parts of the ordinary polylogarithms,
\beq
\text{Cl}_n(\theta) = \left\{\begin{array}{ll}
{1\over2}\left[\text{Li}_n(e^{i\theta})+\text{Li}_n(e^{-i\theta})\right], & n {\rm~ odd,}\\
{1\over2i}\left[\text{Li}_n(e^{i\theta})-\text{Li}_n(e^{-i\theta})\right], & n {\rm~ even.}
\end{array}\right.
\eeq
 All the integrations can now be done very easily using \Eqn{eq:Gonchdef}, and we find,
 \btxtsloppy
\parbox{130mm}{\raggedright\(\displaystyle
 \tilde F(u_1,u_2,u_3) = \)}
\raggedleft\refstepcounter{equation}(\theequation)\label{eq:GRED2}\\
\raggedright\(\displaystyle
G\left(0;u_2\right) G\left(0;u_3\right) G\left(0,\frac{u_2-1}{u_1+u_2-1};1\right)-G\left(1,0;u_2\right) G\left(0,\frac{u_2-1}{u_1+u_2-1};1\right)+\frac{1}{6} \pi ^2 G\left(0,\frac{u_2-1}{u_1+u_2-1};1\right)-G\left(0;u_2\right) G\left(0;u_3\right) G\left(\frac{1}{1-u_1},\frac{u_2-1}{u_1+u_2-1};1\right)+G\left(1,0;u_2\right) G\left(\frac{1}{1-u_1},\frac{u_2-1}{u_1+u_2-1};1\right)-\frac{1}{6} \pi ^2 G\left(\frac{1}{1-u_1},\frac{u_2-1}{u_1+u_2-1};1\right)-G\left(0;u_3\right) G\left(0,1,\frac{1}{u_3};1\right)+G\left(0;u_2\right) G\left(0,\frac{u_2-1}{u_1+u_2-1},0;1\right)+G\left(0;u_3\right) G\left(0,\frac{u_2-1}{u_1+u_2-1},1;1\right)+G\left(0;u_2\right) G\left(0,\frac{u_2-1}{u_1+u_2-1},\frac{1}{1-u_1};1\right)-G\left(0;u_3\right) G\left(0,\frac{u_2-1}{u_1+u_2-1},\frac{1}{1-u_1};1\right)-G\left(0;u_2\right) G\left(0,\frac{u_2-1}{u_1+u_2-1},\frac{u_2-1}{u_1+u_2-1};1\right)+G\left(0;u_3\right) G\left(\frac{1}{1-u_1},1,\frac{1}{u_3};1\right)-G\left(0;u_2\right) G\left(\frac{1}{1-u_1},\frac{u_2-1}{u_1+u_2-1},0;1\right)-G\left(0;u_3\right) G\left(\frac{1}{1-u_1},\frac{u_2-1}{u_1+u_2-1},1;1\right)-G\left(0;u_2\right) G\left(\frac{1}{1-u_1},\frac{u_2-1}{u_1+u_2-1},\frac{1}{1-u_1};1\right)+G\left(0;u_3\right) G\left(\frac{1}{1-u_1},\frac{u_2-1}{u_1+u_2-1},\frac{1}{1-u_1};1\right)+G\left(0;u_2\right) G\left(\frac{1}{1-u_1},\frac{u_2-1}{u_1+u_2-1},\frac{u_2-1}{u_1+u_2-1};1\right)-G\left(0,1,\frac{1}{u_3},0;1\right)+G\left(0,\frac{u_2-1}{u_1+u_2-1},0,1;1\right)-G\left(0,\frac{u_2-1}{u_1+u_2-1},0,\frac{1}{1-u_1};1\right)+G\left(0,\frac{u_2-1}{u_1+u_2-1},1,0;1\right)-G\left(0,\frac{u_2-1}{u_1+u_2-1},\frac{1}{1-u_1},0;1\right)+G\left(0,\frac{u_2-1}{u_1+u_2-1},\frac{1}{1-u_1},1;1\right)-G\left(0,\frac{u_2-1}{u_1+u_2-1},\frac{1}{1-u_1},\frac{1}{1-u_1};1\right)-G\left(0,\frac{u_2-1}{u_1+u_2-1},\frac{u_2-1}{u_1+u_2-1},1;1\right)+G\left(0,\frac{u_2-1}{u_1+u_2-1},\frac{u_2-1}{u_1+u_2-1},\frac{1}{1-u_1};1\right)+G\left(\frac{1}{1-u_1},1,\frac{1}{u_3},0;1\right)-G\left(\frac{1}{1-u_1},\frac{u_2-1}{u_1+u_2-1},0,1;1\right)+G\left(\frac{1}{1-u_1},\frac{u_2-1}{u_1+u_2-1},0,\frac{1}{1-u_1};1\right)-G\left(\frac{1}{1-u_1},\frac{u_2-1}{u_1+u_2-1},1,0;1\right)+G\left(\frac{1}{1-u_1},\frac{u_2-1}{u_1+u_2-1},\frac{1}{1-u_1},0;1\right)-G\left(\frac{1}{1-u_1},\frac{u_2-1}{u_1+u_2-1},\frac{1}{1-u_1},1;1\right)+G\left(\frac{1}{1-u_1},\frac{u_2-1}{u_1+u_2-1},\frac{1}{1-u_1},\frac{1}{1-u_1};1\right)+G\left(\frac{1}{1-u_1},\frac{u_2-1}{u_1+u_2-1},\frac{u_2-1}{u_1+u_2-1},1;1\right)-G\left(\frac{1}{1-u_1},\frac{u_2-1}{u_1+u_2-1},\frac{u_2-1}{u_1+u_2-1},\frac{1}{1-u_1};1\right)-G\left(0;u_2\right) G\left(0;u_3\right) \gcal\left(0,u_{123};1\right)+G\left(1,0;u_2\right) \gcal\left(0,u_{123};1\right)-\frac{1}{6} \pi ^2 \gcal\left(0,u_{123};1\right)+G\left(0;u_2\right) G\left(0;u_3\right) \gcal\left(\frac{1}{1-u_1},u_{123};1\right)-G\left(1,0;u_2\right) \gcal\left(\frac{1}{1-u_1},u_{123};1\right)+\frac{1}{6} \pi ^2 \gcal\left(\frac{1}{1-u_1},u_{123};1\right)-G\left(0;u_2\right) \gcal\left(0,u_{123},0;1\right)-G\left(0;u_3\right) \gcal\left(0,u_{123},1;1\right)-G\left(0;u_2\right) \gcal\left(0,u_{123},\frac{1}{1-u_1};1\right)+G\left(0;u_3\right) \gcal\left(0,u_{123},\frac{1}{1-u_1};1\right)+G\left(0;u_2\right) \gcal\left(0,u_{123},\frac{u_2-1}{u_1+u_2-1};1\right)+G\left(0;u_3\right) \gcal\left(0,u_{123},\frac{1}{u_3};1\right)+G\left(0;u_2\right) \gcal\left(\frac{1}{1-u_1},u_{123},0;1\right)+G\left(0;u_3\right) \gcal\left(\frac{1}{1-u_1},u_{123},1;1\right)+G\left(0;u_2\right) \gcal\left(\frac{1}{1-u_1},u_{123},\frac{1}{1-u_1};1\right)-G\left(0;u_3\right) \gcal\left(\frac{1}{1-u_1},u_{123},\frac{1}{1-u_1};1\right)-G\left(0;u_2\right) \gcal\left(\frac{1}{1-u_1},u_{123},\frac{u_2-1}{u_1+u_2-1};1\right)-G\left(0;u_3\right) \gcal\left(\frac{1}{1-u_1},u_{123},\frac{1}{u_3};1\right)-\gcal\left(0,u_{123},0,1;1\right)+\gcal\left(0,u_{123},0,\frac{1}{1-u_1};1\right)-\gcal\left(0,u_{123},1,0;1\right)+\gcal\left(0,u_{123},\frac{1}{1-u_1},0;1\right)-\gcal\left(0,u_{123},\frac{1}{1-u_1},1;1\right)+\gcal\left(0,u_{123},\frac{1}{1-u_1},\frac{1}{1-u_1};1\right)+\gcal\left(0,u_{123},\frac{u_2-1}{u_1+u_2-1},1;1\right)-\gcal\left(0,u_{123},\frac{u_2-1}{u_1+u_2-1},\frac{1}{1-u_1};1\right)+\gcal\left(0,u_{123},\frac{1}{u_3},0;1\right)+\gcal\left(\frac{1}{1-u_1},u_{123},0,1;1\right)-\gcal\left(\frac{1}{1-u_1},u_{123},0,\frac{1}{1-u_1};1\right)+\gcal\left(\frac{1}{1-u_1},u_{123},1,0;1\right)-\gcal\left(\frac{1}{1-u_1},u_{123},\frac{1}{1-u_1},0;1\right)+\gcal\left(\frac{1}{1-u_1},u_{123},\frac{1}{1-u_1},1;1\right)-\gcal\left(\frac{1}{1-u_1},u_{123},\frac{1}{1-u_1},\frac{1}{1-u_1};1\right)-\gcal\left(\frac{1}{1-u_1},u_{123},\frac{u_2-1}{u_1+u_2-1},1;1\right)+\gcal\left(\frac{1}{1-u_1},u_{123},\frac{u_2-1}{u_1+u_2-1},\frac{1}{1-u_1};1\right)-\gcal\left(\frac{1}{1-u_1},u_{123},\frac{1}{u_3},0;1\right)
 \,.
 \)\etxtsloppy
Note that all the terms in this expression are of uniform transcendental weight four, as expected.
Then $F(u_1,u_2,u_3)$ in \Eqn{eq:f123} is obtained by combining \Eqn{eq:GRED2} with the residues $R_j(u_1,u_2,u_3)$
computed in App.~\ref{app:residues}.


\section{Asymptotic values of the remainder function}
\label{sec:Rasympt}

In this section we study the asymptotic behaviour of the remainder function in various limits. For the sake of simplicity, we exclusively studied strongly ordered limits, \emph{i.e.}, limits where any ratio of conformal cross ratios is either small or large. Note that since the remainder function is completely symmetric in its arguments, it is enough to study the strongly ordered limits for a specific ordering, all other orderings being obtained by symmetry. The technique described in this section to compute the asymptotic behaviour in the various limits can easily be extended to non-strongly ordered limits. In the next section we briefly comment on such limits when all cross ratios are equal.

We start with the limit where all cross ratios are small,
$u_1\ll u_2\ll u_3\ll1$.
We can easily obtain the leading contribution by using {\tt MBasymptotics}. We find,
\beq\bsp
&\lim_{u_1\ll u_2\ll u_3\ll1}\,R_{6,WL}^{(2)}(u_1,u_2,u_3)\\
& = {\pi^2\over 24}\left(\ln u_1\ln u_2 +\ln u_2\ln u_3 + \ln u_3\ln u_1\right) + {17\pi^2\over 1440}+\ord(u_i)\,.
\esp\eeq
In exactly the same way, we can find the asymptotic behaviour when some of the cross ratios are equal to unity and all the others are small,
\beq\bsp
&\lim_{u_1\ll u_2\ll1}\,R_{6,WL}^{(2)}(u_1,u_2,1) = 0\,,\\
&\lim_{u_1\ll1}\,R_{6,WL}^{(2)}(u_1,1,1) = {1\over 2}\zeta_3\,\ln u_1-{\pi^4\over 96} + \ord(u_1)\,.
\esp\eeq
Note that the limit $u_1\ll u_2\ll1$, with $u_3=1$, corresponds to the multi-Regge limit (\ref{eq:mrlimit}).

We now repeat the previous analysis in the limit where the cross ratios are large,
$u_1\gg u_2\gg u_3\gg1$.
Using again {\tt MBasymptotics} to extract the leading behaviour, we find,
\beq\bsp
&\lim_{u_1\gg u_2\gg u_3\gg1}\,R_{6,WL}^{(2)}(u_1,u_2,u_3)= \\
& - \frac{1}{96} \log ^4 \frac{u_1}{u_2u_3} -\frac{5}{48} \pi ^2 \log ^2 \frac{u_1}{u_2u_3} -\frac{1}{2} \zeta_3 \log  \frac{u_1}{u_2u_3}
-\frac{157 \pi ^4}{1440} +\ord(1/u_i)\,.
\esp\eeq
Similarly, for the case were some of the cross ratios are equal to unity, we find,
\beq\bsp
&\lim_{u_1\gg u_2\gg1}\,R_{6,WL}^{(2)}(u_1,u_2,1)=
-\frac{1}{96} \log ^4 \frac{u_1}{u_2} -\frac{1}{12} \pi ^2 \log ^2 \frac{u_1}{u_2}
-\frac{3 \pi ^4}{40} +\ord(1/u_i)\,,\\
&\phantom{a}\\
&\lim_{u_1\gg1}\,R_{6,WL}^{(2)}(u_1,1,1)=
-\frac{1}{96} \log ^4 u_1  -\frac{1}{16} \pi ^2 \log ^2 u_1 +\frac{1}{2} \zeta_3 \log  u_1 -\frac{23 \pi ^4}{480}
 + \ord(1/u_1)\,.
 \esp\eeq



\section{The remainder function for all cross ratios equal}
\label{sec:Ruuu}
In this section we discuss the form of the remainder function in the special case when all the cross ratios are equal, $u_1=u_2=u_3=u$. In \Ref{DelDuca:2009au} several special values were presented for this case. We start by briefly reviewing how these values were obtained and present some additional special values. At the end of this section we give the analytic form of $R_{6,WL}^{(2)}(u,u,u)$ for arbitrary $u$.

In the special case where $u=1$, which corresponds to a regular hexagon~\cite{Brandhuber:2009da, Alday:2009dv},
most of the integrations are easily done using Barnes lemmas and their corollaries, leaving us with at most onefold integrals. Note that some of these integrals involve $\Gamma$ functions with poles in half integer values which lead to multiple binomial sums~\cite{Jegerlehner:2002em, Kalmykov:2007dk}, but all these contributions cancel out when combining all the pieces. Applying this strategy to our integrals, we immediately find the value quoted in \Ref{DelDuca:2009au},
\beq
R_{6,WL}^{(2)}(1,1,1) = -{\pi^4\over 36} \simeq -2.70581...\, .
\eeq
Note that this value agrees with the value conjectured in \Ref{Anastasiou:2009kna}.
The asymptotic behaviour of $R_{6,WL}^{(2)}(u,u,u)$ for $u\to0$ can be obtained in a similar way using {\tt MBasymptotics}, which leaves us with at most trivial onefold integrals. The result is
\beq
\,R_{6,WL}^{(2)}(u,u,u) = {\pi^2\over 8}\ln^2u + {17\pi^4\over 1440} + \ord(u)\,.
\eeq
Finally, the asymptotic value for large $u$ is obtained in exactly the same way. We can perform a rescaling $u\to \lambda^{-1}\,u$ and expand around small values of $\lambda$ using {\tt MBasymptotics}. We find
\beq
\lim_{u\to \infty}\,R_{6,WL}^{(2)}(u,u,u)  = -{\pi^4\over 144} + \ord(1/u)\simeq -0.67645...\,,
\eeq
in very good agreement with the numerical value quoted in \Ref{Anastasiou:2009kna}.

For $u=1/2$, the denominator in \Eqn{eq:F2Euler} drastically simplifies. Repeating the derivation of Sec.~\ref{sec:hardcomp}, we obtain,
\beq\bsp
R_{6,WL}^{(2)}\left({1\over 2},{1\over 2},{1\over 2}\right) &\,= -\frac{105}{64} \zeta_3 \log  2 -\frac{5}{64} \log ^4 2 +\frac{5}{64} \pi ^2 \log ^2 2 -\frac{15}{8} \text{Li}_4\left(\frac{1}{2}\right)+\frac{17 \pi ^4}{2304}\\
&\,\simeq -1.26609\ldots\,.
\esp\eeq

Let us now turn to the generic case where all three cross ratios are equal but they still take generic values.
In this limit it is easy to see that \eqns{eq:sqrt}{eq:sqrt2} reduce to
\beq\label{eq:mudef}
u_{ijk}^{(\pm)} \to \mu^{(\pm)} = {1\pm\sqrt{1-4u}\over 2u}\,,\qquad\quad
v_{ijk}^{(\pm)} \to \nu^{(\pm)} = \pm{1\over\sqrt{1-u}}\,.
\eeq
We can massage the resulting expression and apply the reduction algorithm of App.~\ref{app:Goncharov}
to simplify the expression as much as possible. In particular, we can remove all the dependence on 
$\nu^{(\pm)}$. As regards $\mu^{(\pm)}$, we observe that similar arguments have already been found in the strong coupling case~\cite{Alday:2009dv}\footnote{We are grateful to Paul Heslop for pointing out that
\beq
(1-u)\,\mu^{(+)} = 1+{\mu\over x_\epsilon},\qquad (1-u)\,\mu^{(-)} = 1+{1\over \mu x_\epsilon}\,,\nn
\eeq
where $\mu$ and $x_\eps$ are defined in \Ref{Alday:2009dv}.}.
Note that for $u=1/4$ the square roots in \Eqn{eq:mudef} vanish. This value corresponds to a regular hexagon in a space with a $(2,2)$ signature~\cite{Alday:2009dv}. Using the relations of App.~\ref{app:special_values_Li} and \ref{app:special_values_GPL}, we find,
\beq\bsp\label{eq:r14}
R_{6,WL}^{(2)}&\left({1\over 4},{1\over 4},{1\over 4}\right) =3 \text{Li}_2\left(\frac{1}{3}\right) \log ^2 2 -\frac{9}{2} \text{Li}_2\left(\frac{1}{3}\right) \log ^2 3 -\frac{567}{4} \text{Li}_3\left(\frac{1}{3}\right) \log  2\\
& +\frac{543}{4} \text{Li}_3\left(-\frac{1}{2}\right) \log  2 +\frac{567}{8} \text{Li}_3\left(\frac{1}{3}\right) \log  3 -\frac{567}{4} \text{Li}_3\left(-\frac{1}{2}\right) \log  3 +\frac{1323}{16} \zeta_3 \log  2\\
& +\frac{945}{32} \zeta_3 \log  3 -\frac{39}{32} \log ^4 2 -\frac{257}{64} \log ^4 3 +\frac{173}{8} \log  3  \log ^3 2 +\frac{189}{8} \log ^3 3  \log  2 -\frac{543}{16} \log ^2 3  \log ^2 2\\
& -\frac{63}{16} \pi ^2 \log ^2 2 -\frac{181}{64} \pi ^2 \log ^2 3 +\frac{189}{2} \text{Li}_4\left(\frac{1}{2}\right)+\frac{1701}{8} \text{Li}_4\left(\frac{1}{3}\right)-\frac{543}{16} \text{Li}_4\left(-\frac{1}{3}\right)\\
&+\frac{555}{2} \text{Li}_4\left(-\frac{1}{2}\right)-\frac{9}{2} \text{Li}_2\left(\frac{1}{3}\right)^2-\frac{567}{16} S_{2,2}\left(-\frac{1}{3}\right)-\frac{567}{4} S_{2,2}\left(-\frac{1}{2}\right)-\frac{2123 \pi ^4}{2880}\\
& \simeq 1.08917\ldots\,.
\esp\eeq

Finally, let us turn to the expression for generic values of $u$.
Using the notation introduced in \Eqn{eq:gcaldef} as well as the corresponding one for harmonic polylogarithms,
\beq\bsp\label{eq:hcaldef}
\begin{cal}H\end{cal}(\vec w;1/\mu) &\,= H \left(\vec w;1/\mu^{(+)}\right) +H \left(\vec w;1/\mu^{(-)}\right)\, ,
\esp\eeq
the final answer for the remainder function reads, when all three cross ratios are equal,

 \btxtsloppy
\parbox{130mm}{\raggedright\(\displaystyle
R_{6,WL}^{(2)}(u,u,u)= \)}
\raggedleft\refstepcounter{equation}(\theequation)\label{eq:R62uuu}\\
\raggedright\(\displaystyle
\frac{1}{4} \pi ^2 G\left(1,\frac{1}{2}; u\right)+\frac{1}{8} \pi ^2 G\left(\frac{1}{1-u},\frac{u-1}{2 u-1}; 1\right)-3 G\left(0,1,0,\frac{1}{2}; u\right)-3 G\left(0,1,\frac{1}{2},0; u\right)+\frac{3}{4} G\left(0,\frac{u-1}{2 u-1},0,\frac{1}{1-u}; 1\right)+\frac{3}{4} G\left(0,\frac{u-1}{2 u-1},\frac{1}{1-u},0; 1\right)-\frac{3}{4} G\left(0,\frac{u-1}{2 u-1},\frac{1}{1-u},1; 1\right)+\frac{3}{4} G\left(0,\frac{u-1}{2 u-1},\frac{1}{1-u},\frac{1}{1-u}; 1\right)-\frac{3}{4} G\left(0,\frac{u-1}{2 u-1},\frac{u-1}{2 u-1},\frac{1}{1-u}; 1\right)-6 G\left(1,0,0,\frac{1}{2}; u\right)-3 G\left(1,0,\frac{1}{2},0; u\right)+3 G\left(1,\frac{1}{2},0,0; u\right)-3 G\left(1,\frac{1}{2},1,0; u\right)-\frac{3}{4} G\left(\frac{1}{1-u},1,\frac{1}{u},0; 1\right)+\frac{3}{2} G\left(\frac{1}{1-u},\frac{1}{1-u},1,\frac{1}{1-u}; 1\right)+\frac{3}{4} G\left(\frac{1}{1-u},\frac{u-1}{2 u-1},0,1; 1\right)-\frac{3}{4} G\left(\frac{1}{1-u},\frac{u-1}{2 u-1},0,\frac{1}{1-u}; 1\right)+\frac{3}{4} G\left(\frac{1}{1-u},\frac{u-1}{2 u-1},1,0; 1\right)-\frac{3}{4} G\left(\frac{1}{1-u},\frac{u-1}{2 u-1},\frac{1}{1-u},0; 1\right)+\frac{3}{4} G\left(\frac{1}{1-u},\frac{u-1}{2 u-1},\frac{1}{1-u},1; 1\right)-\frac{3}{4} G\left(\frac{1}{1-u},\frac{u-1}{2 u-1},\frac{1}{1-u},\frac{1}{1-u}; 1\right)-\frac{3}{4} G\left(\frac{1}{1-u},\frac{u-1}{2 u-1},\frac{u-1}{2 u-1},1; 1\right)+\frac{3}{4} G\left(\frac{1}{1-u},\frac{u-1}{2 u-1},\frac{u-1}{2 u-1},\frac{1}{1-u}; 1\right)-\frac{1}{8} \pi ^2 \gcal\left(\frac{1}{1-u},\mu ; 1\right)-\frac{3}{4} \gcal\left(0,\mu ,0,\frac{1}{1-u}; 1\right)-\frac{3}{4} \gcal\left(0,\mu ,\frac{1}{1-u},0; 1\right)+\frac{3}{4} \gcal\left(0,\mu ,\frac{1}{1-u},1; 1\right)-\frac{3}{4} \gcal\left(0,\mu ,\frac{1}{1-u},\frac{1}{1-u}; 1\right)-\frac{3}{4} \gcal\left(0,\mu ,\frac{1}{u},0; 1\right)-\frac{3}{4} \gcal\left(0,\mu ,\frac{u-1}{2 u-1},1; 1\right)+\frac{3}{4} \gcal\left(0,\mu ,\frac{u-1}{2 u-1},\frac{1}{1-u}; 1\right)-\frac{3}{4} \gcal\left(\frac{1}{1-u},\mu ,0,1; 1\right)+\frac{3}{4} \gcal\left(\frac{1}{1-u},\mu ,0,\frac{1}{1-u}; 1\right)-\frac{3}{4} \gcal\left(\frac{1}{1-u},\mu ,1,0; 1\right)+\frac{3}{4} \gcal\left(\frac{1}{1-u},\mu ,\frac{1}{1-u},0; 1\right)-\frac{3}{4} \gcal\left(\frac{1}{1-u},\mu ,\frac{1}{1-u},1; 1\right)+\frac{3}{4} \gcal\left(\frac{1}{1-u},\mu ,\frac{1}{1-u},\frac{1}{1-u}; 1\right)+\frac{3}{4} \gcal\left(\frac{1}{1-u},\mu ,\frac{1}{u},0; 1\right)+\frac{3}{4} \gcal\left(\frac{1}{1-u},\mu ,\frac{u-1}{2 u-1},1; 1\right)-\frac{3}{4} \gcal\left(\frac{1}{1-u},\mu ,\frac{u-1}{2 u-1},\frac{1}{1-u}; 1\right)-\frac{3}{4} G\left(\frac{1}{1-u},1,\frac{1}{u}; 1\right) H(0; u)+\frac{3}{4} G\left(\frac{1}{1-u},\frac{u-1}{2 u-1},0; 1\right) H(0; u)+\frac{3}{4} G\left(\frac{1}{1-u},\frac{u-1}{2 u-1},1; 1\right) H(0; u)-\frac{3}{4} G\left(\frac{1}{1-u},\frac{u-1}{2 u-1},\frac{u-1}{2 u-1}; 1\right) H(0; u)-\frac{3}{4} \gcal\left(0,\mu ,\frac{1}{u}; 1\right) H(0; u)-\frac{3}{4} \gcal\left(0,\mu ,\frac{u-1}{2 u-1}; 1\right) H(0; u)-\frac{3}{4} \gcal\left(\frac{1}{1-u},\mu ,0; 1\right) H(0; u)-\frac{3}{4} \gcal\left(\frac{1}{1-u},\mu ,1; 1\right) H(0; u)+\frac{3}{4} \gcal\left(\frac{1}{1-u},\mu ,\frac{1}{u}; 1\right) H(0; u)+\frac{3}{4} \gcal\left(\frac{1}{1-u},\mu ,\frac{u-1}{2 u-1}; 1\right) H(0; u)+\frac{3}{2} G\left(\frac{1}{1-u},\frac{u-1}{2 u-1}; 1\right) H(0,0; u)-\frac{3}{2} \gcal\left(\frac{1}{1-u},\mu ; 1\right) H(0,0; u)-\frac{1}{8} \pi ^2 H(0,0; u)+3 H(0,0; u) H(0,1; (2 u))+\frac{1}{4} \pi ^2 H(0,1; (2 u))+\frac{3}{2} H(0,0; u) H\left(0,1; \frac{2 u-1}{u-1}\right)-\frac{1}{8} \pi ^2 H\left(0,1; \frac{2 u-1}{u-1}\right)+\frac{3}{4} G\left(\frac{1}{1-u},\frac{u-1}{2 u-1}; 1\right) H(1,0; u)-\frac{3}{4} \gcal\left(\frac{1}{1-u},\mu ; 1\right) H(1,0; u)+3 H(0,1; (2 u)) H(1,0; u)+\frac{3}{4} H\left(0,1; \frac{2 u-1}{u-1}\right) H(1,0; u)+\frac{1}{8} \pi ^2 H(1,0; u)+\frac{1}{8} \pi ^2 H(1,1; u)-6 H(0; u) H(0,0,1; (2 u))-3 H(0; u) H\left(0,0,1; \frac{2 u-1}{u-1}\right)-\frac{3}{4} H(0; u) H\left(1,0,1; \frac{2 u-1}{u-1}\right)-\frac{3}{2} H(0,0,0,0; u)+9 H(0,0,0,1; (2 u))+\frac{9}{2} H\left(0,0,0,1; \frac{2 u-1}{u-1}\right)+\frac{15}{4} H(0,0,1,0; u)-\frac{9}{2} H(0,1,0,0; u)+\frac{3}{2} H\left(0,1,0,1; \frac{2 u-1}{u-1}\right)+\frac{3}{2} H(0,1,1,0; u)-\frac{3}{4} H\left(0,1,1,1; \frac{2 u-1}{u-1}\right)+3 H\left(1,0,0,1; \frac{2 u-1}{u-1}\right)+\frac{15}{4} H(1,0,1,0; u)-3 H(1,1,0,0; u)+\frac{3}{4} H\left(1,1,0,1; \frac{2 u-1}{u-1}\right)+\frac{3}{2} H(1,1,1,0; u)-\frac{1}{8} \pi ^2 H(0; u) \hcal\left(1; \frac{1}{\mu }\right)-\frac{3}{2} H(0,0; u) \hcal\left(0,1; \frac{1}{\mu }\right)-\frac{3}{4} H(1,0; u) \hcal\left(0,1; \frac{1}{\mu }\right)+\frac{1}{8} \pi ^2 \hcal\left(0,1; \frac{1}{\mu }\right)+3 H(0; u) \hcal\left(0,0,1; \frac{1}{\mu }\right)+\frac{3}{4} H(0; u) \hcal\left(0,1,1; \frac{1}{\mu }\right)+\frac{3}{4} H(0; u) \hcal\left(1,0,1; \frac{1}{\mu }\right)-\frac{9}{2} \hcal\left(0,0,0,1; \frac{1}{\mu }\right)-\frac{3}{2} \hcal\left(0,0,1,1; \frac{1}{\mu }\right)-\frac{3}{2} \hcal\left(0,1,0,1; \frac{1}{\mu }\right)-\frac{3}{2} \hcal\left(1,0,0,1; \frac{1}{\mu }\right)+3 H(0; u) \zeta_3+\frac{3}{2} \hcal\left(1; \frac{1}{\mu }\right) \zeta_3+\frac{89 \pi ^4}{1440}
 \,.
\)
\etxtsloppy


\section{Conclusion}
\label{sec:Conclusion}

In this paper, we have given details on the first analytic calculation of the remainder function of the two-loop six-edged Wilson loop
in the Euclidean space in arbitrary kinematics, which we recently performed~\cite{DelDuca:2009au}.
By displaying in detail how the most difficult of the integrals is computed,
we have shown nonetheless how the whole calculation is greatly simplified by exploiting the Regge exactness of the six-edged Wilson loop
in the quasi-multi-Regge kinematics of a pair along the ladder.

The remainder function is given as a combination of Goncharov polylogarithms of uniform transcendental weight four.
The expression we have obtained is very lengthy.
At present, we do not know whether, and if so to what extent, this expression can be further simplified by using some other kinematic limit that leaves the conformal cross ratios unchanged. 
Such a setup is for example found in backward scattering.
Let us consider the physical region in which two gluons undergo a backward scattering. In a $2\to 2$ scattering process,
backward scattering may be obtained from forward scattering by crossing the $t$ and $u$ channels. In a $2\to 4$ scattering process,
we may choose the kinematics in which
$1$ and $2$ are the incoming gluons, with momenta $p_2=(p_2^+/2,0,0,p_2^+/2)$ and $p_1=(p_1^-/2,0,0,-p_1^-/2)$,
and $3, 4, 5, 6$ are the outgoing gluons, with ordering
\begin{equation}
y_3 \ll y_4 \simeq y_5 \ll y_6;\qquad |p_{3\perp}| \simeq |p_{4\perp}|
\simeq |p_{5\perp}| \simeq|p_{6\perp}|\,
.\label{bwqmrk6ptc}
\end{equation}
We may term this the {\em backward quasi-multi-Regge kinematics of a pair along the ladder}.
In the Euclidean region, where the Mandelstam invariants are all negative, the ordering (\ref{bwqmrk6ptc}) entails that
\begin{equation}
-s_{12}, -s_{23}, -s_{61}, -s_{234} \gg -s_{34}, -s_{56}, -s_{345}, -s_{123} \gg  -s_{45} \, .\label{eq:bwmrk2lLip}
\end{equation}
Introducing a parameter $\lambda \gg 1$, the hierarchy above is equivalent to the rescaling
\beq
\{ s_{34}, s_{56}, s_{123}, s_{345} \} = \ord(\lambda)\,, \qquad \{s_{12}, s_{23}, s_{61}, s_{234}\} = \ord(\lambda^2)\,,  \label{eq:bwqmrla}
\eeq
whereas $s_{45}$ is $\ord(1)$. Note that this scaling is equivalent to the scaling of the Mandelstam invariants in the limit where three of the points of the Wilson loop are at infinity, as considered in Ref.~\cite{Alday:2009dv}.
It is easy to see that in the limit (\ref{eq:bwmrk2lLip}) with the rescaling (\ref{eq:bwqmrla})
the cross ratios (\ref{eq:3cross}) do not take trivial limiting values, and thus the six-edged Wilson loop is Regge exact in the
backward QMRK of a pair along the ladder. We could hence repeat our computation in this limit with the hope that the ensuing analytic expression be simpler. Even though we have not performed the full evaluation of the
remainder function in this limit, 
we have examined the diagram $f_H (p_1, p_3, p_5; p_4, p_6, p_2)$ in  \Eqn{eq:w62l} in the six limits obtained from \Eqn{eq:bwmrk2lLip}. We observe that, just
like in Sec.~\ref{sec:2loopwl}, the most complicated hard diagram reduces to a combination of one threefold integral, plus twofold and onefold integrals. Since this kinematic limit leads to the same threefold integral as the standard Regge limit, the result will be expressed in terms of the same functions and thus we expect it to be of similar complexity.

Even though the analytic form of the remainder function is very lengthy, the expression greatly simplifies when considering various limits. In Sec.~\ref{sec:Rasympt} we considered the remainder function in various strongly-ordered limits, and we presented in each case the leading term in the limit where the conformal cross ratios are either large or small. In Sec.~\ref{sec:Ruuu} we also considered the remainder function where all three conformal ratios take equal values, and computed explicitly the value of $R_{6,WL}^{(2)}(u,u,u)$ for $u=1/4, 1/2, 1$ as well as the leading behaviour in the limit of large and small values of $u$. 

The techniques described throughout the paper are generic, and not restricted to the case of a hexagon Wilson loop. In principle they can be applied to the computation of a polygon with an arbitrary number of edges, but in that case the set-up is complicated by the fact that the number of cross ratios grows with the number of edges, giving rise to multiple polylogarithms depending \emph{a priori} on all those cross ratios. However, these techniques could be useful in the computation of special classes of regular polygons where the cross ratios take special values~\cite{Brandhuber:2009da, Alday:2009dv,  Alday:2010vh}. This is currently under investigation.


\section*{Acknowledgements}

We thank Ugo Aglietti, Fernando Alday, Babis Anastasiou, Andi Brandhuber, Nigel Glover, Paul Heslop, Mikhail Kalmykov, Valya Khoze,
Gregory Korchemsky and Gabriele Travaglini for useful discussions and Valentina Forini for a critical reading of the manuscript.
In particular, we thank Nigel Glover for the fruitful collaboration which led to the present work, Paul Heslop and Valya Khoze for providing the numerical value of the two-loop six-point remainder function at several points,
and Alexander Smirnov for providing a version of FIESTA suitable to compute the Wilson-loop integrals. CD thanks the CERN Theory Group 
and CD and VAS thank the Laboratori Nazionali di Frascati for the hospitality during various stages of this work.
This work was partly supported by RFBR, grant 08-02-01451, and
by the EC Marie-Curie Research Training Network ``Tools and Precision
Calculations for Physics Discoveries at Colliders'' under contract MRTN-CT-2006-035505.


\appendix

\section{Nested harmonic sums}\label{app:Ssums}

The nested harmonic sums are defined by~\cite{Vermaseren:1998uu},
\beq\bsp\label{eq:SZSums}
&S_i(n) = Z_i(n) = H_n^{(i)} = \sum_{k=1}^n\,{1\over k^i},\\
& S_{i\vec \jmath}(n) = \sum_{k=1}^n\,{S_{\vec \jmath}(k)\over k^i} {\rm ~~and~~} Z_{i\vec \jmath}(n) = \sum_{k=1}^n\,{Z_{\vec \jmath}(k-1)\over k^i}.
\esp\eeq
The $S$ and $Z$ sums form an algebra. Let us illustrate this on a simple example,
\beq\bsp\label{eq:Ssumalgebra}
S_i(n)S_j(n) &\,= \sum_{k_1=1}^n\,\sum_{k_2=1}^n\,{1\over k_1^i k_2^j}\\
&\,= \sum_{k_1=1}^n\,{1\over k_1^i}\,\sum_{k_2=1}^{k_1}\,{1\over k_2^j}+\sum_{k_2=1}^n\,{1\over k_2^j}\,\sum_{k_1=1}^{k_2}\,{1\over k_1^i} - \sum_{k=1}^n\,{1\over k^{i+j}}\\
&\,=S_{ij}(n)+S_{ji}(n)-S_{i+j}(n).
\esp\eeq
A similar result can be obtained for the $Z$ sums\footnote{Note the sign difference with respect to \Eqn{eq:Ssumalgebra}.},
\beq
Z_i(n)Z_j(n) = Z_{ij}(n)+Z_{ji}(n)+Z_{i+j}(n).
\eeq
For sums of higher weight, a recursive application of the above procedure leads then to the reduction of any product of $S$ or $Z$ sums to a linear combination of those sums. Furthermore, $S$ and $Z$ sums can be interchanged, \emph{e.g.},
\beq\bsp
Z_{11}(n)&\,=\sum_{k=1}^n\,\frac{Z_1(k-1)}{k}=\sum_{k=1}^n\,\frac{1}{k}\,\sum_{\ell=1}^{k-1}\,\frac{1}{\ell}=\sum_{k=1}^n\,\frac{1}{k}\,\left(-\frac{1}{k}+\sum_{\ell=1}^k\,\frac{1}{\ell}\right)\\
&\,=-\sum_{k=1}^n\,\frac{1}{k^2} + \sum_{k=1}^n\,\frac{1}{k}\,S_1(k)=-S_2(n) + S_{11}(n).
\esp\eeq
For $n\rightarrow\infty$, the Euler-Zagier sums converge to multiple zeta values,
\beq
\lim_{n\rightarrow\infty}\,Z_{m_1\ldots m_k}(n) = \zeta(m_k,\ldots,m_1).
\eeq

In Ref.~\cite{Moch:2001zr} generalisations of the $S$ and $Z$ sums were introduced to make them dependent on some variables,
\beq\bsp
&S_i(n;x) = Z_i(n;x) = \sum_{k=1}^n\,{x^k\over k^i},\\
& S_{i\vec \jmath}(n;x_1,\ldots, x_\ell) = \sum_{k=1}^n\,{x_1^k\over k^i}\,S_{\vec \jmath}(k; x_2,\ldots, x_n),\\
&Z_{i\vec \jmath}(n;x_1,\ldots, x_\ell) = \sum_{k=1}^n\,{x_1^k\over k^i}\,Z_{\vec \jmath}(k-1; x_2,\ldots, x_n).
\esp\eeq
Those sums naturally share all the properties of the corresponding number sums introduced in the previous paragraph , \emph{e.g.}, they also form an algebra and the $S$ and $Z$ sums can be interchanged. In Ref.~\cite{Moch:2001zr} several algorithms were derived that allow one to express certain classes of nested sums as linear combinations of $S$ and/or $Z$ sums, and those algorithms are implemented in the {\tt FORM} code {\tt XSummer}~\cite{Moch:2005uc}.
Furthermore, for $n\rightarrow\infty$, the $Z$ sums converge to Goncharov multiple polylogarithms,
\beq
\lim_{n\rightarrow\infty}\,Z_{m_1\ldots m_k}(n;x_1,\ldots,x_k) = \mathrm{Li}_{m_k,\ldots,m_1}(x_k,\ldots,x_1).
\eeq
which are reviewed in the next section.

\section{Goncharov multiple polylogarithms}
\label{app:Goncharov}
\subsection{Definition}
 Let us define~\cite{Goncharov:1998, Goncharov:2001}
 \beq
 \Omega(w) = {\rd t\over t-w},
 \eeq
 and iterated integrations by
 \beq
 \int_0^z \,\Omega(w_1)\circ\ldots\circ\Omega(w_n) = \int_0^z\,{\rd t\over t-w_1}\,\int_0^t\,\Omega(w_2)\circ\ldots\circ\Omega(w_n).
 \eeq
Goncharov multiple polylogarithms can be defined by the iterated integration,
 \beq\label{eq:Gonchdef2}
 G(w_1,\ldots,w_n;z)=\,\int_0^z\,\Omega(w_1)\circ\ldots\circ\Omega(w_n)\,,\\
\eeq
and in the special case where all the $w_i$'s are zero, we define,
\beq
G(\vec 0_n;z) = {1\over n!}\,\ln^n z\,.
\eeq
The vector $\vec w=(w_1, \ldots, w_n)$ is called the weight vector of the polylogarithm and the number of elements in the weight vector is called the weight $w$ of the polylogarithm.
Iterated integrals form a shuffle algebra, and hence we can immediately write,
  \beq\bsp
  G(\vec w_1;z) \, G(\vec w_2;z) &\,=\sum_{\vec w=\vec w_1\uplus\vec w_2}\,G(\vec w;z).\\
      \esp\eeq
      Note that all terms in this equation have the same weight $w_1+w_2$
The algebra properties of the Goncharov polylogarithms imply that not all the $G$ functions are independent, but there must be (polynomial) relations among them.  In particular, we can choose a basis where the rightmost index of all the weight vectors is non zero (apart from objects of the form $G(\vec 0_n;z)$), \emph{e.g.},
\beq\bsp
G(a,0,0;z) &\,= G(0,0;z)\,G(a;z) - G(0,0,a;z) - G(0,a,0;z)\\
&\, = G(0,0;z)\,G(a;z) - G(0,0,a;z) - \left(G(0,a;z)\,G(0;z) - 2G(0,0,a;z)\right)\\
&\,= G(0,0;z)\,G(a;z) + G(0,0,a;z) -G(0,a;z)\,G(0;z)\,.
\esp\eeq
If the rightmost index of a $G$ function is non zero, then the function is invariant under a rescaling of all its arguments,
\beq\label{eq:Gscaling}
G(k\,\vec w;k\, z) = G(\vec w;z)\,.
\eeq
From the definition~(\ref{eq:Gonchdef2}) it is easy to see that,
\beq
\lim_{w_i\to \infty}\,G(\vec w;z) = 0\,.
\eeq
Goncharov multiple polylogarithms can also be represented as multiple nested sums,
\beq
\textrm{Li}_{m_k,\ldots,m_1}(x_k,\ldots,x_1) = \sum_{n_1=1}^\infty{x_1^{n_1}\over n_1^{m_1}}\,\sum_{n_2=1}^{n_1-1}\ldots\sum_{n_k=1}^{n_{k-1}-1}{x_k^{n_k}\over n_k^{m_k}} = Z_{m_1,\ldots,m_k}(\infty;x_1,\ldots,x_k)\,.
\eeq
Since the Li functions are the values at infinity of the $Z$ sums introduced in the previous section, they share all the algebra properties of the $Z$ sums.
The $G$ and Li functions define in fact the same class of function and are related by,
\beq
\textrm{Li}_{m_k,\ldots,m_1}(x_k,\ldots,x_1) = (-1)^k\,G\left(\underbrace{0,\ldots,0}_{m_1-1},{1\over x_1}, \ldots, \underbrace{0,\ldots,0}_{m_k-1}, {1\over x_1\ldots x_k};1\right)\,.
\eeq

\subsection{Special values}
In some cases it is possible to express Goncharov multiple polylogarithms in terms of other functions, \emph{e.g.},
\beq\bsp
G(\vec 0_n;z) = {1\over n!}\,\ln^nz, \qquad &G(\vec a_n;z) = {1\over n!}\,\ln^n\left(1-{z\over a}\right),\\
G(\vec 0_{n-1},a;z) = -\textrm{Li}_n\left({z\over a}\right), \qquad & G(\vec 0_{n},\vec a_{p};z) = (-1)^p\,S_{n,p}\left({z\over a}\right)\,,
\esp\eeq
and in the special case where the elements of the weight vector only take values in the set $\{-1, 0, +1\}$, Goncharov polylogarithms can be expressed in terms of the harmonic polylogarithms introduced by Remiddi and Vermaseren~\cite{Remiddi:1999ew},
\beq
G(\vec w;z) = (-1)^k\,H(\vec w;z)\,,
\eeq
where $k$ is the number of elements in $\vec w$ equal to $(+1)$.

Furthermore, up to weight two, Goncharov polylogarithms can be completely expressed in terms of ordinary logarithms and dilogarithms. In particular, if $a$ and $b$ are non zero, we find,
\beq
G(a,b;z) = \text{Li}_2\left(\frac{b-z}{b-a}\right)-\text{Li}_2\left(\frac{b}{b-a}\right)+\log \left(1-\frac{z}{b}\right) \log \left(\frac{z-a}{b-a}\right)\,.
   \eeq
More special values of Goncharov multiple polylogarithms are presented in App.~\ref{app:special_values_GPL} and \ref{app:GPL_HPL}.

\subsection{Reduction of polylogarithms of the form $G(\vec w(z);1)$}
In this section we present the algorithm used to express a polylogarithm of the form $G(\vec w(z);1)$, where $w$ is a rational function of $z$, as a linear combination of polylogarithms of the form $G(\vec w';z)$, where $\vec w'$ is independent of $z$. This algorithm is a generalisation of the corresponding algorithms described in Refs.~\cite{DelDuca:2009ac, Gehrmann:2001jv, Aglietti:2008fe}.
We start by writing $G(\vec w(z);1)$ as the integral of the derivative,
\beq
G(\vec w(z);1) = G(\vec w(z_0);1) + \int_{z_0}^z\rd t\,{\partial\over\partial t}\,G(\vec w(t);1)\,,
\eeq
where $z_0$ is arbitrary (provided that $G(\vec w(z_0);1)$ exists).
We now carry out the derivative on the integral representation of $G(\vec w(z);1)$,
\beq\label{eq:gred}
G(\vec w(z);1) = G(\vec w(z_0);1) + \int_{z_0}^z\rd t\,\sum_{i=1}^w\int_0^1\Omega(w_1(t))\circ\ldots\circ{\partial\over\partial t}\,\Omega(w_i(t))\circ\ldots\circ\Omega(w_w(t))\,,
\eeq
with
\beq
{\partial\over\partial t}\,\Omega(w_i(t)) = {\rd t_i\over (t_i-w_i(t))^2}\,{\partial\over\partial t} w_i(t).
\eeq
The integrals over the $t_i$ variables are easily performed using partial fractioning and integration by parts. At the end of this procedure, we are left with an integral over $t$ whose integrand is a linear combination (with rational coefficients) of Goncharov polylogarithms of the form $G(\vec w_1(t);1)$, with $w_1 = w-1$. At this point we know recursively how to express these functions in terms of polylogarithms of the form $G(\vec w'_1;t)$ where $\vec w'_1$ is independent of $t$. The last integration is now done using partial fractioning and integration by parts, and since the upper integration limit is $z$, we end up with polylogarithms of the form $G(\vec w';z)$.


\section{Evaluation of the additional residues of $F(u_1,u_2,u_3)$}
\label{app:residues}

\subsection{Evaluation of $R_{-1}(u_1,u_2,u_3)$}
In this appendix we give the details on the computations of the additional residues defined in \Eqn{eq:addres}. For convenience let us start by introducing the definition that $\rcal_j(u_1,u_2,u_3;z_1,z_2)$ is the integrand of $R_{j}(u_1,u_2,u_3)$,
\emph{i.e.}, we define,
\beq
R_{j}(u_1,u_2,u_3) \equiv \mbint \mbint{\rd z_1\over 2\pi i}{\rd z_2\over 2\pi i}\,\rcal_j(u_1,u_2,u_3;z_1,z_2)\,.
\eeq

We now turn to the evaluation of $R_{-1}(u_1,u_2,u_3)$. We close the contours to the right, and take residues in $z_i=n_i$, $n_i\in \mathbb{N}$. We obtain,
\beq\bsp\label{eq:Rm1Res}
R_{-1}&(u_1,u_2,u_3) = - \res_{z_1=0}\,\res_{z_2=0}\,\rcal_{-1}(u_1,u_2,u_3;z_1,z_2) \\
- &\,\sum_{n=1}^\infty\Big(\res_{z_1=n}\,\res_{z_2=0}\,\rcal_{-1}(u_1,u_2,u_3;z_1,z_2)+\res_{z_1=0}\,\res_{z_2=n}\,\rcal_{-1}(u_1,u_2,u_3;z_1,z_2)\Big) \\
-&\, \sum_{n_1=1}^\infty\sum_{n_2=1}^\infty\res_{z_1=n_1}\,\res_{z_2=n_2}\,\rcal_{-1}(u_1,u_2,u_3;z_1,z_2)\,.
\esp\eeq
The first line in \Eqn{eq:Rm1Res} is trivial,
\beq\bsp\label{eq:restrivial}
- &\res_{z_1=0}\,\res_{z_2=0}\,\rcal_{-1}(u_1,u_2,u_3;z_1,z_2) = \\
&\zeta_3 \log u_1+\zeta_3 \log u_2-\frac{1}{4}
   \log ^2u_2 \log ^2u_1-\frac{1}{6} \pi ^2 \log
   ^2u_1-\frac{1}{6} \pi ^2 \log ^2u_2-\frac{1}{6}
   \pi ^2 \log u_2 \log u_1-\frac{17 \pi ^4}{120}
 \esp  \eeq
 The single sum in the second line can be expressed in terms of $S$-sums and the sum can be performed using the algorithms A and B of Refs.~\cite{Moch:2001zr, Moch:2005uc}\footnote{Note that we used a private implementation of these algorithms in Mathematica.},
 \beq\bsp\label{eq:resc4}
 - &\,\sum_{n=1}^\infty\Big(\res_{z_1=n}\,\res_{z_2=0}\,\rcal_{-1}(u_1,u_2,u_3;z_1,z_2)+\res_{z_1=0}\,\res_{z_2=n}\,\rcal_{-1}(u_1,u_2,u_3;z_1,z_2)\Big) \\
 =&\,\sum_{n=1}^\infty\Big(
\frac{u_1^n S_{1,1}(n)}{n^2}+\frac{u_2^n S_{1,1}(n)}{n^2}+\frac{u_1^n S_{1,2}(n)}{n}+\frac{u_1^n S_{2,1}(n)}{n}+\frac{u_2^n S_{1,2}(n)}{n}+\frac{u_2^n S_{2,1}(n)}{n}\\
&-\frac{u_1^n \log u_1 S_{1,1}(n)}{n}-\frac{u_2^n \log u_2 S_{1,1}(n)}{n}-\frac{S_2(n) u_1^n}{n^2}-\frac{S_2(n) u_2^n}{n^2}+\frac{S_1(n) u_1^n \log u_2}{n^2}\\
&+\frac{S_1(n) u_2^n \log u_1}{n^2}+\frac{\pi ^2 u_1^n}{3 n^2}+\frac{\pi ^2 u_2^n}{3 n^2}+\frac{u_1^n \log ^2u_2}{2 n^2}+\frac{u_2^n \log ^2u_1}{2 n^2}-\frac{\pi ^2 S_1(n) u_1^n}{6 n}-\frac{2 S_3(n) u_1^n}{n}\\
&-\frac{\pi ^2 S_1(n) u_2^n}{6 n}-\frac{2 S_3(n) u_2^n}{n}-\frac{S_1(n) u_1^n \log u_1 \log u_2}{n}+\frac{S_2(n) u_1^n \log u_1}{n}+\frac{S_2(n) u_1^n \log u_2}{n}\\
&-\frac{S_1(n) u_2^n \log u_1 \log u_2}{n}+\frac{S_2(n) u_2^n \log u_1}{n}+\frac{S_2(n) u_2^n \log u_2}{n}+\frac{\zeta_3 u_1^n}{n}+\frac{\zeta_3 u_2^n}{n}\\
&-\frac{u_1^n \log u_1 \log ^2u_2}{2 n}-\frac{u_2^n \log ^2u_1 \log u_2}{2 n}-\frac{\pi ^2 u_1^n \log u_1}{3 n}-\frac{\pi ^2 u_1^n \log u_2}{6 n}-\frac{\pi ^2 u_2^n \log u_1}{6 n}\\
&-\frac{\pi ^2 u_2^n \log u_2}{3 n}
 \Big)
 \esp\eeq
 \beq\bsp\nn
 =&\,\zeta_3 H\left(1;u_1\right)+\zeta_3 H\left(1;u_2\right)-\frac{1}{6} \pi ^2 H\left(0;u_2\right) H\left(1;u_1\right)-\frac{1}{6} \pi ^2 H\left(0;u_1\right) H\left(1;u_2\right)\\
 &\,-\frac{1}{6} \pi ^2 H\left(0,1;u_1\right)-\frac{1}{6} \pi ^2 H\left(0,1;u_2\right)-H\left(0,0;u_2\right) H\left(1,0;u_1\right)-\frac{1}{3} \pi ^2 H\left(1,0;u_1\right)\\
 &\,-H\left(0,0;u_1\right) H\left(1,0;u_2\right)-\frac{1}{3} \pi ^2 H\left(1,0;u_2\right)-\frac{1}{6} \pi ^2 H\left(1,1;u_1\right)-\frac{1}{6} \pi ^2 H\left(1,1;u_2\right)\\
 &\,-H\left(0;u_2\right) H\left(0,1,0;u_1\right)-H\left(0;u_1\right) H\left(0,1,0;u_2\right)-H\left(0;u_2\right) H\left(1,1,0;u_1\right)\\
 &\,-H\left(0;u_1\right) H\left(1,1,0;u_2\right)-H\left(0,1,1,0;u_1\right)-H\left(0,1,1,0;u_2\right)-H\left(1,1,1,0;u_1\right)\\
 &\,-H\left(1,1,1,0;u_2\right)\,,
  \esp\eeq
  where $H(\vec w;x)$ denote the standard harmonic polylogarithms of Remiddi and Vermaseren~\cite{Remiddi:1999ew}. The double sum in the third line is rewritten as a nested sum by letting $n=n_1+n_2$,
  \beq\bsp\label{eq:binsum}
 - \sum_{n_1=1}^\infty&\sum_{n_2=1}^\infty\res_{z_1=n_1}\,\res_{z_2=n_2}\,\rcal_{-1}(u_1,u_2,u_3;z_1,z_2) \\
 =\sum_{n=1}^\infty&\sum_{n_1=1}^{n-1}\,\binom{n}{n_1}\Bigg\{
 \frac{\log u_1 u_1^{n_1} u_2^{n-n_1}}{n^2 n_1}-\frac{\log u_1 u_1^{n_1} u_2^{n-n_1}}{n^2 (n_1-n)}+\frac{\log u_1 u_1^{n_1} u_2^{n-n_1}}{n (n_1-n)^2}-\frac{\log u_1 \log u_2 u_1^{n_1} u_2^{n-n_1}}{n n_1}\\
 &+\frac{\log u_1 \log u_2 u_1^{n_1} u_2^{n-n_1}}{n (n_1-n)}+\frac{\log u_2 u_1^{n_1} u_2^{n-n_1}}{n^2 n_1}-\frac{\log u_2 u_1^{n_1} u_2^{n-n_1}}{n^2 (n_1-n)}+\frac{\log u_2 u_1^{n_1} u_2^{n-n_1}}{n n_1^2}\\
 &-\frac{\log u_1 S_1(n) u_1^{n_1} u_2^{n-n_1}}{n n_1}+\frac{\log u_1 S_1(n) u_1^{n_1} u_2^{n-n_1}}{n (n_1-n)}-\frac{\log u_2 S_1(n) u_1^{n_1} u_2^{n-n_1}}{n n_1}
 \\
 &+\frac{\log u_2 S_1(n) u_1^{n_1} u_2^{n-n_1}}{n (n_1-n)}+\frac{2 S_1(n) u_1^{n_1} u_2^{n-n_1}}{n^2 n_1}-\frac{2 S_1(n) u_1^{n_1} u_2^{n-n_1}}{n^2 (n_1-n)}+\frac{S_1(n) u_1^{n_1} u_2^{n-n_1}}{n n_1^2}\\
 &+\frac{S_1(n) u_1^{n_1} u_2^{n-n_1}}{n (n_1-n)^2}+\frac{\log u_1 S_1(n-n_1) u_1^{n_1} u_2^{n-n_1}}{n n_1}-\frac{\log u_1 S_1(n-n_1) u_1^{n_1} u_2^{n-n_1}}{n (n_1-n)}\\
 &+\frac{S_1(n) S_1(n-n_1) u_1^{n_1} u_2^{n-n_1}}{n n_1}-\frac{S_1(n) S_1(n-n_1) u_1^{n_1} u_2^{n-n_1}}{n (n_1-n)}-\frac{S_1(n-n_1) u_1^{n_1} u_2^{n-n_1}}{n^2 n_1}\\
 &+\frac{S_1(n-n_1) u_1^{n_1} u_2^{n-n_1}}{n^2 (n_1-n)}-\frac{S_1(n-n_1) u_1^{n_1} u_2^{n-n_1}}{n n_1^2}+\frac{\log u_2 S_1(n_1) u_1^{n_1} u_2^{n-n_1}}{n n_1}\\
& -\frac{\log u_2 S_1(n_1) u_1^{n_1} u_2^{n-n_1}}{n n_1-n}
 +\frac{S_1(n) S_1(n_1) u_1^{n_1} u_2^{n-n_1}}{n n_1}-\frac{S_1(n) S_1(n_1) u_1^{n_1} u_2^{n-n_1}}{n (n_1-n)}\\
 &-\frac{S_1(n-n_1) S_1(n_1) u_1^{n_1} u_2^{n-n_1}}{n n_1}+\frac{S_1(n-n_1) S_1(n_1) u_1^{n_1} u_2^{n-n_1}}{n (n_1-n)}-\frac{S_1(n_1) u_1^{n_1} u_2^{n-n_1}}{n^2 n_1}\\
 &+\frac{S_1(n_1) u_1^{n_1} u_2^{n-n_1}}{n^2 (n_1-n)}-\frac{S_1(n_1) u_1^{n_1} u_2^{n-n_1}}{n (n_1-n)^2}+\frac{2 S_2(n) u_1^{n_1} u_2^{n-n_1}}{n n_1}-\frac{2 S_2(n) u_1^{n_1} u_2^{n-n_1}}{n (n_1-n)}\\
 &-\frac{2 S_{1,1}(n) u_1^{n_1} u_2^{n-n_1}}{n n_1}+\frac{2 S_{1,1}(n) u_1^{n_1} u_2^{n-n_1}}{n (n_1-n)}-\frac{2 u_1^{n_1} u_2^{n-n_1}}{n^3 n_1}+\frac{2 u_1^{n_1} u_2^{n-n_1}}{n^3 (n_1-n)}-\frac{u_1^{n_1} u_2^{n-n_1}}{n^2 n_1^2}\\
 &-\frac{u_1^{n_1} u_2^{n-n_1}}{n^2 (n_1-n)^2}-\frac{\pi ^2 u_1^{n_1} u_2^{n-n_1}}{6 n n_1}+\frac{\pi ^2 u_1^{n_1} u_2^{n-n_1}}{6 n (n_1-n)}\Bigg\}\,.
 \esp\eeq
 All the sums in this expression can be performed using the algorithms C and D of Refs.~\cite{Moch:2001zr}. With the help of {\tt XSummer}~\cite{Moch:2005uc} we find,
 \btxtsloppy
\parbox{130mm}{\raggedright\(\displaystyle
- \sum_{n_1=1}^\infty\sum_{n_2=1}^\infty\res_{z_1=n_1}\,\res_{z_2=n_2}\,\rcal_{-1}(u_1,u_2,u_3;z_1,z_2)= \)}
\raggedleft\refstepcounter{equation}(\theequation)\label{eq:GRED3}\\
\raggedright\(\displaystyle
-\log u_1 \log u_2 G\left(\frac{1}{u_1},\frac{1}{u_1+u_2};1\right)-\frac{1}{6} \pi ^2 G\left(\frac{1}{u_1},\frac{1}{u_1+u_2};1\right)-\frac{1}{6} \pi ^2 G\left(\frac{1}{u_2},\frac{1}{u_1+u_2};1\right)-2 G\left(0,0,\frac{1}{u_1},\frac{1}{u_1+u_2};1\right)-2 G\left(0,0,\frac{1}{u_2},\frac{1}{u_1+u_2};1\right)+4 G\left(0,0,\frac{1}{u_1+u_2},\frac{1}{u_1};1\right)+4 G\left(0,0,\frac{1}{u_1+u_2},\frac{1}{u_2};1\right)-8 G\left(0,0,\frac{1}{u_1+u_2},\frac{1}{u_1+u_2};1\right)+2 G\left(0,\frac{1}{u_1},0,\frac{1}{u_2};1\right)-2 G\left(0,\frac{1}{u_1},0,\frac{1}{u_1+u_2};1\right)-G\left(0,\frac{1}{u_1},\frac{1}{u_1+u_2},\frac{1}{u_1};1\right)-G\left(0,\frac{1}{u_1},\frac{1}{u_1+u_2},\frac{1}{u_2};1\right)+2 G\left(0,\frac{1}{u_2},0,\frac{1}{u_1};1\right)-2 G\left(0,\frac{1}{u_2},0,\frac{1}{u_1+u_2};1\right)-G\left(0,\frac{1}{u_2},\frac{1}{u_1+u_2},\frac{1}{u_1};1\right)-G\left(0,\frac{1}{u_2},\frac{1}{u_1+u_2},\frac{1}{u_2};1\right)+2 G\left(0,\frac{1}{u_1+u_2},0,\frac{1}{u_1};1\right)+2 G\left(0,\frac{1}{u_1+u_2},0,\frac{1}{u_2};1\right)-4 G\left(0,\frac{1}{u_1+u_2},\frac{1}{u_1+u_2},\frac{1}{u_1+u_2};1\right)+2 G\left(\frac{1}{u_1},0,0,\frac{1}{u_1};1\right)+4 G\left(\frac{1}{u_1},0,0,\frac{1}{u_2};1\right)-2 G\left(\frac{1}{u_1},0,0,\frac{1}{u_1+u_2};1\right)-G\left(\frac{1}{u_1},0,\frac{1}{u_1+u_2},\frac{1}{u_1};1\right)-G\left(\frac{1}{u_1},0,\frac{1}{u_1+u_2},\frac{1}{u_2};1\right)-G\left(\frac{1}{u_1},\frac{1}{u_1+u_2},0,\frac{1}{u_1};1\right)-G\left(\frac{1}{u_1},\frac{1}{u_1+u_2},0,\frac{1}{u_2};1\right)+4 G\left(\frac{1}{u_2},0,0,\frac{1}{u_1};1\right)+2 G\left(\frac{1}{u_2},0,0,\frac{1}{u_2};1\right)-2 G\left(\frac{1}{u_2},0,0,\frac{1}{u_1+u_2};1\right)-G\left(\frac{1}{u_2},0,\frac{1}{u_1+u_2},\frac{1}{u_1};1\right)-G\left(\frac{1}{u_2},0,\frac{1}{u_1+u_2},\frac{1}{u_2};1\right)-G\left(\frac{1}{u_2},\frac{1}{u_1+u_2},0,\frac{1}{u_1};1\right)-G\left(\frac{1}{u_2},\frac{1}{u_1+u_2},0,\frac{1}{u_2};1\right)+\frac{1}{3} \pi ^2 H\left(0,1;u_1\right)+\frac{1}{3} \pi ^2 H\left(0,1;u_2\right)-\frac{1}{3} \pi ^2 H\left(0,1;u_1+u_2\right)+\frac{1}{6} \pi ^2 H\left(1,1;u_1\right)+\frac{1}{6} \pi ^2 H\left(1,1;u_2\right)+12 H\left(0,0,0,1;u_1\right)+12 H\left(0,0,0,1;u_2\right)-12 H\left(0,0,0,1;u_1+u_2\right)-2 H\left(0,0,1,1;u_1\right)-2 H\left(0,0,1,1;u_2\right)+8 H\left(0,0,1,1;u_1+u_2\right)-H\left(0,1,1,1;u_1\right)-H\left(0,1,1,1;u_2\right)-4 H\left(0,1,1,1;u_1+u_2\right)-H\left(1,0,1,1;u_1\right)-H\left(1,0,1,1;u_2\right)-H\left(1,1,0,1;u_1\right)-H\left(1,1,0,1;u_2\right)+G\left(0,\frac{1}{u_1},\frac{1}{u_1+u_2};1\right) \log u_1+G\left(0,\frac{1}{u_2},\frac{1}{u_1+u_2};1\right) \log u_1-2 G\left(0,\frac{1}{u_1+u_2},\frac{1}{u_1};1\right) \log u_1+2 G\left(0,\frac{1}{u_1+u_2},\frac{1}{u_1+u_2};1\right) \log u_1+G\left(\frac{1}{u_1},0,\frac{1}{u_1+u_2};1\right) \log u_1+G\left(\frac{1}{u_1},\frac{1}{u_1+u_2},\frac{1}{u_1};1\right) \log u_1-2 G\left(\frac{1}{u_2},0,\frac{1}{u_1};1\right) \log u_1+G\left(\frac{1}{u_2},0,\frac{1}{u_1+u_2};1\right) \log u_1+G\left(\frac{1}{u_2},\frac{1}{u_1+u_2},\frac{1}{u_1};1\right) \log u_1-4 H\left(0,0,1;u_1\right) \log u_1-4 H\left(0,0,1;u_2\right) \log u_1+4 H\left(0,0,1;u_1+u_2\right) \log u_1+H\left(0,1,1;u_1\right) \log u_1-H\left(0,1,1;u_2\right) \log u_1-2 H\left(0,1,1;u_1+u_2\right) \log u_1-H\left(1,0,1;u_1\right) \log u_1-H\left(1,0,1;u_2\right) \log u_1+H\left(1,1,1;u_1\right) \log u_1+G\left(0,\frac{1}{u_1},\frac{1}{u_1+u_2};1\right) \log u_2+G\left(0,\frac{1}{u_2},\frac{1}{u_1+u_2};1\right) \log u_2-2 G\left(0,\frac{1}{u_1+u_2},\frac{1}{u_2};1\right) \log u_2+2 G\left(0,\frac{1}{u_1+u_2},\frac{1}{u_1+u_2};1\right) \log u_2-2 G\left(\frac{1}{u_1},0,\frac{1}{u_2};1\right) \log u_2+G\left(\frac{1}{u_1},0,\frac{1}{u_1+u_2};1\right) \log u_2+G\left(\frac{1}{u_1},\frac{1}{u_1+u_2},\frac{1}{u_2};1\right) \log u_2+G\left(\frac{1}{u_2},0,\frac{1}{u_1+u_2};1\right) \log u_2+G\left(\frac{1}{u_2},\frac{1}{u_1+u_2},\frac{1}{u_2};1\right) \log u_2-4 H\left(0,0,1;u_1\right) \log u_2-4 H\left(0,0,1;u_2\right) \log u_2+4 H\left(0,0,1;u_1+u_2\right) \log u_2-H\left(0,1,1;u_1\right) \log u_2+H\left(0,1,1;u_2\right) \log u_2-2 H\left(0,1,1;u_1+u_2\right) \log u_2-H\left(1,0,1;u_1\right) \log u_2-H\left(1,0,1;u_2\right) \log u_2+H\left(1,1,1;u_2\right) \log u_2-G\left(\frac{1}{u_2},\frac{1}{u_1+u_2};1\right) \log u_1 \log u_2+2 H\left(0,1;u_1\right) \log u_1 \log u_2+2 H\left(0,1;u_2\right) \log u_1 \log u_2-2 H\left(0,1; u_1+u_2\right) \log u_1 \log u_2+H\left(1,1;u_1\right) \log u_1 \log u_2+H\left(1,1;u_2\right) \log u_1 \log u_2
 \,.
 \)\etxtsloppy

 \subsection{Evaluation of $R_{-1-z_1}(u_1,u_2,u_3)$}
 We now turn to the evaluation of the residue, $R_{-1-z_1}(u_1,u_2,u_3)$, in $z_3 = -1-z_1$ by exchanging the MB integrations with the Euler integration, \Eqn{eq:MBtoEuler}.
Changing the integration variables from $z_2$ to $-z_2$ and shifting the integration contours, we arrive at the following representation for $R_{-1-z_1}(u_1,u_2,u_3)$,
 \beq\bsp\label{eq:Rz1}
 R&_{-1-z_1}(u_1,u_2,u_3) = \tilde R_{-1-z_1}^{(1)}(u_1,u_2,u_3) + \tilde R_{-1-z_1}^{(2)}(u_1,u_2,u_3)\\
 & -\int_{+1/5-i\infty}^{+1/5+i\infty} {\rd z_2\over2\pi i}\, \res_{z_1=0}\,\rcal_{-1-z_1}(u_1,u_2,u_3;z_1,z_2)\,,
\esp \eeq
 where we defined,
 \beq\bsp
\tilde R&_{-1-z_1}^{(1)}(u_1,u_2,u_3) = \int_{-2/3-i\infty}^{-2/3+i\infty} {\rd z_1\over2\pi i}\int_{-1/5-i\infty}^{-1/5+i\infty}   {\rd z_2\over2\pi i}\, z_2\, u_1^{-z_2}\, u_2^{z_1+1}\, u_3^{z_2}\\
&\times\Gamma \left(-z_1-1\right)\, \Gamma
   \left(-z_1\right)\, \Gamma \left(z_1-z_2+1\right)\, \Gamma
   \left(-z_2\right)^2\, \Gamma \left(z_2\right)^2\, \Gamma
   \left(z_1+z_2+1\right)\,,\\
   \tilde R&_{-1-z_1}^{(2)}(u_1,u_2,u_3) = -\int_{-2/3-i\infty}^{-2/3+i\infty} {\rd z_1\over2\pi i}\int_{-1/5-i\infty}^{-1/5+i\infty}   {\rd z_2\over2\pi i}\,
   \frac{z_2^2}{z_1+1}\, u_1^{-z_2}\, u_2^{z_1+1}\, u_3^{z_2}\\
   &\times \Gamma
   \left(-z_1-1\right)\, \Gamma \left(-z_1\right)\, \Gamma \left(z_1-z_2+1\right)\,
   \Gamma \left(-z_2\right)^2\, \Gamma \left(z_2\right)^2\, \Gamma
   \left(z_1+z_2+1\right)\,.
   \esp\eeq
In each case, the integration contours are straight lines and their position is explicitly indicated for each integral. The residue appearing in \Eqn{eq:Rz1} arises from shifting the integration contours\footnote{Note that there are in principle two more non-zero residues coming from crossing the poles in $z_1=-z_2$ and $z_2=-z_1$, but it turns out that these two contributions cancel mutually.} and is given by the onefold integral,
\beq\bsp
\int_{+1/5-i\infty}^{+1/5+i\infty}& {\rd z_2\over2\pi i}\,\res_{z_1=0}\,\rcal_{-1-z_1}(u_1,u_2,u_3;z_1,z_2) \\
=&\,\int_{+1/5-i\infty}^{+1/5+i\infty} {\rd z_2\over2\pi i}\,z_2\, u_1^{-z_2}\, u_3^{z_2}\, \Gamma \left(-z_2\right)^3\, \Gamma
   \left(z_2\right)^3\\
   \times&\, \left(z_2 \log u_2+2 \eug  z_2+z_2
   \psi \left(-z_2\right)+z_2 \psi \left(z_2\right)-1\right)\,,
   \esp\eeq
where $\psi(z) = {\rd\over\rd z}\ln\Gamma(z)$ denotes the digamma function and $\eug$ is the Euler-Mascheroni constant, $\eug = -\psi(1)$. Closing the integration contour to the right and summing up residues in $z_2=n_2 \in \mathbb{N}^*$, we obtain,
\beq\bsp\label{eq:resc10}
\int_{+1/5-i\infty}^{+1/5+i\infty}& {\rd z_2\over2\pi i}\,\res_{z_1=0}\,\rcal_{-1-z_1}(u_1,u_2,u_3;z_1,z_2) \\
=&\,
-\frac{1}{6} \log ^3u_1 H\left(1;-{u_3\over u_1}\right)+\frac{1}{6} \log ^3u_3 H\left(1;-{u_3\over u_1}\right)-\frac{1}{2} \log ^2u_1 H\left(0,1;-{u_3\over u_1}\right)\\
&+\log ^2u_1 H\left(1,1;-{u_3\over u_1}\right)+\frac{1}{2} \log u_2 \log ^2u_1 H\left(1;-{u_3\over u_1}\right)+\frac{1}{2} \log u_3 \log ^2u_1 H\left(1;-{u_3\over u_1}\right)\\
&-\frac{1}{2} \log ^2u_3 \log u_1 H\left(1;-{u_3\over u_1}\right)-\frac{1}{2} \log ^2u_3 H\left(0,1;-{u_3\over u_1}\right)+\log ^2u_3 H\left(1,1;-{u_3\over u_1}\right)\\
&+\frac{1}{2} \log u_2 \log ^2u_3 H\left(1;-{u_3\over u_1}\right)-\frac{1}{2} \pi ^2 \log u_1 H\left(1;-{u_3\over u_1}\right)-\log u_1 H\left(0,0,1;-{u_3\over u_1}\right)\\
&+2 \log u_1 H\left(0,1,1;-{u_3\over u_1}\right)+2 \log u_1 H\left(1,0,1;-{u_3\over u_1}\right)+\log u_2 \log u_1 H\left(0,1;-{u_3\over u_1}\right)\\
&+\log u_3 \log u_1 H\left(0,1;-{u_3\over u_1}\right)-2 \log u_3 \log u_1 H\left(1,1;-{u_3\over u_1}\right)\\
&-\log u_2 \log u_3 \log u_1 H\left(1;-{u_3\over u_1}\right)+\frac{1}{2} \pi ^2 \log u_2 H\left(1;-{u_3\over u_1}\right)+\log u_2 H\left(0,0,1;-{u_3\over u_1}\right)\\
&+\frac{1}{2} \pi ^2 \log u_3 H\left(1;-{u_3\over u_1}\right)+\log u_3 H\left(0,0,1;-{u_3\over u_1}\right)-2 \log u_3 H\left(0,1,1;-{u_3\over u_1}\right)\\
&-2 \log u_3 H\left(1,0,1;-{u_3\over u_1}\right)-\log u_2 \log u_3 H\left(0,1;-{u_3\over u_1}\right)-2 \zeta_3 H\left(1;-{u_3\over u_1}\right)\\
&-\frac{1}{2} \pi ^2 H\left(0,1;-{u_3\over u_1}\right)+\pi ^2 H\left(1,1;-{u_3\over u_1}\right)-H\left(0,0,0,1;-{u_3\over u_1}\right)\\
&+2 H\left(0,0,1,1;-{u_3\over u_1}\right)+2 H\left(0,1,0,1;-{u_3\over u_1}\right)+2 H\left(1,0,0,1;-{u_3\over u_1}\right)
\,.
\esp\eeq

Let us now turn to the computation of $\tilde R_{-1-z_1}^{(1)}(u_1,u_2,u_3)$. We apply \Eqn{eq:MBtoEuler} and we obtain,
\beq\bsp\label{eq:Rz1Euler}
\tilde R_{-1-z_1}^{(1)}&(u_1,u_2,u_3) = \int_{-1/5-i\infty}^{-1/5+i\infty}{\rd z_2\over 2\pi i}\,\res_{z_1=-1}\,\tilde\rcal_{-1-z_1}^{(1)}(u_1,u_2,u_3;z_1, z_2)\\
&+\int_0^1\rd v \int_{-1/5-i\infty}^{-1/5+i\infty}{\rd z_2\over 2\pi i}\,
u_2\, z_2^2\, u_1^{-z_2}\, u_3^{z_2}\, (1-v)^{z_2-1}\, v^{-z_2}\,
   \left(1-\left(1-u_2\right) v\right)^{-z_2-1}\\
   &\times \Gamma
   \left(-z_2\right)^3\, \Gamma \left(z_2\right)^3\,.
   \esp\eeq
 We tacitly exchanged again the MB and Euler integrations, having checked numerically that this operation is allowed.
 The residue in \Eqn{eq:Rz1Euler} comes from the fact that in \Eqn{eq:MBtoEuler} the contour must be such that it separates the poles in $\Gamma(\ldots+z)$ from the poles in $\Gamma(\ldots-z)$. The computation of the residue is trivial,
 \beq\bsp
&\int_{-1/5-i\infty}^{-1/5+i\infty}{\rd z_2\over 2\pi i}\,\res_{z_1=-1}\,\tilde\rcal_{-1-z_1}^{(1)}(u_1,u_2,u_3;z_1, z_2)\\
&=\,
 -\frac{1}{2} \log ^2u_1 H\left(0,1;-{u_3\over u_1}\right)-\frac{1}{2} \log ^2u_3 H\left(0,1;-{u_3\over u_1}\right)-2 \log u_1 H\left(0,0,1;-{u_3\over u_1}\right)\\
 &+\log u_3 \log u_1 H\left(0,1;-{u_3\over u_1}\right)+2 \log u_3 H\left(0,0,1;-{u_3\over u_1}\right)-\frac{1}{2} \pi ^2 H\left(0,1;-{u_3\over u_1}\right)\\
 &-3 H\left(0,0,0,1;-{u_3\over u_1}\right)\,.
 \esp\eeq
 The MB integration in the second term is also trivial and yields
 \beq\bsp\label{eq:Rz1v}
& \int_0^1\rd v\,{u_2\over2\,(1-v)\,(1-(1-u_2)\,v)}\,\Bigg\{
 2\, \text{Li}_3\left(\frac{(v-1) u_3}{v u_1 \left(1-v \left(1-u_2\right)\right)}\right)\\
 &+\text{Li}_2\left(\frac{(v-1) u_3}{v u_1 \left(1-v \left(1-u_2\right)\right)}\right) \left(2 \log \left(1-\left(1-u_2\right) v\right)+2 \log u_1-2 \log u_3-2 \log (1-v)+2 \log v\right)\\
 &-\left(\log ^2\left(1-\left(1-u_2\right) v\right)-2 \log u_1 \log (1-v)-2 \log (1-v) \log \left(1-\left(1-u_2\right) v\right)+2 \log u_3 \log (1-v)\right.\\
 &\left.+2 \log u_1 \log v+2 \log v \log \left(1-\left(1-u_2\right) v\right)+2 \log u_1 \log \left(1-\left(1-u_2\right) v\right)-2 \log u_3 \log v\right.\\
 &\left.-2 \log u_3 \log \left(1-\left(1-u_2\right) v\right)+\log ^2u_1+\log ^2u_3-2 \log u_1 \log u_3+\log ^2(1-v)+\log ^2v\right.\\
 &\left.-2 \log v \log (1-v)+\pi ^2\right) \log \left(1-\frac{u_3 (v-1)}{u_1 v \left(1-\left(1-u_2\right) v\right)}\right)\Bigg\}\,.
 \esp\eeq
 The remaining integration over $v$ can be done in a similar algorithmic way as for $\tilde F(u_1,u_2,u_3)$, so we will be brief on its derivation. We start by expressing the polylogarithms appearing in the integrand of \Eqn{eq:Rz1v} in terms of Goncharov multiple polylogarithms,
 \beq
 \text{Li}_n\left(\frac{(v-1) u_3}{v u_1 \left(1-v \left(1-u_2\right)\right)}\right) = -G\left(\vec 0_{n-1}, \frac{v u_1 \left(1-v \left(1-u_2\right)\right)}{(v-1) u_3};1\right)\,.
 \eeq
Using the algorithm of App.~\ref{app:Goncharov}, we can express the Goncharov polylogarithms in the right-hand side of \Eqn{eq:Rz1v} as a linear combination of Goncharov polylogarithms of the form $G(\ldots;v)$, \emph{e.g.},
 \btxtsloppy
\parbox{130mm}{\raggedright\(\displaystyle
G\left(0,0, \frac{v u_1 \left(1-v \left(1-u_2\right)\right)}{(v-1) u_3};1\right)= \)}
\raggedleft\refstepcounter{equation}(\theequation)\label{eq:GRED4}\\
\raggedright\(\displaystyle
-\log u_1 \log ^2u_2+\log u_3 \log ^2u_2-G\left(0,\frac{1}{1-u_2};1\right) \log u_2-G\left(\frac{1}{1-u_2},0;1\right) \log u_2-G\left(\frac{1}{1-u_2},1;1\right) \log u_2-G\left(\frac{1}{1-u_2},\frac{1}{1-u_2};1\right) \log u_2+\gcal(0,v_{213};1) \log u_2+\gcal(v_{213},1;1) \log u_2+\gcal\left(\frac{1}{1-u_2},v_{213};1\right) \log u_2+G(0;v) \log u_1 \log u_2-G(1;v) \log u_1 \log u_2+G\left(\frac{1}{1-u_2};v\right) \log u_1 \log u_2-G(0;v) \log u_3 \log u_2+G(1;v) \log u_3 \log u_2-G\left(\frac{1}{1-u_2};v\right) \log u_3 \log u_2+\frac{1}{6} \pi ^2 \log u_2-\frac{1}{6} \pi ^2 G(0;v)+\frac{1}{6} \pi ^2 G(1;v)-\frac{1}{6} \pi ^2 G\left(\frac{1}{1-u_2};v\right)+G(0;v) G\left(0,\frac{1}{1-u_2};1\right)-G(1;v) G\left(0,\frac{1}{1-u_2};1\right)+G\left(\frac{1}{1-u_2};v\right) G\left(0,\frac{1}{1-u_2};1\right)+G(0;v) G\left(\frac{1}{1-u_2},0;1\right)-G(1;v) G\left(\frac{1}{1-u_2},0;1\right)+G\left(\frac{1}{1-u_2};v\right) G\left(\frac{1}{1-u_2},0;1\right)+G(0;v) G\left(\frac{1}{1-u_2},1;1\right)-G(1;v) G\left(\frac{1}{1-u_2},1;1\right)+G\left(\frac{1}{1-u_2};v\right) G\left(\frac{1}{1-u_2},1;1\right)+G(0;v) G\left(\frac{1}{1-u_2},\frac{1}{1-u_2};1\right)-G(1;v) G\left(\frac{1}{1-u_2},\frac{1}{1-u_2};1\right)+G\left(\frac{1}{1-u_2};v\right) G\left(\frac{1}{1-u_2},\frac{1}{1-u_2};1\right)-G(0,0,0;v)+G\left(0,0,\frac{1}{1-u_2};1\right)-G\left(0,0,\frac{1}{1-u_2};v\right)+G(0,1,0;v)+G\left(0,1,\frac{1}{1-u_2};v\right)+G\left(0,\frac{1}{1-u_2},0;1\right)-G\left(0,\frac{1}{1-u_2},0;v\right)+G\left(0,\frac{1}{1-u_2},1;1\right)+G\left(0,\frac{1}{1-u_2},\frac{1}{1-u_2};1\right)-G\left(0,\frac{1}{1-u_2},\frac{1}{1-u_2};v\right)+G(1,0,0;v)+G\left(1,0,\frac{1}{1-u_2};v\right)-G(1,1,0;v)-G\left(1,1,\frac{1}{1-u_2};v\right)+G\left(1,\frac{1}{1-u_2},0;v\right)+G\left(1,\frac{1}{1-u_2},\frac{1}{1-u_2};v\right)+G\left(\frac{1}{1-u_2},0,0;1\right)-G\left(\frac{1}{1-u_2},0,0;v\right)+G\left(\frac{1}{1-u_2},0,1;1\right)+G\left(\frac{1}{1-u_2},0,\frac{1}{1-u_2};1\right)-G\left(\frac{1}{1-u_2},0,\frac{1}{1-u_2};v\right)+G\left(\frac{1}{1-u_2},1,0;v\right)+G\left(\frac{1}{1-u_2},1,1;1\right)+G\left(\frac{1}{1-u_2},1,\frac{1}{1-u_2};v\right)+G\left(\frac{1}{1-u_2},\frac{1}{1-u_2},0;1\right)-G\left(\frac{1}{1-u_2},\frac{1}{1-u_2},0;v\right)+G\left(\frac{1}{1-u_2},\frac{1}{1-u_2},1;1\right)+G\left(\frac{1}{1-u_2},\frac{1}{1-u_2},\frac{1}{1-u_2};1\right)-G\left(\frac{1}{1-u_2},\frac{1}{1-u_2},\frac{1}{1-u_2};v\right)-G(0;v) \gcal(0,v_{213};1)+G(1;v) \gcal(0,v_{213};1)-G\left(\frac{1}{1-u_2};v\right) \gcal(0,v_{213};1)-G(0;v) \gcal(v_{213},1;1)+G(1;v) \gcal(v_{213},1;1)-G\left(\frac{1}{1-u_2};v\right) \gcal(v_{213},1;1)-G(0;v) \gcal\left(\frac{1}{1-u_2},v_{213};1\right)+G(1;v) \gcal\left(\frac{1}{1-u_2},v_{213};1\right)-G\left(\frac{1}{1-u_2};v\right) \gcal\left(\frac{1}{1-u_2},v_{213};1\right)-\gcal(0,0,v_{213};1)+\gcal(0,0,v_{213};v)-\gcal(0,1,v_{213};v)-\gcal(0,v_{213},1;1)-\gcal\left(0,\frac{1}{1-u_2},v_{213};1\right)+\gcal\left(0,\frac{1}{1-u_2},v_{213};v\right)-\gcal(1,0,v_{213};v)+\gcal(1,1,v_{213};v)-\gcal\left(1,\frac{1}{1-u_2},v_{213};v\right)-\gcal(v_{213},1,1;1)-\gcal\left(\frac{1}{1-u_2},0,v_{213};1\right)+\gcal\left(\frac{1}{1-u_2},0,v_{213};v\right)-\gcal\left(\frac{1}{1-u_2},1,v_{213};v\right)-\gcal\left(\frac{1}{1-u_2},v_{213},1;1\right)-\gcal\left(\frac{1}{1-u_2},\frac{1}{1-u_2},v_{213};1\right)+\gcal\left(\frac{1}{1-u_2},\frac{1}{1-u_2},v_{213};v\right)-G(0,0;v) \log u_1+G(0,1;v) \log u_1+G\left(0,\frac{1}{1-u_2};1\right) \log u_1-G\left(0,\frac{1}{1-u_2};v\right) \log u_1+G(1,0;v) \log u_1-G(1,1;v) \log u_1+G\left(1,\frac{1}{1-u_2};v\right) \log u_1+G\left(\frac{1}{1-u_2},0;1\right) \log u_1-G\left(\frac{1}{1-u_2},0;v\right) \log u_1+G\left(\frac{1}{1-u_2},1;v\right) \log u_1+G\left(\frac{1}{1-u_2},\frac{1}{1-u_2};1\right) \log u_1-G\left(\frac{1}{1-u_2},\frac{1}{1-u_2};v\right) \log u_1+G(0,0;v) \log u_3-G(0,1;v) \log u_3-G\left(0,\frac{1}{1-u_2};1\right) \log u_3+G\left(0,\frac{1}{1-u_2};v\right) \log u_3-G(1,0;v) \log u_3+G(1,1;v) \log u_3-G\left(1,\frac{1}{1-u_2};v\right) \log u_3-G\left(\frac{1}{1-u_2},0;1\right) \log u_3+G\left(\frac{1}{1-u_2},0;v\right) \log u_3-G\left(\frac{1}{1-u_2},1;v\right) \log u_3-G\left(\frac{1}{1-u_2},\frac{1}{1-u_2};1\right) \log u_3+G\left(\frac{1}{1-u_2},\frac{1}{1-u_2};v\right) \log u_3
\,,
 \)\etxtsloppy
where we introduced the quantity,
\beq\label{eq:sqrt2}
v_{jkl}^{(\pm)} =\frac{u_k-u_l\pm\sqrt{-4 u_j u_k u_l+2 u_k u_l+u_k^2+u_l^2}}{2  \left(1-u_j \right)u_k}\, .
\eeq
Finally, the integration over $v$ is done using the recursive definition of Goncharov polylogarithms, \Eqn{eq:Gonchdef}, and the result reads,
\btxtsloppy
\parbox{130mm}{\raggedright\(\displaystyle
\frac{1}{2} G\left(\frac{1}{1-u_2};1\right) \log ^3u_1+\frac{1}{2} G\left(\frac{1}{1-u_2},0;1\right) \log ^2u_1+\frac{1}{2} G\left(\frac{1}{1-u_2},1;1\right) \log ^2u_1 +\)}
\raggedleft\refstepcounter{equation}(\theequation)\label{eq:triv1}\\
\raggedright\(\displaystyle
\frac{1}{2} G\left(\frac{1}{1-u_2},\frac{1}{1-u_2};1\right) \log ^2u_1-\frac{1}{2} \gcal(v_{213},1;1) \log ^2u_1-\frac{1}{2} \gcal\left(\frac{1}{1-u_2},v_{213};1\right) \log ^2u_1+G\left(\frac{1}{1-u_2};1\right) \log u_2 \log ^2u_1-\frac{3}{2} G\left(\frac{1}{1-u_2};1\right) \log u_3 \log ^2u_1-\frac{1}{12} \pi ^2 \log ^2u_1+G\left(\frac{1}{1-u_2};1\right) \log ^2u_2 \log u_1+\frac{3}{2} G\left(\frac{1}{1-u_2};1\right) \log ^2u_3 \log u_1+\frac{1}{3} \pi ^2 G\left(\frac{1}{1-u_2};1\right) \log u_1+G\left(0,1,\frac{1}{1-u_2};1\right) \log u_1+G\left(0,\frac{1}{1-u_2},1;1\right) \log u_1+G\left(\frac{1}{1-u_2},0,0;1\right) \log u_1+G\left(\frac{1}{1-u_2},0,\frac{1}{1-u_2};1\right) \log u_1+G\left(\frac{1}{1-u_2},1,0;1\right) \log u_1-2 G\left(\frac{1}{1-u_2},1,1;1\right) \log u_1+2 G\left(\frac{1}{1-u_2},1,\frac{1}{1-u_2};1\right) \log u_1+G\left(\frac{1}{1-u_2},\frac{1}{1-u_2},0;1\right) \log u_1+2 G\left(\frac{1}{1-u_2},\frac{1}{1-u_2},1;1\right) \log u_1+G\left(\frac{1}{1-u_2},\frac{1}{1-u_2},\frac{1}{1-u_2};1\right) \log u_1-\gcal\left(0,v_{213},\frac{1}{1-u_2};1\right) \log u_1-\gcal\left(0,\frac{1}{1-u_2},v_{213};1\right) \log u_1-\gcal(v_{213},0,1;1) \log u_1-\gcal(v_{213},1,0;1) \log u_1+2 \gcal(v_{213},1,1;1) \log u_1-2 \gcal\left(v_{213},1,\frac{1}{1-u_2};1\right) \log u_1-2 \gcal\left(v_{213},\frac{1}{1-u_2},1;1\right) \log u_1-\gcal\left(\frac{1}{1-u_2},0,v_{213};1\right) \log u_1-\gcal\left(\frac{1}{1-u_2},v_{213},0;1\right) \log u_1-2 \gcal\left(\frac{1}{1-u_2},v_{213},\frac{1}{1-u_2};1\right) \log u_1-2 \gcal\left(\frac{1}{1-u_2},\frac{1}{1-u_2},v_{213};1\right) \log u_1-G\left(\frac{1}{1-u_2},0;1\right) \log u_3 \log u_1-G\left(\frac{1}{1-u_2},1;1\right) \log u_3 \log u_1-G\left(\frac{1}{1-u_2},\frac{1}{1-u_2};1\right) \log u_3 \log u_1+\gcal(v_{213},1;1) \log u_3 \log u_1+\gcal\left(\frac{1}{1-u_2},v_{213};1\right) \log u_3 \log u_1-2 G\left(\frac{1}{1-u_2};1\right) \log u_2 \log u_3 \log u_1+\frac{1}{6} \pi ^2 \log u_3 \log u_1-\zeta_3 \log u_1-\frac{1}{2} G\left(\frac{1}{1-u_2};1\right) \log ^3u_3+\frac{1}{2} G\left(\frac{1}{1-u_2},0;1\right) \log ^2u_3+\frac{1}{2} G\left(\frac{1}{1-u_2},1;1\right) \log ^2u_3+\frac{1}{2} G\left(\frac{1}{1-u_2},\frac{1}{1-u_2};1\right) \log ^2u_3-\frac{1}{2} \gcal(v_{213},1;1) \log ^2u_3-\frac{1}{2} \gcal\left(\frac{1}{1-u_2},v_{213};1\right) \log ^2u_3+G\left(\frac{1}{1-u_2};1\right) \log u_2 \log ^2u_3-\frac{1}{12} \pi ^2 \log ^2u_3+\frac{1}{2} \pi ^2 G\left(\frac{1}{1-u_2},0;1\right)+\frac{1}{2} \pi ^2 G\left(\frac{1}{1-u_2},1;1\right)+\frac{1}{2} \pi ^2 G\left(\frac{1}{1-u_2},\frac{1}{1-u_2};1\right)+G\left(0,0,1,\frac{1}{1-u_2};1\right)+G\left(0,0,\frac{1}{1-u_2},1;1\right)-2 G\left(0,0,\frac{1}{1-u_2},\frac{1}{1-u_2};1\right)+G\left(0,1,0,\frac{1}{1-u_2};1\right)-2 G\left(0,1,1,\frac{1}{1-u_2};1\right)+G\left(0,1,\frac{1}{1-u_2},0;1\right)-2 G\left(0,1,\frac{1}{1-u_2},1;1\right)+G\left(0,1,\frac{1}{1-u_2},\frac{1}{1-u_2};1\right)+G\left(0,\frac{1}{1-u_2},0,1;1\right)-2 G\left(0,\frac{1}{1-u_2},0,\frac{1}{1-u_2};1\right)+G\left(0,\frac{1}{1-u_2},1,0;1\right)-2 G\left(0,\frac{1}{1-u_2},1,1;1\right)-2 G\left(0,\frac{1}{1-u_2},\frac{1}{1-u_2},0;1\right)-G\left(0,\frac{1}{1-u_2},\frac{1}{1-u_2},1;1\right)-3 G\left(0,\frac{1}{1-u_2},\frac{1}{1-u_2},\frac{1}{1-u_2};1\right)-G\left(\frac{1}{1-u_2},0,0,\frac{1}{1-u_2};1\right)-G\left(\frac{1}{1-u_2},0,1,1;1\right)-G\left(\frac{1}{1-u_2},0,1,\frac{1}{1-u_2};1\right)-G\left(\frac{1}{1-u_2},0,\frac{1}{1-u_2},0;1\right)-2 G\left(\frac{1}{1-u_2},0,\frac{1}{1-u_2},1;1\right)-2 G\left(\frac{1}{1-u_2},0,\frac{1}{1-u_2},\frac{1}{1-u_2};1\right)+G\left(\frac{1}{1-u_2},1,0,0;1\right)-2 G\left(\frac{1}{1-u_2},1,0,1;1\right)+G\left(\frac{1}{1-u_2},1,0,\frac{1}{1-u_2};1\right)-2 G\left(\frac{1}{1-u_2},1,1,0;1\right)+3 G\left(\frac{1}{1-u_2},1,1,1;1\right)-3 G\left(\frac{1}{1-u_2},1,1,\frac{1}{1-u_2};1\right)+G\left(\frac{1}{1-u_2},1,\frac{1}{1-u_2},0;1\right)-3 G\left(\frac{1}{1-u_2},1,\frac{1}{1-u_2},1;1\right)+G\left(\frac{1}{1-u_2},1,\frac{1}{1-u_2},\frac{1}{1-u_2};1\right)-G\left(\frac{1}{1-u_2},\frac{1}{1-u_2},0,0;1\right)-2 G\left(\frac{1}{1-u_2},\frac{1}{1-u_2},0,1;1\right)-2 G\left(\frac{1}{1-u_2},\frac{1}{1-u_2},0,\frac{1}{1-u_2};1\right)-3 G\left(\frac{1}{1-u_2},\frac{1}{1-u_2},1,1;1\right)-G\left(\frac{1}{1-u_2},\frac{1}{1-u_2},1,\frac{1}{1-u_2};1\right)-2 G\left(\frac{1}{1-u_2},\frac{1}{1-u_2},\frac{1}{1-u_2},0;1\right)-3 G\left(\frac{1}{1-u_2},\frac{1}{1-u_2},\frac{1}{1-u_2},1;1\right)-3 G\left(\frac{1}{1-u_2},\frac{1}{1-u_2},\frac{1}{1-u_2},\frac{1}{1-u_2};1\right)-\frac{1}{2} \pi ^2 \gcal(v_{213},1;1)-\frac{1}{2} \pi ^2 \gcal\left(\frac{1}{1-u_2},v_{213};1\right)+\gcal\left(0,0,v_{213},\frac{1}{1-u_2};1\right)+\gcal\left(0,0,\frac{1}{1-u_2},v_{213};1\right)+\gcal\left(0,v_{213},1,\frac{1}{1-u_2};1\right)+\gcal\left(0,v_{213},\frac{1}{1-u_2},1;1\right)+\gcal\left(0,\frac{1}{1-u_2},0,v_{213};1\right)+\gcal\left(0,\frac{1}{1-u_2},v_{213},1;1\right)+\gcal\left(0,\frac{1}{1-u_2},v_{213},\frac{1}{1-u_2};1\right)+2 \gcal\left(0,\frac{1}{1-u_2},\frac{1}{1-u_2},v_{213};1\right)-\gcal(v_{213},0,0,1;1)-\gcal(v_{213},0,1,0;1)+2 \gcal(v_{213},0,1,1;1)-\gcal\left(v_{213},0,1,\frac{1}{1-u_2};1\right)-\gcal\left(v_{213},0,\frac{1}{1-u_2},1;1\right)-\gcal(v_{213},1,0,0;1)+2 \gcal(v_{213},1,0,1;1)-\gcal\left(v_{213},1,0,\frac{1}{1-u_2};1\right)+2 \gcal(v_{213},1,1,0;1)-3 \gcal(v_{213},1,1,1;1)+3 \gcal\left(v_{213},1,1,\frac{1}{1-u_2};1\right)-\gcal\left(v_{213},1,\frac{1}{1-u_2},0;1\right)+3 \gcal\left(v_{213},1,\frac{1}{1-u_2},1;1\right)-\gcal\left(v_{213},1,\frac{1}{1-u_2},\frac{1}{1-u_2};1\right)-\gcal\left(v_{213},\frac{1}{1-u_2},0,1;1\right)-\gcal\left(v_{213},\frac{1}{1-u_2},1,0;1\right)+3 \gcal\left(v_{213},\frac{1}{1-u_2},1,1;1\right)-\gcal\left(v_{213},\frac{1}{1-u_2},1,\frac{1}{1-u_2};1\right)-\gcal\left(v_{213},\frac{1}{1-u_2},\frac{1}{1-u_2},1;1\right)+\gcal\left(\frac{1}{1-u_2},0,0,v_{213};1\right)+\gcal\left(\frac{1}{1-u_2},0,v_{213},1;1\right)+\gcal\left(\frac{1}{1-u_2},0,v_{213},\frac{1}{1-u_2};1\right)+2 \gcal\left(\frac{1}{1-u_2},0,\frac{1}{1-u_2},v_{213};1\right)-\gcal\left(\frac{1}{1-u_2},v_{213},0,0;1\right)+\gcal\left(\frac{1}{1-u_2},v_{213},0,1;1\right)-\gcal\left(\frac{1}{1-u_2},v_{213},0,\frac{1}{1-u_2};1\right)+\gcal\left(\frac{1}{1-u_2},v_{213},1,0;1\right)+2 \gcal\left(\frac{1}{1-u_2},v_{213},1,\frac{1}{1-u_2};1\right)-\gcal\left(\frac{1}{1-u_2},v_{213},\frac{1}{1-u_2},0;1\right)+2 \gcal\left(\frac{1}{1-u_2},v_{213},\frac{1}{1-u_2},1;1\right)-\gcal\left(\frac{1}{1-u_2},v_{213},\frac{1}{1-u_2},\frac{1}{1-u_2};1\right)+2 \gcal\left(\frac{1}{1-u_2},\frac{1}{1-u_2},0,v_{213};1\right)+2 \gcal\left(\frac{1}{1-u_2},\frac{1}{1-u_2},v_{213},1;1\right)+\gcal\left(\frac{1}{1-u_2},\frac{1}{1-u_2},v_{213},\frac{1}{1-u_2};1\right)+3 \gcal\left(\frac{1}{1-u_2},\frac{1}{1-u_2},\frac{1}{1-u_2},v_{213};1\right)+H\left(0,0,0,1,1-u_2\right)-\frac{1}{6} \pi ^2 G\left(\frac{1}{1-u_2};1\right) \log u_2+2 G\left(0,\frac{1}{1-u_2},\frac{1}{1-u_2};1\right) \log u_2+2 G\left(\frac{1}{1-u_2},0,\frac{1}{1-u_2};1\right) \log u_2+G\left(\frac{1}{1-u_2},1,\frac{1}{1-u_2};1\right) \log u_2+2 G\left(\frac{1}{1-u_2},\frac{1}{1-u_2},0;1\right) \log u_2+2 G\left(\frac{1}{1-u_2},\frac{1}{1-u_2},1;1\right) \log u_2+3 G\left(\frac{1}{1-u_2},\frac{1}{1-u_2},\frac{1}{1-u_2};1\right) \log u_2-\gcal\left(0,v_{213},\frac{1}{1-u_2};1\right) \log u_2-\gcal\left(0,\frac{1}{1-u_2},v_{213};1\right) \log u_2-\gcal\left(v_{213},1,\frac{1}{1-u_2};1\right) \log u_2-\gcal\left(v_{213},\frac{1}{1-u_2},1;1\right) \log u_2-\gcal\left(\frac{1}{1-u_2},0,v_{213};1\right) \log u_2-\gcal\left(\frac{1}{1-u_2},v_{213},1;1\right) \log u_2-\gcal\left(\frac{1}{1-u_2},v_{213},\frac{1}{1-u_2};1\right) \log u_2-2 \gcal\left(\frac{1}{1-u_2},\frac{1}{1-u_2},v_{213};1\right) \log u_2-G\left(\frac{1}{1-u_2};1\right) \log ^2u_2 \log u_3-\frac{1}{3} \pi ^2 G\left(\frac{1}{1-u_2};1\right) \log u_3-G\left(0,1,\frac{1}{1-u_2};1\right) \log u_3-G\left(0,\frac{1}{1-u_2},1;1\right) \log u_3-G\left(\frac{1}{1-u_2},0,0;1\right) \log u_3-G\left(\frac{1}{1-u_2},0,\frac{1}{1-u_2};1\right) \log u_3-G\left(\frac{1}{1-u_2},1,0;1\right) \log u_3+2 G\left(\frac{1}{1-u_2},1,1;1\right) \log u_3-2 G\left(\frac{1}{1-u_2},1,\frac{1}{1-u_2};1\right) \log u_3-G\left(\frac{1}{1-u_2},\frac{1}{1-u_2},0;1\right) \log u_3-2 G\left(\frac{1}{1-u_2},\frac{1}{1-u_2},1;1\right) \log u_3-G\left(\frac{1}{1-u_2},\frac{1}{1-u_2},\frac{1}{1-u_2};1\right) \log u_3+\gcal\left(0,v_{213},\frac{1}{1-u_2};1\right) \log u_3+\gcal\left(0,\frac{1}{1-u_2},v_{213};1\right) \log u_3+\gcal(v_{213},0,1;1) \log u_3+\gcal(v_{213},1,0;1) \log u_3-2 \gcal(v_{213},1,1;1) \log u_3+2 \gcal\left(v_{213},1,\frac{1}{1-u_2};1\right) \log u_3+2 \gcal\left(v_{213},\frac{1}{1-u_2},1;1\right) \log u_3+\gcal\left(\frac{1}{1-u_2},0,v_{213};1\right) \log u_3+\gcal\left(\frac{1}{1-u_2},v_{213},0;1\right) \log u_3+2 \gcal\left(\frac{1}{1-u_2},v_{213},\frac{1}{1-u_2};1\right) \log u_3+2 \gcal\left(\frac{1}{1-u_2},\frac{1}{1-u_2},v_{213};1\right) \log u_3+\log u_3 \zeta_3-\frac{11 \pi ^4}{90}\,.\)
\etxtsloppy

We now turn to the evaluation of $\tilde R_{-1-z_1}^{(2)}(u_1,u_2,u_3)$. The computation follows the same lines as for $\tilde R_{-1-z_1}^{(1)}(u_1,u_2,u_3)$, with a slight complication coming from the denominator in the integrand. To get rid of the denominator, we rewrite $\tilde R_{-1-z_1}^{(2)}(u_1,u_2,u_3)$ as the integral of the derivative with respect to $u_2$,
\beq\label{eq:Rz12}
\tilde R_{-1-z_1}^{(2)}(u_1,u_2,u_3) = \tilde R_{-1-z_1}^{(2)}(u_1,1,u_3) + \int_1^{u_2}\rd u\,{\partial\over\partial u}\,\tilde R_{-1-z_1}^{(2)}(u_1,u,u_3)\,.
\eeq
Let us start with the first term in \Eqn{eq:Rz12}.
The value for $u_2=1$ is easily obtained by applying Barnes lemmas, by means of which one of the two integrations can be performed. This leaves us with a onefold integral trivial to compute,
\beq\bsp\label{eq:triv}
\tilde R_{-1-z_1}^{(2)}&(u_1,1,u_3) = \int_{+1/5-i\infty}^{+1/5+i\infty}{\rd z_2\over 2\pi i}\,
z_2^2\, u_1^{-z_2}\, u_3^{z_2}\, \Gamma \left(-z_2\right)^3\, \Gamma
   \left(z_2\right)^3\, \left(\psi\left(-z_2\right)+\psi\left(z_2\right)+2 \gamma \right)\\
   =\,&-\frac{1}{6} \log ^3u_1 H\left(1;-{u_3\over u_1}\right)+\frac{1}{6} \log ^3u_3 H\left(1;-{u_3\over u_1}\right)+\log ^2u_1 H\left(1,1;-{u_3\over u_1}\right)\\
   &+\frac{1}{2} \log u_3 \log ^2u_1 H\left(1;-{u_3\over u_1}\right)-\frac{1}{2} \log ^2u_3 \log u_1 H\left(1;-{u_3\over u_1}\right)+\log ^2u_3 H\left(1,1;-{u_3\over u_1}\right)\\
   &-\frac{1}{2} \pi ^2 \log u_1 H\left(1;-{u_3\over u_1}\right)+\log u_1 H\left(0,0,1;-{u_3\over u_1}\right)+2 \log u_1 H\left(0,1,1;-{u_3\over u_1}\right)\\
   &+2 \log u_1 H\left(1,0,1;-{u_3\over u_1}\right)-2 \log u_3 \log u_1 H\left(1,1;-{u_3\over u_1}\right)+\frac{1}{2} \pi ^2 \log u_3 H\left(1;-{u_3\over u_1}\right)\\
   &-\log u_3 H\left(0,0,1;-{u_3\over u_1}\right)-2 \log u_3 H\left(0,1,1;-{u_3\over u_1}\right)-2 \log u_3 H\left(1,0,1;-{u_3\over u_1}\right)\\
   &-2 \zeta_3 H\left(1;-{u_3\over u_1}\right)+\pi ^2 H\left(1,1;-{u_3\over u_1}\right)+2 H\left(0,0,0,1;-{u_3\over u_1}\right)\\
   &+2 H\left(0,0,1,1;-{u_3\over u_1}\right)+2 H\left(0,1,0,1;-{u_3\over u_1}\right)+2 H\left(1,0,0,1;-{u_3\over u_1}\right)\,.
   \esp\eeq
The remaining term is again computed by exchanging one of the MB integrations with an Euler integration. Shifting the contours such that they satisfy the assumptions underlying \Eqn{eq:MBtoEuler} introduces an additional residue of the form,
\beq\bsp
\int_1^{u_2}&\rd u\,{\partial\over\partial u}\,\res_{z_1=-1}\,\rcal_{-1-z_1}^{(2)}(u_1,u,u_3;z_1,z_2)\\
 =&\, \int_1^{u_2}\rd u\,{\partial\over\partial u}\,
\int_{+1/5-i\infty}^{+1/5+i\infty}{\rd z_2\over 2\pi i}\,
z_2^2\, u_1^{-z_2}\, u_3^{z_2} \Gamma \left(-z_2\right)^3\, \Gamma
   \left(z_2\right)^3\,u^{-1}\\
      =&\,\int_{+1/5-i\infty}^{+1/5+i\infty}{\rd z_2\over 2\pi i}\,
z_2^2\, u_1^{-z_2}\, u_3^{z_2} \Gamma \left(-z_2\right)^3\, \Gamma
   \left(z_2\right)^3\,\ln u_2\\
      =&\,\frac{1}{2} \log u_2 \log ^2u_1 H\left(1;-{u_3\over u_1}\right)+\frac{1}{2} \log u_2 \log ^2u_3 H\left(1;-{u_3\over u_1}\right)+\log u_2 \log u_1 H\left(0,1;-{u_3\over u_1}\right)\\
      &-\log u_2 \log u_3 \log u_1 H\left(1;-{u_3\over u_1}\right)+\frac{1}{2} \pi ^2 \log u_2 H\left(1;-{u_3\over u_1}\right)+\log u_2 H\left(0,0,1;-{u_3\over u_1}\right)\\
      &-\log u_2 \log u_3 H\left(0,1;-{u_3\over u_1}\right)\,.
\esp\eeq
After inserting an Euler integral for the integral over $z_1$ in the second term of \Eqn{eq:Rz12} and exchanging the Euler integration and the integration over $z_2$, the MB integral is trivial and can be performed by closing the contour and summing up residues,
\beq\bsp
\int_0^1&\rd v\int_1^{u_2}\rd u\,{u_3\over2}\,\left(u v-v+1\right)^{-1} \left(-u_1 v^2+u_1 u v^2+u_1 v-u_3 v+u_3\right)^{-1}\\
&\times\Bigg\{\log ^2\left(\left(u-1\right) v+1\right)-2 \log u_1 \log (1-v)-2 \log (1-v) \log \left(\left(u-1\right) v+1\right)\\
&+2 \log u_3 \log (1-v)+2 \log u_1 \log v+2 \log v \log \left(\left(u-1\right) v+1\right)+2 \log u_1 \log \left(\left(u-1\right) v+1\right)\\
&-2 \log u_3 \log v-2 \log u_3 \log \left(\left(u-1\right) v+1\right)+\log ^2u_1+\log ^2u_3-2 \log u_1 \log u_3+\log ^2(1-v)\\
&+\log ^2v-2 \log v \log (1-v)+\pi ^2\Bigg\}\,,
\esp\eeq
and the integration over $u$ can be done easily in terms of ordinary polylogarithms, yielding,
\btxtsloppy
\parbox{130mm}{\raggedright\(\displaystyle
\int_0^1\rd v\,{1\over 6}\,v^{-1}\,(1-v)^{-1}\,\Bigg\{
\log ^3\left(v \left(u_2-1\right)+1\right)-3 \log (1-v) \log ^2\left(v \left(u_2-1\right)+1\right)+\)}
\raggedleft\refstepcounter{equation}(\theequation)\label{eq:vint}\\
\raggedright\(\displaystyle
3 \log v \log ^2\left(v \left(u_2-1\right)+1\right) +3 \log u_1 \log ^2\left(v \left(u_2-1\right)+1\right)-3 \log u_3 \log ^2\left(v \left(u_2-1\right)+1\right) -3 \log \left(-\frac{u_1 \left(u_2-1\right) v^2+\left(u_1-u_3\right) v+u_3}{(v-1) u_3}\right) \log ^2\left(v \left(u_2-1\right)+1\right) +3 {\log ^2(1-v)} \log \left(v \left(u_2-1\right)+1\right) +3 \log ^2v \log \left(v \left(u_2-1\right)+1\right) +3 \log ^2u_1 \log \left(v \left(u_2-1\right)+1\right) +3 \log ^2u_3 \log \left(v \left(u_2-1\right)+1\right) -6 \log (1-v) \log v \log \left(v \left(u_2-1\right)+1\right) -6 {\log (1-v) \log u_1} \log \left(v \left(u_2-1\right)+1\right) +6 \log v \log u_1 \log \left(v \left(u_2-1\right)+1\right) +6 {\log (1-v) \log u_3} \log \left(v \left(u_2-1\right)+1\right) -6 \log v \log u_3 \log \left(v \left(u_2-1\right)+1\right) -6 \log u_1 \log u_3 \log \left(v \left(u_2-1\right)+1\right) +6 {\log (1-v) \log \left(-\frac{u_1 \left(u_2-1\right) v^2+\left(u_1-u_3\right) v+u_3}{(v-1) u_3}\right) \log \left(v \left(u_2-1\right)+1\right) }-6 \log v \log \left(-\frac{u_1 \left(u_2-1\right) v^2+\left(u_1-u_3\right) v+u_3}{(v-1) u_3}\right) \log \left(v \left(u_2-1\right)+1\right) -6 \log u_1 \log \left(-\frac{u_1 \left(u_2-1\right) v^2+\left(u_1-u_3\right) v+u_3}{(v-1) u_3}\right) \log \left(v \left(u_2-1\right)+1\right) +6 \log u_3 \log \left(-\frac{u_1 \left(u_2-1\right) v^2+\left(u_1-u_3\right) v+u_3}{(v-1) u_3}\right) \log \left(v \left(u_2-1\right)+1\right) -6 \text{Li}_2\left(\frac{v u_1 \left(v \left(u_2-1\right)+1\right)}{(v-1) u_3}\right) \log \left(v \left(u_2-1\right)+1\right) +3 \pi ^2 \log \left(v \left(u_2-1\right)+1\right) -3 {\log ^2(1-v) \log \left(u_1 \left(u_2-1\right) v^2+\left(u_1-u_3\right) v+u_3\right)} -3 \log ^2v \log \left(u_1 \left(u_2-1\right) v^2+\left(u_1-u_3\right) v+u_3\right) -3 \log ^2u_1 \log \left(u_1 \left(u_2-1\right) v^2+\left(u_1-u_3\right) v+u_3\right) -3 \log ^2u_3 \log \left(u_1 \left(u_2-1\right) v^2+\left(u_1-u_3\right) v+u_3\right) +6 {\log (1-v) \log v \log \left(u_1 \left(u_2-1\right) v^2+\left(u_1-u_3\right) v+u_3\right)} +6 {\log (1-v) \log u_1 \log \left(u_1 \left(u_2-1\right) v^2+\left(u_1-u_3\right) v+u_3\right)} -6 \log v \log u_1 \log \left(u_1 \left(u_2-1\right) v^2+\left(u_1-u_3\right) v+u_3\right)-6 {\log (1-v) \log u_3 \log \left(u_1 \left(u_2-1\right) v^2+\left(u_1-u_3\right) v+u_3\right)} +6 \log v \log u_3 \log \left(u_1 \left(u_2-1\right) v^2+\left(u_1-u_3\right) v+u_3\right) +6 \log u_1 \log u_3 \log \left(u_1 \left(u_2-1\right) v^2+\left(u_1-u_3\right) v+u_3\right) -3 \pi ^2 \log \left(u_1 \left(u_2-1\right) v^2+\left(u_1-u_3\right) v+u_3\right) +3 \log ^2(1-v) \log \left(v u_1-v u_3+u_3\right) +3 \log ^2v \log \left(v u_1-v u_3+u_3\right) +3 \log ^2u_1 \log \left(v u_1-v u_3+u_3\right)  +3 \log ^2u_3 \log \left(v u_1-v u_3+u_3\right) -6 \log (1-v) \log v \log \left(v u_1-v u_3+u_3\right) -6 \log (1-v) \log u_1 \log \left(v u_1-v u_3+u_3\right) +6 \log v \log u_1 \log \left(v u_1-v u_3+u_3\right) +6 \log (1-v) \log u_3 \log \left(v u_1-v u_3+u_3\right) -6 \log v \log u_3 \log \left(v u_1-v u_3+u_3\right)-6 \log u_1 \log u_3 \log \left(v u_1-v u_3+u_3\right) +3 \pi ^2 \log \left(v u_1-v u_3+u_3\right)-6 \log (1-v) \text{Li}_2\left(\frac{v u_1}{(v-1) u_3}\right) +6 \log v \text{Li}_2\left(\frac{v u_1}{(v-1) u_3}\right) +6 \log u_1 \text{Li}_2\left(\frac{v u_1}{(v-1) u_3}\right) -6 \log u_3 \text{Li}_2\left(\frac{v u_1}{(v-1) u_3}\right) +6 \log (1-v) \text{Li}_2\left(\frac{v u_1 \left(v \left(u_2-1\right)+1\right)}{(v-1) u_3}\right)-6 \log v \text{Li}_2\left(\frac{v u_1 \left(v \left(u_2-1\right)+1\right)}{(v-1) u_3}\right) -6 \log u_1 \text{Li}_2\left(\frac{v u_1 \left(v \left(u_2-1\right)+1\right)}{(v-1) u_3}\right) +6 \log u_3 \text{Li}_2\left(\frac{v u_1 \left(v \left(u_2-1\right)+1\right)}{(v-1) u_3}\right) -6 \text{Li}_3\left(\frac{v u_1}{(v-1) u_3}\right) +6 \text{Li}_3\left(\frac{v u_1 \left(v \left(u_2-1\right)+1\right)}{(v-1) u_3}\right)
\Bigg\}\,.\)
\etxtsloppy
We are thus left with only the integration over $v$ to be done. We proceed in the by now usual way by converting all the polylogarithms in \Eqn{eq:vint} into Goncharov polylogarithms using the algorithm of App.~\ref{app:Goncharov} and then perform the integration over $v$ using the recursive definition of the $G$-function. At the end of this procedure we find,
\btxtsloppy
\parbox{130mm}{\raggedright\(\displaystyle
\frac{1}{2} G\left(0,\frac{1}{1-u_2};1\right) \log ^2u_1+\frac{1}{2} G\left(0,\frac{1}{1-\frac{u_1}{u_3}};1\right) \log ^2u_1+\)}
\raggedleft\refstepcounter{equation}(\theequation)\label{eq:triv2}\\
\raggedright\(\displaystyle
\frac{1}{2} G\left(\frac{1}{1-u_2},1;1\right) \log ^2u_1+
\frac{1}{2} G\left(-\frac{u_3}{u_1-u_3},1;1\right) \log ^2u_1-\frac{1}{2} \gcal(0,v_{213};1) \log ^2u_1-\frac{1}{2} \gcal(v_{213},1;1) \log ^2u_1+G\left(0,0,\frac{1}{1-u_2};1\right) \log u_1+G\left(0,1,\frac{1}{1-u_2};1\right) \log u_1+G\left(0,\frac{1}{1-u_2},0;1\right) \log u_1+G\left(0,\frac{1}{1-u_2},\frac{1}{1-u_2};1\right) \log u_1+G\left(0,\frac{1}{1-\frac{u_1}{u_3}},0;1\right) \log u_1-G\left(0,\frac{1}{1-\frac{u_1}{u_3}},1;1\right) \log u_1+G\left(\frac{1}{1-u_2},0,1;1\right) \log u_1+G\left(\frac{1}{1-u_2},1,0;1\right) \log u_1-2 G\left(\frac{1}{1-u_2},1,1;1\right) \log u_1+G\left(\frac{1}{1-u_2},1,\frac{1}{1-u_2};1\right) \log u_1+G\left(\frac{1}{1-u_2},\frac{1}{1-u_2},1;1\right) \log u_1+G\left(\frac{1}{1-\frac{u_1}{u_3}},0,1;1\right) \log u_1+G\left(\frac{1}{1-\frac{u_1}{u_3}},1,0;1\right) \log u_1-2 G\left(\frac{1}{1-\frac{u_1}{u_3}},1,1;1\right) \log u_1-\gcal(0,v_{213},0;1) \log u_1+\gcal(0,v_{213},1;1) \log u_1-\gcal\left(0,v_{213},\frac{1}{1-u_2};1\right) \log u_1-\gcal(v_{213},0,1;1) \log u_1-\gcal(v_{213},1,0;1) \log u_1+2 \gcal(v_{213},1,1;1) \log u_1-\gcal\left(v_{213},1,\frac{1}{1-u_2};1\right) \log u_1-\gcal\left(v_{213},\frac{1}{1-u_2},1;1\right) \log u_1-G\left(0,\frac{1}{1-u_2};1\right) \log u_3 \log u_1-G\left(0,\frac{1}{1-\frac{u_1}{u_3}};1\right) \log u_3 \log u_1-G\left(\frac{1}{1-u_2},1;1\right) \log u_3 \log u_1-G\left(\frac{1}{1-\frac{u_1}{u_3}},1;1\right) \log u_3 \log u_1+\gcal(0,v_{213};1) \log u_3 \log u_1+\gcal(v_{213},1;1) \log u_3 \log u_1+\frac{1}{2} G\left(0,\frac{1}{1-u_2};1\right) \log ^2u_3+\frac{1}{2} G\left(0,\frac{1}{1-\frac{u_1}{u_3}};1\right) \log ^2u_3+\frac{1}{2} G\left(\frac{1}{1-u_2},1;1\right) \log ^2u_3+\frac{1}{2} G\left(-\frac{u_3}{u_1-u_3},1;1\right) \log ^2u_3-\frac{1}{2} \gcal(0,v_{213};1) \log ^2u_3-\frac{1}{2} \gcal(v_{213},1;1) \log ^2u_3+\frac{1}{2} \pi ^2 G\left(0,\frac{1}{1-u_2};1\right)+\frac{1}{2} \pi ^2 G\left(0,\frac{1}{1-\frac{u_1}{u_3}};1\right)+\frac{1}{2} \pi ^2 G\left(\frac{1}{1-u_2},1;1\right)+\frac{1}{2} \pi ^2 G\left(-\frac{u_3}{u_1-u_3},1;1\right)+G\left(0,0,0,\frac{1}{1-u_2};1\right)+G\left(0,0,\frac{1}{1-u_2},0;1\right)+G\left(0,0,\frac{1}{1-u_2},\frac{1}{1-u_2};1\right)+G\left(0,1,0,\frac{1}{1-u_2};1\right)-2 G\left(0,1,1,\frac{1}{1-u_2};1\right)+G\left(0,1,\frac{1}{1-u_2},0;1\right)-2 G\left(0,1,\frac{1}{1-u_2},1;1\right)+G\left(0,1,\frac{1}{1-u_2},\frac{1}{1-u_2};1\right)+G\left(0,\frac{1}{1-u_2},0,0;1\right)+G\left(0,\frac{1}{1-u_2},0,\frac{1}{1-u_2};1\right)-G\left(0,\frac{1}{1-u_2},1,1;1\right)+G\left(0,\frac{1}{1-u_2},\frac{1}{1-u_2},0;1\right)+G\left(0,\frac{1}{1-u_2},\frac{1}{1-u_2},\frac{1}{1-u_2};1\right)+G\left(0,\frac{1}{1-\frac{u_1}{u_3}},0,0;1\right)-G\left(0,\frac{1}{1-\frac{u_1}{u_3}},0,1;1\right)-G\left(0,\frac{1}{1-\frac{u_1}{u_3}},1,0;1\right)+G\left(0,\frac{1}{1-\frac{u_1}{u_3}},1,1;1\right)+G\left(\frac{1}{1-u_2},0,0,1;1\right)+G\left(\frac{1}{1-u_2},0,1,0;1\right)-2 G\left(\frac{1}{1-u_2},0,1,1;1\right)+G\left(\frac{1}{1-u_2},0,1,\frac{1}{1-u_2};1\right)+G\left(\frac{1}{1-u_2},0,\frac{1}{1-u_2},1;1\right)+G\left(\frac{1}{1-u_2},1,0,0;1\right)-2 G\left(\frac{1}{1-u_2},1,0,1;1\right)+G\left(\frac{1}{1-u_2},1,0,\frac{1}{1-u_2};1\right)-2 G\left(\frac{1}{1-u_2},1,1,0;1\right)+3 G\left(\frac{1}{1-u_2},1,1,1;1\right)-2 G\left(\frac{1}{1-u_2},1,1,\frac{1}{1-u_2};1\right)+G\left(\frac{1}{1-u_2},1,\frac{1}{1-u_2},0;1\right)-2 G\left(\frac{1}{1-u_2},1,\frac{1}{1-u_2},1;1\right)+G\left(\frac{1}{1-u_2},1,\frac{1}{1-u_2},\frac{1}{1-u_2};1\right)+G\left(\frac{1}{1-u_2},\frac{1}{1-u_2},0,1;1\right)+G\left(\frac{1}{1-u_2},\frac{1}{1-u_2},1,0;1\right)-2 G\left(\frac{1}{1-u_2},\frac{1}{1-u_2},1,1;1\right)+G\left(\frac{1}{1-u_2},\frac{1}{1-u_2},1,\frac{1}{1-u_2};1\right)+G\left(\frac{1}{1-u_2},\frac{1}{1-u_2},\frac{1}{1-u_2},1;1\right)+G\left(\frac{1}{1-\frac{u_1}{u_3}},0,0,1;1\right)+G\left(\frac{1}{1-\frac{u_1}{u_3}},0,1,0;1\right)-2 G\left(\frac{1}{1-\frac{u_1}{u_3}},0,1,1;1\right)+G\left(\frac{1}{1-\frac{u_1}{u_3}},1,0,0;1\right)-2 G\left(\frac{1}{1-\frac{u_1}{u_3}},1,0,1;1\right)-2 G\left(\frac{1}{1-\frac{u_1}{u_3}},1,1,0;1\right)+3 G\left(\frac{1}{1-\frac{u_1}{u_3}},1,1,1;1\right)-\frac{1}{2} \pi ^2 \gcal(0,v_{213};1)-\frac{1}{2} \pi ^2 \gcal(v_{213},1;1)-\gcal(0,v_{213},0,0;1)+\gcal(0,v_{213},0,1;1)-\gcal\left(0,v_{213},0,\frac{1}{1-u_2};1\right)+\gcal(0,v_{213},1,0;1)-\gcal(0,v_{213},1,1;1)+\gcal\left(0,v_{213},1,\frac{1}{1-u_2};1\right)-\gcal\left(0,v_{213},\frac{1}{1-u_2},0;1\right)+\gcal\left(0,v_{213},\frac{1}{1-u_2},1;1\right)-\gcal\left(0,v_{213},\frac{1}{1-u_2},\frac{1}{1-u_2};1\right)-\gcal(v_{213},0,0,1;1)-\gcal(v_{213},0,1,0;1)+2 \gcal(v_{213},0,1,1;1)-\gcal\left(v_{213},0,1,\frac{1}{1-u_2};1\right)-\gcal\left(v_{213},0,\frac{1}{1-u_2},1;1\right)-\gcal(v_{213},1,0,0;1)+2 \gcal(v_{213},1,0,1;1)-\gcal\left(v_{213},1,0,\frac{1}{1-u_2};1\right)+2 \gcal(v_{213},1,1,0;1)-3 \gcal(v_{213},1,1,1;1)+2 \gcal\left(v_{213},1,1,\frac{1}{1-u_2};1\right)-\gcal\left(v_{213},1,\frac{1}{1-u_2},0;1\right)+2 \gcal\left(v_{213},1,\frac{1}{1-u_2},1;1\right)-\gcal\left(v_{213},1,\frac{1}{1-u_2},\frac{1}{1-u_2};1\right)-\gcal\left(v_{213},\frac{1}{1-u_2},0,1;1\right)-\gcal\left(v_{213},\frac{1}{1-u_2},1,0;1\right)+2 \gcal\left(v_{213},\frac{1}{1-u_2},1,1;1\right)-\gcal\left(v_{213},\frac{1}{1-u_2},1,\frac{1}{1-u_2};1\right)-\gcal\left(v_{213},\frac{1}{1-u_2},\frac{1}{1-u_2},1;1\right)-G\left(0,0,\frac{1}{1-u_2};1\right) \log u_3-G\left(0,1,\frac{1}{1-u_2};1\right) \log u_3-G\left(0,\frac{1}{1-u_2},0;1\right) \log u_3-G\left(0,\frac{1}{1-u_2},\frac{1}{1-u_2};1\right) \log u_3-G\left(0,\frac{1}{1-\frac{u_1}{u_3}},0;1\right) \log u_3+G\left(0,\frac{1}{1-\frac{u_1}{u_3}},1;1\right) \log u_3-G\left(\frac{1}{1-u_2},0,1;1\right) \log u_3-G\left(\frac{1}{1-u_2},1,0;1\right) \log u_3+2 G\left(\frac{1}{1-u_2},1,1;1\right) \log u_3-G\left(\frac{1}{1-u_2},1,\frac{1}{1-u_2};1\right) \log u_3-G\left(\frac{1}{1-u_2},\frac{1}{1-u_2},1;1\right) \log u_3-G\left(\frac{1}{1-\frac{u_1}{u_3}},0,1;1\right) \log u_3-G\left(\frac{1}{1-\frac{u_1}{u_3}},1,0;1\right) \log u_3+2 G\left(\frac{1}{1-\frac{u_1}{u_3}},1,1;1\right) \log u_3+\gcal(0,v_{213},0;1) \log u_3-\gcal(0,v_{213},1;1) \log u_3+\gcal\left(0,v_{213},\frac{1}{1-u_2};1\right) \log u_3+\gcal(v_{213},0,1;1) \log u_3+\gcal(v_{213},1,0;1) \log u_3-2 \gcal(v_{213},1,1;1) \log u_3+\gcal\left(v_{213},1,\frac{1}{1-u_2};1\right) \log u_3+\gcal\left(v_{213},\frac{1}{1-u_2},1;1\right) \log u_3
\,.\)
\etxtsloppy


\section{Limits of the remainder function}
\label{sec:limits}

\subsection{Multi-Regge limits}
In this section we discuss the behaviour of the remainder function in the multi-Regge kinematics. These are defined by requiring the final-state gluons to be strongly ordered in rapidity while having comparable transverse momentum. If we choose the gluons 3 and 4 as incoming, this implies the hierarchy of scales,
\beq\label{eq:MRK}
-s_{34}\gg -s_{234},\, -s_{345} \gg -s_{56},\,-s_{61},\,-s_{12} \gg -s_{45},\, -s_{456},\,-s_{23}\,.
\eeq
Introducing a parameter $\lambda\ll 1$, the above hierarchy is equivalent to a rescaling,
\beq
\{s_{234}, s_{345}\} = \ord(\lambda)\,,\qquad
\{s_{56}, s_{61}, s_{12}\} = \ord(\lambda^2)\,,\qquad
\{s_{45}, s_{456}, s_{23}\} = \ord(\lambda^3)\,,
\eeq
whereas $s_{34}$ is $\ord(1)$. In this limit all three conformal cross ratios take limiting 
values~\cite{DelDuca:2008jg,Brower:2008nm, Brower:2008ia},
\beq
u_1 = \ord(\lambda),\qquad u_2 = \ord(\lambda), \qquad u_3 = 1+\ord(\lambda),
\label{eq:mrlimit}
\eeq
and it was shown that in the Euclidean region the remainder function must vanish.

As a consistency check of our computation, we computed the leading behaviour of the remainder function in the multi-Regge limit. Apart from
$f_H (p_1, p_3, p_5; p_4, p_6, p_2)$, the sums of all the other terms in \Eqn{eq:w62l} is expressed in terms of harmonic polylogarithms, thus
we can directly expand the harmonic polylogarithms in the scaling parameter $\lambda$ and only keep the leading term. However,
$f_H (p_1, p_3, p_5; p_4, p_6, p_2)$ is expressed in terms of Goncharov polylogarithms whose arguments are complicated functions of the conformal cross ratios. It is therefore easier to compute $f_H (p_1, p_3, p_5; p_4, p_6, p_2)$ from scratch in the limit under consideration. In what follows, this technique is described on the example of the threefold contribution to $f_H (p_1, p_3, p_5; p_4, p_6, p_2)$ presented in \Eqn{eq:3foldH4}.

Defining a quantity $\bar u_3$ by $u_3=1+\bar u_3$, we can reformulate the problem as finding the leading behaviour in the limit $\lambda\to0$ of the integral
\beq\bsp\label{eq:3foldH4ubar}
 -{1\over4}&\,\mbint\mbint\mbint{\rd z_1\over 2\pi i}\,{\rd z_2\over 2\pi i}\,{\rd z_3\over 2\pi i}\,(z_1\,z_2+z_2\,z_3+z_3\,z_1)\,\lambda^{z_1+z_2}\, u_1^{z_1}\, u_2^{z_2}\, (1+\lambda\,\bar u_3)^{z_3}\\
 &\times \Gamma \left(-z_1\right)^2\, \Gamma
   \left(-z_2\right)^2\, \Gamma \left(-z_3\right)^2\, \Gamma
   \left(z_1+z_2\right)\, \Gamma \left(z_2+z_3\right)\, \Gamma
   \left(z_3+z_1\right)\,.
  \esp \eeq
The code {\tt MBasymptotics} allows us to extract the leading behaviour for $\lambda\to0$ of MB integrals of the form
\beq
\mbint\ldots\mbint\left(\prod_i {\rd z_i\over 2\pi i}\right) \lambda^{P(z_i)}\,f(z_i)\,,
\eeq
where $P(z_i)$ is a polynomial in the integration variables, and $f(z_i)$ denotes a function independent of $\lambda$. However, the integral in \Eqn{eq:3foldH4ubar} violates this form. We can bring \Eqn{eq:3foldH4ubar} into the desired form to the price of introducing an additional MB integral by applying \Eqn{MB} to the term $(1+\lambda \bar u_3)^{z_3}$. \Eqn{eq:3foldH4ubar} now takes the form,
\beq\bsp\label{eq:3foldH4ubar2}
 -{1\over4}&\,\mbint\mbint\mbint\mbint{\rd z_1\over 2\pi i}\,{\rd z_2\over 2\pi i}\,{\rd z_3\over 2\pi i}\,{\rd z\over 2\pi i}\,
 \left(z_2 z_3+z_1 z_2+z_1z_3\right) \, \lambda^{z_1+z_2+z}\,u_1^{z_1}
   u_2^{z_2}\, \bar{u}_3^z\\
&   \times \Gamma (-z)\, \Gamma \left(-z_1\right)^2\, \Gamma
   \left(-z_2\right)^2\, \Gamma \left(z_1+z_2\right)\, \Gamma
   \left(z-z_3\right)\, \Gamma \left(-z_3\right)\, \Gamma \left(z_1+z_3\right)\,
   \Gamma \left(z_2+z_3\right)\\
         -{1\over4}&\,\mbint\mbint\mbint{\rd z_1\over 2\pi i}\,{\rd z_2\over 2\pi i}\,{\rd z_3\over 2\pi i}\,
   \left(z_2 z_3+z_1z_2+z_1z_3\right)\, \lambda^{z_1+z_2}\, u_1^{z_1}\,
   u_2^{z_2}\\
  &\times \Gamma \left(-z_1\right)^2\,
   \Gamma \left(-z_2\right)^2\, \Gamma \left(z_1+z_2\right)\, \Gamma
   \left(-z_3\right)^2\, \Gamma \left(z_1+z_3\right)\, \Gamma
   \left(z_2+z_3\right)\\
      -{1\over4}&\,\mbint\mbint\mbint{\rd z_1\over 2\pi i}\,{\rd z_2\over 2\pi i}\,{\rd z_3\over 2\pi i}\,
   \left(z_2 z_3+z_1z_2+z_1z_3\right)\, \lambda^{z_1+z_2+1}\, u_1^{z_1}\,
   u_2^{z_2}\,\bar{u}_3 z_3\\
&   \times \Gamma \left(-z_1\right)^2\,
   \Gamma \left(-z_2\right)^2\, \Gamma \left(z_1+z_2\right)\, \Gamma
   \left(-z_3\right)^2\, \Gamma \left(z_1+z_3\right)\, \Gamma
   \left(z_2+z_3\right)\,,
   \esp\eeq
where the contours for the integrations over $z_i$, $i=1,2,3$, are given in \Eqn{eq:contours} and the contour for the integral over $z$ is a straight vertical line with $\re(z)=+{3\over2}$. The threefold contributions arise when shifting the $z$ contour from the form required by \Eqn{MB} to a straight line. All the integrals in \Eqn{eq:3foldH4ubar2} match precisely the form required by {\tt MBasymptotics}, and we find,
\beq\bsp
&-\frac{1}{4} \zeta_3 \log (u_1u_2)
  -\frac{1}{192} \log ^4\frac{u_1}{u_2}
   -\frac{5}{96} \pi ^2 \log^2\frac{u_1}{u_2}
  -\frac{137 \pi ^4}{2880}\\
&+\frac{1}{4}\int_{+1/3-i\infty}^{+1/3+i\infty}{\rd z_1\over 2\pi i}\, z_1^2\, \left(u_1^{z_1}\,u_2^{-z_1}+u_1^{-z_1}\,u_2^{z_1}\right)\, \Gamma \left(-z_1\right)^4\, \Gamma \left(z_1\right)^2\,
   \Gamma \left(2 z_1\right)\\
&-\frac{1}{24}\int_{-1/5-i\infty}^{-1/5+i\infty} {\rd z_3\over 2\pi i}\,z_3\, \Gamma \left(-z_3\right)^2\, \Gamma \left(z_3\right)^2\,
   \Big(6 \left(\log u_1+\log u_2 +2 \eug
   \right) \psi \left(z_3\right)+6 \eug  \log u_2\\
   &\qquad\qquad+6
   \log u_1 \left(\log u_2+\eug \right)+6 \psi
   \left(z_3\right)^2+\pi ^2+6 \eug ^2\Big)\,.
   \esp\eeq
The integral over $z_3$ can be evaluated in terms of harmonic polylogarithms in the usual way by closing the contour to the right and summing up residues. The integral over $z_1$ is more special, because it involves poles in half-integer values of the $\Gamma$ function. Summing up the tower of residues leads to multiple binomial sums~\cite{Jegerlehner:2002em, Kalmykov:2007dk}. However, we observe that this contribution cancels against similar contributions coming from the twofold contributions to $f_H (p_1, p_3, p_5; p_4, p_6, p_2)$, so we do not discuss this issue further. Finally, combining all the contributions, \Eqn{eq:w62l}, we find that the remainder function vanishes in the multi-Regge limit~(\ref{eq:MRK}),
\beq
\lim_{\lambda\to 0}R_6^{(2)}(\lambda\,u_1, \lambda\,u_2, 1+\lambda\,\bar u_3) = 0\,.
\eeq
Note that there are five more ways in which we could have defined the multi-Regge limit, corresponding to the five cyclic permutations of the indices in \Eqn{eq:MRK}. It is clear that in terms of conformal cross ratios the six different ways of taking the multi-Regge limit are equivalent pairwise, and we have checked that our expression satisfies
\beq
\lim_{\lambda\to 0}R_6^{(2)}(1+\lambda\,\bar u_1, \lambda\,u_2, \lambda\, u_3) = 0 {\rm ~~and~~} \lim_{\lambda\to 0}R_6^{(2)}(\lambda\, u_1,1+ \lambda\,\bar u_2, \lambda\, u_3) = 0\,,
\eeq
Therefore, our result has the correct behaviour in all the multi-Regge limits.


\subsection{Collinear limits}
In this section we compute the remainder function in collinear kinematics. If the momenta of two external particles, say 1 and 2, become collinear, then the conformal cross ratios take the particular values,
\beq
u_1 \to 0,\qquad u_2 \to (1-z)\,{s_{56}\over s_{234}},\qquad u_3\to 1-u_2\,,
\eeq
and the remainder function must vanish in this limit, {\it i.e.}
\beq\label{eq:R62col}
\lim_{\lambda\to0}R_6^{(2)}(\lambda\, u_1,u_2,1-u_2) = 0, \qquad \forall u_2\,.
\eeq
We proceed in a similar way to the multi-Regge limit, and we again only discuss here the case of the threefold contribution to $f_H (p_1, p_3, p_5; p_4, p_6, p_2)$. Using {\tt MBasymptotics} we obtain the leading behaviour of \Eqn{eq:3foldH4} in the limit $\lambda\to0$
\beq\bsp
\frac{1}{4} &\mbint\mbint{\rd z_2\over2\pi i}\,{\rd z_3\over2\pi i}\,
\left(1-u_2\right)^{z_3}\, u_2^{z_2}\, \Gamma
   \left(-z_2\right)^2\, \Gamma \left(z_2\right)\, \Gamma \left(-z_3\right)^2\,
   \Gamma \left(z_3\right)\, \Gamma \left(z_2+z_3\right)\\
   &\qquad\times \left(z_3 z_2 \log
   u_1+2 \eug\,  z_3 z_2+z_2+z_3+z_3 z_2\, \psi
   \left(z_2\right)+z_3 z_2\, \psi\left(z_3\right)\right)\,,
   \esp\eeq
with the integration contours given by \Eqn{eq:contours}. Closing the integration contours to the right and taking residues results in binomial nested sums, similar to the ones encountered in \Eqn{eq:binsum}. All the sums can hence be done using {\tt XSummer} and result in polylogarithms. Since the expression is rather lengthy and does not add anything new to the discussion, we do not show the result explicitly. Finally, combining this result with the contributions coming from the other contributions in \Eqn{eq:w62l} we see that our expression for the remainder function satisfies \Eqn{eq:R62col}.

Similarly to the multi-Regge case, we could have defined five additional collinear limits obtained by cyclic permutations of the external momenta. The six limits one obtains in this way are again equivalent pairwise at the level of the reminder function (up to redefining, \emph{e.g.}, $u_2\to1-u_3$ in \Eqn{eq:R62col}), and we checked explicitly that our result does not only satisfy the constraint (\ref{eq:R62col}) but also
\beq
\lim_{\lambda\to0}R_6^{(2)}(u_1,\lambda\,u_2,1-u_1) = 0 {\rm ~~and~~} \lim_{\lambda\to0}R_6^{(2)}(u_1,1-u_1,\lambda u_3) = 0,
\quad\forall u_1\,.
\eeq

\section{Special values of ordinary and harmonic polylogarithms}
\label{app:special_values_Li}
In this appendix we present several special values of polylogarithms up to weight four we encountered throughout our computation. All the identities of this section were obtained either using the {\tt PSLQ} algorithm~\cite{pslq1, pslq2} or using the {\tt HPL} package~\cite{Maitre:2005uu}. The question whether a given transcendental number can be expressed as a polynomial with rational coefficients of other transcendental numbers, \emph{i.e.}, the problem of finding a basis in the space of transcendental numbers, is an open mathematical problem, and we must therefore make an \emph{a priori} choice for our basis. Our choice consists in monomials in the following transcendental numbers:
\begin{itemize}
\item weight one: $\log 2$, $\log 3$,
\item weight two: $\pi^2$, $\text{Li}_2(1/3)$,
\item weight three: $\zeta_3$, $\text{Li}_3(1/3)$, $\text{Li}_3(-1/2)$,
\item weight four: $\text{Li}_4(1/2)$, $\text{Li}_4(-1/2)$, $\text{Li}_4(1/3)$, $\text{Li}_4(-1/3)$, $S_{2,2}(-1/2)$, $S_{2,2}(-1/3)$.
\end{itemize}
Note that the values of harmonic polylogarithms in $1/2$ presented in this appendix are sufficient to obtain all harmonic polylogarithms in $1/2$ up to weight four. Furthermore, as the space of harmonic polylogarithms is closed under the transformations $x \to (1-x)/(1+x)$ and $x\to -x$, these values are at the same time sufficient to construct all harmonic polylogarithms up to weight four in $-1/2$ and $\pm1/3$, and we have hence proved at the same time that all harmonic polylogarithms up to weight four in these values can be expressed completely in the basis we just defined.

\subsection{Polylogarithms of weight two}
\begin{align}
\text{Li}_2(-8) =&\,  3 \log ^2 3 -6 \log  2  \log  3 +6 \text{Li}_2\left(\frac{1}{3}\right)-\frac{\pi ^2}{2}\\
\text{Li}_2(-3) =&\,  -\log ^2 3 -2 \text{Li}_2\left(\frac{1}{3}\right)\\
\text{Li}_2(-2) =&\,  \frac{1}{2} \log ^2 3 -\log  2  \log  3 +\text{Li}_2\left(\frac{1}{3}\right)-\frac{\pi ^2}{6}\\
\text{Li}_2\left(-\frac{1}{2}\right) =&\,  -\frac{1}{2} \log ^2 2 -\frac{1}{2} \log ^2 3 +\log  3  \log  2 -\text{Li}_2\left(\frac{1}{3}\right)\\
\text{Li}_2\left(-\frac{1}{3}\right) =&\,  \frac{1}{2} \log ^2 3 +2 \text{Li}_2\left(\frac{1}{3}\right)-\frac{\pi ^2}{6}\\
\text{Li}_2\left(-\frac{1}{8}\right) =&\,  -\frac{9}{2} \log ^2 2 -3 \log ^2 3 +6 \log  3  \log  2 -6 \text{Li}_2\left(\frac{1}{3}\right)+\frac{\pi ^2}{3}\\
\text{Li}_2\left(\frac{1}{4}\right) =&\,  -2 \log ^2 2 -\log ^2 3 +2 \log  3  \log  2 -2 \text{Li}_2\left(\frac{1}{3}\right)+\frac{\pi ^2}{6}\\
\text{Li}_2\left(\frac{2}{3}\right) =&\,  -\log ^2 3 +\log  2  \log  3 -\text{Li}_2\left(\frac{1}{3}\right)+\frac{\pi ^2}{6}\\
\text{Li}_2\left(\frac{2}{3}\right) =&\,  -\log ^2 3 +\log  2  \log  3 -\text{Li}_2\left(\frac{1}{3}\right)+\frac{\pi ^2}{6}\\
\text{Li}_2\left(\frac{3}{4}\right) =&\,  -2 \log ^2 2 +\log ^2 3 +2 \text{Li}_2\left(\frac{1}{3}\right)
\end{align}

\subsection{Polylogarithms of weight three}
\begin{align}
\text{Li}_3(-8) =&\,  -3 \log ^3 2 -\frac{3}{2} \pi ^2 \log  2 +18 \text{Li}_3\left(-\frac{1}{2}\right)+\frac{49 \zeta_3}{4}\\
\text{Li}_3(-3) =&\,  -\frac{1}{3} \log ^3 3 +2 \text{Li}_3\left(\frac{1}{3}\right)-\frac{13 \zeta_3}{6}\\
\text{Li}_3(-2) =&\,  -\frac{1}{6} \log ^3 2 -\frac{1}{6} \pi ^2 \log  2 +\text{Li}_3\left(-\frac{1}{2}\right)\\
\text{Li}_3\left(-\frac{1}{3}\right) =&\,  -\frac{1}{6} \log ^3 3 +\frac{1}{6} \pi ^2 \log  3 +2 \text{Li}_3\left(\frac{1}{3}\right)-\frac{13 \zeta_3}{6}\\
\text{Li}_3\left(-\frac{1}{8}\right) =&\,  \frac{3}{2} \log ^3 2 -\pi ^2 \log  2 +18 \text{Li}_3\left(-\frac{1}{2}\right)+\frac{49 \zeta_3}{4}\\
\text{Li}_3\left(\frac{1}{4}\right) =&\,  \frac{2}{3} \log ^3 2 -\frac{1}{3} \pi ^2 \log  2 +4 \text{Li}_3\left(-\frac{1}{2}\right)+\frac{7 \zeta_3}{2}\\
\text{Li}_3\left(\frac{2}{3}\right) =&\, \frac{1}{6} \log ^3 2 +\frac{1}{3} \log ^3 3 -\frac{1}{2} \log ^2 3  \log  2 +\frac{1}{6} \pi ^2 \log  2 -\frac{1}{6} \pi ^2 \log  3 -\text{Li}_3\left(\frac{1}{3}\right)\\
&\nn-\text{Li}_3\left(-\frac{1}{2}\right)+\zeta_3\\
\text{Li}_3\left(\frac{3}{4}\right) =&\,  2 \log ^3 2 +\frac{1}{3} \log ^3 3 -2 \log  3  \log ^2 2 -2 \text{Li}_3\left(\frac{1}{3}\right)-4 \text{Li}_3\left(-\frac{1}{2}\right)-\frac{\zeta_3}{3}
\end{align}

\subsection{Polylogarithms of weight four}
\begin{align}
\text{Li}_4&(-8) =  -4 \log ^4 2 -\frac{1}{2} \pi ^2 \log ^2 2 -42 \text{Li}_4\left(\frac{1}{2}\right)-54 \text{Li}_4\left(-\frac{1}{2}\right)-\frac{3 \pi ^4}{40}\\
\text{Li}_4&(-3) =  -\frac{1}{24} \log ^4 3 -\frac{1}{12} \pi ^2 \log ^2 3 -\text{Li}_4\left(-\frac{1}{3}\right)-\frac{7 \pi ^4}{360}\\
\text{Li}_4&(-2) =  -\frac{1}{24} \log ^4 2 -\frac{1}{12} \pi ^2 \log ^2 2 -\text{Li}_4\left(-\frac{1}{2}\right)-\frac{7 \pi ^4}{360}\\
\text{Li}_4&\left(-\frac{1}{8}\right) =  \frac{5}{8} \log ^4 2 -\frac{1}{4} \pi ^2 \log ^2 2 +42 \text{Li}_4\left(\frac{1}{2}\right)+54 \text{Li}_4\left(-\frac{1}{2}\right)+\frac{\pi ^4}{18}\\
\text{Li}_4&\left(\frac{1}{9}\right) =  8 \text{Li}_4\left(-\frac{1}{3}\right)+8 \text{Li}_4\left(\frac{1}{3}\right)\\
\text{Li}_4&\left(\frac{1}{4}\right) =  8 \text{Li}_4\left(-\frac{1}{2}\right)+8 \text{Li}_4\left(\frac{1}{2}\right)\\
\text{Li}_4&\left(\frac{2}{3}\right) =  -\text{Li}_3\left(-\frac{1}{2}\right) \log  2 +\text{Li}_3\left(-\frac{1}{2}\right) \log  3 +\zeta_3 \log  2 -\zeta_3 \log  3 +\frac{1}{12} \log ^4 2 \\
&\nn-\frac{1}{12} \log ^4 3 -\frac{1}{6} \log  3  \log ^3 2 +\frac{1}{6} \log ^3 3  \log  2 +\frac{1}{12} \pi ^2 \log ^2 2 +\frac{1}{12} \pi ^2 \log ^2 3 -\frac{1}{6} \pi ^2 \log  3  \log  2\\
&\nn -\text{Li}_4\left(\frac{1}{3}\right)-\text{Li}_4\left(-\frac{1}{2}\right)+S_{2,2}\left(-\frac{1}{2}\right)+\frac{\pi ^4}{90}
\end{align}
\begin{align}
\text{Li}_4&\left(\frac{3}{4}\right) =  4 \text{Li}_3\left(\frac{1}{3}\right) \log  2 -2 \text{Li}_3\left(\frac{1}{3}\right) \log  3 -\frac{19}{3} \zeta_3 \log  2 +\frac{19}{6} \zeta_3 \log  3 -\frac{4}{3} \log ^4 2 \\
&\nn+\frac{1}{4} \log ^4 3 +\frac{4}{3} \log  3  \log ^3 2 -\frac{2}{3} \log ^3 3  \log  2 +\frac{1}{3} \pi ^2 \log ^2 2 -\frac{1}{12} \pi ^2 \log ^2 3 -8 \text{Li}_4\left(\frac{1}{2}\right)\\
&\nn-\text{Li}_4\left(-\frac{1}{3}\right)-8 \text{Li}_4\left(-\frac{1}{2}\right)+S_{2,2}\left(-\frac{1}{3}\right)+\frac{\pi ^4}{90}\\
\text{Li}_4&\left(\frac{8}{9}\right) =  -36 \text{Li}_3\left(\frac{1}{3}\right) \log  2 +18 \text{Li}_3\left(\frac{1}{3}\right) \log  3 +57 \zeta_3 \log  2 -\frac{57}{2} \zeta_3 \log  3 +\frac{16}{3} \log ^4 2\\
&\nn -\frac{77}{24} \log ^4 3 +10 \log ^3 3  \log  2 -9 \log ^2 3  \log ^2 2 -\frac{5}{6} \pi ^2 \log ^2 2 +\frac{11}{6} \pi ^2 \log ^2 3 -3 \pi ^2 \log  3  \log  2 \\
&\nn+74 \text{Li}_4\left(\frac{1}{2}\right)-8 \text{Li}_4\left(\frac{1}{3}\right)+10 \text{Li}_4\left(-\frac{1}{3}\right)+54 \text{Li}_4\left(-\frac{1}{2}\right)-9 S_{2,2}\left(-\frac{1}{3}\right)-\frac{11 \pi ^4}{120}\\
S_{2,2}&\left(-\frac{1}{8}\right) =  -36 \text{Li}_3\left(\frac{1}{3}\right) \log  2 +54 \text{Li}_3\left(-\frac{1}{2}\right) \log  2 +18 \text{Li}_3\left(\frac{1}{3}\right) \log  3\\
&\nn -36 \text{Li}_3\left(-\frac{1}{2}\right) \log  3 +\frac{363}{4} \zeta_3 \log  2 -51 \zeta_3 \log  3 +\frac{89}{24} \log ^4 2 -\frac{15}{8} \log ^4 3 +6 \log  3  \log ^3 2\\
&\nn +6 \log ^3 3  \log  2 -9 \log ^2 3  \log ^2 2 -\frac{29}{6} \pi ^2 \log ^2 2 +\frac{3}{2} \pi ^2 \log ^2 3 +116 \text{Li}_4\left(\frac{1}{2}\right)+18 \text{Li}_4\left(-\frac{1}{3}\right)\\
&\nn+108 \text{Li}_4\left(-\frac{1}{2}\right)-9 S_{2,2}\left(-\frac{1}{3}\right)-\frac{17 \pi ^4}{360}\\
S_{2,2}&\left(\frac{1}{3}\right) =  -\text{Li}_3\left(\frac{1}{3}\right) \log  2 +\text{Li}_3\left(-\frac{1}{2}\right) \log  2 +\text{Li}_3\left(\frac{1}{3}\right) \log  3 -\text{Li}_3\left(-\frac{1}{2}\right) \log  3\\
&\nn -\frac{1}{24} \log ^4 2 -\frac{1}{24} \log ^4 3 +\frac{1}{6} \log  3  \log ^3 2 +\frac{1}{6} \log ^3 3  \log  2 -\frac{1}{4} \log ^2 3  \log ^2 2 +2 \text{Li}_4\left(\frac{1}{3}\right)\\
&\nn+2 \text{Li}_4\left(-\frac{1}{2}\right)-S_{2,2}\left(-\frac{1}{2}\right)\\
S_{2,2}&\left(\frac{1}{2}\right) =  -\frac{1}{8} \zeta_3 \log  2 +\frac{1}{24} \log ^4 2 +\frac{\pi ^4}{720}\\
S_{2,2}&\left(\frac{2}{3}\right) =  -\text{Li}_3\left(-\frac{1}{2}\right) \log  2 -\text{Li}_3\left(\frac{1}{3}\right) \log  3 +\zeta_3 \log  2 -\zeta_3 \log  3 +\frac{1}{24} \log ^4 2 +\frac{1}{8} \log ^4 3\\
&\nn -\frac{1}{6} \log ^3 3  \log  2 -2 \text{Li}_4\left(\frac{1}{3}\right)-2 \text{Li}_4\left(-\frac{1}{2}\right)+S_{2,2}\left(-\frac{1}{2}\right)+\frac{\pi ^4}{360}\\
S_{2,2}&\left(\frac{3}{4}\right) =  -8 \text{Li}_3\left(-\frac{1}{2}\right) \log  2 -2 \text{Li}_3\left(\frac{1}{3}\right) \log  3 -9 \zeta_3 \log  2 +\frac{19}{6} \zeta_3 \log  3 +\frac{2}{3} \log ^4 2 \\
&\nn+\frac{5}{24} \log ^4 3 -\frac{4}{3} \log  3  \log ^3 2 +\frac{2}{3} \pi ^2 \log ^2 2 -\frac{1}{6} \pi ^2 \log ^2 3 -16 \text{Li}_4\left(\frac{1}{2}\right)-2 \text{Li}_4\left(-\frac{1}{3}\right)\\
&\nn-16 \text{Li}_4\left(-\frac{1}{2}\right)+S_{2,2}\left(-\frac{1}{3}\right)+\frac{\pi ^4}{360}\\
S_{2,2}&\left(\frac{8}{9}\right) =  -36 \text{Li}_3\left(\frac{1}{3}\right) \log  2 -6 \text{Li}_3\left(\frac{1}{3}\right) \log  3 -36 \text{Li}_3\left(-\frac{1}{2}\right) \log  3 +57 \zeta_3 \log  2\\
&\nn -\frac{107}{3} \zeta_3 \log  3 +\frac{4}{3} \log ^4 2 +\frac{35}{24} \log ^4 3 +6 \log  3  \log ^3 2 +2 \log ^3 3  \log  2 -9 \log ^2 3  \log ^2 2 -\frac{4}{3} \pi ^2 \log ^2 2\\
&\nn +\frac{1}{6} \pi ^2 \log ^2 3 +32 \text{Li}_4\left(\frac{1}{2}\right)-16 \text{Li}_4\left(\frac{1}{3}\right)+2 \text{Li}_4\left(-\frac{1}{3}\right)-9 S_{2,2}\left(-\frac{1}{3}\right)-\frac{7 \pi ^4}{45}
\end{align}
\begin{align}
H&\left(-1,1,-1,1; \frac{1}{2}\right) =  -\text{Li}_2\left(\frac{1}{3}\right) \log ^2 2 +2 \text{Li}_2\left(\frac{1}{3}\right) \log ^2 3 -4 \text{Li}_3\left(\frac{1}{3}\right) \log  2\\
&\nn +8 \text{Li}_3\left(-\frac{1}{2}\right) \log  2 -2 \text{Li}_2\left(\frac{1}{3}\right) \log  3  \log  2 +4 \text{Li}_3\left(\frac{1}{3}\right) \log  3 +\frac{181}{12} \zeta_3 \log  2 -\frac{73}{12} \zeta_3 \log  3 \\
&\nn+\frac{11}{8} \log ^4 2 +\frac{1}{12} \log ^4 3 +\frac{5}{6} \log  3  \log ^3 2 -\frac{1}{3} \log ^3 3  \log  2 -\frac{1}{2} \log ^2 3  \log ^2 2 -\frac{23}{24} \pi ^2 \log ^2 2\\
&\nn +\frac{1}{4} \pi ^2 \log ^2 3 -\frac{1}{12} \pi ^2 \log  3  \log  2 +32 \text{Li}_4\left(\frac{1}{2}\right)+4 \text{Li}_4\left(-\frac{1}{3}\right)+32 \text{Li}_4\left(-\frac{1}{2}\right)+2 \text{Li}_2\left(\frac{1}{3}\right)^2\\
&\nn-\frac{\pi ^2 \text{Li}_2\left(\frac{1}{3}\right)}{6}-2 S_{2,2}\left(-\frac{1}{3}\right)+\frac{\pi ^4}{1440}\\
H&\left(0,-1,-1,1; \frac{1}{2}\right) =  -\frac{1}{2} \text{Li}_2\left(\frac{1}{3}\right) \log ^2 2 +\text{Li}_2\left(\frac{1}{3}\right) \log ^2 3 +9 \text{Li}_3\left(\frac{1}{3}\right) \log  2\\
&\nn -7 \text{Li}_3\left(-\frac{1}{2}\right) \log  2 -\text{Li}_2\left(\frac{1}{3}\right) \log  3  \log  2 -4 \text{Li}_3\left(\frac{1}{3}\right) \log  3 +8 \text{Li}_3\left(-\frac{1}{2}\right) \log  3 -\frac{115}{24} \zeta_3 \log  2\\
&\nn -\frac{5}{3} \zeta_3 \log  3 -\frac{1}{2} \log ^4 2 +\frac{5}{12} \log ^4 3 -\frac{5}{6} \log  3  \log ^3 2 -2 \log ^3 3  \log  2 +\frac{9}{4} \log ^2 3  \log ^2 2 +\frac{7}{24} \pi ^2 \log ^2 2\\
&\nn +\frac{1}{8} \pi ^2 \log ^2 3 +\frac{1}{12} \pi ^2 \log  3  \log  2 -9 \text{Li}_4\left(\frac{1}{2}\right)-12 \text{Li}_4\left(\frac{1}{3}\right)+2 \text{Li}_4\left(-\frac{1}{3}\right)-22 \text{Li}_4\left(-\frac{1}{2}\right)\\
&\nn+\text{Li}_2\left(\frac{1}{3}\right)^2-\frac{\pi ^2 \text{Li}_2\left(\frac{1}{3}\right)}{12}+2 S_{2,2}\left(-\frac{1}{3}\right)+8 S_{2,2}\left(-\frac{1}{2}\right)+\frac{7 \pi ^4}{240}\\
H&\left(0,-1,1,-1; \frac{1}{2}\right) =  -\frac{1}{2} \text{Li}_2\left(\frac{1}{3}\right) \log ^2 2 -\text{Li}_2\left(\frac{1}{3}\right) \log ^2 3 -12 \text{Li}_3\left(\frac{1}{3}\right) \log  2\\
&\nn +12 \text{Li}_3\left(-\frac{1}{2}\right) \log  2 +2 \text{Li}_2\left(\frac{1}{3}\right) \log  3  \log  2 +6 \text{Li}_3\left(\frac{1}{3}\right) \log  3 -12 \text{Li}_3\left(-\frac{1}{2}\right) \log  3\\
&\nn +\frac{95}{8} \zeta_3 \log  2 -\frac{19}{8} \zeta_3 \log  3 +\frac{7}{12} \log ^4 2 -\frac{19}{32} \log ^4 3 +\frac{5}{2} \log  3  \log ^3 2 +3 \log ^3 3  \log  2 -\frac{17}{4} \log ^2 3  \log ^2 2\\
&\nn -\frac{17}{24} \pi ^2 \log ^2 2 -\frac{1}{48} \pi ^2 \log ^2 3 -\frac{1}{12} \pi ^2 \log  3  \log  2 +20 \text{Li}_4\left(\frac{1}{2}\right)+\frac{27 \text{Li}_4\left(\frac{1}{3}\right)}{2}-\frac{3 \text{Li}_4\left(-\frac{1}{3}\right)}{4}\\
&\nn+35 \text{Li}_4\left(-\frac{1}{2}\right)-\text{Li}_2\left(\frac{1}{3}\right)^2+\frac{\pi ^2 \text{Li}_2\left(\frac{1}{3}\right)}{12}-3 S_{2,2}\left(-\frac{1}{3}\right)-8 S_{2,2}\left(-\frac{1}{2}\right)-\frac{7 \pi ^4}{240}\\
H&\left(0,-1,1,1; \frac{1}{2}\right) =  \frac{1}{2} \text{Li}_2\left(\frac{1}{3}\right) \log ^2 2 -4 \text{Li}_3\left(-\frac{1}{2}\right) \log  2 -\frac{7}{2} \zeta_3 \log  2 -\frac{1}{6} \log ^4 2\\
&\nn -\frac{1}{2} \log  3  \log ^3 2 +\frac{1}{4} \log ^2 3  \log ^2 2 +\frac{1}{3} \pi ^2 \log ^2 2 -9 \text{Li}_4\left(\frac{1}{2}\right)-12 \text{Li}_4\left(-\frac{1}{2}\right)-\frac{17 \pi ^4}{1440}\\
H&\left(0,0,-1,1; \frac{1}{2}\right) =  -\text{Li}_3\left(-\frac{1}{2}\right) \log  2 -\frac{3}{4} \zeta_3 \log  2 -\frac{5}{24} \log ^4 2 +\frac{1}{8} \pi ^2 \log ^2 2\\
&\nn -2 \text{Li}_4\left(\frac{1}{2}\right)-6 \text{Li}_4\left(-\frac{1}{2}\right)-\frac{31 \pi ^4}{1440}\\
H&\left(0,0,1,-1; \frac{1}{2}\right) =  \text{Li}_3\left(\frac{1}{3}\right) \log  2 -\text{Li}_3\left(-\frac{1}{2}\right) \log  2 -\frac{1}{2} \text{Li}_3\left(\frac{1}{3}\right) \log  3\\
&\nn +\text{Li}_3\left(-\frac{1}{2}\right) \log  3 -\frac{53}{24} \zeta_3 \log  2 +\frac{17}{12} \zeta_3 \log  3 +\frac{1}{12} \log ^4 2 +\frac{5}{96} \log ^4 3 -\frac{1}{6} \log  3  \log ^3 2 -\frac{1}{6} \log ^3 3  \log  2\\
&\nn +\frac{1}{4} \log ^2 3  \log ^2 2 +\frac{1}{24} \pi ^2 \log ^2 2 -\frac{1}{24} \pi ^2 \log ^2 3 -2 \text{Li}_4\left(\frac{1}{2}\right)-\frac{\text{Li}_4\left(-\frac{1}{3}\right)}{2}+2 \text{Li}_4\left(-\frac{1}{2}\right)\\
&\nn+\frac{S_{2,2}\left(-\frac{1}{3}\right)}{4}+S_{2,2}\left(-\frac{1}{2}\right)+\frac{11 \pi ^4}{480}
\end{align}
\begin{align}
H&\left(0,1,-1,-1; \frac{1}{2}\right) =  3 \text{Li}_3\left(\frac{1}{3}\right) \log  2 -3 \text{Li}_3\left(-\frac{1}{2}\right) \log  2 -\frac{3}{2} \text{Li}_3\left(\frac{1}{3}\right) \log  3 \\
&\nn+3 \text{Li}_3\left(-\frac{1}{2}\right) \log  3 -\frac{107}{24} \zeta_3 \log  2 +\frac{25}{12} \zeta_3 \log  3 -\frac{1}{3} \log ^4 2 +\frac{11}{96} \log ^4 3 -\frac{1}{2} \log  3  \log ^3 2 -\frac{1}{2} \log ^3 3  \log  2 \\
&\nn+\frac{3}{4} \log ^2 3  \log ^2 2 +\frac{7}{24} \pi ^2 \log ^2 2 -\frac{1}{24} \pi ^2 \log ^2 3 -8 \text{Li}_4\left(\frac{1}{2}\right)-2 \text{Li}_4\left(\frac{1}{3}\right)-\frac{\text{Li}_4\left(-\frac{1}{3}\right)}{2}\\
&\nn-11 \text{Li}_4\left(-\frac{1}{2}\right)+\frac{3 S_{2,2}\left(-\frac{1}{3}\right)}{4}-\frac{\pi ^4}{1440}\\
H&\left(0,1,-1,1; \frac{1}{2}\right) =  -7 \text{Li}_3\left(\frac{1}{3}\right) \log  2 +10 \text{Li}_3\left(-\frac{1}{2}\right) \log  2 +\frac{7}{2} \text{Li}_3\left(\frac{1}{3}\right) \log  3\\
&\nn -7 \text{Li}_3\left(-\frac{1}{2}\right) \log  3 +\frac{77}{12} \zeta_3 \log  2 +\frac{35}{24} \zeta_3 \log  3 +\frac{17}{24} \log ^4 2 -\frac{7}{48} \log ^4 3 +\frac{7}{6} \log  3  \log ^3 2 +\frac{7}{6} \log ^3 3  \log  2 \\
&\nn-\frac{7}{4} \log ^2 3  \log ^2 2 -\frac{7}{12} \pi ^2 \log ^2 2 -\frac{7}{48} \pi ^2 \log ^2 3 +17 \text{Li}_4\left(\frac{1}{2}\right)+\frac{21 \text{Li}_4\left(\frac{1}{3}\right)}{2}-\frac{7 \text{Li}_4\left(-\frac{1}{3}\right)}{4}\\
&\nn+32 \text{Li}_4\left(-\frac{1}{2}\right)-\frac{7 S_{2,2}\left(-\frac{1}{3}\right)}{4}-7 S_{2,2}\left(-\frac{1}{2}\right)-\frac{\pi ^4}{240}\\
H&\left(0,1,0,-1; \frac{1}{2}\right) =  2 \text{Li}_3\left(\frac{1}{3}\right) \log  2 -2 \text{Li}_3\left(-\frac{1}{2}\right) \log  2 -\text{Li}_3\left(\frac{1}{3}\right) \log  3\\
&\nn +2 \text{Li}_3\left(-\frac{1}{2}\right) \log  3 +\frac{25}{12} \zeta_3 \log  2 -\frac{11}{3} \zeta_3 \log  3 +\frac{5}{24} \log ^4 2 -\frac{1}{48} \log ^4 3 -\frac{1}{3} \log  3  \log ^3 2 -\frac{1}{3} \log ^3 3  \log  2 \\
&\nn+\frac{1}{2} \log ^2 3  \log ^2 2 -\frac{1}{8} \pi ^2 \log ^2 2 +\frac{1}{6} \pi ^2 \log ^2 3 +3 \text{Li}_4\left(\frac{1}{2}\right)-6 \text{Li}_4\left(\frac{1}{3}\right)+2 \text{Li}_4\left(-\frac{1}{3}\right)\\
&\nn-3 \text{Li}_4\left(-\frac{1}{2}\right)+\frac{S_{2,2}\left(-\frac{1}{3}\right)}{2}+2 S_{2,2}\left(-\frac{1}{2}\right)+\frac{3 \pi ^4}{160}\\
H&\left(0,1,1,-1; \frac{1}{2}\right) =  8 \text{Li}_3\left(\frac{1}{3}\right) \log  2 -8 \text{Li}_3\left(-\frac{1}{2}\right) \log  2 -4 \text{Li}_3\left(\frac{1}{3}\right) \log  3\\
&\nn +8 \text{Li}_3\left(-\frac{1}{2}\right) \log  3 -\frac{14}{3} \zeta_3 \log  2 -\frac{5}{3} \zeta_3 \log  3 -\frac{1}{3} \log ^4 2 +\frac{1}{6} \log ^4 3 -\frac{4}{3} \log  3  \log ^3 2 -\frac{4}{3} \log ^3 3  \log  2\\
&\nn +2 \log ^2 3  \log ^2 2 +\frac{1}{3} \pi ^2 \log ^2 2 +\frac{1}{6} \pi ^2 \log ^2 3 -10 \text{Li}_4\left(\frac{1}{2}\right)-12 \text{Li}_4\left(\frac{1}{3}\right)+2 \text{Li}_4\left(-\frac{1}{3}\right)\\
&\nn-23 \text{Li}_4\left(-\frac{1}{2}\right)+2 S_{2,2}\left(-\frac{1}{3}\right)+8 S_{2,2}\left(-\frac{1}{2}\right)+\frac{19 \pi ^4}{720}
\end{align}

\subsection{Polylogarithms of complex arguments}
\begin{align}
\text{Li}_3\left(\frac{1}{2}+\frac{i}{2}\right)+ \text{Li}_3\left(\frac{1}{2}-\frac{i}{2}\right) &\,= 
\frac{1}{24} \log ^3 2 -\frac{5}{96} \pi ^2 \log  2 +\frac{35}{32} \zeta_3\\
\text{Li}_4\left(\frac{1}{2}+\frac{i}{2}\right)+ \text{Li}_4\left(\frac{1}{2}-\frac{i}{2}\right) &\,=
\frac{1}{48} \log ^4 2 -\frac{5}{384} \pi ^2 \log ^2 2 +\frac{5 \text{Li}_4\left(\frac{1}{2}\right)}{8}+\frac{343 \pi ^4}{46080}
\end{align}

\section{Special values of Goncharov multiple polylogarithms}
\label{app:special_values_GPL}

In this appendix we present special values of Goncharov multiple polylogarithms that we encountered throughout our computation. The results are expressed in the same transcendental basis as in App.~\ref{app:special_values_Li}. All the expressions have been derived using the reduction algorithm of App.~\ref{app:Goncharov}. Note that we present here only values for polylogarithms of the form $G(\vec w;1)$. It is very easy to extend this list to other types of polylogarithms by applying \Eqn{eq:Gscaling}.

\subsection{Polylogarithms of weight two}
\begin{align}
G&\left(\frac{4}{3},1; 1\right) =  \log ^2 3 +2 \text{Li}_2\left(\frac{1}{3}\right)\\
G&\left(\frac{3}{2},1; 1\right) =  -\frac{1}{2} \log ^2 3 +\log  2  \log  3 -\text{Li}_2\left(\frac{1}{3}\right)+\frac{\pi ^2}{6}\\
G&(2,1; 1) =  \frac{\pi ^2}{12}
\end{align}

\subsection{Polylogarithms of weight three}
\begin{align}
G&\left(-\frac{1}{2},-\frac{1}{3},-1; 1\right) =  4 \text{Li}_2\left(\frac{1}{3}\right) \log  2 +\text{Li}_2\left(\frac{1}{3}\right) \log  3 +\frac{1}{2} \log ^3 2 -\frac{7}{6} \log ^3 3 -\log  3  \log ^2 2 \\
&\,\nn+2 \log ^2 3  \log  2 -\frac{1}{6} \pi ^2 \log  2 +\frac{2}{3} \pi ^2 \log  3 +10 \text{Li}_3\left(\frac{1}{3}\right)-3 \text{Li}_3\left(-\frac{1}{2}\right)-10 \zeta_3\\
G&\left(-\frac{1}{2},-\frac{1}{3},0; 1\right) =  \text{Li}_2\left(\frac{1}{3}\right) \log  3 +\frac{1}{6} \log ^3 2 -\frac{1}{4} \log ^3 3 +\frac{1}{6} \pi ^2 \log  2 -\frac{7}{12} \pi ^2 \log  3\\
&\,\nn -\text{Li}_3\left(\frac{1}{3}\right)+\frac{11 \text{Li}_3\left(-\frac{1}{3}\right)}{2}-\text{Li}_3\left(-\frac{1}{2}\right)+\frac{17 \zeta_3}{4}\\
G&\left(-\frac{1}{2},0,-1; 1\right) =  \text{Li}_2\left(\frac{1}{3}\right) \log  2 +\frac{1}{2} \log ^3 2 -\frac{1}{24} \log ^3 3 -\log  3  \log ^2 2 +\frac{1}{2} \log ^2 3  \log  2 \\
&\,\nn+\frac{1}{12} \pi ^2 \log  2 +\frac{1}{8} \pi ^2 \log  3 +\frac{\text{Li}_3\left(\frac{1}{3}\right)}{2}-\frac{\text{Li}_3\left(-\frac{1}{3}\right)}{4}-3 \text{Li}_3\left(-\frac{1}{2}\right)-\frac{23 \zeta_3}{8}\\
G&\left(-\frac{1}{3},-\frac{1}{2},-1; 1\right) =  -8 \text{Li}_2\left(\frac{1}{3}\right) \log  2 -\frac{19}{6} \log ^3 2 +\frac{1}{3} \log ^3 3 +\frac{13}{2} \log  3  \log ^2 2\\
&\,\nn -4 \log ^2 3  \log  2-\frac{7}{12} \pi ^2 \log  2 -\frac{1}{6} \pi ^2 \log  3 -2 \text{Li}_3\left(\frac{1}{3}\right)+19 \text{Li}_3\left(-\frac{1}{2}\right)+\frac{121 \zeta_3}{8}\\
G&\left(-\frac{1}{3},-\frac{1}{2},-\frac{1}{3}; 1\right) =  -14 \text{Li}_2\left(\frac{1}{3}\right) \log  2 -\frac{17}{3} \log ^3 2 +14 \log  3  \log ^2 2 -7 \log ^2 3  \log  2 \\
&\,\nn-\frac{4}{3} \pi ^2 \log  2 +34 \text{Li}_3\left(-\frac{1}{2}\right)+\frac{49 \zeta_3}{2}\\
G&\left(-\frac{1}{3},-\frac{1}{2},1; 1\right) =  -10 \text{Li}_2\left(\frac{1}{3}\right) \log  2 -3 \log ^3 2 -\frac{1}{6} \log ^3 3 +6 \log  3  \log ^2 2 -\frac{7}{2} \log ^2 3  \log  2 \\
&\,\nn-\frac{5}{6} \pi ^2 \log  2 +\frac{1}{3} \pi ^2 \log  3 +\text{Li}_3\left(\frac{1}{3}\right)+18 \text{Li}_3\left(-\frac{1}{2}\right)+\frac{45 \zeta_3}{4}
\end{align}
\begin{align}
G&\left(-\frac{1}{3},-\frac{1}{3},-\frac{1}{2}; 1\right) =  2 \text{Li}_2\left(\frac{1}{3}\right) \log  2 +\frac{17}{6} \log ^3 2 -4 \log  3  \log ^2 2 +\log ^2 3  \log  2 +\pi ^2 \log  2\\
&\,\nn -17 \text{Li}_3\left(-\frac{1}{2}\right)-\frac{49 \zeta_3}{4}\\
G&\left(-\frac{1}{3},1,-\frac{1}{2}; 1\right) =  10 \text{Li}_2\left(\frac{1}{3}\right) \log  2 -6 \text{Li}_2\left(\frac{1}{3}\right) \log  3 +\frac{25}{6} \log ^3 2 -\frac{2}{3} \log ^3 3\\
&\,\nn -8 \log  3  \log ^2 2 +5 \log ^2 3  \log  2 +\pi ^2 \log  2 -\frac{2}{3} \pi ^2 \log  3 -14 \text{Li}_3\left(\frac{1}{3}\right)-25 \text{Li}_3\left(-\frac{1}{2}\right)-\frac{77 \zeta_3}{12}\\
G&(0,-2,-8; 1) =  -9 \text{Li}_2\left(\frac{1}{3}\right) \log  2 +6 \text{Li}_2\left(\frac{1}{3}\right) \log  3 -\frac{11}{6} \log ^3 2 +\frac{7}{6} \log ^3 3 \\
&\,\nn+5 \log  3  \log ^2 2 -5 \log ^2 3  \log  2 +\frac{1}{6} \pi ^2 \log  2 +\frac{1}{3} \pi ^2 \log  3 +11 \text{Li}_3\left(\frac{1}{3}\right)+4 \text{Li}_3\left(-\frac{1}{2}\right)-\frac{31 \zeta_3}{6}\\
G&(0,-2,1; 1) =  -\frac{1}{6} \log ^3 2 +\frac{1}{24} \log ^3 3 +\frac{1}{2} \log  3  \log ^2 2 -\frac{1}{2} \log ^2 3  \log  2 -\frac{1}{6} \pi ^2 \log  2\\
&\,\nn +\frac{7}{24} \pi ^2 \log  3 +\frac{\text{Li}_3\left(\frac{1}{3}\right)}{2}-\frac{3 \text{Li}_3\left(-\frac{1}{3}\right)}{4}+\text{Li}_3\left(-\frac{1}{2}\right)-\frac{13 \zeta_3}{8}\\
G&(0,1,2; 1) =  \frac{1}{6} \pi ^2 \log  2 -\frac{5 \zeta_3}{8}\\
G&(0,2,1; 1) =  \frac{13 \zeta_3}{8}-\frac{1}{4} \pi ^2 \log  2 \\
G&(0,3,-1; 1) =  \text{Li}_2\left(\frac{1}{3}\right) \log  2 +\frac{1}{2} \log ^3 2 -\frac{5}{24} \log ^3 3 -\frac{3}{2} \log  3  \log ^2 2 +\frac{3}{2} \log ^2 3  \log  2 \\
&\,\nn-\frac{1}{6} \pi ^2 \log  2 +\frac{1}{24} \pi ^2 \log  3 +\frac{\text{Li}_3\left(\frac{1}{3}\right)}{2}+\frac{3 \text{Li}_3\left(-\frac{1}{3}\right)}{4}-3 \text{Li}_3\left(-\frac{1}{2}\right)-\frac{9 \zeta_3}{8}\\
G&(0,4,-2; 1) =  -\text{Li}_2\left(\frac{1}{3}\right) \log  2 +\text{Li}_2\left(\frac{1}{3}\right) \log  3 -\frac{5}{6} \log ^3 2 +\frac{1}{2} \log ^3 3 +\frac{5}{2} \log  3  \log ^2 2 \\
&\,\nn-\frac{3}{2} \log ^2 3  \log  2 -\frac{1}{6} \pi ^2 \log  2 -\frac{1}{6} \pi ^2 \log  3 +4 \text{Li}_3\left(-\frac{1}{2}\right)+\frac{7 \zeta_3}{2}\\
G&\left(\frac{4}{3},1,4; 1\right) =  6 \text{Li}_2\left(\frac{1}{3}\right) \log  3 -4 \text{Li}_2\left(\frac{1}{3}\right) \log  2 +\frac{2}{3} \log ^3 3 -2 \log  2  \log ^2 3 +\frac{2}{3} \pi ^2 \log  3\\
&\,\nn +14 \text{Li}_3\left(\frac{1}{3}\right)-\frac{65 \zeta_3}{6}\\
G&\left(\frac{4}{3},\frac{4}{3},1; 1\right) =  -4 \text{Li}_2\left(\frac{1}{3}\right) \log  2 -\frac{2}{3} \log ^3 2 +2 \log  3  \log ^2 2 -2 \log ^2 3  \log  2 +4 \text{Li}_3\left(-\frac{1}{2}\right)\\
&\,\nn+\frac{5 \zeta_3}{2}\\
G&\left(\frac{4}{3},\frac{3}{2},1; 1\right) =  4 \text{Li}_2\left(\frac{1}{3}\right) \log  2 +\frac{5}{3} \log ^3 2 -\frac{5}{12} \log ^3 3 -4 \log  3  \log ^2 2 +2 \log ^2 3  \log  2 \\
&\,\nn+\frac{1}{2} \pi ^2 \log  2 +\frac{1}{12} \pi ^2 \log  3 +3 \text{Li}_3\left(\frac{1}{3}\right)-\frac{\text{Li}_3\left(-\frac{1}{3}\right)}{2}-10 \text{Li}_3\left(-\frac{1}{2}\right)-\frac{19 \zeta_3}{2}\\
G&\left(\frac{4}{3},2,1; 1\right) =  -2 \text{Li}_2\left(\frac{1}{3}\right) \log  2 -\frac{2}{3} \log ^3 2 -\frac{1}{3} \log ^3 3 +2 \log  3  \log ^2 2 -\log ^2 3  \log  2 \\
&\,\nn-\frac{1}{12} \pi ^2 \log  2 +2 \text{Li}_3\left(\frac{1}{3}\right)+4 \text{Li}_3\left(-\frac{1}{2}\right)+\frac{47 \zeta_3}{24}
\end{align}
\begin{align}
G&(2,0,1; 1) =  \frac{1}{4} \pi ^2 \log  2 -\zeta_3\\
G&(2,1,0; 1) =  -\frac{5 \zeta_3}{8}\\
G&(2,1,1; 1) =  -\frac{3 \zeta_3}{4}\\
G&(2,1,2; 1) =  \frac{\zeta_3}{4}-\frac{1}{12} \pi ^2 \log  2 \\
G&(2,2,1; 1) =  -\frac{\zeta_3}{8}\\
G&(3,0,-1; 1) =  \text{Li}_2\left(\frac{1}{3}\right) \log  2 -2 \text{Li}_2\left(\frac{1}{3}\right) \log  3 +\frac{1}{6} \log ^3 2 -\frac{5}{8} \log ^3 3 -\frac{1}{2} \log  3  \log ^2 2\\
&\,\nn +\frac{1}{2} \log ^2 3  \log  2 +\frac{1}{12} \pi ^2 \log  2 +\frac{1}{8} \pi ^2 \log  3 -\frac{\text{Li}_3\left(\frac{1}{3}\right)}{2}-\frac{7 \text{Li}_3\left(-\frac{1}{3}\right)}{4}-\text{Li}_3\left(-\frac{1}{2}\right)-\frac{3 \zeta_3}{2}\\
G&(4,0,-2; 1) =  3 \text{Li}_2\left(\frac{1}{3}\right) \log  2 -2 \text{Li}_2\left(\frac{1}{3}\right) \log  3 +\frac{5}{6} \log ^3 2 -\frac{19}{24} \log ^3 3 -\frac{5}{2} \log  3  \log ^2 2 \\
&\,\nn+2 \log ^2 3  \log  2 +\frac{1}{12} \pi ^2 \log  2 +\frac{7}{24} \pi ^2 \log  3 +\frac{\text{Li}_3\left(\frac{1}{3}\right)}{2}-\frac{7 \text{Li}_3\left(-\frac{1}{3}\right)}{4}-6 \text{Li}_3\left(-\frac{1}{2}\right)-\frac{23 \zeta_3}{4}
\end{align}

\subsection{Polylogarithms of weight four}
\begin{align}
G&\left(-\frac{1}{2},0,-1,-1; 1\right) =  \frac{1}{2} \text{Li}_2\left(\frac{1}{3}\right) \log ^2 2 +8 \text{Li}_3\left(\frac{1}{3}\right) \log  2 -11 \text{Li}_3\left(-\frac{1}{2}\right) \log  2 \\
&\nn-4 \text{Li}_3\left(\frac{1}{3}\right) \log  3 +8 \text{Li}_3\left(-\frac{1}{2}\right) \log  3 -\frac{35}{8} \zeta_3 \log  2 -\frac{37}{24} \zeta_3 \log  3 +\frac{5}{24} \log ^4 2 +\frac{1}{6} \log ^4 3\\
&\nn -\frac{11}{6} \log  3  \log ^3 2 -\frac{4}{3} \log ^3 3  \log  2 +\frac{9}{4} \log ^2 3  \log ^2 2 +\frac{7}{24} \pi ^2 \log ^2 2 +\frac{1}{6} \pi ^2 \log ^2 3 -8 \text{Li}_4\left(\frac{1}{2}\right)\\
&\nn-12 \text{Li}_4\left(\frac{1}{3}\right)+2 \text{Li}_4\left(-\frac{1}{3}\right)-23 \text{Li}_4\left(-\frac{1}{2}\right)+2 S_{2,2}\left(-\frac{1}{3}\right)+8 S_{2,2}\left(-\frac{1}{2}\right)+\frac{\pi ^4}{480}\\
G&\left(-\frac{1}{2},0,-\frac{1}{2},-1; 1\right) =  \text{Li}_2\left(\frac{1}{3}\right) \log ^2 3 +17 \text{Li}_3\left(\frac{1}{3}\right) \log  2 -19 \text{Li}_3\left(-\frac{1}{2}\right) \log  2\\
&\nn -\text{Li}_2\left(\frac{1}{3}\right) \log  3  \log  2 -\frac{15}{2} \text{Li}_3\left(\frac{1}{3}\right) \log  3 +19 \text{Li}_3\left(-\frac{1}{2}\right) \log  3 -\frac{13}{4} \zeta_3 \log  2 -\frac{31}{4} \zeta_3 \log  3\\
&\nn +\frac{13}{24} \log ^4 2 +\frac{13}{32} \log ^4 3 -\frac{19}{6} \log  3  \log ^3 2 -\frac{10}{3} \log ^3 3  \log  2 +\frac{19}{4} \log ^2 3  \log ^2 2 +\frac{1}{4} \pi ^2 \log ^2 2\\
&\nn +\frac{7}{12} \pi ^2 \log ^2 3 -\frac{1}{6} \pi ^2 \log  3  \log  2 -6 \text{Li}_4\left(\frac{1}{2}\right)-30 \text{Li}_4\left(\frac{1}{3}\right)+\frac{15 \text{Li}_4\left(-\frac{1}{3}\right)}{2}-38 \text{Li}_4\left(-\frac{1}{2}\right)\\
&\nn+\text{Li}_2\left(\frac{1}{3}\right)^2-\frac{\pi ^2 \text{Li}_2\left(\frac{1}{3}\right)}{12}+\frac{15 S_{2,2}\left(-\frac{1}{3}\right)}{4}+19 S_{2,2}\left(-\frac{1}{2}\right)+\frac{\pi ^4}{16}
\end{align}
\begin{align}
G&\left(-\frac{1}{3},-1,-\frac{1}{2},0; 1\right) =  -\frac{7}{2} \text{Li}_2\left(\frac{1}{3}\right) \log ^2 3 +15 \text{Li}_3\left(\frac{1}{3}\right) \log  2 -19 \text{Li}_3\left(-\frac{1}{2}\right) \log  2\\
&\nn +3 \text{Li}_2\left(\frac{1}{3}\right) \log  3  \log  2 -\frac{21}{2} \text{Li}_3\left(\frac{1}{3}\right) \log  3 +11 \text{Li}_3\left(-\frac{1}{2}\right) \log  3 -\frac{293}{12} \zeta_3 \log  2 +\frac{45}{8} \zeta_3 \log  3\\
&\nn -\frac{31}{12} \log ^4 2 -\frac{1}{6} \log ^4 3 -\frac{11}{6} \log  3  \log ^3 2 -\log ^3 3  \log  2 +\frac{11}{4} \log ^2 3  \log ^2 2 +\frac{23}{24} \pi ^2 \log ^2 2 -\frac{1}{48} \pi ^2 \log ^2 3\\
&\nn +\frac{1}{3} \pi ^2 \log  3  \log  2 -57 \text{Li}_4\left(\frac{1}{2}\right)-\frac{33 \text{Li}_4\left(\frac{1}{3}\right)}{2}-\frac{13 \text{Li}_4\left(-\frac{1}{3}\right)}{4}-70 \text{Li}_4\left(-\frac{1}{2}\right)-\frac{7 \text{Li}_2\left(\frac{1}{3}\right)^2}{2}\\
&\nn+\frac{\pi ^2 \text{Li}_2\left(\frac{1}{3}\right)}{2}+\frac{21 S_{2,2}\left(-\frac{1}{3}\right)}{4}+11 S_{2,2}\left(-\frac{1}{2}\right)+\frac{41 \pi ^4}{720}\\
G&\left(-\frac{1}{3},-\frac{1}{2},-\frac{1}{3},-1; 1\right) =  -12 \text{Li}_2\left(\frac{1}{3}\right) \log ^2 2 +\frac{11}{2} \text{Li}_2\left(\frac{1}{3}\right) \log ^2 3 \\
&\nn-123 \text{Li}_3\left(\frac{1}{3}\right) \log  2 +195 \text{Li}_3\left(-\frac{1}{2}\right) \log  2 +3 \text{Li}_2\left(\frac{1}{3}\right) \log  3  \log  2 +\frac{123}{2} \text{Li}_3\left(\frac{1}{3}\right) \log  3\\
&\nn -157 \text{Li}_3\left(-\frac{1}{2}\right) \log  3 +\frac{249}{4} \zeta_3 \log  2 +\frac{477}{8} \zeta_3 \log  3 -\frac{73}{12} \log ^4 2 -\frac{1}{16} \log ^4 3 +\frac{229}{6} \log  3  \log ^3 2\\
&\nn +22 \log ^3 3  \log  2 -\frac{181}{4} \log ^2 3  \log ^2 2 -5 \pi ^2 \log ^2 2 -\frac{85}{16} \pi ^2 \log ^2 3 +\frac{7}{3} \pi ^2 \log  3  \log  2 +112 \text{Li}_4\left(\frac{1}{2}\right)\\
&\nn+\frac{477 \text{Li}_4\left(\frac{1}{3}\right)}{2}-\frac{231 \text{Li}_4\left(-\frac{1}{3}\right)}{4}+365 \text{Li}_4\left(-\frac{1}{2}\right)+\frac{11 \text{Li}_2\left(\frac{1}{3}\right)^2}{2}-\pi ^2 \text{Li}_2\left(\frac{1}{3}\right)-\frac{123 S_{2,2}\left(-\frac{1}{3}\right)}{4}\\
&\nn-157 S_{2,2}\left(-\frac{1}{2}\right)-\frac{397 \pi ^4}{720}\\
G&\left(-\frac{1}{3},-\frac{1}{2},-\frac{1}{3},0; 1\right) =  \frac{9}{2} \text{Li}_2\left(\frac{1}{3}\right) \log ^2 3 -30 \text{Li}_3\left(\frac{1}{3}\right) \log  2 +64 \text{Li}_3\left(-\frac{1}{2}\right) \log  2\\
&\nn +3 \text{Li}_2\left(\frac{1}{3}\right) \log  3  \log  2 +14 \text{Li}_3\left(\frac{1}{3}\right) \log  3 -62 \text{Li}_3\left(-\frac{1}{2}\right) \log  3 -\frac{47}{3} \zeta_3 \log  2 +\frac{239}{6} \zeta_3 \log  3\\
&\nn -\frac{37}{24} \log ^4 2 +\frac{5}{3} \log ^4 3 +\frac{31}{3} \log  3  \log ^3 2 +\frac{13}{2} \log ^3 3  \log  2 -\frac{31}{2} \log ^2 3  \log ^2 2 -\frac{1}{12} \pi ^2 \log ^2 2 -\frac{8}{3} \pi ^2 \log ^2 3 \\
&\nn+\pi ^2 \log  3  \log  2 +18 \text{Li}_4\left(\frac{1}{2}\right)+96 \text{Li}_4\left(\frac{1}{3}\right)-34 \text{Li}_4\left(-\frac{1}{3}\right)+139 \text{Li}_4\left(-\frac{1}{2}\right)+\frac{9 \text{Li}_2\left(\frac{1}{3}\right)^2}{2}\\
&\nn+\frac{\pi ^2 \text{Li}_2\left(\frac{1}{3}\right)}{3}-7 S_{2,2}\left(-\frac{1}{3}\right)-62 S_{2,2}\left(-\frac{1}{2}\right)-\frac{61 \pi ^4}{360}\\
G&\left(-\frac{1}{3},-\frac{1}{2},1,1; 1\right) =  -12 \text{Li}_2\left(\frac{1}{3}\right) \log ^2 2 +6 \text{Li}_2\left(\frac{1}{3}\right) \log ^2 3 +15 \text{Li}_3\left(\frac{1}{3}\right) \log  2 \\
&\nn+15 \text{Li}_3\left(-\frac{1}{2}\right) \log  2 +\frac{7}{2} \text{Li}_3\left(\frac{1}{3}\right) \log  3 +19 \text{Li}_3\left(-\frac{1}{2}\right) \log  3 +\frac{149}{4} \zeta_3 \log  2 -\frac{589}{24} \zeta_3 \log  3\\
&\nn -\frac{5}{12} \log ^4 2 +\frac{5}{12} \log ^4 3 +\frac{37}{6} \log  3  \log ^3 2 -2 \log ^3 3  \log  2 -\frac{11}{4} \log ^2 3  \log ^2 2 -\frac{3}{2} \pi ^2 \log ^2 2 +\frac{71}{48} \pi ^2 \log ^2 3 \\
&\nn-\frac{5}{6} \pi ^2 \log  3  \log  2 +52 \text{Li}_4\left(\frac{1}{2}\right)-\frac{59 \text{Li}_4\left(\frac{1}{3}\right)}{2}+\frac{63 \text{Li}_4\left(-\frac{1}{3}\right)}{4}+17 \text{Li}_4\left(-\frac{1}{2}\right)+6 \text{Li}_2\left(\frac{1}{3}\right)^2\\
&\nn-\frac{7 S_{2,2}\left(-\frac{1}{3}\right)}{4}+19 S_{2,2}\left(-\frac{1}{2}\right)+\frac{\pi ^4}{40}
\end{align}
\begin{align}
G&\left(-\frac{1}{3},-\frac{1}{3},-\frac{1}{2},-1; 1\right) =  -\text{Li}_2\left(\frac{1}{3}\right) \log ^2 2 -2 \text{Li}_2\left(\frac{1}{3}\right) \log ^2 3 +25 \text{Li}_3\left(\frac{1}{3}\right) \log  2 \\
&\nn-55 \text{Li}_3\left(-\frac{1}{2}\right) \log  2 -\frac{31}{2} \text{Li}_3\left(\frac{1}{3}\right) \log  3 +31 \text{Li}_3\left(-\frac{1}{2}\right) \log  3 -\frac{123}{4} \zeta_3 \log  2 -\frac{155}{24} \zeta_3 \log  3\\
&\nn -3 \log ^4 2 +\frac{7}{48} \log ^4 3 -\frac{16}{3} \log  3  \log ^3 2 -\frac{25}{6} \log ^3 3  \log  2 +\frac{29}{4} \log ^2 3  \log ^2 2 +\frac{17}{6} \pi ^2 \log ^2 2\\
&\nn +\frac{13}{16} \pi ^2 \log ^2 3 -\frac{1}{2} \pi ^2 \log  3  \log  2 -101 \text{Li}_4\left(\frac{1}{2}\right)-\frac{93 \text{Li}_4\left(\frac{1}{3}\right)}{2}+\frac{31 \text{Li}_4\left(-\frac{1}{3}\right)}{4}-160 \text{Li}_4\left(-\frac{1}{2}\right)\\
&\nn-2 \text{Li}_2\left(\frac{1}{3}\right)^2+\frac{\pi ^2 \text{Li}_2\left(\frac{1}{3}\right)}{3}+\frac{31 S_{2,2}\left(-\frac{1}{3}\right)}{4}+31 S_{2,2}\left(-\frac{1}{2}\right)+\frac{13 \pi ^4}{180}\\
G&\left(-\frac{1}{3},-\frac{1}{3},-\frac{1}{2},1; 1\right) =  2 \text{Li}_2\left(\frac{1}{3}\right) \log ^2 2 -6 \text{Li}_2\left(\frac{1}{3}\right) \log ^2 3 -15 \text{Li}_3\left(\frac{1}{3}\right) \log  2\\
&\nn -18 \text{Li}_3\left(-\frac{1}{2}\right) \log  2 -\frac{7}{2} \text{Li}_3\left(\frac{1}{3}\right) \log  3 -18 \text{Li}_3\left(-\frac{1}{2}\right) \log  3 -\frac{145}{4} \zeta_3 \log  2 +\frac{565}{24} \zeta_3 \log  3\\
&\nn -\frac{25}{8} \log ^4 2 -\frac{13}{24} \log ^4 3 -\frac{5}{3} \log  3  \log ^3 2 +\frac{5}{2} \log ^3 3  \log  2 -\frac{5}{4} \log ^2 3  \log ^2 2 +\frac{13}{6} \pi ^2 \log ^2 2\\
&\nn -\frac{71}{48} \pi ^2 \log ^2 3 +\pi ^2 \log  3  \log  2 -94 \text{Li}_4\left(\frac{1}{2}\right)+\frac{57 \text{Li}_4\left(\frac{1}{3}\right)}{2}-\frac{71 \text{Li}_4\left(-\frac{1}{3}\right)}{4}-71 \text{Li}_4\left(-\frac{1}{2}\right)\\
&\nn-6 \text{Li}_2\left(\frac{1}{3}\right)^2+\frac{7 S_{2,2}\left(-\frac{1}{3}\right)}{4}-18 S_{2,2}\left(-\frac{1}{2}\right)-\frac{4 \pi ^4}{45}\\
G&\left(-\frac{1}{3},0,-3,1; 1\right) =  8 \text{Li}_2\left(\frac{1}{3}\right) \log ^2 3 -28 \text{Li}_3\left(\frac{1}{3}\right) \log  2 +8 \text{Li}_3\left(-\frac{1}{2}\right) \log  2\\
&\nn -8 \text{Li}_2\left(\frac{1}{3}\right) \log  3  \log  2 +26 \text{Li}_3\left(\frac{1}{3}\right) \log  3 -8 \text{Li}_3\left(-\frac{1}{2}\right) \log  3 +57 \zeta_3 \log  2 -\frac{199}{6} \zeta_3 \log  3 +3 \log ^4 2\\
&\nn -\frac{5}{8} \log ^4 3 -\frac{4}{3} \log  3  \log ^3 2 +\frac{2}{3} \log ^3 3  \log  2 -\frac{4}{3} \pi ^2 \log ^2 2 +\frac{3}{2} \pi ^2 \log ^2 3 -\frac{4}{3} \pi ^2 \log  3  \log  2 +64 \text{Li}_4\left(\frac{1}{2}\right)\\
&\nn+6 \text{Li}_4\left(\frac{1}{3}\right)+22 \text{Li}_4\left(-\frac{1}{3}\right)+32 \text{Li}_4\left(-\frac{1}{2}\right)+8 \text{Li}_2\left(\frac{1}{3}\right)^2-\frac{2 \pi ^2 \text{Li}_2\left(\frac{1}{3}\right)}{3}-13 S_{2,2}\left(-\frac{1}{3}\right)\\
&\nn-8 S_{2,2}\left(-\frac{1}{2}\right)-\frac{9 \pi ^4}{40}\\
G&\left(-\frac{1}{3},0,-\frac{1}{2},1; 1\right) =  4 \text{Li}_2\left(\frac{1}{3}\right) \log ^2 2 -5 \text{Li}_2\left(\frac{1}{3}\right) \log ^2 3 -29 \text{Li}_3\left(\frac{1}{3}\right) \log  2\\
&\nn +20 \text{Li}_3\left(-\frac{1}{2}\right) \log  2 +3 \text{Li}_2\left(\frac{1}{3}\right) \log  3  \log  2 +\frac{9}{2} \text{Li}_3\left(\frac{1}{3}\right) \log  3 -40 \text{Li}_3\left(-\frac{1}{2}\right) \log  3 -\frac{93}{4} \zeta_3 \log  2\\
&\nn +\frac{263}{8} \zeta_3 \log  3 -\frac{49}{24} \log ^4 2 -\frac{29}{48} \log ^4 3 +\frac{8}{3} \log  3  \log ^3 2 +\frac{19}{3} \log ^3 3  \log  2 -\frac{27}{4} \log ^2 3  \log ^2 2 +\frac{1}{2} \pi ^2 \log ^2 2 \\
&\nn-\frac{101}{48} \pi ^2 \log ^2 3 +\frac{4}{3} \pi ^2 \log  3  \log  2 -38 \text{Li}_4\left(\frac{1}{2}\right)+\frac{129 \text{Li}_4\left(\frac{1}{3}\right)}{2}-\frac{97 \text{Li}_4\left(-\frac{1}{3}\right)}{4}+29 \text{Li}_4\left(-\frac{1}{2}\right)\\
&\nn-\frac{7 \text{Li}_2\left(\frac{1}{3}\right)^2}{2}+\frac{\pi ^2 \text{Li}_2\left(\frac{1}{3}\right)}{2}-\frac{9 S_{2,2}\left(-\frac{1}{3}\right)}{4}-40 S_{2,2}\left(-\frac{1}{2}\right)-\frac{131 \pi ^4}{720}\\
G&(0,0,1,2; 1) =  \frac{1}{8} \zeta_3 \log  2 -\frac{1}{24} \log ^4 2 +\frac{1}{24} \pi ^2 \log ^2 2 -\text{Li}_4\left(\frac{1}{2}\right)+\frac{\pi ^4}{288}
\end{align}
\begin{align}
G&(0,0,2,1; 1) =  \frac{1}{8} \log ^4 2 +3 \text{Li}_4\left(\frac{1}{2}\right)-\frac{11 \pi ^4}{720}\\
G&(0,1,0,2; 1) =  -\frac{1}{4} \zeta_3 \log  2 -\frac{1}{24} \log ^4 2 -\frac{1}{24} \pi ^2 \log ^2 2 -\text{Li}_4\left(\frac{1}{2}\right)+\frac{19 \pi ^4}{1440}\\
G&(0,1,1,2; 1) =  \frac{3}{4} \zeta_3 \log  2 +\frac{1}{12} \log ^4 2 -\frac{1}{12} \pi ^2 \log ^2 2 +2 \text{Li}_4\left(\frac{1}{2}\right)-\frac{\pi ^4}{60}\\
G&(0,1,2,0; 1) =  \frac{1}{8} \log ^4 2 -\frac{1}{24} \pi ^2 \log ^2 2 +3 \text{Li}_4\left(\frac{1}{2}\right)-\frac{29 \pi ^4}{1440}\\
G&(0,1,2,1; 1) =  -\frac{\pi ^4}{480}\\
G&(0,1,2,2; 1) =  \frac{5}{8} \zeta_3 \log  2 -\frac{1}{12} \pi ^2 \log ^2 2 -\frac{\pi ^4}{480}\\
G&(0,2,0,1; 1) =  -\frac{1}{8} \log ^4 2 -\frac{1}{8} \pi ^2 \log ^2 2 -3 \text{Li}_4\left(\frac{1}{2}\right)+\frac{7 \pi ^4}{288}\\
G&(0,2,1,0; 1) =  -\frac{1}{8} \log ^4 2 +\frac{1}{8} \pi ^2 \log ^2 2 -3 \text{Li}_4\left(\frac{1}{2}\right)+\frac{\pi ^4}{160}\\
G&(0,2,1,1; 1) =  -\frac{1}{12} \log ^4 2 +\frac{1}{12} \pi ^2 \log ^2 2 -2 \text{Li}_4\left(\frac{1}{2}\right)+\frac{\pi ^4}{180}\\
G&(0,2,1,2; 1) =  \zeta_3 \log  2 +\frac{1}{6} \log ^4 2 +\frac{1}{12} \pi ^2 \log ^2 2 +4 \text{Li}_4\left(\frac{1}{2}\right)-\frac{5 \pi ^4}{144}\\
G&(0,2,2,1; 1) =  -\frac{1}{12} \log ^4 2 -\frac{1}{24} \pi ^2 \log ^2 2 -2 \text{Li}_4\left(\frac{1}{2}\right)+\frac{\pi ^4}{80}\\
G&\left(\frac{4}{3},\frac{3}{2},\frac{3}{2},0; 1\right) =  \text{Li}_2\left(\frac{1}{3}\right) \log ^2 3 +30 \text{Li}_3\left(\frac{1}{3}\right) \log  2 -42 \text{Li}_3\left(-\frac{1}{2}\right) \log  2\\
&\nn -\text{Li}_2\left(\frac{1}{3}\right) \log  3  \log  2 -3 \text{Li}_3\left(\frac{1}{3}\right) \log  3 +42 \text{Li}_3\left(-\frac{1}{2}\right) \log  3 +\frac{79}{6} \zeta_3 \log  2 -\frac{149}{4} \zeta_3 \log  3\\
&\nn +\frac{23}{12} \log ^4 2 -\log ^4 3 -7 \log  3  \log ^3 2 -\frac{31}{6} \log ^3 3  \log  2 +10 \log ^2 3  \log ^2 2 -\frac{1}{6} \pi ^2 \log ^2 2 +\frac{21}{8} \pi ^2 \log ^2 3\\
&\nn -\frac{5}{6} \pi ^2 \log  3  \log  2 +4 \text{Li}_4\left(\frac{1}{2}\right)-68 \text{Li}_4\left(\frac{1}{3}\right)+\frac{49 \text{Li}_4\left(-\frac{1}{3}\right)}{2}-84 \text{Li}_4\left(-\frac{1}{2}\right)+\frac{\text{Li}_2\left(\frac{1}{3}\right)^2}{2}\\
&\nn-\frac{\pi ^2 \text{Li}_2\left(\frac{1}{3}\right)}{6}+\frac{3 S_{2,2}\left(-\frac{1}{3}\right)}{2}+42 S_{2,2}\left(-\frac{1}{2}\right)+\frac{\pi ^4}{15}\\
G&\left(\frac{4}{3},\frac{3}{2},\frac{3}{2},1; 1\right) =  -2 \text{Li}_2\left(\frac{1}{3}\right) \log ^2 3 +13 \text{Li}_3\left(\frac{1}{3}\right) \log  2 -23 \text{Li}_3\left(-\frac{1}{2}\right) \log  2 \\
&\nn-\frac{13}{2} \text{Li}_3\left(\frac{1}{3}\right) \log  3 +23 \text{Li}_3\left(-\frac{1}{2}\right) \log  3 +\frac{29}{12} \zeta_3 \log  2 -\frac{305}{24} \zeta_3 \log  3 +\frac{17}{24} \log ^4 2 -\frac{1}{2} \log ^4 3\\
&\nn -\frac{23}{6} \log  3  \log ^3 2 -\frac{13}{6} \log ^3 3  \log  2 +\frac{21}{4} \log ^2 3  \log ^2 2 +\frac{1}{4} \pi ^2 \log ^2 2 +\frac{17}{16} \pi ^2 \log ^2 3 -\frac{2}{3} \pi ^2 \log  3  \log  2\\
&\nn -6 \text{Li}_4\left(\frac{1}{2}\right)-\frac{69 \text{Li}_4\left(\frac{1}{3}\right)}{2}+\frac{47 \text{Li}_4\left(-\frac{1}{3}\right)}{4}-46 \text{Li}_4\left(-\frac{1}{2}\right)-2 \text{Li}_2\left(\frac{1}{3}\right)^2+\frac{\pi ^2 \text{Li}_2\left(\frac{1}{3}\right)}{6}\\
&\nn+\frac{13 S_{2,2}\left(-\frac{1}{3}\right)}{4}+23 S_{2,2}\left(-\frac{1}{2}\right)+\frac{13 \pi ^4}{180}
\end{align}
\begin{align}
G&\left(\frac{4}{3},\frac{3}{2},2,0; 1\right) =  \frac{15}{2} \text{Li}_2\left(\frac{1}{3}\right) \log ^2 2 +\text{Li}_2\left(\frac{1}{3}\right) \log ^2 3 -3 \text{Li}_3\left(\frac{1}{3}\right) \log  2\\
&\nn -27 \text{Li}_3\left(-\frac{1}{2}\right) \log  2 -2 \text{Li}_2\left(\frac{1}{3}\right) \log  3  \log  2 +\frac{5}{2} \text{Li}_3\left(\frac{1}{3}\right) \log  3 -5 \text{Li}_3\left(-\frac{1}{2}\right) \log  3 -\frac{283}{12} \zeta_3 \log  2\\
&\nn +\frac{25}{24} \zeta_3 \log  3 +\frac{3}{2} \log ^4 2 +\frac{7}{32} \log ^4 3 -7 \log  3  \log ^3 2 -\frac{1}{2} \log ^3 3  \log  2 +\frac{7}{2} \log ^2 3  \log ^2 2 +\frac{4}{3} \pi ^2 \log ^2 2\\
&\nn -\frac{3}{8} \pi ^2 \log ^2 3 +\frac{5}{6} \pi ^2 \log  3  \log  2 -38 \text{Li}_4\left(\frac{1}{2}\right)+9 \text{Li}_4\left(\frac{1}{3}\right)-\text{Li}_4\left(-\frac{1}{3}\right)-39 \text{Li}_4\left(-\frac{1}{2}\right)\\
&\nn+\text{Li}_2\left(\frac{1}{3}\right)^2-\frac{7 \pi ^2 \text{Li}_2\left(\frac{1}{3}\right)}{12}-\frac{5 S_{2,2}\left(-\frac{1}{3}\right)}{4}-5 S_{2,2}\left(-\frac{1}{2}\right)-\frac{13 \pi ^4}{480}\\
G&\left(\frac{4}{3},2,0,0; 1\right) =  2 \text{Li}_2\left(\frac{1}{3}\right) \log ^2 2 -\text{Li}_2\left(\frac{1}{3}\right) \log ^2 3 -6 \text{Li}_3\left(\frac{1}{3}\right) \log  2\\
&\nn +6 \text{Li}_3\left(-\frac{1}{2}\right) \log  2 +2 \text{Li}_3\left(\frac{1}{3}\right) \log  3 -8 \text{Li}_3\left(-\frac{1}{2}\right) \log  3 +\frac{1}{3} \zeta_3 \log  2 +\frac{29}{6} \zeta_3 \log  3 +\frac{1}{24} \log ^4 2 \\
&\nn-\frac{1}{6} \log ^4 3 +\frac{2}{3} \log  3  \log ^3 2 +\log ^3 3  \log  2 -\log ^2 3  \log ^2 2 -\frac{1}{6} \pi ^2 \log ^2 2 -\frac{1}{4} \pi ^2 \log ^2 3 +12 \text{Li}_4\left(\frac{1}{3}\right)\\
&\nn-3 \text{Li}_4\left(-\frac{1}{3}\right)+14 \text{Li}_4\left(-\frac{1}{2}\right)-\text{Li}_2\left(\frac{1}{3}\right)^2-S_{2,2}\left(-\frac{1}{3}\right)-8 S_{2,2}\left(-\frac{1}{2}\right)-\frac{7 \pi ^4}{360}\\
G&\left(\frac{4}{3},2,\frac{4}{3},1; 1\right) =  \frac{1}{2} \text{Li}_2\left(\frac{1}{3}\right) \log ^2 2 +3 \text{Li}_2\left(\frac{1}{3}\right) \log ^2 3 +9 \text{Li}_3\left(\frac{1}{3}\right) \log  2\\
&\nn -5 \text{Li}_3\left(-\frac{1}{2}\right) \log  2 -\frac{3}{2} \text{Li}_3\left(\frac{1}{3}\right) \log  3 +3 \text{Li}_3\left(-\frac{1}{2}\right) \log  3 -\frac{29}{6} \zeta_3 \log  2 -\frac{5}{8} \zeta_3 \log  3 -\frac{1}{12} \log ^4 2\\
&\nn +\frac{13}{16} \log ^4 3 -\frac{1}{3} \log  3  \log ^3 2 -\frac{3}{2} \log ^3 3  \log  2 +\log ^2 3  \log ^2 2 +\frac{1}{6} \pi ^2 \log ^2 2 -\frac{1}{48} \pi ^2 \log ^2 3 +\frac{1}{6} \pi ^2 \log  3  \log  2\\
&\nn -6 \text{Li}_4\left(\frac{1}{2}\right)-\frac{9 \text{Li}_4\left(\frac{1}{3}\right)}{2}+\frac{3 \text{Li}_4\left(-\frac{1}{3}\right)}{4}-13 \text{Li}_4\left(-\frac{1}{2}\right)+3 \text{Li}_2\left(\frac{1}{3}\right)^2-\frac{\pi ^2 \text{Li}_2\left(\frac{1}{3}\right)}{6}\\
&\nn+\frac{3 S_{2,2}\left(-\frac{1}{3}\right)}{4}+3 S_{2,2}\left(-\frac{1}{2}\right)-\frac{11 \pi ^4}{720}\\
G&\left(\frac{4}{3},2,\frac{3}{2},1; 1\right) =  -\frac{1}{2} \text{Li}_2\left(\frac{1}{3}\right) \log ^2 2 +\frac{3}{2} \text{Li}_2\left(\frac{1}{3}\right) \log ^2 3 -15 \text{Li}_3\left(\frac{1}{3}\right) \log  2\\
&\nn +27 \text{Li}_3\left(-\frac{1}{2}\right) \log  2 -\text{Li}_2\left(\frac{1}{3}\right) \log  3  \log  2 +\frac{21}{2} \text{Li}_3\left(\frac{1}{3}\right) \log  3 -11 \text{Li}_3\left(-\frac{1}{2}\right) \log  3 +\frac{377}{12} \zeta_3 \log  2\\
&\nn -\frac{45}{8} \zeta_3 \log  3 +\frac{5}{4} \log ^4 2 -\frac{1}{3} \log ^4 3 +\frac{7}{3} \log  3  \log ^3 2 +2 \log ^3 3  \log  2 -3 \log ^2 3  \log ^2 2 -\frac{41}{24} \pi ^2 \log ^2 2\\
&\nn +\frac{13}{48} \pi ^2 \log ^2 3 -\frac{1}{2} \pi ^2 \log  3  \log  2 +57 \text{Li}_4\left(\frac{1}{2}\right)+\frac{33 \text{Li}_4\left(\frac{1}{3}\right)}{2}+\frac{13 \text{Li}_4\left(-\frac{1}{3}\right)}{4}+70 \text{Li}_4\left(-\frac{1}{2}\right)\\
&\nn+\frac{3 \text{Li}_2\left(\frac{1}{3}\right)^2}{2}-\frac{21 S_{2,2}\left(-\frac{1}{3}\right)}{4}-11 S_{2,2}\left(-\frac{1}{2}\right)-\frac{61 \pi ^4}{720}\\
G&\left(\frac{4}{3},2,4,0; 1\right) =  -2 \text{Li}_2\left(\frac{1}{3}\right) \log ^2 2 -5 \text{Li}_2\left(\frac{1}{3}\right) \log ^2 3 +17 \text{Li}_3\left(\frac{1}{3}\right) \log  2 \\
&\nn-5 \text{Li}_3\left(-\frac{1}{2}\right) \log  2 +6 \text{Li}_2\left(\frac{1}{3}\right) \log  3  \log  2 -\frac{23}{2} \text{Li}_3\left(\frac{1}{3}\right) \log  3 +3 \text{Li}_3\left(-\frac{1}{2}\right) \log  3 -\frac{385}{12} \zeta_3 \log  2\\
&\nn +\frac{365}{24} \zeta_3 \log  3 -\frac{23}{24} \log ^4 2 -\frac{1}{24} \log ^4 3 -\frac{7}{6} \log  3  \log ^3 2 +\frac{1}{6} \log ^3 3  \log  2 -\frac{1}{4} \log ^2 3  \log ^2 2 +\frac{31}{24} \pi ^2 \log ^2 2
\end{align}
\begin{align}
&\nn -\frac{9}{16} \pi ^2 \log ^2 3 +\frac{2}{3} \pi ^2 \log  3  \log  2 -17 \text{Li}_4\left(\frac{1}{2}\right)-\frac{3 \text{Li}_4\left(\frac{1}{3}\right)}{2}-\frac{39 \text{Li}_4\left(-\frac{1}{3}\right)}{4}-2 \text{Li}_4\left(-\frac{1}{2}\right)\\
&\nn-5 \text{Li}_2\left(\frac{1}{3}\right)^2+\frac{\pi ^2 \text{Li}_2\left(\frac{1}{3}\right)}{2}+\frac{23 S_{2,2}\left(-\frac{1}{3}\right)}{4}+3 S_{2,2}\left(-\frac{1}{2}\right)+\frac{59 \pi ^4}{720}\\
G&\left(\frac{3}{2},1,2,0; 1\right) =  7 \text{Li}_3\left(\frac{1}{3}\right) \log  2 -4 \text{Li}_3\left(-\frac{1}{2}\right) \log  2 -\frac{7}{2} \text{Li}_3\left(\frac{1}{3}\right) \log  3\\
&\nn +7 \text{Li}_3\left(-\frac{1}{2}\right) \log  3 -\frac{85}{12} \zeta_3 \log  2 +\frac{5}{4} \zeta_3 \log  3 -\frac{5}{24} \log ^4 2 +\frac{19}{96} \log ^4 3 -\frac{7}{6} \log  3  \log ^3 2 -\frac{7}{6} \log ^3 3  \log  2\\
&\nn +\frac{7}{4} \log ^2 3  \log ^2 2 +\frac{3}{8} \pi ^2 \log ^2 2 +\frac{1}{24} \pi ^2 \log ^2 3 -7 \text{Li}_4\left(\frac{1}{2}\right)-8 \text{Li}_4\left(\frac{1}{3}\right)+\frac{\text{Li}_4\left(-\frac{1}{3}\right)}{2}\\
&\nn-10 \text{Li}_4\left(-\frac{1}{2}\right)+\frac{7 S_{2,2}\left(-\frac{1}{3}\right)}{4}+4 S_{2,2}\left(-\frac{1}{2}\right)+\frac{47 \pi ^4}{720}\\
G&\left(\frac{3}{2},2,0,1; 1\right) =  -\text{Li}_2\left(\frac{1}{3}\right) \log ^2 3 +4 \text{Li}_3\left(\frac{1}{3}\right) \log  2 -8 \text{Li}_3\left(-\frac{1}{2}\right) \log  2\\
&\nn +2 \text{Li}_2\left(\frac{1}{3}\right) \log  3  \log  2 -2 \text{Li}_3\left(\frac{1}{3}\right) \log  3 +4 \text{Li}_3\left(-\frac{1}{2}\right) \log  3 +\frac{5}{3} \zeta_3 \log  2 -\frac{137}{24} \zeta_3 \log  3 +\frac{5}{24} \log ^4 2\\
&\nn -\frac{25}{96} \log ^4 3 -\frac{2}{3} \log  3  \log ^3 2 +\frac{1}{3} \log ^3 3  \log  2 +\frac{1}{4} \pi ^2 \log ^2 2 +\frac{29}{48} \pi ^2 \log ^2 3 -\frac{2}{3} \pi ^2 \log  3  \log  2 -4 \text{Li}_4\left(\frac{1}{2}\right)\\
&\nn-\frac{21 \text{Li}_4\left(\frac{1}{3}\right)}{2}+\frac{13 \text{Li}_4\left(-\frac{1}{3}\right)}{4}-15 \text{Li}_4\left(-\frac{1}{2}\right)-\text{Li}_2\left(\frac{1}{3}\right)^2+\frac{2 \pi ^2 \text{Li}_2\left(\frac{1}{3}\right)}{3}+S_{2,2}\left(-\frac{1}{3}\right)\\
&\nn+8 S_{2,2}\left(-\frac{1}{2}\right)-\frac{\pi ^4}{90}\\
G&\left(\frac{3}{2},2,1,0; 1\right) =  -3 \text{Li}_3\left(\frac{1}{3}\right) \log  2 +4 \text{Li}_3\left(-\frac{1}{2}\right) \log  2 +\frac{3}{2} \text{Li}_3\left(\frac{1}{3}\right) \log  3\\
&\nn -3 \text{Li}_3\left(-\frac{1}{2}\right) \log  3 +\frac{3}{4} \zeta_3 \log  2 +\frac{67}{24} \zeta_3 \log  3 -\frac{1}{24} \log ^4 2 -\frac{1}{48} \log ^4 3 +\frac{1}{2} \log  3  \log ^3 2 +\frac{1}{2} \log ^3 3  \log  2\\
&\nn -\frac{3}{4} \log ^2 3  \log ^2 2 -\frac{7}{48} \pi ^2 \log ^2 3 +2 \text{Li}_4\left(\frac{1}{2}\right)+\frac{13 \text{Li}_4\left(\frac{1}{3}\right)}{2}-\frac{7 \text{Li}_4\left(-\frac{1}{3}\right)}{4}+9 \text{Li}_4\left(-\frac{1}{2}\right)\\
&\nn-\frac{3 S_{2,2}\left(-\frac{1}{3}\right)}{4}-4 S_{2,2}\left(-\frac{1}{2}\right)-\frac{3 \pi ^4}{160}\\
G&(2,0,0,1; 1) =  \frac{7}{8} \zeta_3 \log  2 +\frac{1}{24} \log ^4 2 +\frac{1}{12} \pi ^2 \log ^2 2 +\text{Li}_4\left(\frac{1}{2}\right)-\frac{\pi ^4}{80}\\
G&(2,0,1,0; 1) =  -\frac{7}{4} \zeta_3 \log  2 +\frac{1}{24} \log ^4 2 -\frac{1}{24} \pi ^2 \log ^2 2 +\text{Li}_4\left(\frac{1}{2}\right)+\frac{\pi ^4}{1440}\\
G&(2,0,1,1; 1) =  \frac{19 \pi ^4}{1440}-\frac{7}{4} \zeta_3 \log  2 \\
G&(2,0,1,2; 1) =  -\frac{13}{8} \zeta_3 \log  2 -\frac{1}{6} \log ^4 2 -\frac{1}{12} \pi ^2 \log ^2 2 -4 \text{Li}_4\left(\frac{1}{2}\right)+\frac{7 \pi ^4}{180}\\
G&(2,0,2,1; 1) =  -\frac{21}{8} \zeta_3 \log  2 +\frac{1}{4} \pi ^2 \log ^2 2 +\frac{7 \pi ^4}{720}\\
G&(2,1,0,0; 1) =  \frac{7}{8} \zeta_3 \log  2 +\frac{1}{24} \log ^4 2 -\frac{1}{24} \pi ^2 \log ^2 2 +\text{Li}_4\left(\frac{1}{2}\right)-\frac{\pi ^4}{288}\\
G&(2,1,0,1; 1) =  \frac{7}{2} \zeta_3 \log  2 +\frac{1}{6} \log ^4 2 -\frac{1}{6} \pi ^2 \log ^2 2 +4 \text{Li}_4\left(\frac{1}{2}\right)-\frac{71 \pi ^4}{1440}
\end{align}
\begin{align}
G&(2,1,0,2; 1) =  \frac{13}{4} \zeta_3 \log  2 +\frac{1}{6} \log ^4 2 -\frac{1}{8} \pi ^2 \log ^2 2 +4 \text{Li}_4\left(\frac{1}{2}\right)-\frac{11 \pi ^4}{240}\\
G&(2,1,1,0; 1) =  -\frac{7}{4} \zeta_3 \log  2 -\frac{1}{12} \log ^4 2 +\frac{1}{12} \pi ^2 \log ^2 2 -2 \text{Li}_4\left(\frac{1}{2}\right)+\frac{11 \pi ^4}{360}\\
G&(2,1,1,1; 1) =  \frac{7 \pi ^4}{720}\\
G&(2,1,1,2; 1) =  \frac{3}{4} \zeta_3 \log  2 -\frac{\pi ^4}{288}\\
G&(2,1,2,0; 1) =  -\frac{21}{8} \zeta_3 \log  2 -\frac{1}{6} \log ^4 2 +\frac{1}{8} \pi ^2 \log ^2 2 -4 \text{Li}_4\left(\frac{1}{2}\right)+\frac{\pi ^4}{24}\\
G&(2,1,2,1; 1) =  -\frac{7}{2} \zeta_3 \log  2 -\frac{1}{6} \log ^4 2 +\frac{1}{6} \pi ^2 \log ^2 2 -4 \text{Li}_4\left(\frac{1}{2}\right)+\frac{13 \pi ^4}{288}\\
G&(2,1,2,2; 1) =  -\frac{23}{8} \zeta_3 \log  2 -\frac{1}{8} \log ^4 2 +\frac{1}{6} \pi ^2 \log ^2 2 -3 \text{Li}_4\left(\frac{1}{2}\right)+\frac{\pi ^4}{30}\\
G&(2,2,0,1; 1) =  \frac{21}{8} \zeta_3 \log  2 +\frac{1}{12} \log ^4 2 -\frac{5}{24} \pi ^2 \log ^2 2 +2 \text{Li}_4\left(\frac{1}{2}\right)-\frac{7 \pi ^4}{288}\\
G&(2,2,1,0; 1) =  \frac{\pi ^4}{480}\\
G&(2,2,1,1; 1) =  \frac{7}{4} \zeta_3 \log  2 +\frac{1}{12} \log ^4 2 -\frac{1}{12} \pi ^2 \log ^2 2 +2 \text{Li}_4\left(\frac{1}{2}\right)-\frac{\pi ^4}{48}\\
G&(2,2,1,2; 1) =  \frac{11}{4} \zeta_3 \log  2 +\frac{1}{8} \log ^4 2 -\frac{1}{8} \pi ^2 \log ^2 2 +3 \text{Li}_4\left(\frac{1}{2}\right)-\frac{\pi ^4}{30}\\
G&(2,2,2,1; 1) =  -\frac{7}{8} \zeta_3 \log  2 -\frac{1}{24} \log ^4 2 +\frac{1}{24} \pi ^2 \log ^2 2 -\text{Li}_4\left(\frac{1}{2}\right)+\frac{\pi ^4}{90}
\end{align}


\section{Goncharov multiple polylogarithm in terms of harmonic polylogarithms}
\label{app:GPL_HPL}
If we consider Goncharov polylogarithms of the form $G(\vec w(a);1)$, it is sometimes possible to reexpress this function in terms of harmonic polylogarithms in $a$. In the following we present the identities we encountered throughout our computation.

\subsection{Polylogarithms of weight one}
\beq
G(a;1)  =  -H\left(1;{1\over a}\right)
\eeq

\subsection{Polylogarithms of weight two}
\begin{align}
G&(0,a; 1) = 
-H\left(0,1; \frac{1}{a}\right)\\
G&(-1,a; 1) = 
\log  2\,  H\left(-1; \frac{1}{a}\right)+H\left(-1,1; \frac{1}{a}\right)-H\left(0,1; \frac{1}{a}\right)\\
G&(a,0; 1) = 
H\left(0,1; \frac{1}{a}\right)\\
G&(-a,-1; 1) = 
\log  2\,  H\left(-1; \frac{1}{a}\right)+\log  2\,  H\left(1; \frac{1}{a}\right)-H\left(0,-1; \frac{1}{a}\right)-H\left(1,-1; \frac{1}{a}\right)
\end{align}
\begin{align}
G&(a,-1; 1) = 
-\log  2\,  H\left(-1; \frac{1}{a}\right)-\log  2\,  H\left(1; \frac{1}{a}\right)-H\left(-1,1; \frac{1}{a}\right)+H\left(0,1; \frac{1}{a}\right)\\
G&(a,1; 1) = 
H\left(0,1; \frac{1}{a}\right)+H\left(1,1; \frac{1}{a}\right)\\
G&\left(a^2,-a; 1\right) = 
H\left(-1,-1; \frac{1}{a}\right)-H\left(-1,1; \frac{1}{a}\right)-H\left(0,-1; \frac{1}{a}\right)+H\left(0,1; \frac{1}{a}\right)\\
&\nn-H\left(1,-1; \frac{1}{a}\right)\\
G&\left(a^2,a; 1\right) = 
-H\left(-1,1; \frac{1}{a}\right)-H\left(0,-1; \frac{1}{a}\right)+H\left(0,1; \frac{1}{a}\right)-H\left(1,-1; \frac{1}{a}\right)\\
&\nn+H\left(1,1; \frac{1}{a}\right)\\
G&(-a,a; 1) = 
-H\left(-1,1; \frac{1}{a}\right)\\
G&(a,-a; 1) = 
-H\left(1,-1; \frac{1}{a}\right)\\
G&(a,a; 1) = 
H\left(1,1; \frac{1}{a}\right)\\
G&\left(-a,a^2; 1\right) = 
H\left(-1,-1; \frac{1}{a}\right)+H\left(0,-1; \frac{1}{a}\right)-H\left(0,1; \frac{1}{a}\right)\\
G&\left(a,a^2; 1\right) = 
H\left(0,-1; \frac{1}{a}\right)-H\left(0,1; \frac{1}{a}\right)+H\left(1,1; \frac{1}{a}\right)
\end{align}

\subsection{Polylogarithms of weight three}
\begin{align}
G&(0,0,a; 1) = 
-H\left(0,0,1; \frac{1}{a}\right)\\
G&(0,1,a; 1) = 
\frac{1}{6} \pi ^2 H\left(1; \frac{1}{a}\right)-H\left(0,0,1; \frac{1}{a}\right)-H\left(1,0,1; \frac{1}{a}\right)\\
G&(0,a,0; 1) = 
2 H\left(0,0,1; \frac{1}{a}\right)\\
G&(0,-a,-1; 1) = 
\log  2\,  H\left(0,-1; \frac{1}{a}\right)+\log  2\,  H\left(0,1; \frac{1}{a}\right)+\frac{1}{12} \pi ^2 H\left(1; \frac{1}{a}\right)\\
&\nn-2 H\left(0,0,-1; \frac{1}{a}\right)-H\left(0,1,-1; \frac{1}{a}\right)-H\left(1,0,-1; \frac{1}{a}\right)\\
G&(0,a,-1; 1) = 
-\log  2\,  H\left(0,-1; \frac{1}{a}\right)-\log  2\,  H\left(0,1; \frac{1}{a}\right)-\frac{1}{12} \pi ^2 H\left(-1; \frac{1}{a}\right)\\
&\nn-H\left(-1,0,1; \frac{1}{a}\right)-H\left(0,-1,1; \frac{1}{a}\right)+2 H\left(0,0,1; \frac{1}{a}\right)\\
G&(0,a,1; 1) = 
-\frac{1}{6} \pi ^2 H\left(1; \frac{1}{a}\right)+2 H\left(0,0,1; \frac{1}{a}\right)+H\left(0,1,1; \frac{1}{a}\right)\\
&\nn+H\left(1,0,1; \frac{1}{a}\right)
\end{align}
\begin{align}
G&\left(0,a^2,-a; 1\right) = 
H\left(-1,0,-1; \frac{1}{a}\right)-2 H\left(-1,0,1; \frac{1}{a}\right)+2 H\left(0,-1,-1; \frac{1}{a}\right)\\
&\nn-2 H\left(0,-1,1; \frac{1}{a}\right)-4 H\left(0,0,-1; \frac{1}{a}\right)+4 H\left(0,0,1; \frac{1}{a}\right)-2 H\left(0,1,-1; \frac{1}{a}\right)\\
G&\left(0,a^2,a; 1\right) = 
-2 H\left(0,-1,1; \frac{1}{a}\right)-4 H\left(0,0,-1; \frac{1}{a}\right)+4 H\left(0,0,1; \frac{1}{a}\right)\\
&\nn-2 H\left(0,1,-1; \frac{1}{a}\right)+2 H\left(0,1,1; \frac{1}{a}\right)-2 H\left(1,0,-1; \frac{1}{a}\right)+H\left(1,0,1; \frac{1}{a}\right)\\
G&(0,-a,a; 1) = 
-H\left(0,-1,1; \frac{1}{a}\right)\\
G&(0,a,-a; 1) = 
-H\left(0,1,-1; \frac{1}{a}\right)\\
G&(0,a,a; 1) = 
H\left(0,1,1; \frac{1}{a}\right)\\
G&\left(0,-a,a^2; 1\right) = 
-H\left(-1,0,-1; \frac{1}{a}\right)+2 H\left(-1,0,1; \frac{1}{a}\right)+H\left(0,-1,-1; \frac{1}{a}\right)\\
&\nn+3 H\left(0,0,-1; \frac{1}{a}\right)-3 H\left(0,0,1; \frac{1}{a}\right)\\
G&\left(0,a,a^2; 1\right) = 
3 H\left(0,0,-1; \frac{1}{a}\right)-3 H\left(0,0,1; \frac{1}{a}\right)+H\left(0,1,1; \frac{1}{a}\right)\\
&\nn+2 H\left(1,0,-1; \frac{1}{a}\right)-H\left(1,0,1; \frac{1}{a}\right)\\
G&(-a,0,-1; 1) = 
-\log  2\,  H\left(0,-1; \frac{1}{a}\right)-\log  2\,  H\left(0,1; \frac{1}{a}\right)+\frac{1}{12} \pi ^2 H\left(-1; \frac{1}{a}\right)\\
&\nn+H\left(0,0,-1; \frac{1}{a}\right)+H\left(0,1,-1; \frac{1}{a}\right)\\
G&(a,0,-1; 1) = 
\log  2\,  H\left(0,-1; \frac{1}{a}\right)+\log  2\,  H\left(0,1; \frac{1}{a}\right)-\frac{1}{12} \pi ^2 H\left(1; \frac{1}{a}\right)\\
&\nn+H\left(0,-1,1; \frac{1}{a}\right)-H\left(0,0,1; \frac{1}{a}\right)\\
G&(a,0,1; 1) = 
\frac{1}{6} \pi ^2 H\left(1; \frac{1}{a}\right)-H\left(0,0,1; \frac{1}{a}\right)-H\left(0,1,1; \frac{1}{a}\right)\\
G&(-a,0,a; 1) = 
-H\left(-1,0,1; \frac{1}{a}\right)\\
G&(a,0,-a; 1) = 
-H\left(1,0,-1; \frac{1}{a}\right)\\
G&(a,0,a; 1) = 
H\left(1,0,1; \frac{1}{a}\right)\\
G&\left(-a,0,a^2; 1\right) = 
2 H\left(-1,0,-1; \frac{1}{a}\right)-2 H\left(-1,0,1; \frac{1}{a}\right)+H\left(0,-1,-1; \frac{1}{a}\right)\\
&\nn+H\left(0,0,-1; \frac{1}{a}\right)-H\left(0,0,1; \frac{1}{a}\right)
\end{align}
\begin{align}
G&\left(a,0,a^2; 1\right) = 
H\left(0,0,-1; \frac{1}{a}\right)-H\left(0,0,1; \frac{1}{a}\right)+H\left(0,1,1; \frac{1}{a}\right)\\
&\nn-2 H\left(1,0,-1; \frac{1}{a}\right)+2 H\left(1,0,1; \frac{1}{a}\right)\\
G&(a,1,0; 1) = 
-H\left(0,0,1; \frac{1}{a}\right)-H\left(1,0,1; \frac{1}{a}\right)\\
G&(a,1,1; 1) = 
-H\left(0,0,1; \frac{1}{a}\right)-H\left(0,1,1; \frac{1}{a}\right)-H\left(1,0,1; \frac{1}{a}\right)-H\left(1,1,1; \frac{1}{a}\right)\\
G&(-a,1,-a; 1) = 
H\left(-1,-1,-1; \frac{1}{a}\right)-H\left(-1,0,-1; \frac{1}{a}\right)\\
G&(-a,1,a; 1) = 
-H\left(-1,-1,1; \frac{1}{a}\right)-H\left(1,-1,-1; \frac{1}{a}\right)-H\left(1,-1,1; \frac{1}{a}\right)\\
&\nn+H\left(1,0,-1; \frac{1}{a}\right)\\
G&(a,1,-a; 1) = 
H\left(-1,0,1; \frac{1}{a}\right)+H\left(-1,1,-1; \frac{1}{a}\right)+H\left(-1,1,1; \frac{1}{a}\right)\\
&\nn+H\left(1,1,-1; \frac{1}{a}\right)\\
G&(a,1,a; 1) = 
-H\left(1,0,1; \frac{1}{a}\right)-H\left(1,1,1; \frac{1}{a}\right)\\
G&\left(-a,1,a^2; 1\right) = 
H\left(-1,-1,-1; \frac{1}{a}\right)-H\left(-1,0,-1; \frac{1}{a}\right)+H\left(0,-1,-1; \frac{1}{a}\right)\\
&\nn+H\left(0,0,-1; \frac{1}{a}\right)-H\left(0,0,1; \frac{1}{a}\right)+2 H\left(1,0,-1; \frac{1}{a}\right)-H\left(1,0,1; \frac{1}{a}\right)\\
G&\left(a,1,a^2; 1\right) = 
-H\left(-1,0,-1; \frac{1}{a}\right)+2 H\left(-1,0,1; \frac{1}{a}\right)+H\left(0,0,-1; \frac{1}{a}\right)\\
&\nn-H\left(0,0,1; \frac{1}{a}\right)+H\left(0,1,1; \frac{1}{a}\right)-H\left(1,0,1; \frac{1}{a}\right)-H\left(1,1,1; \frac{1}{a}\right)\\
G&\left(a^2,0,-a; 1\right) = 
H\left(-1,0,-1; \frac{1}{a}\right)-H\left(0,-1,-1; \frac{1}{a}\right)+H\left(0,-1,1; \frac{1}{a}\right)\\
&\nn+H\left(0,0,-1; \frac{1}{a}\right)-H\left(0,0,1; \frac{1}{a}\right)+H\left(0,1,-1; \frac{1}{a}\right)-H\left(1,0,-1; \frac{1}{a}\right)\\
G&\left(a^2,0,a; 1\right) = 
-H\left(-1,0,1; \frac{1}{a}\right)+H\left(0,-1,1; \frac{1}{a}\right)+H\left(0,0,-1; \frac{1}{a}\right)\\
&\nn-H\left(0,0,1; \frac{1}{a}\right)+H\left(0,1,-1; \frac{1}{a}\right)-H\left(0,1,1; \frac{1}{a}\right)+H\left(1,0,1; \frac{1}{a}\right)\\
G&\left(a^2,-a,0; 1\right) = 
-2 H\left(-1,0,-1; \frac{1}{a}\right)+2 H\left(-1,0,1; \frac{1}{a}\right)-H\left(0,-1,-1; \frac{1}{a}\right)\\
&\nn+H\left(0,-1,1; \frac{1}{a}\right)+3 H\left(0,0,-1; \frac{1}{a}\right)-3 H\left(0,0,1; \frac{1}{a}\right)+H\left(0,1,-1; \frac{1}{a}\right)\\
&\nn+H\left(1,0,-1; \frac{1}{a}\right)
\end{align}
\begin{align}
G&\left(a^2,a,0; 1\right) = 
H\left(-1,0,1; \frac{1}{a}\right)+H\left(0,-1,1; \frac{1}{a}\right)+3 H\left(0,0,-1; \frac{1}{a}\right)\\
&\nn-3 H\left(0,0,1; \frac{1}{a}\right)+H\left(0,1,-1; \frac{1}{a}\right)-H\left(0,1,1; \frac{1}{a}\right)+2 H\left(1,0,-1; \frac{1}{a}\right)\\
&\nn-2 H\left(1,0,1; \frac{1}{a}\right)\\
G&\left(a^2,-a,1; 1\right) = 
H\left(-1,-1,-1; \frac{1}{a}\right)-H\left(-1,-1,1; \frac{1}{a}\right)-H\left(-1,0,-1; \frac{1}{a}\right)\\
&\nn+H\left(-1,0,1; \frac{1}{a}\right)-H\left(-1,1,-1; \frac{1}{a}\right)-2 H\left(0,-1,-1; \frac{1}{a}\right)+2 H\left(0,-1,1; \frac{1}{a}\right)\\
&\nn+3 H\left(0,0,-1; \frac{1}{a}\right)-3 H\left(0,0,1; \frac{1}{a}\right)+2 H\left(0,1,-1; \frac{1}{a}\right)-H\left(0,1,1; \frac{1}{a}\right)\\
&\nn-H\left(1,-1,-1; \frac{1}{a}\right)+H\left(1,0,-1; \frac{1}{a}\right)\\
G&\left(a^2,a,1; 1\right) = 
H\left(-1,0,1; \frac{1}{a}\right)+H\left(-1,1,1; \frac{1}{a}\right)-H\left(0,-1,-1; \frac{1}{a}\right)\\
&\nn+2 H\left(0,-1,1; \frac{1}{a}\right)+3 H\left(0,0,-1; \frac{1}{a}\right)-3 H\left(0,0,1; \frac{1}{a}\right)+2 H\left(0,1,-1; \frac{1}{a}\right)\\
&\nn-2 H\left(0,1,1; \frac{1}{a}\right)+H\left(1,-1,1; \frac{1}{a}\right)+H\left(1,0,-1; \frac{1}{a}\right)-H\left(1,0,1; \frac{1}{a}\right)\\
&\nn+H\left(1,1,-1; \frac{1}{a}\right)-H\left(1,1,1; \frac{1}{a}\right)\\
G&\left(a^2,a^2,-a; 1\right) = 
H\left(-1,-1,-1; \frac{1}{a}\right)-H\left(-1,-1,1; \frac{1}{a}\right)-H\left(-1,1,-1; \frac{1}{a}\right)\\
&\nn+H\left(-1,1,1; \frac{1}{a}\right)-H\left(0,-1,-1; \frac{1}{a}\right)+H\left(0,-1,1; \frac{1}{a}\right)+H\left(0,1,-1; \frac{1}{a}\right)\\
&\nn-H\left(0,1,1; \frac{1}{a}\right)-H\left(1,-1,-1; \frac{1}{a}\right)+H\left(1,-1,1; \frac{1}{a}\right)+H\left(1,0,-1; \frac{1}{a}\right)\\
&\nn-H\left(1,0,1; \frac{1}{a}\right)+H\left(1,1,-1; \frac{1}{a}\right)\\
G&\left(a^2,a^2,a; 1\right) = 
-H\left(-1,-1,1; \frac{1}{a}\right)-H\left(-1,0,-1; \frac{1}{a}\right)+H\left(-1,0,1; \frac{1}{a}\right)\\
&\nn-H\left(-1,1,-1; \frac{1}{a}\right)+H\left(-1,1,1; \frac{1}{a}\right)-H\left(0,-1,-1; \frac{1}{a}\right)+H\left(0,-1,1; \frac{1}{a}\right)\\
&\nn+H\left(0,1,-1; \frac{1}{a}\right)-H\left(0,1,1; \frac{1}{a}\right)-H\left(1,-1,-1; \frac{1}{a}\right)+H\left(1,-1,1; \frac{1}{a}\right)\\
&\nn+H\left(1,1,-1; \frac{1}{a}\right)-H\left(1,1,1; \frac{1}{a}\right)\\
G&\left(a^2,-a,-a; 1\right) = 
H\left(-1,-1,-1; \frac{1}{a}\right)-H\left(-1,-1,1; \frac{1}{a}\right)-H\left(-1,0,-1; \frac{1}{a}\right)\\
&\nn+H\left(-1,0,1; \frac{1}{a}\right)-H\left(-1,1,-1; \frac{1}{a}\right)-H\left(0,-1,-1; \frac{1}{a}\right)+H\left(0,-1,1; \frac{1}{a}\right)\\
&\nn+H\left(0,0,-1; \frac{1}{a}\right)-H\left(0,0,1; \frac{1}{a}\right)+H\left(0,1,-1; \frac{1}{a}\right)-H\left(1,-1,-1; \frac{1}{a}\right)
\end{align}
\begin{align}
G&\left(a^2,-a,a; 1\right) = 
-H\left(-1,-1,1; \frac{1}{a}\right)-H\left(-1,0,-1; \frac{1}{a}\right)+H\left(-1,0,1; \frac{1}{a}\right)\\
&\nn-H\left(-1,1,-1; \frac{1}{a}\right)+H\left(-1,1,1; \frac{1}{a}\right)+H\left(0,-1,1; \frac{1}{a}\right)+H\left(0,0,-1; \frac{1}{a}\right)\\
&\nn-H\left(0,0,1; \frac{1}{a}\right)+H\left(0,1,-1; \frac{1}{a}\right)-H\left(0,1,1; \frac{1}{a}\right)+H\left(1,-1,1; \frac{1}{a}\right)\\
G&\left(a^2,a,-a; 1\right) = 
-H\left(-1,1,-1; \frac{1}{a}\right)-H\left(0,-1,-1; \frac{1}{a}\right)+H\left(0,-1,1; \frac{1}{a}\right)\\
&\nn+H\left(0,0,-1; \frac{1}{a}\right)-H\left(0,0,1; \frac{1}{a}\right)+H\left(0,1,-1; \frac{1}{a}\right)-H\left(1,-1,-1; \frac{1}{a}\right)\\
&\nn+H\left(1,-1,1; \frac{1}{a}\right)+H\left(1,0,-1; \frac{1}{a}\right)-H\left(1,0,1; \frac{1}{a}\right)+H\left(1,1,-1; \frac{1}{a}\right)\\
G&\left(a^2,a,a; 1\right) = 
H\left(-1,1,1; \frac{1}{a}\right)+H\left(0,-1,1; \frac{1}{a}\right)+H\left(0,0,-1; \frac{1}{a}\right)\\
&\nn-H\left(0,0,1; \frac{1}{a}\right)+H\left(0,1,-1; \frac{1}{a}\right)-H\left(0,1,1; \frac{1}{a}\right)+H\left(1,-1,1; \frac{1}{a}\right)\\
&\nn+H\left(1,0,-1; \frac{1}{a}\right)-H\left(1,0,1; \frac{1}{a}\right)+H\left(1,1,-1; \frac{1}{a}\right)-H\left(1,1,1; \frac{1}{a}\right)\\
G&\left(a^2,-a,a^2; 1\right) = 
H\left(-1,-1,-1; \frac{1}{a}\right)-H\left(-1,-1,1; \frac{1}{a}\right)-H\left(-1,0,-1; \frac{1}{a}\right)\\
&\nn+H\left(-1,0,1; \frac{1}{a}\right)-H\left(-1,1,-1; \frac{1}{a}\right)-H\left(1,-1,-1; \frac{1}{a}\right)-H\left(1,0,-1; \frac{1}{a}\right)\\
&\nn+H\left(1,0,1; \frac{1}{a}\right)\\
G&\left(a^2,a,a^2; 1\right) = 
H\left(-1,0,-1; \frac{1}{a}\right)-H\left(-1,0,1; \frac{1}{a}\right)+H\left(-1,1,1; \frac{1}{a}\right)\\
&\nn+H\left(1,-1,1; \frac{1}{a}\right)+H\left(1,0,-1; \frac{1}{a}\right)-H\left(1,0,1; \frac{1}{a}\right)+H\left(1,1,-1; \frac{1}{a}\right)\\
&\nn-H\left(1,1,1; \frac{1}{a}\right)\\
G&(a,a,0; 1) = 
-H\left(0,1,1; \frac{1}{a}\right)-H\left(1,0,1; \frac{1}{a}\right)\\
G&(-a,-a,1; 1) = 
H\left(-1,-1,-1; \frac{1}{a}\right)-H\left(0,-1,-1; \frac{1}{a}\right)\\
G&(-a,a,1; 1) = 
H\left(-1,0,1; \frac{1}{a}\right)+H\left(-1,1,1; \frac{1}{a}\right)+H\left(0,-1,1; \frac{1}{a}\right)\\
&\nn+H\left(1,-1,-1; \frac{1}{a}\right)+H\left(1,-1,1; \frac{1}{a}\right)-H\left(1,0,-1; \frac{1}{a}\right)\\
G&(a,-a,1; 1) = 
-H\left(-1,0,1; \frac{1}{a}\right)-H\left(-1,1,-1; \frac{1}{a}\right)-H\left(-1,1,1; \frac{1}{a}\right)\\
&\nn+H\left(0,1,-1; \frac{1}{a}\right)-H\left(1,-1,-1; \frac{1}{a}\right)+H\left(1,0,-1; \frac{1}{a}\right)
\end{align}
\begin{align}
G&(a,a,1; 1) = 
-H\left(0,1,1; \frac{1}{a}\right)-H\left(1,1,1; \frac{1}{a}\right)\\
G&\left(-a,a^2,0; 1\right) = 
-H\left(-1,0,-1; \frac{1}{a}\right)-2 H\left(0,-1,-1; \frac{1}{a}\right)-4 H\left(0,0,-1; \frac{1}{a}\right)\\
&\nn+4 H\left(0,0,1; \frac{1}{a}\right)\\
G&\left(a,a^2,0; 1\right) = 
-4 H\left(0,0,-1; \frac{1}{a}\right)+4 H\left(0,0,1; \frac{1}{a}\right)-2 H\left(0,1,1; \frac{1}{a}\right)\\
&\nn-H\left(1,0,1; \frac{1}{a}\right)\\
G&\left(-a,a^2,1; 1\right) = 
H\left(-1,-1,-1; \frac{1}{a}\right)+H\left(-1,0,-1; \frac{1}{a}\right)-H\left(-1,0,1; \frac{1}{a}\right)\\
&\nn-H\left(0,-1,-1; \frac{1}{a}\right)-H\left(0,-1,1; \frac{1}{a}\right)-4 H\left(0,0,-1; \frac{1}{a}\right)+4 H\left(0,0,1; \frac{1}{a}\right)\\
&\nn-H\left(0,1,-1; \frac{1}{a}\right)+H\left(0,1,1; \frac{1}{a}\right)-2 H\left(1,0,-1; \frac{1}{a}\right)+H\left(1,0,1; \frac{1}{a}\right)\\
G&\left(a,a^2,1; 1\right) = 
H\left(-1,0,-1; \frac{1}{a}\right)-2 H\left(-1,0,1; \frac{1}{a}\right)+H\left(0,-1,-1; \frac{1}{a}\right)\\
&\nn-H\left(0,-1,1; \frac{1}{a}\right)-4 H\left(0,0,-1; \frac{1}{a}\right)+4 H\left(0,0,1; \frac{1}{a}\right)-H\left(0,1,-1; \frac{1}{a}\right)\\
&\nn-H\left(0,1,1; \frac{1}{a}\right)-H\left(1,0,-1; \frac{1}{a}\right)+H\left(1,0,1; \frac{1}{a}\right)-H\left(1,1,1; \frac{1}{a}\right)\\
G&\left(-a,a^2,-a; 1\right) = 
H\left(-1,-1,-1; \frac{1}{a}\right)+H\left(-1,0,-1; \frac{1}{a}\right)-H\left(-1,0,1; \frac{1}{a}\right)\\
&\nn-H\left(0,-1,1; \frac{1}{a}\right)-2 H\left(0,0,-1; \frac{1}{a}\right)+2 H\left(0,0,1; \frac{1}{a}\right)-H\left(0,1,-1; \frac{1}{a}\right)\\
G&\left(-a,a^2,a; 1\right) = 
-H\left(-1,-1,1; \frac{1}{a}\right)-H\left(0,-1,-1; \frac{1}{a}\right)-H\left(0,-1,1; \frac{1}{a}\right)\\
&\nn-2 H\left(0,0,-1; \frac{1}{a}\right)+2 H\left(0,0,1; \frac{1}{a}\right)-H\left(0,1,-1; \frac{1}{a}\right)+H\left(0,1,1; \frac{1}{a}\right)\\
&\nn-H\left(1,-1,-1; \frac{1}{a}\right)-H\left(1,-1,1; \frac{1}{a}\right)-H\left(1,0,-1; \frac{1}{a}\right)+H\left(1,0,1; \frac{1}{a}\right)\\
G&\left(a,a^2,-a; 1\right) = 
H\left(-1,0,-1; \frac{1}{a}\right)-H\left(-1,0,1; \frac{1}{a}\right)+H\left(-1,1,-1; \frac{1}{a}\right)\\
&\nn+H\left(-1,1,1; \frac{1}{a}\right)+H\left(0,-1,-1; \frac{1}{a}\right)-H\left(0,-1,1; \frac{1}{a}\right)-2 H\left(0,0,-1; \frac{1}{a}\right)\\
&\nn+2 H\left(0,0,1; \frac{1}{a}\right)-H\left(0,1,-1; \frac{1}{a}\right)-H\left(0,1,1; \frac{1}{a}\right)+H\left(1,1,-1; \frac{1}{a}\right)\\
G&\left(a,a^2,a; 1\right) = 
-H\left(0,-1,1; \frac{1}{a}\right)-2 H\left(0,0,-1; \frac{1}{a}\right)+2 H\left(0,0,1; \frac{1}{a}\right)\\
&\nn-H\left(0,1,-1; \frac{1}{a}\right)-H\left(1,0,-1; \frac{1}{a}\right)+H\left(1,0,1; \frac{1}{a}\right)-H\left(1,1,1; \frac{1}{a}\right)
\end{align}
\begin{align}
G&\left(-a,a^2,a^2; 1\right) = 
H\left(-1,-1,-1; \frac{1}{a}\right)+H\left(-1,0,-1; \frac{1}{a}\right)-H\left(-1,0,1; \frac{1}{a}\right)\\
&\nn+H\left(0,-1,-1; \frac{1}{a}\right)-H\left(0,-1,1; \frac{1}{a}\right)-H\left(0,1,-1; \frac{1}{a}\right)+H\left(0,1,1; \frac{1}{a}\right)\\
G&\left(a,a^2,a^2; 1\right) = 
H\left(0,-1,-1; \frac{1}{a}\right)-H\left(0,-1,1; \frac{1}{a}\right)-H\left(0,1,-1; \frac{1}{a}\right)\\
&\nn+H\left(0,1,1; \frac{1}{a}\right)-H\left(1,0,-1; \frac{1}{a}\right)+H\left(1,0,1; \frac{1}{a}\right)-H\left(1,1,1; \frac{1}{a}\right)\\
G&(-a,-a,a; 1) = 
-H\left(-1,-1,1; \frac{1}{a}\right)\\
G&(-a,a,-a; 1) = 
-H\left(-1,1,-1; \frac{1}{a}\right)\\
G&(-a,a,a; 1) = 
H\left(-1,1,1; \frac{1}{a}\right)\\
G&(a,-a,-a; 1) = 
-H\left(1,-1,-1; \frac{1}{a}\right)\\
G&(a,-a,a; 1) = 
H\left(1,-1,1; \frac{1}{a}\right)\\
G&(a,a,-a; 1) = 
H\left(1,1,-1; \frac{1}{a}\right)\\
G&(a,a,a; 1) = 
-H\left(1,1,1; \frac{1}{a}\right)\\
G&\left(-a,-a,a^2; 1\right) = 
H\left(-1,-1,-1; \frac{1}{a}\right)+H\left(0,-1,-1; \frac{1}{a}\right)+H\left(0,0,-1; \frac{1}{a}\right)\\
&\nn-H\left(0,0,1; \frac{1}{a}\right)\\
G&\left(-a,a,a^2; 1\right) = 
H\left(-1,0,-1; \frac{1}{a}\right)-H\left(-1,0,1; \frac{1}{a}\right)+H\left(-1,1,1; \frac{1}{a}\right)\\
&\nn+H\left(0,-1,-1; \frac{1}{a}\right)+H\left(0,0,-1; \frac{1}{a}\right)-H\left(0,0,1; \frac{1}{a}\right)+H\left(1,-1,-1; \frac{1}{a}\right)\\
&\nn+H\left(1,-1,1; \frac{1}{a}\right)+H\left(1,0,-1; \frac{1}{a}\right)-H\left(1,0,1; \frac{1}{a}\right)\\
G&\left(a,-a,a^2; 1\right) = 
-H\left(-1,0,-1; \frac{1}{a}\right)+H\left(-1,0,1; \frac{1}{a}\right)-H\left(-1,1,-1; \frac{1}{a}\right)\\
&\nn-H\left(-1,1,1; \frac{1}{a}\right)+H\left(0,0,-1; \frac{1}{a}\right)-H\left(0,0,1; \frac{1}{a}\right)+H\left(0,1,1; \frac{1}{a}\right)\\
&\nn-H\left(1,-1,-1; \frac{1}{a}\right)-H\left(1,0,-1; \frac{1}{a}\right)+H\left(1,0,1; \frac{1}{a}\right)\\
G&\left(a,a,a^2; 1\right) = 
H\left(0,0,-1; \frac{1}{a}\right)-H\left(0,0,1; \frac{1}{a}\right)+H\left(0,1,1; \frac{1}{a}\right)\\
&\nn-H\left(1,1,1; \frac{1}{a}\right)
\end{align}

\subsection{Polylogarithms of weight four}
\begin{align}
G&(0,0,0,a; 1) = 
-H\left(0,0,0,1; \frac{1}{a}\right)\\
G&(0,0,1,a; 1) = 
\zeta_3 H\left(1; \frac{1}{a}\right)-H\left(0,0,0,1; \frac{1}{a}\right)-H\left(1,0,0,1; \frac{1}{a}\right)\\
G&(0,0,-a,0; 1) = 
-3 H\left(0,0,0,-1; \frac{1}{a}\right)\\
G&(0,0,a,0; 1) = 
3 H\left(0,0,0,1; \frac{1}{a}\right)\\
G&(0,0,-a,-1; 1) = 
\log  2  H\left(0,0,-1; \frac{1}{a}\right)+\log  2  H\left(0,0,1; \frac{1}{a}\right)+\frac{3}{4} \zeta_3 H\left(1; \frac{1}{a}\right)\\
&\nn+\frac{1}{12} \pi ^2 H\left(0,1; \frac{1}{a}\right)-3 H\left(0,0,0,-1; \frac{1}{a}\right)-H\left(0,0,1,-1; \frac{1}{a}\right)-H\left(0,1,0,-1; \frac{1}{a}\right)\\
&\nn-H\left(1,0,0,-1; \frac{1}{a}\right)\\
G&(0,0,a,-1; 1) = 
-\log  2  H\left(0,0,-1; \frac{1}{a}\right)-\log  2  H\left(0,0,1; \frac{1}{a}\right)-\frac{3}{4} \zeta_3 H\left(-1; \frac{1}{a}\right)\\
&\nn-\frac{1}{12} \pi ^2 H\left(0,-1; \frac{1}{a}\right)-H\left(-1,0,0,1; \frac{1}{a}\right)-H\left(0,-1,0,1; \frac{1}{a}\right)-H\left(0,0,-1,1; \frac{1}{a}\right)\\
&\nn+3 H\left(0,0,0,1; \frac{1}{a}\right)\\
G&(0,0,a,1; 1) = 
-\zeta_3 H\left(1; \frac{1}{a}\right)-\frac{1}{6} \pi ^2 H\left(0,1; \frac{1}{a}\right)+3 H\left(0,0,0,1; \frac{1}{a}\right)\\
&\nn+H\left(0,0,1,1; \frac{1}{a}\right)+H\left(0,1,0,1; \frac{1}{a}\right)+H\left(1,0,0,1; \frac{1}{a}\right)\\
G&\left(0,0,a^2,-a; 1\right) = 
3 H\left(-1,0,0,-1; \frac{1}{a}\right)-4 H\left(-1,0,0,1; \frac{1}{a}\right)\\
&\nn+2 H\left(0,-1,0,-1; \frac{1}{a}\right)-4 H\left(0,-1,0,1; \frac{1}{a}\right)+4 H\left(0,0,-1,-1; \frac{1}{a}\right)\\
&\nn-4 H\left(0,0,-1,1; \frac{1}{a}\right)-12 H\left(0,0,0,-1; \frac{1}{a}\right)+12 H\left(0,0,0,1; \frac{1}{a}\right)-4 H\left(0,0,1,-1; \frac{1}{a}\right)\\
G&\left(0,0,a^2,a; 1\right) = 
-4 H\left(0,0,-1,1; \frac{1}{a}\right)-12 H\left(0,0,0,-1; \frac{1}{a}\right)\\
&\nn+12 H\left(0,0,0,1; \frac{1}{a}\right)-4 H\left(0,0,1,-1; \frac{1}{a}\right)+4 H\left(0,0,1,1; \frac{1}{a}\right)-4 H\left(0,1,0,-1; \frac{1}{a}\right)\\
&\nn+2 H\left(0,1,0,1; \frac{1}{a}\right)-4 H\left(1,0,0,-1; \frac{1}{a}\right)+3 H\left(1,0,0,1; \frac{1}{a}\right)\\
G&(0,0,-a,a; 1) = 
-H\left(0,0,-1,1; \frac{1}{a}\right)\\
G&(0,0,a,-a; 1) = 
-H\left(0,0,1,-1; \frac{1}{a}\right)\\
G&(0,0,a,a; 1) = 
H\left(0,0,1,1; \frac{1}{a}\right)
\end{align}
\begin{align}
G&\left(0,0,-a,a^2; 1\right) = 
-3 H\left(-1,0,0,-1; \frac{1}{a}\right)+4 H\left(-1,0,0,1; \frac{1}{a}\right)\\
&\nn-H\left(0,-1,0,-1; \frac{1}{a}\right)+2 H\left(0,-1,0,1; \frac{1}{a}\right)+H\left(0,0,-1,-1; \frac{1}{a}\right)\\
&\nn+7 H\left(0,0,0,-1; \frac{1}{a}\right)-7 H\left(0,0,0,1; \frac{1}{a}\right)\\
G&\left(0,0,a,a^2; 1\right) = 
7 H\left(0,0,0,-1; \frac{1}{a}\right)-7 H\left(0,0,0,1; \frac{1}{a}\right)+H\left(0,0,1,1; \frac{1}{a}\right)\\
&\nn+2 H\left(0,1,0,-1; \frac{1}{a}\right)-H\left(0,1,0,1; \frac{1}{a}\right)+4 H\left(1,0,0,-1; \frac{1}{a}\right)-3 H\left(1,0,0,1; \frac{1}{a}\right)\\
G&(0,1,0,a; 1) = 
\frac{1}{6} \pi ^2 H\left(0,1; \frac{1}{a}\right)-H\left(0,0,0,1; \frac{1}{a}\right)-H\left(0,1,0,1; \frac{1}{a}\right)\\
G&(0,1,1,a; 1) = 
-\zeta_3 H\left(1; \frac{1}{a}\right)+\frac{1}{6} \pi ^2 H\left(0,1; \frac{1}{a}\right)+\frac{1}{6} \pi ^2 H\left(1,1; \frac{1}{a}\right)\\
&\nn-H\left(0,0,0,1; \frac{1}{a}\right)-H\left(0,1,0,1; \frac{1}{a}\right)-H\left(1,0,0,1; \frac{1}{a}\right)-H\left(1,1,0,1; \frac{1}{a}\right)\\
G&(0,1,a,1; 1) = 
-\frac{1}{3} \pi ^2 H\left(0,1; \frac{1}{a}\right)-\frac{1}{3} \pi ^2 H\left(1,1; \frac{1}{a}\right)+3 H\left(0,0,0,1; \frac{1}{a}\right)\\
&\nn+H\left(0,0,1,1; \frac{1}{a}\right)+2 H\left(0,1,0,1; \frac{1}{a}\right)+3 H\left(1,0,0,1; \frac{1}{a}\right)+H\left(1,0,1,1; \frac{1}{a}\right)\\
&\nn+2 H\left(1,1,0,1; \frac{1}{a}\right)\\
G&(0,1,a,a; 1) = 
-\frac{1}{6} \pi ^2 H\left(1,1; \frac{1}{a}\right)+H\left(0,0,1,1; \frac{1}{a}\right)+H\left(1,0,0,1; \frac{1}{a}\right)\\
&\nn+H\left(1,0,1,1; \frac{1}{a}\right)+H\left(1,1,0,1; \frac{1}{a}\right)\\
G&(0,-a,0,-1; 1) = 
-2 \log  2  H\left(0,0,-1; \frac{1}{a}\right)-2 \log  2  H\left(0,0,1; \frac{1}{a}\right)\\
&\nn+\frac{1}{12} \pi ^2 H\left(0,-1; \frac{1}{a}\right)-\frac{1}{12} \pi ^2 H\left(0,1; \frac{1}{a}\right)+3 H\left(0,0,0,-1; \frac{1}{a}\right)+2 H\left(0,0,1,-1; \frac{1}{a}\right)\\
&\nn+H\left(0,1,0,-1; \frac{1}{a}\right)\\
G&(0,a,0,-1; 1) = 
2 \log  2  H\left(0,0,-1; \frac{1}{a}\right)+2 \log  2  H\left(0,0,1; \frac{1}{a}\right)\\
&\nn+\frac{1}{12} \pi ^2 H\left(0,-1; \frac{1}{a}\right)-\frac{1}{12} \pi ^2 H\left(0,1; \frac{1}{a}\right)+H\left(0,-1,0,1; \frac{1}{a}\right)+2 H\left(0,0,-1,1; \frac{1}{a}\right)\\
&\nn-3 H\left(0,0,0,1; \frac{1}{a}\right)\\
G&(0,a,0,1; 1) = 
\frac{1}{3} \pi ^2 H\left(0,1; \frac{1}{a}\right)-3 H\left(0,0,0,1; \frac{1}{a}\right)-2 H\left(0,0,1,1; \frac{1}{a}\right)\\
&\nn-H\left(0,1,0,1; \frac{1}{a}\right)\\
G&(0,-a,0,a; 1) = 
-H\left(0,-1,0,1; \frac{1}{a}\right)
\end{align}
\begin{align}
G&(0,a,0,-a; 1) = 
-H\left(0,1,0,-1; \frac{1}{a}\right)\\
G&(0,a,0,a; 1) = 
H\left(0,1,0,1; \frac{1}{a}\right)\\
G&\left(0,-a,0,a^2; 1\right) = 
H\left(0,-1,0,-1; \frac{1}{a}\right)+2 H\left(0,0,-1,-1; \frac{1}{a}\right)\\
&\nn+4 H\left(0,0,0,-1; \frac{1}{a}\right)-4 H\left(0,0,0,1; \frac{1}{a}\right)\\
G&\left(0,a,0,a^2; 1\right) = 
4 H\left(0,0,0,-1; \frac{1}{a}\right)-4 H\left(0,0,0,1; \frac{1}{a}\right)+2 H\left(0,0,1,1; \frac{1}{a}\right)\\
&\nn+H\left(0,1,0,1; \frac{1}{a}\right)\\
G&(0,a,1,0; 1) = 
2 \zeta_3 H\left(1; \frac{1}{a}\right)-3 H\left(0,0,0,1; \frac{1}{a}\right)-H\left(0,1,0,1; \frac{1}{a}\right)\\
&\nn-2 H\left(1,0,0,1; \frac{1}{a}\right)\\
G&(0,a,1,1; 1) = 
\zeta_3 H\left(1; \frac{1}{a}\right)+\frac{1}{6} \pi ^2 H\left(0,1; \frac{1}{a}\right)+\frac{1}{6} \pi ^2 H\left(1,1; \frac{1}{a}\right)\\
&\nn-3 H\left(0,0,0,1; \frac{1}{a}\right)-2 H\left(0,0,1,1; \frac{1}{a}\right)-2 H\left(0,1,0,1; \frac{1}{a}\right)-H\left(0,1,1,1; \frac{1}{a}\right)\\
&\nn-2 H\left(1,0,0,1; \frac{1}{a}\right)-H\left(1,0,1,1; \frac{1}{a}\right)-H\left(1,1,0,1; \frac{1}{a}\right)\\
G&(0,a,1,a; 1) = 
\frac{1}{3} \pi ^2 H\left(1,1; \frac{1}{a}\right)-H\left(0,1,0,1; \frac{1}{a}\right)-H\left(0,1,1,1; \frac{1}{a}\right)\\
&\nn-3 H\left(1,0,0,1; \frac{1}{a}\right)-H\left(1,0,1,1; \frac{1}{a}\right)-2 H\left(1,1,0,1; \frac{1}{a}\right)\\
G&\left(0,-a,1,a^2; 1\right) = 
\frac{1}{3} \pi ^2 H\left(-1,-1; \frac{1}{a}\right)-\frac{1}{6} \pi ^2 H\left(-1,1; \frac{1}{a}\right)\\
&\nn-\frac{1}{6} \pi ^2 H\left(1,-1; \frac{1}{a}\right)+3 H\left(-1,-1,0,-1; \frac{1}{a}\right)-2 H\left(-1,-1,0,1; \frac{1}{a}\right)\\
&\nn+H\left(-1,0,-1,-1; \frac{1}{a}\right)-6 H\left(-1,0,0,-1; \frac{1}{a}\right)+4 H\left(-1,0,0,1; \frac{1}{a}\right)\\
&\nn-2 H\left(-1,1,0,-1; \frac{1}{a}\right)+2 H\left(-1,1,0,1; \frac{1}{a}\right)+H\left(0,-1,-1,-1; \frac{1}{a}\right)\\
&\nn-2 H\left(0,-1,0,-1; \frac{1}{a}\right)+2 H\left(0,-1,0,1; \frac{1}{a}\right)+2 H\left(0,0,-1,-1; \frac{1}{a}\right)\\
&\nn+4 H\left(0,0,0,-1; \frac{1}{a}\right)-4 H\left(0,0,0,1; \frac{1}{a}\right)+2 H\left(0,1,0,-1; \frac{1}{a}\right)-H\left(0,1,0,1; \frac{1}{a}\right)\\
&\nn-2 H\left(1,-1,0,-1; \frac{1}{a}\right)+2 H\left(1,-1,0,1; \frac{1}{a}\right)+5 H\left(1,0,0,-1; \frac{1}{a}\right)-3 H\left(1,0,0,1; \frac{1}{a}\right)
\end{align}
\begin{align}
G&\left(0,a,1,a^2; 1\right) = 
-\frac{1}{6} \pi ^2 H\left(-1,1; \frac{1}{a}\right)-\frac{1}{6} \pi ^2 H\left(1,-1; \frac{1}{a}\right)+\frac{1}{3} \pi ^2 H\left(1,1; \frac{1}{a}\right)\\
&\nn-3 H\left(-1,0,0,-1; \frac{1}{a}\right)+5 H\left(-1,0,0,1; \frac{1}{a}\right)-2 H\left(-1,1,0,-1; \frac{1}{a}\right)\\
&\nn+2 H\left(-1,1,0,1; \frac{1}{a}\right)-H\left(0,-1,0,-1; \frac{1}{a}\right)+2 H\left(0,-1,0,1; \frac{1}{a}\right)\\
&\nn+4 H\left(0,0,0,-1; \frac{1}{a}\right)-4 H\left(0,0,0,1; \frac{1}{a}\right)+2 H\left(0,0,1,1; \frac{1}{a}\right)+2 H\left(0,1,0,-1; \frac{1}{a}\right)\\
&\nn-2 H\left(0,1,0,1; \frac{1}{a}\right)-H\left(0,1,1,1; \frac{1}{a}\right)-2 H\left(1,-1,0,-1; \frac{1}{a}\right)+2 H\left(1,-1,0,1; \frac{1}{a}\right)\\
&\nn+4 H\left(1,0,0,-1; \frac{1}{a}\right)-6 H\left(1,0,0,1; \frac{1}{a}\right)-H\left(1,0,1,1; \frac{1}{a}\right)+2 H\left(1,1,0,-1; \frac{1}{a}\right)\\
&\nn-3 H\left(1,1,0,1; \frac{1}{a}\right)\\
G&\left(0,a^2,0,-a; 1\right) = 
H\left(0,-1,0,-1; \frac{1}{a}\right)+2 H\left(0,-1,0,1; \frac{1}{a}\right)\\
&\nn-4 H\left(0,0,-1,-1; \frac{1}{a}\right)+4 H\left(0,0,-1,1; \frac{1}{a}\right)+6 H\left(0,0,0,-1; \frac{1}{a}\right)-6 H\left(0,0,0,1; \frac{1}{a}\right)\\
&\nn+4 H\left(0,0,1,-1; \frac{1}{a}\right)-2 H\left(0,1,0,-1; \frac{1}{a}\right)\\
G&\left(0,a^2,0,a; 1\right) = 
-2 H\left(0,-1,0,1; \frac{1}{a}\right)+4 H\left(0,0,-1,1; \frac{1}{a}\right)\\
&\nn+6 H\left(0,0,0,-1; \frac{1}{a}\right)-6 H\left(0,0,0,1; \frac{1}{a}\right)+4 H\left(0,0,1,-1; \frac{1}{a}\right)-4 H\left(0,0,1,1; \frac{1}{a}\right)\\
&\nn+2 H\left(0,1,0,-1; \frac{1}{a}\right)+H\left(0,1,0,1; \frac{1}{a}\right)\\
G&\left(0,a^2,-a,0; 1\right) = 
-6 H\left(-1,0,0,-1; \frac{1}{a}\right)+8 H\left(-1,0,0,1; \frac{1}{a}\right)\\
&\nn-5 H\left(0,-1,0,-1; \frac{1}{a}\right)+6 H\left(0,-1,0,1; \frac{1}{a}\right)-4 H\left(0,0,-1,-1; \frac{1}{a}\right)\\
&\nn+4 H\left(0,0,-1,1; \frac{1}{a}\right)+18 H\left(0,0,0,-1; \frac{1}{a}\right)-18 H\left(0,0,0,1; \frac{1}{a}\right)\\
&\nn+4 H\left(0,0,1,-1; \frac{1}{a}\right)+2 H\left(0,1,0,-1; \frac{1}{a}\right)\\
G&\left(0,a^2,a,0; 1\right) = 
2 H\left(0,-1,0,1; \frac{1}{a}\right)+4 H\left(0,0,-1,1; \frac{1}{a}\right)\\
&\nn+18 H\left(0,0,0,-1; \frac{1}{a}\right)-18 H\left(0,0,0,1; \frac{1}{a}\right)+4 H\left(0,0,1,-1; \frac{1}{a}\right)-4 H\left(0,0,1,1; \frac{1}{a}\right)\\
&\nn+6 H\left(0,1,0,-1; \frac{1}{a}\right)-5 H\left(0,1,0,1; \frac{1}{a}\right)+8 H\left(1,0,0,-1; \frac{1}{a}\right)-6 H\left(1,0,0,1; \frac{1}{a}\right)
\end{align}
\begin{align}
G&\left(0,a^2,-a,1; 1\right) = 
-\frac{1}{6} \pi ^2 H\left(-1,-1; \frac{1}{a}\right)-\frac{1}{6} \pi ^2 H\left(0,-1; \frac{1}{a}\right)\\
&\nn+\frac{1}{6} \pi ^2 H\left(0,1; \frac{1}{a}\right)-2 H\left(-1,-1,0,1; \frac{1}{a}\right)+H\left(-1,0,-1,-1; \frac{1}{a}\right)\\
&\nn-2 H\left(-1,0,-1,1; \frac{1}{a}\right)-2 H\left(-1,0,0,-1; \frac{1}{a}\right)+4 H\left(-1,0,0,1; \frac{1}{a}\right)\\
&\nn-2 H\left(-1,0,1,-1; \frac{1}{a}\right)+2 H\left(0,-1,-1,-1; \frac{1}{a}\right)-2 H\left(0,-1,-1,1; \frac{1}{a}\right)\\
&\nn-5 H\left(0,-1,0,-1; \frac{1}{a}\right)+6 H\left(0,-1,0,1; \frac{1}{a}\right)-2 H\left(0,-1,1,-1; \frac{1}{a}\right)\\
&\nn-8 H\left(0,0,-1,-1; \frac{1}{a}\right)+8 H\left(0,0,-1,1; \frac{1}{a}\right)+18 H\left(0,0,0,-1; \frac{1}{a}\right)\\
&\nn-18 H\left(0,0,0,1; \frac{1}{a}\right)+8 H\left(0,0,1,-1; \frac{1}{a}\right)-4 H\left(0,0,1,1; \frac{1}{a}\right)\\
&\nn-2 H\left(0,1,-1,-1; \frac{1}{a}\right)+4 H\left(0,1,0,-1; \frac{1}{a}\right)-2 H\left(0,1,0,1; \frac{1}{a}\right)\\
G&\left(0,a^2,a,1; 1\right) = 
-\frac{1}{6} \pi ^2 H\left(0,-1; \frac{1}{a}\right)+\frac{1}{6} \pi ^2 H\left(0,1; \frac{1}{a}\right)-\frac{1}{6} \pi ^2 H\left(1,1; \frac{1}{a}\right)\\
&\nn-2 H\left(0,-1,0,-1; \frac{1}{a}\right)+4 H\left(0,-1,0,1; \frac{1}{a}\right)+2 H\left(0,-1,1,1; \frac{1}{a}\right)\\
&\nn-4 H\left(0,0,-1,-1; \frac{1}{a}\right)+8 H\left(0,0,-1,1; \frac{1}{a}\right)+18 H\left(0,0,0,-1; \frac{1}{a}\right)\\
&\nn-18 H\left(0,0,0,1; \frac{1}{a}\right)+8 H\left(0,0,1,-1; \frac{1}{a}\right)-8 H\left(0,0,1,1; \frac{1}{a}\right)+2 H\left(0,1,-1,1; \frac{1}{a}\right)\\
&\nn+6 H\left(0,1,0,-1; \frac{1}{a}\right)-5 H\left(0,1,0,1; \frac{1}{a}\right)+2 H\left(0,1,1,-1; \frac{1}{a}\right)-2 H\left(0,1,1,1; \frac{1}{a}\right)\\
&\nn+2 H\left(1,0,-1,1; \frac{1}{a}\right)+4 H\left(1,0,0,-1; \frac{1}{a}\right)-2 H\left(1,0,0,1; \frac{1}{a}\right)+2 H\left(1,0,1,-1; \frac{1}{a}\right)\\
&\nn-H\left(1,0,1,1; \frac{1}{a}\right)+2 H\left(1,1,0,-1; \frac{1}{a}\right)\\
G&\left(0,a^2,a^2,-a; 1\right) = 
-H\left(-1,-1,0,-1; \frac{1}{a}\right)+2 H\left(-1,-1,0,1; \frac{1}{a}\right)\\
&\nn+4 H\left(-1,0,0,-1; \frac{1}{a}\right)-4 H\left(-1,0,0,1; \frac{1}{a}\right)+2 H\left(-1,0,1,1; \frac{1}{a}\right)\\
&\nn+2 H\left(0,-1,-1,-1; \frac{1}{a}\right)-2 H\left(0,-1,-1,1; \frac{1}{a}\right)-2 H\left(0,-1,1,-1; \frac{1}{a}\right)\\
&\nn+2 H\left(0,-1,1,1; \frac{1}{a}\right)-4 H\left(0,0,-1,-1; \frac{1}{a}\right)+4 H\left(0,0,-1,1; \frac{1}{a}\right)\\
&\nn+4 H\left(0,0,1,-1; \frac{1}{a}\right)-4 H\left(0,0,1,1; \frac{1}{a}\right)-2 H\left(0,1,-1,-1; \frac{1}{a}\right)\\
&\nn+2 H\left(0,1,-1,1; \frac{1}{a}\right)+2 H\left(0,1,0,-1; \frac{1}{a}\right)-2 H\left(0,1,0,1; \frac{1}{a}\right)+2 H\left(0,1,1,-1; \frac{1}{a}\right)
\end{align}
\begin{align}
G&\left(0,a^2,a^2,a; 1\right) = 
-2 H\left(0,-1,-1,1; \frac{1}{a}\right)-2 H\left(0,-1,0,-1; \frac{1}{a}\right)\\
&\nn+2 H\left(0,-1,0,1; \frac{1}{a}\right)-2 H\left(0,-1,1,-1; \frac{1}{a}\right)+2 H\left(0,-1,1,1; \frac{1}{a}\right)\\
&\nn-4 H\left(0,0,-1,-1; \frac{1}{a}\right)+4 H\left(0,0,-1,1; \frac{1}{a}\right)+4 H\left(0,0,1,-1; \frac{1}{a}\right)\\
&\nn-4 H\left(0,0,1,1; \frac{1}{a}\right)-2 H\left(0,1,-1,-1; \frac{1}{a}\right)+2 H\left(0,1,-1,1; \frac{1}{a}\right)\\
&\nn+2 H\left(0,1,1,-1; \frac{1}{a}\right)-2 H\left(0,1,1,1; \frac{1}{a}\right)-2 H\left(1,0,-1,-1; \frac{1}{a}\right)\\
&\nn-4 H\left(1,0,0,-1; \frac{1}{a}\right)+4 H\left(1,0,0,1; \frac{1}{a}\right)-2 H\left(1,1,0,-1; \frac{1}{a}\right)+H\left(1,1,0,1; \frac{1}{a}\right)\\
G&\left(0,a^2,-a,a^2; 1\right) = 
2 H\left(-1,-1,0,-1; \frac{1}{a}\right)-4 H\left(-1,-1,0,1; \frac{1}{a}\right)\\
&\nn+H\left(-1,0,-1,-1; \frac{1}{a}\right)-2 H\left(-1,0,-1,1; \frac{1}{a}\right)-7 H\left(-1,0,0,-1; \frac{1}{a}\right)\\
&\nn+7 H\left(-1,0,0,1; \frac{1}{a}\right)-2 H\left(-1,0,1,-1; \frac{1}{a}\right)+2 H\left(0,-1,-1,-1; \frac{1}{a}\right)\\
&\nn-2 H\left(0,-1,-1,1; \frac{1}{a}\right)-2 H\left(0,-1,0,-1; \frac{1}{a}\right)+2 H\left(0,-1,0,1; \frac{1}{a}\right)\\
&\nn-2 H\left(0,-1,1,-1; \frac{1}{a}\right)-2 H\left(0,1,-1,-1; \frac{1}{a}\right)-2 H\left(0,1,0,-1; \frac{1}{a}\right)\\
&\nn+2 H\left(0,1,0,1; \frac{1}{a}\right)\\
G&\left(0,a^2,a,a^2; 1\right) = 
2 H\left(0,-1,0,-1; \frac{1}{a}\right)-2 H\left(0,-1,0,1; \frac{1}{a}\right)\\
&\nn+2 H\left(0,-1,1,1; \frac{1}{a}\right)+2 H\left(0,1,-1,1; \frac{1}{a}\right)+2 H\left(0,1,0,-1; \frac{1}{a}\right)\\
&\nn-2 H\left(0,1,0,1; \frac{1}{a}\right)+2 H\left(0,1,1,-1; \frac{1}{a}\right)-2 H\left(0,1,1,1; \frac{1}{a}\right)\\
&\nn+2 H\left(1,0,-1,1; \frac{1}{a}\right)+7 H\left(1,0,0,-1; \frac{1}{a}\right)-7 H\left(1,0,0,1; \frac{1}{a}\right)\\
&\nn+2 H\left(1,0,1,-1; \frac{1}{a}\right)-H\left(1,0,1,1; \frac{1}{a}\right)+4 H\left(1,1,0,-1; \frac{1}{a}\right)\\
&\nn-2 H\left(1,1,0,1; \frac{1}{a}\right)\\
G&(0,a,a,0; 1) = 
-2 H\left(0,0,1,1; \frac{1}{a}\right)-H\left(0,1,0,1; \frac{1}{a}\right)\\
G&(0,a,a,1; 1) = 
-\frac{1}{6} \pi ^2 H\left(1,1; \frac{1}{a}\right)-2 H\left(0,0,1,1; \frac{1}{a}\right)-H\left(0,1,1,1; \frac{1}{a}\right)\\
&\nn+2 H\left(1,0,0,1; \frac{1}{a}\right)+H\left(1,1,0,1; \frac{1}{a}\right)
\end{align}
\begin{align}
G&\left(0,-a,a^2,0; 1\right) = 
6 H\left(-1,0,0,-1; \frac{1}{a}\right)-8 H\left(-1,0,0,1; \frac{1}{a}\right)\\
&\nn+H\left(0,-1,0,-1; \frac{1}{a}\right)-4 H\left(0,-1,0,1; \frac{1}{a}\right)-4 H\left(0,0,-1,-1; \frac{1}{a}\right)\\
&\nn-18 H\left(0,0,0,-1; \frac{1}{a}\right)+18 H\left(0,0,0,1; \frac{1}{a}\right)\\
G&\left(0,a,a^2,0; 1\right) = 
-18 H\left(0,0,0,-1; \frac{1}{a}\right)+18 H\left(0,0,0,1; \frac{1}{a}\right)\\
&\nn-4 H\left(0,0,1,1; \frac{1}{a}\right)-4 H\left(0,1,0,-1; \frac{1}{a}\right)+H\left(0,1,0,1; \frac{1}{a}\right)\\
&\nn-8 H\left(1,0,0,-1; \frac{1}{a}\right)+6 H\left(1,0,0,1; \frac{1}{a}\right)\\
G&\left(0,-a,a^2,1; 1\right) = 
-\frac{1}{6} \pi ^2 H\left(-1,-1; \frac{1}{a}\right)+\frac{1}{6} \pi ^2 H\left(-1,1; \frac{1}{a}\right)\\
&\nn+\frac{1}{6} \pi ^2 H\left(0,-1; \frac{1}{a}\right)-\frac{1}{6} \pi ^2 H\left(0,1; \frac{1}{a}\right)+\frac{1}{6} \pi ^2 H\left(1,-1; \frac{1}{a}\right)\\
&\nn-3 H\left(-1,-1,0,-1; \frac{1}{a}\right)+4 H\left(-1,-1,0,1; \frac{1}{a}\right)-H\left(-1,0,-1,-1; \frac{1}{a}\right)\\
&\nn+2 H\left(-1,0,-1,1; \frac{1}{a}\right)+11 H\left(-1,0,0,-1; \frac{1}{a}\right)-11 H\left(-1,0,0,1; \frac{1}{a}\right)\\
&\nn+2 H\left(-1,0,1,-1; \frac{1}{a}\right)-2 H\left(-1,0,1,1; \frac{1}{a}\right)+2 H\left(-1,1,0,-1; \frac{1}{a}\right)\\
&\nn-2 H\left(-1,1,0,1; \frac{1}{a}\right)+H\left(0,-1,-1,-1; \frac{1}{a}\right)+5 H\left(0,-1,0,-1; \frac{1}{a}\right)\\
&\nn-7 H\left(0,-1,0,1; \frac{1}{a}\right)-H\left(0,0,-1,-1; \frac{1}{a}\right)-3 H\left(0,0,-1,1; \frac{1}{a}\right)\\
&\nn-18 H\left(0,0,0,-1; \frac{1}{a}\right)+18 H\left(0,0,0,1; \frac{1}{a}\right)-3 H\left(0,0,1,-1; \frac{1}{a}\right)\\
&\nn+3 H\left(0,0,1,1; \frac{1}{a}\right)-4 H\left(0,1,0,-1; \frac{1}{a}\right)+3 H\left(0,1,0,1; \frac{1}{a}\right)\\
&\nn+2 H\left(1,-1,0,-1; \frac{1}{a}\right)-2 H\left(1,-1,0,1; \frac{1}{a}\right)-5 H\left(1,0,0,-1; \frac{1}{a}\right)\\
&\nn+3 H\left(1,0,0,1; \frac{1}{a}\right)
\end{align}
\begin{align}
G&\left(0,a,a^2,1; 1\right) = 
\frac{1}{6} \pi ^2 H\left(-1,1; \frac{1}{a}\right)+\frac{1}{6} \pi ^2 H\left(0,-1; \frac{1}{a}\right)\\
&\nn-\frac{1}{6} \pi ^2 H\left(0,1; \frac{1}{a}\right)+\frac{1}{6} \pi ^2 H\left(1,-1; \frac{1}{a}\right)-\frac{1}{6} \pi ^2 H\left(1,1; \frac{1}{a}\right)\\
&\nn+3 H\left(-1,0,0,-1; \frac{1}{a}\right)-5 H\left(-1,0,0,1; \frac{1}{a}\right)+2 H\left(-1,1,0,-1; \frac{1}{a}\right)\\
&\nn-2 H\left(-1,1,0,1; \frac{1}{a}\right)+3 H\left(0,-1,0,-1; \frac{1}{a}\right)-4 H\left(0,-1,0,1; \frac{1}{a}\right)\\
&\nn+3 H\left(0,0,-1,-1; \frac{1}{a}\right)-3 H\left(0,0,-1,1; \frac{1}{a}\right)-18 H\left(0,0,0,-1; \frac{1}{a}\right)\\
&\nn+18 H\left(0,0,0,1; \frac{1}{a}\right)-3 H\left(0,0,1,-1; \frac{1}{a}\right)-H\left(0,0,1,1; \frac{1}{a}\right)\\
&\nn-7 H\left(0,1,0,-1; \frac{1}{a}\right)+5 H\left(0,1,0,1; \frac{1}{a}\right)-H\left(0,1,1,1; \frac{1}{a}\right)\\
&\nn+2 H\left(1,-1,0,-1; \frac{1}{a}\right)-2 H\left(1,-1,0,1; \frac{1}{a}\right)+2 H\left(1,0,-1,-1; \frac{1}{a}\right)\\
&\nn-2 H\left(1,0,-1,1; \frac{1}{a}\right)-11 H\left(1,0,0,-1; \frac{1}{a}\right)+11 H\left(1,0,0,1; \frac{1}{a}\right)\\
&\nn-2 H\left(1,0,1,-1; \frac{1}{a}\right)+H\left(1,0,1,1; \frac{1}{a}\right)-4 H\left(1,1,0,-1; \frac{1}{a}\right)\\
&\nn+3 H\left(1,1,0,1; \frac{1}{a}\right)\\
G&\left(0,-a,a^2,a^2; 1\right) = 
-H\left(-1,-1,0,-1; \frac{1}{a}\right)+2 H\left(-1,-1,0,1; \frac{1}{a}\right)\\
&\nn-H\left(-1,0,-1,-1; \frac{1}{a}\right)+2 H\left(-1,0,-1,1; \frac{1}{a}\right)+3 H\left(-1,0,0,-1; \frac{1}{a}\right)\\
&\nn-3 H\left(-1,0,0,1; \frac{1}{a}\right)+2 H\left(-1,0,1,-1; \frac{1}{a}\right)-2 H\left(-1,0,1,1; \frac{1}{a}\right)\\
&\nn+H\left(0,-1,-1,-1; \frac{1}{a}\right)+H\left(0,-1,0,-1; \frac{1}{a}\right)-H\left(0,-1,0,1; \frac{1}{a}\right)\\
&\nn+3 H\left(0,0,-1,-1; \frac{1}{a}\right)-3 H\left(0,0,-1,1; \frac{1}{a}\right)-3 H\left(0,0,1,-1; \frac{1}{a}\right)\\
&\nn+3 H\left(0,0,1,1; \frac{1}{a}\right)\\
G&\left(0,a,a^2,a^2; 1\right) = 
3 H\left(0,0,-1,-1; \frac{1}{a}\right)-3 H\left(0,0,-1,1; \frac{1}{a}\right)\\
&\nn-3 H\left(0,0,1,-1; \frac{1}{a}\right)+3 H\left(0,0,1,1; \frac{1}{a}\right)-H\left(0,1,0,-1; \frac{1}{a}\right)\\
&\nn+H\left(0,1,0,1; \frac{1}{a}\right)-H\left(0,1,1,1; \frac{1}{a}\right)+2 H\left(1,0,-1,-1; \frac{1}{a}\right)\\
&\nn-2 H\left(1,0,-1,1; \frac{1}{a}\right)-3 H\left(1,0,0,-1; \frac{1}{a}\right)+3 H\left(1,0,0,1; \frac{1}{a}\right)\\
&\nn-2 H\left(1,0,1,-1; \frac{1}{a}\right)+H\left(1,0,1,1; \frac{1}{a}\right)-2 H\left(1,1,0,-1; \frac{1}{a}\right)+H\left(1,1,0,1; \frac{1}{a}\right)
\end{align}
\begin{align}
G&(0,-a,-a,a; 1) = 
-H\left(0,-1,-1,1; \frac{1}{a}\right)\\
G&(0,-a,a,-a; 1) = 
-H\left(0,-1,1,-1; \frac{1}{a}\right)\\
G&(0,-a,a,a; 1) = 
H\left(0,-1,1,1; \frac{1}{a}\right)\\
G&(0,a,-a,-a; 1) = 
-H\left(0,1,-1,-1; \frac{1}{a}\right)\\
G&(0,a,-a,a; 1) = 
H\left(0,1,-1,1; \frac{1}{a}\right)\\
G&(0,a,a,-a; 1) = 
H\left(0,1,1,-1; \frac{1}{a}\right)\\
G&(0,a,a,a; 1) = 
-H\left(0,1,1,1; \frac{1}{a}\right)\\
G&(-a,0,0,-1; 1) = 
\log  2  H\left(0,0,-1; \frac{1}{a}\right)+\log  2  H\left(0,0,1; \frac{1}{a}\right)+\frac{3}{4} \zeta_3 H\left(-1; \frac{1}{a}\right)\\
&\nn-\frac{1}{12} \pi ^2 H\left(0,-1; \frac{1}{a}\right)-H\left(0,0,0,-1; \frac{1}{a}\right)-H\left(0,0,1,-1; \frac{1}{a}\right)\\
G&(a,0,0,-1; 1) = 
-\log  2  H\left(0,0,-1; \frac{1}{a}\right)-\log  2  H\left(0,0,1; \frac{1}{a}\right)-\frac{3}{4} \zeta_3 H\left(1; \frac{1}{a}\right)\\
&\nn+\frac{1}{12} \pi ^2 H\left(0,1; \frac{1}{a}\right)-H\left(0,0,-1,1; \frac{1}{a}\right)+H\left(0,0,0,1; \frac{1}{a}\right)\\
G&(a,0,0,1; 1) = 
\zeta_3 H\left(1; \frac{1}{a}\right)-\frac{1}{6} \pi ^2 H\left(0,1; \frac{1}{a}\right)+H\left(0,0,0,1; \frac{1}{a}\right)\\
&\nn+H\left(0,0,1,1; \frac{1}{a}\right)\\
G&(-a,0,0,a; 1) = 
-H\left(-1,0,0,1; \frac{1}{a}\right)\\
G&(a,0,0,-a; 1) = 
-H\left(1,0,0,-1; \frac{1}{a}\right)\\
G&(a,0,0,a; 1) = 
H\left(1,0,0,1; \frac{1}{a}\right)\\
G&\left(-a,0,0,a^2; 1\right) = 
4 H\left(-1,0,0,-1; \frac{1}{a}\right)-4 H\left(-1,0,0,1; \frac{1}{a}\right)\\
&\nn+2 H\left(0,-1,0,-1; \frac{1}{a}\right)-2 H\left(0,-1,0,1; \frac{1}{a}\right)+H\left(0,0,-1,-1; \frac{1}{a}\right)\\
&\nn+H\left(0,0,0,-1; \frac{1}{a}\right)-H\left(0,0,0,1; \frac{1}{a}\right)\\
G&(a,0,1,0; 1) = 
-2 \zeta_3 H\left(1; \frac{1}{a}\right)+H\left(0,0,0,1; \frac{1}{a}\right)+H\left(0,1,0,1; \frac{1}{a}\right)\\
G&(a,0,1,1; 1) = 
-\zeta_3 H\left(1; \frac{1}{a}\right)+H\left(0,0,0,1; \frac{1}{a}\right)+H\left(0,0,1,1; \frac{1}{a}\right)\\
&\nn+H\left(0,1,0,1; \frac{1}{a}\right)+H\left(0,1,1,1; \frac{1}{a}\right)
\end{align}
\begin{align}
G&(a,0,1,a; 1) = 
-\frac{1}{3} \pi ^2 H\left(1,1; \frac{1}{a}\right)+2 H\left(0,1,0,1; \frac{1}{a}\right)+H\left(0,1,1,1; \frac{1}{a}\right)\\
&\nn+2 H\left(1,0,0,1; \frac{1}{a}\right)+H\left(1,0,1,1; \frac{1}{a}\right)+2 H\left(1,1,0,1; \frac{1}{a}\right)\\
G&\left(-a,0,1,a^2; 1\right) = 
-\frac{1}{3} \pi ^2 H\left(-1,-1; \frac{1}{a}\right)+\frac{1}{6} \pi ^2 H\left(-1,1; \frac{1}{a}\right)\\
&\nn+\frac{1}{6} \pi ^2 H\left(1,-1; \frac{1}{a}\right)\nn-4 H\left(-1,-1,0,-1; \frac{1}{a}\right)+4 H\left(-1,-1,0,1; \frac{1}{a}\right)\\
&\nn-2 H\left(-1,0,-1,-1; \frac{1}{a}\right)+4 H\left(-1,0,0,-1; \frac{1}{a}\right)-3 H\left(-1,0,0,1; \frac{1}{a}\right)\\
&\nn+2 H\left(-1,1,0,-1; \frac{1}{a}\right)-2 H\left(-1,1,0,1; \frac{1}{a}\right)-H\left(0,-1,-1,-1; \frac{1}{a}\right)\\
&\nn+5 H\left(0,-1,0,-1; \frac{1}{a}\right)-4 H\left(0,-1,0,1; \frac{1}{a}\right)+H\left(0,0,-1,-1; \frac{1}{a}\right)\\
&\nn+H\left(0,0,0,-1; \frac{1}{a}\right)-H\left(0,0,0,1; \frac{1}{a}\right)-2 H\left(0,1,0,-1; \frac{1}{a}\right)\\
&\nn+H\left(0,1,0,1; \frac{1}{a}\right)+2 H\left(1,-1,0,-1; \frac{1}{a}\right)-2 H\left(1,-1,0,1; \frac{1}{a}\right)\\
&\nn+2 H\left(1,0,-1,-1; \frac{1}{a}\right)-H\left(1,0,0,1; \frac{1}{a}\right)\\
G&\left(a,0,1,a^2; 1\right) = 
\frac{1}{6} \pi ^2 H\left(-1,1; \frac{1}{a}\right)+\frac{1}{6} \pi ^2 H\left(1,-1; \frac{1}{a}\right)-\frac{1}{3} \pi ^2 H\left(1,1; \frac{1}{a}\right)\\
&\nn-H\left(-1,0,0,-1; \frac{1}{a}\right)-2 H\left(-1,0,1,1; \frac{1}{a}\right)+2 H\left(-1,1,0,-1; \frac{1}{a}\right)\\
&\nn-2 H\left(-1,1,0,1; \frac{1}{a}\right)+H\left(0,-1,0,-1; \frac{1}{a}\right)-2 H\left(0,-1,0,1; \frac{1}{a}\right)\\
&\nn+H\left(0,0,0,-1; \frac{1}{a}\right)-H\left(0,0,0,1; \frac{1}{a}\right)+H\left(0,0,1,1; \frac{1}{a}\right)\\
&\nn-4 H\left(0,1,0,-1; \frac{1}{a}\right)+5 H\left(0,1,0,1; \frac{1}{a}\right)+H\left(0,1,1,1; \frac{1}{a}\right)\\
&\nn+2 H\left(1,-1,0,-1; \frac{1}{a}\right)-2 H\left(1,-1,0,1; \frac{1}{a}\right)-3 H\left(1,0,0,-1; \frac{1}{a}\right)\\
&\nn+4 H\left(1,0,0,1; \frac{1}{a}\right)+2 H\left(1,0,1,1; \frac{1}{a}\right)-4 H\left(1,1,0,-1; \frac{1}{a}\right)\\
&\nn+4 H\left(1,1,0,1; \frac{1}{a}\right)\\
G&(a,0,a,0; 1) = 
-H\left(0,1,0,1; \frac{1}{a}\right)-2 H\left(1,0,0,1; \frac{1}{a}\right)\\
G&(a,0,a,1; 1) = 
\frac{1}{3} \pi ^2 H\left(1,1; \frac{1}{a}\right)-H\left(0,1,0,1; \frac{1}{a}\right)-3 H\left(1,0,0,1; \frac{1}{a}\right)\\
&\nn-2 H\left(1,0,1,1; \frac{1}{a}\right)-2 H\left(1,1,0,1; \frac{1}{a}\right)
\end{align}
\begin{align}
G&\left(-a,0,a^2,1; 1\right) = 
\frac{1}{3} \pi ^2 H\left(-1,-1; \frac{1}{a}\right)-\frac{1}{6} \pi ^2 H\left(-1,1; \frac{1}{a}\right)\\
&\nn-\frac{1}{6} \pi ^2 H\left(1,-1; \frac{1}{a}\right)+4 H\left(-1,-1,0,-1; \frac{1}{a}\right)-4 H\left(-1,-1,0,1; \frac{1}{a}\right)\\
&\nn+4 H\left(-1,0,-1,-1; \frac{1}{a}\right)-2 H\left(-1,0,-1,1; \frac{1}{a}\right)-8 H\left(-1,0,0,-1; \frac{1}{a}\right)\\
&\nn+7 H\left(-1,0,0,1; \frac{1}{a}\right)-2 H\left(-1,0,1,-1; \frac{1}{a}\right)+2 H\left(-1,0,1,1; \frac{1}{a}\right)\\
&\nn-2 H\left(-1,1,0,-1; \frac{1}{a}\right)+2 H\left(-1,1,0,1; \frac{1}{a}\right)+H\left(0,-1,-1,-1; \frac{1}{a}\right)\\
&\nn-3 H\left(0,-1,0,-1; \frac{1}{a}\right)+3 H\left(0,-1,0,1; \frac{1}{a}\right)-3 H\left(0,0,-1,-1; \frac{1}{a}\right)\\
&\nn-H\left(0,0,-1,1; \frac{1}{a}\right)-6 H\left(0,0,0,-1; \frac{1}{a}\right)+6 H\left(0,0,0,1; \frac{1}{a}\right)\\
&\nn-H\left(0,0,1,-1; \frac{1}{a}\right)+H\left(0,0,1,1; \frac{1}{a}\right)-2 H\left(0,1,0,-1; \frac{1}{a}\right)\\
&\nn+H\left(0,1,0,1; \frac{1}{a}\right)-2 H\left(1,-1,0,-1; \frac{1}{a}\right)+2 H\left(1,-1,0,1; \frac{1}{a}\right)\\
&\nn-2 H\left(1,0,-1,-1; \frac{1}{a}\right)+H\left(1,0,0,1; \frac{1}{a}\right)\\
G&\left(a,0,a^2,1; 1\right) = 
-\frac{1}{6} \pi ^2 H\left(-1,1; \frac{1}{a}\right)-\frac{1}{6} \pi ^2 H\left(1,-1; \frac{1}{a}\right)+\frac{1}{3} \pi ^2 H\left(1,1; \frac{1}{a}\right)\\
&\nn+H\left(-1,0,0,-1; \frac{1}{a}\right)+2 H\left(-1,0,1,1; \frac{1}{a}\right)-2 H\left(-1,1,0,-1; \frac{1}{a}\right)\\
&\nn+2 H\left(-1,1,0,1; \frac{1}{a}\right)+H\left(0,-1,0,-1; \frac{1}{a}\right)-2 H\left(0,-1,0,1; \frac{1}{a}\right)\\
&\nn+H\left(0,0,-1,-1; \frac{1}{a}\right)-H\left(0,0,-1,1; \frac{1}{a}\right)-6 H\left(0,0,0,-1; \frac{1}{a}\right)\\
&\nn+6 H\left(0,0,0,1; \frac{1}{a}\right)-H\left(0,0,1,-1; \frac{1}{a}\right)-3 H\left(0,0,1,1; \frac{1}{a}\right)\\
&\nn+3 H\left(0,1,0,-1; \frac{1}{a}\right)-3 H\left(0,1,0,1; \frac{1}{a}\right)-H\left(0,1,1,1; \frac{1}{a}\right)\\
&\nn-2 H\left(1,-1,0,-1; \frac{1}{a}\right)+2 H\left(1,-1,0,1; \frac{1}{a}\right)-2 H\left(1,0,-1,-1; \frac{1}{a}\right)\\
&\nn+2 H\left(1,0,-1,1; \frac{1}{a}\right)+7 H\left(1,0,0,-1; \frac{1}{a}\right)-8 H\left(1,0,0,1; \frac{1}{a}\right)\\
&\nn+2 H\left(1,0,1,-1; \frac{1}{a}\right)-4 H\left(1,0,1,1; \frac{1}{a}\right)+4 H\left(1,1,0,-1; \frac{1}{a}\right)\\
&\nn-4 H\left(1,1,0,1; \frac{1}{a}\right)\\
G&(-a,0,-a,a; 1) = 
-H\left(-1,0,-1,1; \frac{1}{a}\right)\\
G&(-a,0,a,-a; 1) = 
-H\left(-1,0,1,-1; \frac{1}{a}\right)
\end{align}
\begin{align}
G&(-a,0,a,a; 1) = 
H\left(-1,0,1,1; \frac{1}{a}\right)\\
G&(a,0,-a,-a; 1) = 
-H\left(1,0,-1,-1; \frac{1}{a}\right)\\
G&(a,0,-a,a; 1) = 
H\left(1,0,-1,1; \frac{1}{a}\right)\\
G&(a,0,a,-a; 1) = 
H\left(1,0,1,-1; \frac{1}{a}\right)\\
G&(a,0,a,a; 1) = 
-H\left(1,0,1,1; \frac{1}{a}\right)\\
G&(a,1,0,0; 1) = 
H\left(0,0,0,1; \frac{1}{a}\right)+H\left(1,0,0,1; \frac{1}{a}\right)\\
G&(a,1,0,1; 1) = 
-\frac{1}{6} \pi ^2 H\left(0,1; \frac{1}{a}\right)-\frac{1}{6} \pi ^2 H\left(1,1; \frac{1}{a}\right)+H\left(0,0,0,1; \frac{1}{a}\right)\\
&\nn+H\left(0,0,1,1; \frac{1}{a}\right)+H\left(1,0,0,1; \frac{1}{a}\right)+H\left(1,0,1,1; \frac{1}{a}\right)\\
G&(a,1,0,a; 1) = 
-2 H\left(0,1,0,1; \frac{1}{a}\right)-H\left(0,1,1,1; \frac{1}{a}\right)-H\left(1,1,0,1; \frac{1}{a}\right)\\
G&\left(-a,1,0,a^2; 1\right) = 
2 H\left(-1,-1,0,-1; \frac{1}{a}\right)-2 H\left(-1,-1,0,1; \frac{1}{a}\right)\\
&\nn+H\left(-1,0,-1,-1; \frac{1}{a}\right)+H\left(-1,0,0,-1; \frac{1}{a}\right)-H\left(-1,0,0,1; \frac{1}{a}\right)\\
&\nn+2 H\left(0,-1,-1,-1; \frac{1}{a}\right)-4 H\left(0,-1,0,-1; \frac{1}{a}\right)+2 H\left(0,-1,0,1; \frac{1}{a}\right)\\
&\nn+H\left(0,0,-1,-1; \frac{1}{a}\right)+H\left(0,0,0,-1; \frac{1}{a}\right)-H\left(0,0,0,1; \frac{1}{a}\right)\\
&\nn+4 H\left(0,1,0,-1; \frac{1}{a}\right)-2 H\left(0,1,0,1; \frac{1}{a}\right)\\
G&\left(a,1,0,a^2; 1\right) = 
-2 H\left(0,-1,0,-1; \frac{1}{a}\right)+4 H\left(0,-1,0,1; \frac{1}{a}\right)\\
&\nn+H\left(0,0,0,-1; \frac{1}{a}\right)-H\left(0,0,0,1; \frac{1}{a}\right)+H\left(0,0,1,1; \frac{1}{a}\right)+2 H\left(0,1,0,-1; \frac{1}{a}\right)\\
&\nn-4 H\left(0,1,0,1; \frac{1}{a}\right)-2 H\left(0,1,1,1; \frac{1}{a}\right)-H\left(1,0,0,-1; \frac{1}{a}\right)+H\left(1,0,0,1; \frac{1}{a}\right)\\
&\nn-H\left(1,0,1,1; \frac{1}{a}\right)+2 H\left(1,1,0,-1; \frac{1}{a}\right)-2 H\left(1,1,0,1; \frac{1}{a}\right)\\
G&(a,1,1,0; 1) = 
H\left(0,0,0,1; \frac{1}{a}\right)+H\left(0,1,0,1; \frac{1}{a}\right)+H\left(1,0,0,1; \frac{1}{a}\right)\\
&\nn+H\left(1,1,0,1; \frac{1}{a}\right)
\end{align}
\begin{align}
G&(a,1,1,1; 1) = 
H\left(0,0,0,1; \frac{1}{a}\right)+H\left(0,0,1,1; \frac{1}{a}\right)+H\left(0,1,0,1; \frac{1}{a}\right)\\
&\nn+H\left(0,1,1,1; \frac{1}{a}\right)+H\left(1,0,0,1; \frac{1}{a}\right)+H\left(1,0,1,1; \frac{1}{a}\right)+H\left(1,1,0,1; \frac{1}{a}\right)\\
&\nn+H\left(1,1,1,1; \frac{1}{a}\right)\\
G&(a,1,1,a; 1) = 
H\left(1,0,0,1; \frac{1}{a}\right)+H\left(1,0,1,1; \frac{1}{a}\right)+H\left(1,1,0,1; \frac{1}{a}\right)\\
&\nn+H\left(1,1,1,1; \frac{1}{a}\right)\\
G&\left(-a,1,1,a^2; 1\right) = 
H\left(-1,-1,-1,-1; \frac{1}{a}\right)-H\left(-1,-1,0,-1; \frac{1}{a}\right)\\
&\nn-H\left(-1,0,-1,-1; \frac{1}{a}\right)+H\left(-1,0,0,-1; \frac{1}{a}\right)+H\left(0,-1,-1,-1; \frac{1}{a}\right)\\
&\nn-H\left(0,-1,0,-1; \frac{1}{a}\right)+H\left(0,0,-1,-1; \frac{1}{a}\right)+H\left(0,0,0,-1; \frac{1}{a}\right)\\
&\nn-H\left(0,0,0,1; \frac{1}{a}\right)+2 H\left(0,1,0,-1; \frac{1}{a}\right)-H\left(0,1,0,1; \frac{1}{a}\right)\\
&\nn+2 H\left(1,0,-1,-1; \frac{1}{a}\right)-H\left(1,0,0,1; \frac{1}{a}\right)+2 H\left(1,1,0,-1; \frac{1}{a}\right)\\
&\nn-H\left(1,1,0,1; \frac{1}{a}\right)\\
G&\left(a,1,1,a^2; 1\right) = 
H\left(-1,-1,0,-1; \frac{1}{a}\right)-2 H\left(-1,-1,0,1; \frac{1}{a}\right)\\
&\nn-H\left(-1,0,0,-1; \frac{1}{a}\right)-2 H\left(-1,0,1,1; \frac{1}{a}\right)-H\left(0,-1,0,-1; \frac{1}{a}\right)\\
&\nn+2 H\left(0,-1,0,1; \frac{1}{a}\right)+H\left(0,0,0,-1; \frac{1}{a}\right)-H\left(0,0,0,1; \frac{1}{a}\right)\\
&\nn+H\left(0,0,1,1; \frac{1}{a}\right)-H\left(0,1,0,1; \frac{1}{a}\right)-H\left(0,1,1,1; \frac{1}{a}\right)\\
&\nn+H\left(1,0,0,1; \frac{1}{a}\right)+H\left(1,0,1,1; \frac{1}{a}\right)+H\left(1,1,0,1; \frac{1}{a}\right)\\
&\nn+H\left(1,1,1,1; \frac{1}{a}\right)\\
G&(a,1,a,0; 1) = 
H\left(0,1,0,1; \frac{1}{a}\right)+H\left(0,1,1,1; \frac{1}{a}\right)+H\left(1,0,0,1; \frac{1}{a}\right)\\
&\nn+H\left(1,1,0,1; \frac{1}{a}\right)\\
G&(a,1,a,1; 1) = 
H\left(0,1,0,1; \frac{1}{a}\right)+H\left(0,1,1,1; \frac{1}{a}\right)+H\left(1,1,0,1; \frac{1}{a}\right)\\
&\nn+H\left(1,1,1,1; \frac{1}{a}\right)
\end{align}
\begin{align}
G&\left(-a,1,a^2,0; 1\right) = 
-H\left(-1,-1,0,-1; \frac{1}{a}\right)+H\left(-1,0,0,-1; \frac{1}{a}\right)\\
&\nn-2 H\left(0,-1,-1,-1; \frac{1}{a}\right)+H\left(0,-1,0,-1; \frac{1}{a}\right)-4 H\left(0,0,-1,-1; \frac{1}{a}\right)\\
&\nn-6 H\left(0,0,0,-1; \frac{1}{a}\right)+6 H\left(0,0,0,1; \frac{1}{a}\right)-4 H\left(0,1,0,-1; \frac{1}{a}\right)\\
&\nn+2 H\left(0,1,0,1; \frac{1}{a}\right)-2 H\left(1,0,-1,-1; \frac{1}{a}\right)-5 H\left(1,0,0,-1; \frac{1}{a}\right)\\
&\nn+4 H\left(1,0,0,1; \frac{1}{a}\right)\\
G&\left(a,1,a^2,0; 1\right) = 
4 H\left(-1,0,0,-1; \frac{1}{a}\right)-5 H\left(-1,0,0,1; \frac{1}{a}\right)\\
&\nn+2 H\left(-1,0,1,1; \frac{1}{a}\right)+2 H\left(0,-1,0,-1; \frac{1}{a}\right)-4 H\left(0,-1,0,1; \frac{1}{a}\right)\\
&\nn-6 H\left(0,0,0,-1; \frac{1}{a}\right)+6 H\left(0,0,0,1; \frac{1}{a}\right)-4 H\left(0,0,1,1; \frac{1}{a}\right)\\
&\nn+H\left(0,1,0,1; \frac{1}{a}\right)+2 H\left(0,1,1,1; \frac{1}{a}\right)+H\left(1,0,0,1; \frac{1}{a}\right)\\
&\nn+H\left(1,1,0,1; \frac{1}{a}\right)\\
G&\left(-a,1,a^2,1; 1\right) = 
H\left(-1,-1,-1,-1; \frac{1}{a}\right)-H\left(-1,-1,0,-1; \frac{1}{a}\right)\\
&\nn+H\left(-1,0,-1,-1; \frac{1}{a}\right)+H\left(-1,0,0,-1; \frac{1}{a}\right)-H\left(-1,0,0,1; \frac{1}{a}\right)\\
&\nn+2 H\left(-1,1,0,-1; \frac{1}{a}\right)-H\left(-1,1,0,1; \frac{1}{a}\right)-H\left(0,-1,-1,-1; \frac{1}{a}\right)\\
&\nn+3 H\left(0,-1,0,-1; \frac{1}{a}\right)-H\left(0,-1,0,1; \frac{1}{a}\right)-3 H\left(0,0,-1,-1; \frac{1}{a}\right)\\
&\nn-H\left(0,0,-1,1; \frac{1}{a}\right)-6 H\left(0,0,0,-1; \frac{1}{a}\right)+6 H\left(0,0,0,1; \frac{1}{a}\right)\\
&\nn-H\left(0,0,1,-1; \frac{1}{a}\right)+H\left(0,0,1,1; \frac{1}{a}\right)-6 H\left(0,1,0,-1; \frac{1}{a}\right)\\
&\nn+3 H\left(0,1,0,1; \frac{1}{a}\right)+2 H\left(1,-1,0,-1; \frac{1}{a}\right)-H\left(1,-1,0,1; \frac{1}{a}\right)\\
&\nn-2 H\left(1,0,-1,-1; \frac{1}{a}\right)-H\left(1,0,-1,1; \frac{1}{a}\right)-5 H\left(1,0,0,-1; \frac{1}{a}\right)\\
&\nn+5 H\left(1,0,0,1; \frac{1}{a}\right)-H\left(1,0,1,-1; \frac{1}{a}\right)+H\left(1,0,1,1; \frac{1}{a}\right)\\
&\nn-4 H\left(1,1,0,-1; \frac{1}{a}\right)+2 H\left(1,1,0,1; \frac{1}{a}\right)
\end{align}
\begin{align}
G&\left(a,1,a^2,1; 1\right) = 
-2 H\left(-1,-1,0,-1; \frac{1}{a}\right)+4 H\left(-1,-1,0,1; \frac{1}{a}\right)\\
&\nn-H\left(-1,0,-1,-1; \frac{1}{a}\right)+H\left(-1,0,-1,1; \frac{1}{a}\right)+5 H\left(-1,0,0,-1; \frac{1}{a}\right)\\
&\nn-5 H\left(-1,0,0,1; \frac{1}{a}\right)+H\left(-1,0,1,-1; \frac{1}{a}\right)+2 H\left(-1,0,1,1; \frac{1}{a}\right)\\
&\nn+H\left(-1,1,0,-1; \frac{1}{a}\right)-2 H\left(-1,1,0,1; \frac{1}{a}\right)+3 H\left(0,-1,0,-1; \frac{1}{a}\right)\\
&\nn-6 H\left(0,-1,0,1; \frac{1}{a}\right)+H\left(0,0,-1,-1; \frac{1}{a}\right)-H\left(0,0,-1,1; \frac{1}{a}\right)\\
&\nn-6 H\left(0,0,0,-1; \frac{1}{a}\right)+6 H\left(0,0,0,1; \frac{1}{a}\right)-H\left(0,0,1,-1; \frac{1}{a}\right)\\
&\nn-3 H\left(0,0,1,1; \frac{1}{a}\right)-H\left(0,1,0,-1; \frac{1}{a}\right)+3 H\left(0,1,0,1; \frac{1}{a}\right)\\
&\nn+H\left(0,1,1,1; \frac{1}{a}\right)+H\left(1,-1,0,-1; \frac{1}{a}\right)-2 H\left(1,-1,0,1; \frac{1}{a}\right)\\
&\nn-H\left(1,0,0,-1; \frac{1}{a}\right)+H\left(1,0,0,1; \frac{1}{a}\right)-H\left(1,0,1,1; \frac{1}{a}\right)\\
&\nn+H\left(1,1,0,1; \frac{1}{a}\right)+H\left(1,1,1,1; \frac{1}{a}\right)\\
G&\left(-a,1,a^2,a^2; 1\right) = 
H\left(-1,-1,-1,-1; \frac{1}{a}\right)-H\left(-1,-1,0,-1; \frac{1}{a}\right)\\
&\nn+H\left(-1,0,-1,-1; \frac{1}{a}\right)+H\left(-1,0,0,-1; \frac{1}{a}\right)-H\left(-1,0,0,1; \frac{1}{a}\right)\\
&\nn+2 H\left(-1,1,0,-1; \frac{1}{a}\right)-H\left(-1,1,0,1; \frac{1}{a}\right)+H\left(0,-1,-1,-1; \frac{1}{a}\right)\\
&\nn+H\left(0,-1,0,-1; \frac{1}{a}\right)-H\left(0,-1,0,1; \frac{1}{a}\right)+H\left(0,0,-1,-1; \frac{1}{a}\right)\\
&\nn-H\left(0,0,-1,1; \frac{1}{a}\right)-H\left(0,0,1,-1; \frac{1}{a}\right)+H\left(0,0,1,1; \frac{1}{a}\right)\\
&\nn+2 H\left(1,-1,0,-1; \frac{1}{a}\right)-H\left(1,-1,0,1; \frac{1}{a}\right)-H\left(1,0,-1,1; \frac{1}{a}\right)\\
&\nn-H\left(1,0,0,-1; \frac{1}{a}\right)+H\left(1,0,0,1; \frac{1}{a}\right)-H\left(1,0,1,-1; \frac{1}{a}\right)\\
&\nn+H\left(1,0,1,1; \frac{1}{a}\right)-2 H\left(1,1,0,-1; \frac{1}{a}\right)+H\left(1,1,0,1; \frac{1}{a}\right)
\end{align}
\begin{align}
G&\left(a,1,a^2,a^2; 1\right) = 
-H\left(-1,-1,0,-1; \frac{1}{a}\right)+2 H\left(-1,-1,0,1; \frac{1}{a}\right)\\
&\nn-H\left(-1,0,-1,-1; \frac{1}{a}\right)+H\left(-1,0,-1,1; \frac{1}{a}\right)+H\left(-1,0,0,-1; \frac{1}{a}\right)\\
&\nn-H\left(-1,0,0,1; \frac{1}{a}\right)+H\left(-1,0,1,-1; \frac{1}{a}\right)+H\left(-1,1,0,-1; \frac{1}{a}\right)\\
&\nn-2 H\left(-1,1,0,1; \frac{1}{a}\right)+H\left(0,0,-1,-1; \frac{1}{a}\right)-H\left(0,0,-1,1; \frac{1}{a}\right)\\
&\nn-H\left(0,0,1,-1; \frac{1}{a}\right)+H\left(0,0,1,1; \frac{1}{a}\right)-H\left(0,1,0,-1; \frac{1}{a}\right)\\
&\nn+H\left(0,1,0,1; \frac{1}{a}\right)-H\left(0,1,1,1; \frac{1}{a}\right)+H\left(1,-1,0,-1; \frac{1}{a}\right)\\
&\nn-2 H\left(1,-1,0,1; \frac{1}{a}\right)-H\left(1,0,0,-1; \frac{1}{a}\right)+H\left(1,0,0,1; \frac{1}{a}\right)\\
&\nn-H\left(1,0,1,1; \frac{1}{a}\right)+H\left(1,1,0,1; \frac{1}{a}\right)+H\left(1,1,1,1; \frac{1}{a}\right)\\
G&(a,1,a,a; 1) = 
H\left(1,1,0,1; \frac{1}{a}\right)+H\left(1,1,1,1; \frac{1}{a}\right)\\
G&\left(a^2,0,0,-a; 1\right) = 
H\left(-1,0,0,-1; \frac{1}{a}\right)-H\left(0,-1,0,-1; \frac{1}{a}\right)\\
&\nn+H\left(0,0,-1,-1; \frac{1}{a}\right)-H\left(0,0,-1,1; \frac{1}{a}\right)-H\left(0,0,0,-1; \frac{1}{a}\right)+H\left(0,0,0,1; \frac{1}{a}\right)\\
&\nn-H\left(0,0,1,-1; \frac{1}{a}\right)+H\left(0,1,0,-1; \frac{1}{a}\right)-H\left(1,0,0,-1; \frac{1}{a}\right)\\
G&\left(a^2,0,0,a; 1\right) = 
-H\left(-1,0,0,1; \frac{1}{a}\right)+H\left(0,-1,0,1; \frac{1}{a}\right)-H\left(0,0,-1,1; \frac{1}{a}\right)\\
&\nn-H\left(0,0,0,-1; \frac{1}{a}\right)+H\left(0,0,0,1; \frac{1}{a}\right)-H\left(0,0,1,-1; \frac{1}{a}\right)+H\left(0,0,1,1; \frac{1}{a}\right)\\
&\nn-H\left(0,1,0,1; \frac{1}{a}\right)+H\left(1,0,0,1; \frac{1}{a}\right)\\
G&\left(a^2,0,-a,0; 1\right) = 
-2 H\left(-1,0,0,-1; \frac{1}{a}\right)+H\left(0,-1,0,-1; \frac{1}{a}\right)\\
&\nn-2 H\left(0,-1,0,1; \frac{1}{a}\right)+2 H\left(0,0,-1,-1; \frac{1}{a}\right)-2 H\left(0,0,-1,1; \frac{1}{a}\right)\\
&\nn-4 H\left(0,0,0,-1; \frac{1}{a}\right)+4 H\left(0,0,0,1; \frac{1}{a}\right)-2 H\left(0,0,1,-1; \frac{1}{a}\right)+2 H\left(1,0,0,-1; \frac{1}{a}\right)\\
G&\left(a^2,0,a,0; 1\right) = 
2 H\left(-1,0,0,1; \frac{1}{a}\right)-2 H\left(0,0,-1,1; \frac{1}{a}\right)\\
&\nn-4 H\left(0,0,0,-1; \frac{1}{a}\right)+4 H\left(0,0,0,1; \frac{1}{a}\right)-2 H\left(0,0,1,-1; \frac{1}{a}\right)\\
&\nn+2 H\left(0,0,1,1; \frac{1}{a}\right)-2 H\left(0,1,0,-1; \frac{1}{a}\right)+H\left(0,1,0,1; \frac{1}{a}\right)-2 H\left(1,0,0,1; \frac{1}{a}\right)
\end{align}
\begin{align}
G&\left(a^2,0,-a,1; 1\right) = 
\frac{1}{3} \pi ^2 H\left(-1,-1; \frac{1}{a}\right)-\frac{1}{6} \pi ^2 H\left(-1,1; \frac{1}{a}\right)\\
&\nn-\frac{1}{6} \pi ^2 H\left(1,-1; \frac{1}{a}\right)+2 H\left(-1,-1,0,-1; \frac{1}{a}\right)+2 H\left(-1,0,-1,-1; \frac{1}{a}\right)\\
&\nn-H\left(-1,0,-1,1; \frac{1}{a}\right)-5 H\left(-1,0,0,-1; \frac{1}{a}\right)+3 H\left(-1,0,0,1; \frac{1}{a}\right)\\
&\nn-H\left(-1,0,1,-1; \frac{1}{a}\right)+2 H\left(-1,0,1,1; \frac{1}{a}\right)-H\left(-1,1,0,-1; \frac{1}{a}\right)\\
&\nn-H\left(0,-1,-1,-1; \frac{1}{a}\right)+H\left(0,-1,-1,1; \frac{1}{a}\right)-H\left(0,-1,0,1; \frac{1}{a}\right)\\
&\nn+H\left(0,-1,1,-1; \frac{1}{a}\right)+3 H\left(0,0,-1,-1; \frac{1}{a}\right)-3 H\left(0,0,-1,1; \frac{1}{a}\right)\\
&\nn-4 H\left(0,0,0,-1; \frac{1}{a}\right)+4 H\left(0,0,0,1; \frac{1}{a}\right)-3 H\left(0,0,1,-1; \frac{1}{a}\right)+H\left(0,0,1,1; \frac{1}{a}\right)\\
&\nn+H\left(0,1,-1,-1; \frac{1}{a}\right)-H\left(1,-1,0,-1; \frac{1}{a}\right)-H\left(1,0,-1,-1; \frac{1}{a}\right)+2 H\left(1,0,0,-1; \frac{1}{a}\right)\\
G&\left(a^2,0,a,1; 1\right) = 
-\frac{1}{6} \pi ^2 H\left(-1,1; \frac{1}{a}\right)-\frac{1}{6} \pi ^2 H\left(1,-1; \frac{1}{a}\right)+\frac{1}{3} \pi ^2 H\left(1,1; \frac{1}{a}\right)\\
&\nn+2 H\left(-1,0,0,1; \frac{1}{a}\right)+H\left(-1,0,1,1; \frac{1}{a}\right)+H\left(-1,1,0,1; \frac{1}{a}\right)\\
&\nn-H\left(0,-1,1,1; \frac{1}{a}\right)+H\left(0,0,-1,-1; \frac{1}{a}\right)-3 H\left(0,0,-1,1; \frac{1}{a}\right)\\
&\nn-4 H\left(0,0,0,-1; \frac{1}{a}\right)+4 H\left(0,0,0,1; \frac{1}{a}\right)-3 H\left(0,0,1,-1; \frac{1}{a}\right)\\
&\nn+3 H\left(0,0,1,1; \frac{1}{a}\right)-H\left(0,1,-1,1; \frac{1}{a}\right)-H\left(0,1,0,-1; \frac{1}{a}\right)\\
&\nn-H\left(0,1,1,-1; \frac{1}{a}\right)+H\left(0,1,1,1; \frac{1}{a}\right)+H\left(1,-1,0,1; \frac{1}{a}\right)\\
&\nn-2 H\left(1,0,-1,-1; \frac{1}{a}\right)+H\left(1,0,-1,1; \frac{1}{a}\right)+3 H\left(1,0,0,-1; \frac{1}{a}\right)\\
&\nn-5 H\left(1,0,0,1; \frac{1}{a}\right)+H\left(1,0,1,-1; \frac{1}{a}\right)-2 H\left(1,0,1,1; \frac{1}{a}\right)-2 H\left(1,1,0,1; \frac{1}{a}\right)\\
G&\left(a^2,0,a^2,-a; 1\right) = 
2 H\left(-1,-1,0,-1; \frac{1}{a}\right)-4 H\left(-1,-1,0,1; \frac{1}{a}\right)\\
&\nn+3 H\left(-1,0,-1,-1; \frac{1}{a}\right)-3 H\left(-1,0,-1,1; \frac{1}{a}\right)-5 H\left(-1,0,0,-1; \frac{1}{a}\right)\\
&\nn+5 H\left(-1,0,0,1; \frac{1}{a}\right)-3 H\left(-1,0,1,-1; \frac{1}{a}\right)-H\left(-1,1,0,-1; \frac{1}{a}\right)\\
&\nn+2 H\left(-1,1,0,1; \frac{1}{a}\right)-2 H\left(0,-1,0,-1; \frac{1}{a}\right)+2 H\left(0,-1,0,1; \frac{1}{a}\right)\\
&\nn+2 H\left(0,1,0,-1; \frac{1}{a}\right)-2 H\left(0,1,0,1; \frac{1}{a}\right)-H\left(1,-1,0,-1; \frac{1}{a}\right)
\end{align}

\begin{align}
&\nn+2 H\left(1,-1,0,1; \frac{1}{a}\right)-2 H\left(1,0,-1,-1; \frac{1}{a}\right)+2 H\left(1,0,-1,1; \frac{1}{a}\right)\\
&\nn+4 H\left(1,0,0,-1; \frac{1}{a}\right)-4 H\left(1,0,0,1; \frac{1}{a}\right)+2 H\left(1,0,1,-1; \frac{1}{a}\right)\\
G&\left(a^2,0,a^2,a; 1\right) = 
-2 H\left(-1,0,-1,1; \frac{1}{a}\right)-4 H\left(-1,0,0,-1; \frac{1}{a}\right)\\
&\nn+4 H\left(-1,0,0,1; \frac{1}{a}\right)-2 H\left(-1,0,1,-1; \frac{1}{a}\right)+2 H\left(-1,0,1,1; \frac{1}{a}\right)\\
&\nn-2 H\left(-1,1,0,-1; \frac{1}{a}\right)+H\left(-1,1,0,1; \frac{1}{a}\right)-2 H\left(0,-1,0,-1; \frac{1}{a}\right)\\
&\nn+2 H\left(0,-1,0,1; \frac{1}{a}\right)+2 H\left(0,1,0,-1; \frac{1}{a}\right)-2 H\left(0,1,0,1; \frac{1}{a}\right)\\
&\nn-2 H\left(1,-1,0,-1; \frac{1}{a}\right)+H\left(1,-1,0,1; \frac{1}{a}\right)+3 H\left(1,0,-1,1; \frac{1}{a}\right)\\
&\nn+5 H\left(1,0,0,-1; \frac{1}{a}\right)-5 H\left(1,0,0,1; \frac{1}{a}\right)+3 H\left(1,0,1,-1; \frac{1}{a}\right)\\
&\nn-3 H\left(1,0,1,1; \frac{1}{a}\right)+4 H\left(1,1,0,-1; \frac{1}{a}\right)-2 H\left(1,1,0,1; \frac{1}{a}\right)\\
G&\left(a^2,0,-a,a^2; 1\right) = 
-2 H\left(-1,-1,0,-1; \frac{1}{a}\right)+4 H\left(-1,-1,0,1; \frac{1}{a}\right)\\
&\nn+H\left(-1,0,-1,1; \frac{1}{a}\right)+4 H\left(-1,0,0,-1; \frac{1}{a}\right)-4 H\left(-1,0,0,1; \frac{1}{a}\right)\\
&\nn+H\left(-1,0,1,-1; \frac{1}{a}\right)+H\left(-1,1,0,-1; \frac{1}{a}\right)-2 H\left(-1,1,0,1; \frac{1}{a}\right)\\
&\nn-H\left(0,-1,-1,-1; \frac{1}{a}\right)+H\left(0,-1,-1,1; \frac{1}{a}\right)+3 H\left(0,-1,0,-1; \frac{1}{a}\right)\\
&\nn-3 H\left(0,-1,0,1; \frac{1}{a}\right)+H\left(0,-1,1,-1; \frac{1}{a}\right)+H\left(0,1,-1,-1; \frac{1}{a}\right)\\
&\nn-H\left(0,1,0,-1; \frac{1}{a}\right)+H\left(0,1,0,1; \frac{1}{a}\right)+H\left(1,-1,0,-1; \frac{1}{a}\right)-2 H\left(1,-1,0,1; \frac{1}{a}\right)\\
&\nn-H\left(1,0,-1,-1; \frac{1}{a}\right)-3 H\left(1,0,0,-1; \frac{1}{a}\right)+3 H\left(1,0,0,1; \frac{1}{a}\right)\\
G&\left(a^2,0,a,a^2; 1\right) = 
3 H\left(-1,0,0,-1; \frac{1}{a}\right)-3 H\left(-1,0,0,1; \frac{1}{a}\right)\\
&\nn+H\left(-1,0,1,1; \frac{1}{a}\right)+2 H\left(-1,1,0,-1; \frac{1}{a}\right)-H\left(-1,1,0,1; \frac{1}{a}\right)\\
&\nn+H\left(0,-1,0,-1; \frac{1}{a}\right)-H\left(0,-1,0,1; \frac{1}{a}\right)-H\left(0,-1,1,1; \frac{1}{a}\right)+2 H\left(1,1,0,1; \frac{1}{a}\right)\\
&\nn-H\left(0,1,-1,1; \frac{1}{a}\right)-3 H\left(0,1,0,-1; \frac{1}{a}\right)+3 H\left(0,1,0,1; \frac{1}{a}\right)-H\left(0,1,1,-1; \frac{1}{a}\right)\\
&\nn+H\left(0,1,1,1; \frac{1}{a}\right)+2 H\left(1,-1,0,-1; \frac{1}{a}\right)-H\left(1,-1,0,1; \frac{1}{a}\right)-H\left(1,0,-1,1; \frac{1}{a}\right)\\
&\nn-4 H\left(1,0,0,-1; \frac{1}{a}\right)+4 H\left(1,0,0,1; \frac{1}{a}\right)-H\left(1,0,1,-1; \frac{1}{a}\right)-4 H\left(1,1,0,-1; \frac{1}{a}\right)
\end{align}
\begin{align}
G&\left(a^2,-a,0,0; 1\right) = 
4 H\left(-1,0,0,-1; \frac{1}{a}\right)-4 H\left(-1,0,0,1; \frac{1}{a}\right)\\
&\nn+2 H\left(0,-1,0,-1; \frac{1}{a}\right)-2 H\left(0,-1,0,1; \frac{1}{a}\right)+H\left(0,0,-1,-1; \frac{1}{a}\right)\\
&\nn-H\left(0,0,-1,1; \frac{1}{a}\right)-7 H\left(0,0,0,-1; \frac{1}{a}\right)+7 H\left(0,0,0,1; \frac{1}{a}\right)\\
&\nn-H\left(0,0,1,-1; \frac{1}{a}\right)-H\left(0,1,0,-1; \frac{1}{a}\right)-H\left(1,0,0,-1; \frac{1}{a}\right)\\
G&\left(a^2,a,0,0; 1\right) = 
-H\left(-1,0,0,1; \frac{1}{a}\right)-H\left(0,-1,0,1; \frac{1}{a}\right)-H\left(0,0,-1,1; \frac{1}{a}\right)\\
&\nn-7 H\left(0,0,0,-1; \frac{1}{a}\right)+7 H\left(0,0,0,1; \frac{1}{a}\right)-H\left(0,0,1,-1; \frac{1}{a}\right)\\
&\nn+H\left(0,0,1,1; \frac{1}{a}\right)-2 H\left(0,1,0,-1; \frac{1}{a}\right)+2 H\left(0,1,0,1; \frac{1}{a}\right)\\
&\nn-4 H\left(1,0,0,-1; \frac{1}{a}\right)+4 H\left(1,0,0,1; \frac{1}{a}\right)\\
G&\left(a^2,-a,0,1; 1\right) = 
-\frac{1}{6} \pi ^2 H\left(-1,-1; \frac{1}{a}\right)+\frac{1}{6} \pi ^2 H\left(-1,1; \frac{1}{a}\right)\\
&\nn+\frac{1}{6} \pi ^2 H\left(0,-1; \frac{1}{a}\right)-\frac{1}{6} \pi ^2 H\left(0,1; \frac{1}{a}\right)+\frac{1}{6} \pi ^2 H\left(1,-1; \frac{1}{a}\right)-2 H\left(-1,0,-1,-1; \frac{1}{a}\right)\\
&\nn+2 H\left(-1,0,-1,1; \frac{1}{a}\right)+4 H\left(-1,0,0,-1; \frac{1}{a}\right)-4 H\left(-1,0,0,1; \frac{1}{a}\right)\\
&\nn+2 H\left(-1,0,1,-1; \frac{1}{a}\right)-2 H\left(-1,0,1,1; \frac{1}{a}\right)-H\left(0,-1,-1,-1; \frac{1}{a}\right)\\
&\nn+H\left(0,-1,-1,1; \frac{1}{a}\right)+H\left(0,-1,0,-1; \frac{1}{a}\right)-H\left(0,-1,0,1; \frac{1}{a}\right)\\
&\nn+H\left(0,-1,1,-1; \frac{1}{a}\right)+4 H\left(0,0,-1,-1; \frac{1}{a}\right)-4 H\left(0,0,-1,1; \frac{1}{a}\right)\\
&\nn-7 H\left(0,0,0,-1; \frac{1}{a}\right)+7 H\left(0,0,0,1; \frac{1}{a}\right)-4 H\left(0,0,1,-1; \frac{1}{a}\right)\\
&\nn+3 H\left(0,0,1,1; \frac{1}{a}\right)+H\left(0,1,-1,-1; \frac{1}{a}\right)-H\left(0,1,0,-1; \frac{1}{a}\right)\\
&\nn+H\left(1,0,-1,-1; \frac{1}{a}\right)-H\left(1,0,0,-1; \frac{1}{a}\right)\\
G&\left(a^2,a,0,1; 1\right) = 
\frac{1}{6} \pi ^2 H\left(-1,1; \frac{1}{a}\right)+\frac{1}{6} \pi ^2 H\left(0,-1; \frac{1}{a}\right)-\frac{1}{6} \pi ^2 H\left(0,1; \frac{1}{a}\right)\\
&\nn+\frac{1}{6} \pi ^2 H\left(1,-1; \frac{1}{a}\right)-\frac{1}{6} \pi ^2 H\left(1,1; \frac{1}{a}\right)-H\left(-1,0,0,1; \frac{1}{a}\right)-H\left(-1,0,1,1; \frac{1}{a}\right)\\
&\nn-H\left(0,-1,0,1; \frac{1}{a}\right)-H\left(0,-1,1,1; \frac{1}{a}\right)+3 H\left(0,0,-1,-1; \frac{1}{a}\right)-4 H\left(0,0,-1,1; \frac{1}{a}\right)\\
&\nn-7 H\left(0,0,0,-1; \frac{1}{a}\right)+7 H\left(0,0,0,1; \frac{1}{a}\right)-4 H\left(0,0,1,-1; \frac{1}{a}\right)+4 H\left(0,0,1,1; \frac{1}{a}\right)\\
&\nn-H\left(0,1,-1,1; \frac{1}{a}\right)-H\left(0,1,0,-1; \frac{1}{a}\right)+H\left(0,1,0,1; \frac{1}{a}\right)-H\left(0,1,1,-1; \frac{1}{a}\right)
\end{align}

\begin{align}
&\nn+H\left(0,1,1,1; \frac{1}{a}\right)+2 H\left(1,0,-1,-1; \frac{1}{a}\right)-2 H\left(1,0,-1,1; \frac{1}{a}\right)-4 H\left(1,0,0,-1; \frac{1}{a}\right)\\
&\nn+4 H\left(1,0,0,1; \frac{1}{a}\right)-2 H\left(1,0,1,-1; \frac{1}{a}\right)+2 H\left(1,0,1,1; \frac{1}{a}\right)\\
G&\left(a^2,-a,0,a^2; 1\right) = 
2 H\left(-1,-1,0,-1; \frac{1}{a}\right)-2 H\left(-1,-1,0,1; \frac{1}{a}\right)\\
&\nn-2 H\left(-1,1,0,-1; \frac{1}{a}\right)+2 H\left(-1,1,0,1; \frac{1}{a}\right)+H\left(0,-1,-1,-1; \frac{1}{a}\right)\\
&\nn-H\left(0,-1,-1,1; \frac{1}{a}\right)-3 H\left(0,-1,0,-1; \frac{1}{a}\right)+3 H\left(0,-1,0,1; \frac{1}{a}\right)\\
&\nn-H\left(0,-1,1,-1; \frac{1}{a}\right)-H\left(0,1,-1,-1; \frac{1}{a}\right)+H\left(0,1,0,-1; \frac{1}{a}\right)\\
&\nn-H\left(0,1,0,1; \frac{1}{a}\right)-2 H\left(1,-1,0,-1; \frac{1}{a}\right)+2 H\left(1,-1,0,1; \frac{1}{a}\right)\\
&\nn-H\left(1,0,-1,-1; \frac{1}{a}\right)-H\left(1,0,0,-1; \frac{1}{a}\right)+H\left(1,0,0,1; \frac{1}{a}\right)\\
G&\left(a^2,a,0,a^2; 1\right) = 
H\left(-1,0,0,-1; \frac{1}{a}\right)-H\left(-1,0,0,1; \frac{1}{a}\right)\\
&\nn+H\left(-1,0,1,1; \frac{1}{a}\right)-2 H\left(-1,1,0,-1; \frac{1}{a}\right)+2 H\left(-1,1,0,1; \frac{1}{a}\right)\\
&\nn-H\left(0,-1,0,-1; \frac{1}{a}\right)+H\left(0,-1,0,1; \frac{1}{a}\right)+H\left(0,-1,1,1; \frac{1}{a}\right)+H\left(0,1,-1,1; \frac{1}{a}\right)\\
&\nn+3 H\left(0,1,0,-1; \frac{1}{a}\right)-3 H\left(0,1,0,1; \frac{1}{a}\right)+H\left(0,1,1,-1; \frac{1}{a}\right)-H\left(0,1,1,1; \frac{1}{a}\right)\\
&\nn-2 H\left(1,-1,0,-1; \frac{1}{a}\right)+2 H\left(1,-1,0,1; \frac{1}{a}\right)+2 H\left(1,1,0,-1; \frac{1}{a}\right)-2 H\left(1,1,0,1; \frac{1}{a}\right)\\
G&\left(a^2,-a,1,0; 1\right) = 
-2 H\left(-1,-1,0,-1; \frac{1}{a}\right)+2 H\left(-1,-1,0,1; \frac{1}{a}\right)\\
&\nn-H\left(-1,0,-1,-1; \frac{1}{a}\right)+H\left(-1,0,-1,1; \frac{1}{a}\right)+3 H\left(-1,0,0,-1; \frac{1}{a}\right)\\
&\nn-3 H\left(-1,0,0,1; \frac{1}{a}\right)+H\left(-1,0,1,-1; \frac{1}{a}\right)+H\left(-1,1,0,-1; \frac{1}{a}\right)\\
&\nn+4 H\left(0,-1,0,-1; \frac{1}{a}\right)-4 H\left(0,-1,0,1; \frac{1}{a}\right)+H\left(0,0,-1,-1; \frac{1}{a}\right)\\
&\nn-H\left(0,0,-1,1; \frac{1}{a}\right)-7 H\left(0,0,0,-1; \frac{1}{a}\right)+7 H\left(0,0,0,1; \frac{1}{a}\right)-H\left(0,0,1,-1; \frac{1}{a}\right)\\
&\nn-3 H\left(0,1,0,-1; \frac{1}{a}\right)+2 H\left(0,1,0,1; \frac{1}{a}\right)+H\left(1,-1,0,-1; \frac{1}{a}\right)-H\left(1,0,0,-1; \frac{1}{a}\right)
\end{align}
\begin{align}
G&\left(a^2,a,1,0; 1\right) = 
-H\left(-1,0,0,1; \frac{1}{a}\right)-H\left(-1,1,0,1; \frac{1}{a}\right)\\
&\nn+2 H\left(0,-1,0,-1; \frac{1}{a}\right)-3 H\left(0,-1,0,1; \frac{1}{a}\right)-H\left(0,0,-1,1; \frac{1}{a}\right)\\
&\nn-7 H\left(0,0,0,-1; \frac{1}{a}\right)+7 H\left(0,0,0,1; \frac{1}{a}\right)-H\left(0,0,1,-1; \frac{1}{a}\right)\\
&\nn+H\left(0,0,1,1; \frac{1}{a}\right)-4 H\left(0,1,0,-1; \frac{1}{a}\right)+4 H\left(0,1,0,1; \frac{1}{a}\right)\\
&\nn-H\left(1,-1,0,1; \frac{1}{a}\right)-H\left(1,0,-1,1; \frac{1}{a}\right)-3 H\left(1,0,0,-1; \frac{1}{a}\right)\\
&\nn+3 H\left(1,0,0,1; \frac{1}{a}\right)-H\left(1,0,1,-1; \frac{1}{a}\right)+H\left(1,0,1,1; \frac{1}{a}\right)\\
&\nn-2 H\left(1,1,0,-1; \frac{1}{a}\right)+2 H\left(1,1,0,1; \frac{1}{a}\right)\\
G&\left(a^2,-a,1,a^2; 1\right) = 
H\left(-1,-1,-1,-1; \frac{1}{a}\right)-H\left(-1,-1,-1,1; \frac{1}{a}\right)\\
&\nn-H\left(-1,-1,0,-1; \frac{1}{a}\right)+H\left(-1,-1,0,1; \frac{1}{a}\right)-H\left(-1,-1,1,-1; \frac{1}{a}\right)\\
&\nn-2 H\left(-1,0,-1,-1; \frac{1}{a}\right)+2 H\left(-1,0,-1,1; \frac{1}{a}\right)+3 H\left(-1,0,0,-1; \frac{1}{a}\right)\\
&\nn-3 H\left(-1,0,0,1; \frac{1}{a}\right)+2 H\left(-1,0,1,-1; \frac{1}{a}\right)-H\left(-1,0,1,1; \frac{1}{a}\right)\\
&\nn-H\left(-1,1,-1,-1; \frac{1}{a}\right)+H\left(-1,1,0,-1; \frac{1}{a}\right)+H\left(0,-1,0,-1; \frac{1}{a}\right)\\
&\nn-H\left(0,-1,0,1; \frac{1}{a}\right)-H\left(0,-1,1,1; \frac{1}{a}\right)-H\left(0,1,-1,1; \frac{1}{a}\right)\\
&\nn-3 H\left(0,1,0,-1; \frac{1}{a}\right)+3 H\left(0,1,0,1; \frac{1}{a}\right)-H\left(0,1,1,-1; \frac{1}{a}\right)\\
&\nn+H\left(0,1,1,1; \frac{1}{a}\right)-H\left(1,-1,-1,-1; \frac{1}{a}\right)+H\left(1,-1,0,-1; \frac{1}{a}\right)\\
&\nn+H\left(1,0,-1,-1; \frac{1}{a}\right)-2 H\left(1,0,-1,1; \frac{1}{a}\right)-4 H\left(1,0,0,-1; \frac{1}{a}\right)\\
&\nn+4 H\left(1,0,0,1; \frac{1}{a}\right)-2 H\left(1,0,1,-1; \frac{1}{a}\right)+H\left(1,0,1,1; \frac{1}{a}\right)\\
&\nn-4 H\left(1,1,0,-1; \frac{1}{a}\right)+2 H\left(1,1,0,1; \frac{1}{a}\right)\\
G&\left(a^2,a,1,a^2; 1\right) = 
-2 H\left(-1,-1,0,-1; \frac{1}{a}\right)+4 H\left(-1,-1,0,1; \frac{1}{a}\right)\\
&\nn-H\left(-1,0,-1,-1; \frac{1}{a}\right)+2 H\left(-1,0,-1,1; \frac{1}{a}\right)+4 H\left(-1,0,0,-1; \frac{1}{a}\right)\\
&\nn-4 H\left(-1,0,0,1; \frac{1}{a}\right)+2 H\left(-1,0,1,-1; \frac{1}{a}\right)-H\left(-1,0,1,1; \frac{1}{a}\right)\\
&\nn-H\left(-1,1,0,1; \frac{1}{a}\right)-H\left(-1,1,1,1; \frac{1}{a}\right)-H\left(0,-1,-1,-1; \frac{1}{a}\right)
\end{align}

\begin{align}
&\nn+H\left(0,-1,-1,1; \frac{1}{a}\right)+3 H\left(0,-1,0,-1; \frac{1}{a}\right)-3 H\left(0,-1,0,1; \frac{1}{a}\right)\\
&\nn+H\left(0,-1,1,-1; \frac{1}{a}\right)+H\left(0,1,-1,-1; \frac{1}{a}\right)-H\left(0,1,0,-1; \frac{1}{a}\right)\\
&\nn+H\left(0,1,0,1; \frac{1}{a}\right)-H\left(1,-1,0,1; \frac{1}{a}\right)-H\left(1,-1,1,1; \frac{1}{a}\right)\\
&\nn+H\left(1,0,-1,-1; \frac{1}{a}\right)-2 H\left(1,0,-1,1; \frac{1}{a}\right)-3 H\left(1,0,0,-1; \frac{1}{a}\right)\\
&\nn+3 H\left(1,0,0,1; \frac{1}{a}\right)-2 H\left(1,0,1,-1; \frac{1}{a}\right)+2 H\left(1,0,1,1; \frac{1}{a}\right)\\
&\nn-H\left(1,1,-1,1; \frac{1}{a}\right)-H\left(1,1,0,-1; \frac{1}{a}\right)+H\left(1,1,0,1; \frac{1}{a}\right)\\
&\nn-H\left(1,1,1,-1; \frac{1}{a}\right)+H\left(1,1,1,1; \frac{1}{a}\right)\\
G&\left(a^2,a^2,0,-a; 1\right) = 
H\left(-1,-1,0,-1; \frac{1}{a}\right)-H\left(-1,0,-1,-1; \frac{1}{a}\right)\\
&\nn+H\left(-1,0,-1,1; \frac{1}{a}\right)+H\left(-1,0,0,-1; \frac{1}{a}\right)-H\left(-1,0,0,1; \frac{1}{a}\right)\\
&\nn+H\left(-1,0,1,-1; \frac{1}{a}\right)-H\left(-1,1,0,-1; \frac{1}{a}\right)-H\left(0,-1,-1,-1; \frac{1}{a}\right)\\
&\nn+H\left(0,-1,-1,1; \frac{1}{a}\right)+H\left(0,-1,1,-1; \frac{1}{a}\right)-H\left(0,-1,1,1; \frac{1}{a}\right)\\
&\nn+H\left(0,0,-1,-1; \frac{1}{a}\right)-H\left(0,0,-1,1; \frac{1}{a}\right)-H\left(0,0,1,-1; \frac{1}{a}\right)\\
&\nn+H\left(0,0,1,1; \frac{1}{a}\right)+H\left(0,1,-1,-1; \frac{1}{a}\right)-H\left(0,1,-1,1; \frac{1}{a}\right)\\
&\nn-H\left(0,1,0,-1; \frac{1}{a}\right)+H\left(0,1,0,1; \frac{1}{a}\right)-H\left(0,1,1,-1; \frac{1}{a}\right)\\
&\nn-H\left(1,-1,0,-1; \frac{1}{a}\right)+H\left(1,0,-1,-1; \frac{1}{a}\right)-H\left(1,0,-1,1; \frac{1}{a}\right)\\
&\nn-H\left(1,0,0,-1; \frac{1}{a}\right)+H\left(1,0,0,1; \frac{1}{a}\right)-H\left(1,0,1,-1; \frac{1}{a}\right)+H\left(1,1,0,-1; \frac{1}{a}\right)\\
G&\left(a^2,a^2,0,a; 1\right) = 
-H\left(-1,-1,0,1; \frac{1}{a}\right)+H\left(-1,0,-1,1; \frac{1}{a}\right)\\
&\nn+H\left(-1,0,0,-1; \frac{1}{a}\right)-H\left(-1,0,0,1; \frac{1}{a}\right)+H\left(-1,0,1,-1; \frac{1}{a}\right)\\
&\nn-H\left(-1,0,1,1; \frac{1}{a}\right)+H\left(-1,1,0,1; \frac{1}{a}\right)+H\left(0,-1,-1,1; \frac{1}{a}\right)\\
&\nn+H\left(0,-1,0,-1; \frac{1}{a}\right)-H\left(0,-1,0,1; \frac{1}{a}\right)+H\left(0,-1,1,-1; \frac{1}{a}\right)\\
&\nn-H\left(0,-1,1,1; \frac{1}{a}\right)+H\left(0,0,-1,-1; \frac{1}{a}\right)-H\left(0,0,-1,1; \frac{1}{a}\right)
\end{align}

\begin{align}
&\nn-H\left(0,0,1,-1; \frac{1}{a}\right)+H\left(0,0,1,1; \frac{1}{a}\right)+H\left(0,1,-1,-1; \frac{1}{a}\right)\\
&\nn-H\left(0,1,-1,1; \frac{1}{a}\right)-H\left(0,1,1,-1; \frac{1}{a}\right)+H\left(0,1,1,1; \frac{1}{a}\right)\\
&\nn+H\left(1,-1,0,1; \frac{1}{a}\right)-H\left(1,0,-1,1; \frac{1}{a}\right)-H\left(1,0,0,-1; \frac{1}{a}\right)\\
&\nn+H\left(1,0,0,1; \frac{1}{a}\right)-H\left(1,0,1,-1; \frac{1}{a}\right)+H\left(1,0,1,1; \frac{1}{a}\right)-H\left(1,1,0,1; \frac{1}{a}\right)\\
G&\left(a^2,a^2,1,-a; 1\right) = 
H\left(-1,-1,-1,-1; \frac{1}{a}\right)-H\left(-1,-1,-1,1; \frac{1}{a}\right)\\
&\nn-H\left(-1,-1,1,-1; \frac{1}{a}\right)+H\left(-1,-1,1,1; \frac{1}{a}\right)-2 H\left(-1,0,-1,-1; \frac{1}{a}\right)\\
&\nn+2 H\left(-1,0,-1,1; \frac{1}{a}\right)+2 H\left(-1,0,1,-1; \frac{1}{a}\right)-2 H\left(-1,0,1,1; \frac{1}{a}\right)\\
&\nn-H\left(-1,1,-1,-1; \frac{1}{a}\right)+H\left(-1,1,-1,1; \frac{1}{a}\right)+H\left(-1,1,1,-1; \frac{1}{a}\right)\\
&\nn-H\left(-1,1,1,1; \frac{1}{a}\right)-H\left(0,-1,-1,-1; \frac{1}{a}\right)+H\left(0,-1,-1,1; \frac{1}{a}\right)\\
&\nn+H\left(0,-1,1,-1; \frac{1}{a}\right)-H\left(0,-1,1,1; \frac{1}{a}\right)+H\left(0,0,-1,-1; \frac{1}{a}\right)\\
&\nn-H\left(0,0,-1,1; \frac{1}{a}\right)-H\left(0,0,1,-1; \frac{1}{a}\right)+H\left(0,0,1,1; \frac{1}{a}\right)\\
&\nn+H\left(0,1,-1,-1; \frac{1}{a}\right)-H\left(0,1,-1,1; \frac{1}{a}\right)-H\left(0,1,0,-1; \frac{1}{a}\right)\\
&\nn+H\left(0,1,0,1; \frac{1}{a}\right)-H\left(0,1,1,-1; \frac{1}{a}\right)-H\left(1,-1,-1,-1; \frac{1}{a}\right)\\
&\nn+H\left(1,-1,-1,1; \frac{1}{a}\right)+H\left(1,-1,1,-1; \frac{1}{a}\right)-H\left(1,-1,1,1; \frac{1}{a}\right)\\
&\nn+H\left(1,0,-1,-1; \frac{1}{a}\right)-H\left(1,0,-1,1; \frac{1}{a}\right)-H\left(1,0,1,-1; \frac{1}{a}\right)\\
&\nn+H\left(1,0,1,1; \frac{1}{a}\right)+H\left(1,1,-1,-1; \frac{1}{a}\right)-H\left(1,1,-1,1; \frac{1}{a}\right)\\
&\nn-H\left(1,1,0,-1; \frac{1}{a}\right)+H\left(1,1,0,1; \frac{1}{a}\right)-H\left(1,1,1,-1; \frac{1}{a}\right)\\
G&\left(a^2,a^2,1,a; 1\right) = 
-H\left(-1,-1,-1,1; \frac{1}{a}\right)-H\left(-1,-1,0,-1; \frac{1}{a}\right)\\
&\nn+H\left(-1,-1,0,1; \frac{1}{a}\right)-H\left(-1,-1,1,-1; \frac{1}{a}\right)+H\left(-1,-1,1,1; \frac{1}{a}\right)\\
&\nn-H\left(-1,0,-1,-1; \frac{1}{a}\right)+H\left(-1,0,-1,1; \frac{1}{a}\right)+H\left(-1,0,1,-1; \frac{1}{a}\right)\\
&\nn-H\left(-1,0,1,1; \frac{1}{a}\right)-H\left(-1,1,-1,-1; \frac{1}{a}\right)+H\left(-1,1,-1,1; \frac{1}{a}\right)
\end{align}

\begin{align}
&\nn+H\left(-1,1,1,-1; \frac{1}{a}\right)-H\left(-1,1,1,1; \frac{1}{a}\right)+H\left(0,-1,-1,1; \frac{1}{a}\right)\\
&\nn+H\left(0,-1,0,-1; \frac{1}{a}\right)-H\left(0,-1,0,1; \frac{1}{a}\right)+H\left(0,-1,1,-1; \frac{1}{a}\right)\\
&\nn-H\left(0,-1,1,1; \frac{1}{a}\right)+H\left(0,0,-1,-1; \frac{1}{a}\right)-H\left(0,0,-1,1; \frac{1}{a}\right)\\
&\nn-H\left(0,0,1,-1; \frac{1}{a}\right)+H\left(0,0,1,1; \frac{1}{a}\right)+H\left(0,1,-1,-1; \frac{1}{a}\right)\\
&\nn-H\left(0,1,-1,1; \frac{1}{a}\right)-H\left(0,1,1,-1; \frac{1}{a}\right)+H\left(0,1,1,1; \frac{1}{a}\right)\\
&\nn-H\left(1,-1,-1,-1; \frac{1}{a}\right)+H\left(1,-1,-1,1; \frac{1}{a}\right)+H\left(1,-1,1,-1; \frac{1}{a}\right)\\
&\nn-H\left(1,-1,1,1; \frac{1}{a}\right)+2 H\left(1,0,-1,-1; \frac{1}{a}\right)-2 H\left(1,0,-1,1; \frac{1}{a}\right)\\
&\nn-2 H\left(1,0,1,-1; \frac{1}{a}\right)+2 H\left(1,0,1,1; \frac{1}{a}\right)+H\left(1,1,-1,-1; \frac{1}{a}\right)\\
&\nn-H\left(1,1,-1,1; \frac{1}{a}\right)-H\left(1,1,1,-1; \frac{1}{a}\right)+H\left(1,1,1,1; \frac{1}{a}\right)\\
G&\left(a^2,a^2,-a,0; 1\right) = 
-2 H\left(-1,-1,0,-1; \frac{1}{a}\right)+2 H\left(-1,-1,0,1; \frac{1}{a}\right)\\
&\nn-2 H\left(-1,0,-1,-1; \frac{1}{a}\right)+2 H\left(-1,0,-1,1; \frac{1}{a}\right)+2 H\left(-1,0,1,-1; \frac{1}{a}\right)\\
&\nn-2 H\left(-1,0,1,1; \frac{1}{a}\right)+2 H\left(-1,1,0,-1; \frac{1}{a}\right)-2 H\left(-1,1,0,1; \frac{1}{a}\right)\\
&\nn-H\left(0,-1,-1,-1; \frac{1}{a}\right)+H\left(0,-1,-1,1; \frac{1}{a}\right)+2 H\left(0,-1,0,-1; \frac{1}{a}\right)\\
&\nn-2 H\left(0,-1,0,1; \frac{1}{a}\right)+H\left(0,-1,1,-1; \frac{1}{a}\right)-H\left(0,-1,1,1; \frac{1}{a}\right)\\
&\nn+3 H\left(0,0,-1,-1; \frac{1}{a}\right)-3 H\left(0,0,-1,1; \frac{1}{a}\right)-3 H\left(0,0,1,-1; \frac{1}{a}\right)\\
&\nn+3 H\left(0,0,1,1; \frac{1}{a}\right)+H\left(0,1,-1,-1; \frac{1}{a}\right)-H\left(0,1,-1,1; \frac{1}{a}\right)\\
&\nn-3 H\left(0,1,0,-1; \frac{1}{a}\right)+3 H\left(0,1,0,1; \frac{1}{a}\right)-H\left(0,1,1,-1; \frac{1}{a}\right)\\
&\nn+2 H\left(1,-1,0,-1; \frac{1}{a}\right)-2 H\left(1,-1,0,1; \frac{1}{a}\right)+H\left(1,0,-1,-1; \frac{1}{a}\right)\\
&\nn-H\left(1,0,-1,1; \frac{1}{a}\right)-3 H\left(1,0,0,-1; \frac{1}{a}\right)+3 H\left(1,0,0,1; \frac{1}{a}\right)\\
&\nn-H\left(1,0,1,-1; \frac{1}{a}\right)-H\left(1,1,0,-1; \frac{1}{a}\right)
\end{align}

\begin{align}
G&\left(a^2,a^2,a,0; 1\right) = 
H\left(-1,-1,0,1; \frac{1}{a}\right)+H\left(-1,0,-1,1; \frac{1}{a}\right)\\
&\nn+3 H\left(-1,0,0,-1; \frac{1}{a}\right)-3 H\left(-1,0,0,1; \frac{1}{a}\right)+H\left(-1,0,1,-1; \frac{1}{a}\right)\\
&\nn-H\left(-1,0,1,1; \frac{1}{a}\right)+2 H\left(-1,1,0,-1; \frac{1}{a}\right)-2 H\left(-1,1,0,1; \frac{1}{a}\right)\\
&\nn+H\left(0,-1,-1,1; \frac{1}{a}\right)+3 H\left(0,-1,0,-1; \frac{1}{a}\right)-3 H\left(0,-1,0,1; \frac{1}{a}\right)\\
&\nn+H\left(0,-1,1,-1; \frac{1}{a}\right)-H\left(0,-1,1,1; \frac{1}{a}\right)+3 H\left(0,0,-1,-1; \frac{1}{a}\right)\\
&\nn-3 H\left(0,0,-1,1; \frac{1}{a}\right)-3 H\left(0,0,1,-1; \frac{1}{a}\right)+3 H\left(0,0,1,1; \frac{1}{a}\right)\\
&\nn+H\left(0,1,-1,-1; \frac{1}{a}\right)-H\left(0,1,-1,1; \frac{1}{a}\right)-2 H\left(0,1,0,-1; \frac{1}{a}\right)\\
&\nn+2 H\left(0,1,0,1; \frac{1}{a}\right)-H\left(0,1,1,-1; \frac{1}{a}\right)+H\left(0,1,1,1; \frac{1}{a}\right)\\
&\nn+2 H\left(1,-1,0,-1; \frac{1}{a}\right)-2 H\left(1,-1,0,1; \frac{1}{a}\right)+2 H\left(1,0,-1,-1; \frac{1}{a}\right)-2 H\left(1,0,-1,1; \frac{1}{a}\right)\\
&\nn-2 H\left(1,0,1,-1; \frac{1}{a}\right)+2 H\left(1,0,1,1; \frac{1}{a}\right)-2 H\left(1,1,0,-1; \frac{1}{a}\right)+2 H\left(1,1,0,1; \frac{1}{a}\right)\\
G&\left(a^2,a^2,-a,1; 1\right) = 
H\left(-1,-1,-1,-1; \frac{1}{a}\right)-H\left(-1,-1,-1,1; \frac{1}{a}\right)\\
&\nn-H\left(-1,-1,1,-1; \frac{1}{a}\right)+H\left(-1,-1,1,1; \frac{1}{a}\right)-H\left(-1,0,-1,-1; \frac{1}{a}\right)\\
&\nn+H\left(-1,0,-1,1; \frac{1}{a}\right)+H\left(-1,0,1,-1; \frac{1}{a}\right)-H\left(-1,0,1,1; \frac{1}{a}\right)\\
&\nn-H\left(-1,1,-1,-1; \frac{1}{a}\right)+H\left(-1,1,-1,1; \frac{1}{a}\right)+H\left(-1,1,0,-1; \frac{1}{a}\right)\\
&\nn-H\left(-1,1,0,1; \frac{1}{a}\right)+H\left(-1,1,1,-1; \frac{1}{a}\right)-2 H\left(0,-1,-1,-1; \frac{1}{a}\right)\\
&\nn+2 H\left(0,-1,-1,1; \frac{1}{a}\right)+2 H\left(0,-1,1,-1; \frac{1}{a}\right)-2 H\left(0,-1,1,1; \frac{1}{a}\right)\\
&\nn+3 H\left(0,0,-1,-1; \frac{1}{a}\right)-3 H\left(0,0,-1,1; \frac{1}{a}\right)-3 H\left(0,0,1,-1; \frac{1}{a}\right)+3 H\left(0,0,1,1; \frac{1}{a}\right)\\
&\nn+2 H\left(0,1,-1,-1; \frac{1}{a}\right)-2 H\left(0,1,-1,1; \frac{1}{a}\right)-H\left(0,1,0,-1; \frac{1}{a}\right)+H\left(0,1,0,1; \frac{1}{a}\right)\\
&\nn-2 H\left(0,1,1,-1; \frac{1}{a}\right)+H\left(0,1,1,1; \frac{1}{a}\right)-H\left(1,-1,-1,-1; \frac{1}{a}\right)+H\left(1,-1,-1,1; \frac{1}{a}\right)\\
&\nn+H\left(1,-1,0,-1; \frac{1}{a}\right)-H\left(1,-1,0,1; \frac{1}{a}\right)+H\left(1,-1,1,-1; \frac{1}{a}\right)\\
&\nn+2 H\left(1,0,-1,-1; \frac{1}{a}\right)-2 H\left(1,0,-1,1; \frac{1}{a}\right)-3 H\left(1,0,0,-1; \frac{1}{a}\right)+3 H\left(1,0,0,1; \frac{1}{a}\right)\\
&\nn-2 H\left(1,0,1,-1; \frac{1}{a}\right)+H\left(1,0,1,1; \frac{1}{a}\right)+H\left(1,1,-1,-1; \frac{1}{a}\right)-H\left(1,1,0,-1; \frac{1}{a}\right)
\end{align}
\begin{align}
G&\left(a^2,a^2,a,1; 1\right) = 
H\left(-1,-1,0,1; \frac{1}{a}\right)+H\left(-1,-1,1,1; \frac{1}{a}\right)\\
&\nn-H\left(-1,0,-1,-1; \frac{1}{a}\right)+2 H\left(-1,0,-1,1; \frac{1}{a}\right)+3 H\left(-1,0,0,-1; \frac{1}{a}\right)\\
&\nn-3 H\left(-1,0,0,1; \frac{1}{a}\right)+2 H\left(-1,0,1,-1; \frac{1}{a}\right)-2 H\left(-1,0,1,1; \frac{1}{a}\right)\\
&\nn+H\left(-1,1,-1,1; \frac{1}{a}\right)+H\left(-1,1,0,-1; \frac{1}{a}\right)-H\left(-1,1,0,1; \frac{1}{a}\right)+H\left(-1,1,1,-1; \frac{1}{a}\right)\\
&\nn-H\left(-1,1,1,1; \frac{1}{a}\right)-H\left(0,-1,-1,-1; \frac{1}{a}\right)+2 H\left(0,-1,-1,1; \frac{1}{a}\right)\\
&\nn+H\left(0,-1,0,-1; \frac{1}{a}\right)-H\left(0,-1,0,1; \frac{1}{a}\right)+2 H\left(0,-1,1,-1; \frac{1}{a}\right)-2 H\left(0,-1,1,1; \frac{1}{a}\right)\\
&\nn+3 H\left(0,0,-1,-1; \frac{1}{a}\right)-3 H\left(0,0,-1,1; \frac{1}{a}\right)-3 H\left(0,0,1,-1; \frac{1}{a}\right)+3 H\left(0,0,1,1; \frac{1}{a}\right)\\
&\nn+2 H\left(0,1,-1,-1; \frac{1}{a}\right)-2 H\left(0,1,-1,1; \frac{1}{a}\right)-2 H\left(0,1,1,-1; \frac{1}{a}\right)+2 H\left(0,1,1,1; \frac{1}{a}\right)\\
&\nn+H\left(1,-1,-1,1; \frac{1}{a}\right)+H\left(1,-1,0,-1; \frac{1}{a}\right)-H\left(1,-1,0,1; \frac{1}{a}\right)\\
&\nn+H\left(1,-1,1,-1; \frac{1}{a}\right)-H\left(1,-1,1,1; \frac{1}{a}\right)+H\left(1,0,-1,-1; \frac{1}{a}\right)-H\left(1,0,-1,1; \frac{1}{a}\right)\\
&\nn-H\left(1,0,1,-1; \frac{1}{a}\right)+H\left(1,0,1,1; \frac{1}{a}\right)+H\left(1,1,-1,-1; \frac{1}{a}\right)-H\left(1,1,-1,1; \frac{1}{a}\right)\\
&\nn-H\left(1,1,1,-1; \frac{1}{a}\right)+H\left(1,1,1,1; \frac{1}{a}\right)\\
G&\left(a^2,a^2,a^2,-a; 1\right) = 
H\left(-1,-1,-1,-1; \frac{1}{a}\right)-H\left(-1,-1,-1,1; \frac{1}{a}\right)\\
&\nn-H\left(-1,-1,1,-1; \frac{1}{a}\right)+H\left(-1,-1,1,1; \frac{1}{a}\right)-H\left(-1,1,-1,-1; \frac{1}{a}\right)\\
&\nn+H\left(-1,1,-1,1; \frac{1}{a}\right)+H\left(-1,1,1,-1; \frac{1}{a}\right)-H\left(-1,1,1,1; \frac{1}{a}\right)\\
&\nn-H\left(0,-1,-1,-1; \frac{1}{a}\right)+H\left(0,-1,-1,1; \frac{1}{a}\right)+H\left(0,-1,1,-1; \frac{1}{a}\right)\\
&\nn-H\left(0,-1,1,1; \frac{1}{a}\right)+H\left(0,1,-1,-1; \frac{1}{a}\right)-H\left(0,1,-1,1; \frac{1}{a}\right)-H\left(0,1,1,-1; \frac{1}{a}\right)\\
&\nn+H\left(0,1,1,1; \frac{1}{a}\right)-H\left(1,-1,-1,-1; \frac{1}{a}\right)+H\left(1,-1,-1,1; \frac{1}{a}\right)\\
&\nn+H\left(1,-1,1,-1; \frac{1}{a}\right)-H\left(1,-1,1,1; \frac{1}{a}\right)+H\left(1,0,-1,-1; \frac{1}{a}\right)-H\left(1,0,-1,1; \frac{1}{a}\right)\\
&\nn-H\left(1,0,1,-1; \frac{1}{a}\right)+H\left(1,0,1,1; \frac{1}{a}\right)+H\left(1,1,-1,-1; \frac{1}{a}\right)-H\left(1,1,-1,1; \frac{1}{a}\right)\\
&\nn-H\left(1,1,0,-1; \frac{1}{a}\right)+H\left(1,1,0,1; \frac{1}{a}\right)-H\left(1,1,1,-1; \frac{1}{a}\right)
\end{align}
\begin{align}
G&\left(a^2,a^2,a^2,a; 1\right) = 
-H\left(-1,-1,-1,1; \frac{1}{a}\right)-H\left(-1,-1,0,-1; \frac{1}{a}\right)\\
&\nn+H\left(-1,-1,0,1; \frac{1}{a}\right)-H\left(-1,-1,1,-1; \frac{1}{a}\right)+H\left(-1,-1,1,1; \frac{1}{a}\right)\\
&\nn-H\left(-1,0,-1,-1; \frac{1}{a}\right)+H\left(-1,0,-1,1; \frac{1}{a}\right)+H\left(-1,0,1,-1; \frac{1}{a}\right)\\
&\nn-H\left(-1,0,1,1; \frac{1}{a}\right)-H\left(-1,1,-1,-1; \frac{1}{a}\right)+H\left(-1,1,-1,1; \frac{1}{a}\right)\\
&\nn+H\left(-1,1,1,-1; \frac{1}{a}\right)-H\left(-1,1,1,1; \frac{1}{a}\right)-H\left(0,-1,-1,-1; \frac{1}{a}\right)\\
&\nn+H\left(0,-1,-1,1; \frac{1}{a}\right)+H\left(0,-1,1,-1; \frac{1}{a}\right)-H\left(0,-1,1,1; \frac{1}{a}\right)\\
&\nn+H\left(0,1,-1,-1; \frac{1}{a}\right)-H\left(0,1,-1,1; \frac{1}{a}\right)-H\left(0,1,1,-1; \frac{1}{a}\right)\\
&\nn+H\left(0,1,1,1; \frac{1}{a}\right)-H\left(1,-1,-1,-1; \frac{1}{a}\right)+H\left(1,-1,-1,1; \frac{1}{a}\right)\\
&\nn+H\left(1,-1,1,-1; \frac{1}{a}\right)-H\left(1,-1,1,1; \frac{1}{a}\right)+H\left(1,1,-1,-1; \frac{1}{a}\right)\\
&\nn-H\left(1,1,-1,1; \frac{1}{a}\right)-H\left(1,1,1,-1; \frac{1}{a}\right)+H\left(1,1,1,1; \frac{1}{a}\right)\\
G&\left(a^2,a^2,-a,a^2; 1\right) = 
H\left(-1,-1,-1,-1; \frac{1}{a}\right)-H\left(-1,-1,-1,1; \frac{1}{a}\right)\\
&\nn-H\left(-1,-1,1,-1; \frac{1}{a}\right)+H\left(-1,-1,1,1; \frac{1}{a}\right)-H\left(-1,0,-1,-1; \frac{1}{a}\right)\\
&\nn+H\left(-1,0,-1,1; \frac{1}{a}\right)+H\left(-1,0,1,-1; \frac{1}{a}\right)-H\left(-1,0,1,1; \frac{1}{a}\right)\\
&\nn-H\left(-1,1,-1,-1; \frac{1}{a}\right)+H\left(-1,1,-1,1; \frac{1}{a}\right)+H\left(-1,1,0,-1; \frac{1}{a}\right)\\
&\nn-H\left(-1,1,0,1; \frac{1}{a}\right)+H\left(-1,1,1,-1; \frac{1}{a}\right)-H\left(1,-1,-1,-1; \frac{1}{a}\right)\\
&\nn+H\left(1,-1,-1,1; \frac{1}{a}\right)+H\left(1,-1,0,-1; \frac{1}{a}\right)-H\left(1,-1,0,1; \frac{1}{a}\right)\\
&\nn+H\left(1,-1,1,-1; \frac{1}{a}\right)+H\left(1,1,-1,-1; \frac{1}{a}\right)+H\left(1,1,0,-1; \frac{1}{a}\right)-H\left(1,1,0,1; \frac{1}{a}\right)\\
G&\left(a^2,a^2,a,a^2; 1\right) = 
H\left(-1,-1,0,-1; \frac{1}{a}\right)-H\left(-1,-1,0,1; \frac{1}{a}\right)\\
&\nn+H\left(-1,-1,1,1; \frac{1}{a}\right)+H\left(-1,1,-1,1; \frac{1}{a}\right)+H\left(-1,1,0,-1; \frac{1}{a}\right)-H\left(-1,1,0,1; \frac{1}{a}\right)\\
&\nn+H\left(-1,1,1,-1; \frac{1}{a}\right)-H\left(-1,1,1,1; \frac{1}{a}\right)+H\left(1,-1,-1,1; \frac{1}{a}\right)+H\left(1,-1,0,-1; \frac{1}{a}\right)\\
&\nn-H\left(1,-1,0,1; \frac{1}{a}\right)+H\left(1,-1,1,-1; \frac{1}{a}\right)-H\left(1,-1,1,1; \frac{1}{a}\right)+H\left(1,0,-1,-1; \frac{1}{a}\right)\\
&\nn-H\left(1,0,-1,1; \frac{1}{a}\right)-H\left(1,0,1,-1; \frac{1}{a}\right)+H\left(1,0,1,1; \frac{1}{a}\right)+H\left(1,1,-1,-1; \frac{1}{a}\right)
\end{align}

\begin{align}
&\nn
-H\left(1,1,-1,1; \frac{1}{a}\right)-H\left(1,1,1,-1; \frac{1}{a}\right)+H\left(1,1,1,1; \frac{1}{a}\right)\\
G&\left(a^2,-a,a^2,0; 1\right) = 
-2 H\left(-1,-1,0,-1; \frac{1}{a}\right)+2 H\left(-1,-1,0,1; \frac{1}{a}\right)\\
&\nn-H\left(-1,0,-1,-1; \frac{1}{a}\right)+H\left(-1,0,-1,1; \frac{1}{a}\right)+3 H\left(-1,0,0,-1; \frac{1}{a}\right)\\
&\nn-3 H\left(-1,0,0,1; \frac{1}{a}\right)+H\left(-1,0,1,-1; \frac{1}{a}\right)+H\left(-1,1,0,-1; \frac{1}{a}\right)\\
&\nn-2 H\left(0,-1,-1,-1; \frac{1}{a}\right)+2 H\left(0,-1,-1,1; \frac{1}{a}\right)+2 H\left(0,-1,0,-1; \frac{1}{a}\right)\\
&\nn-2 H\left(0,-1,0,1; \frac{1}{a}\right)+2 H\left(0,-1,1,-1; \frac{1}{a}\right)+2 H\left(0,1,-1,-1; \frac{1}{a}\right)\\
&\nn+2 H\left(0,1,0,-1; \frac{1}{a}\right)-2 H\left(0,1,0,1; \frac{1}{a}\right)+H\left(1,-1,0,-1; \frac{1}{a}\right)\\
&\nn+2 H\left(1,0,-1,-1; \frac{1}{a}\right)+4 H\left(1,0,0,-1; \frac{1}{a}\right)-4 H\left(1,0,0,1; \frac{1}{a}\right)\\
G&\left(a^2,a,a^2,0; 1\right) = 
-4 H\left(-1,0,0,-1; \frac{1}{a}\right)+4 H\left(-1,0,0,1; \frac{1}{a}\right)\\
&\nn-2 H\left(-1,0,1,1; \frac{1}{a}\right)-H\left(-1,1,0,1; \frac{1}{a}\right)-2 H\left(0,-1,0,-1; \frac{1}{a}\right)\\
&\nn+2 H\left(0,-1,0,1; \frac{1}{a}\right)-2 H\left(0,-1,1,1; \frac{1}{a}\right)-2 H\left(0,1,-1,1; \frac{1}{a}\right)\\
&\nn-2 H\left(0,1,0,-1; \frac{1}{a}\right)+2 H\left(0,1,0,1; \frac{1}{a}\right)-2 H\left(0,1,1,-1; \frac{1}{a}\right)\\
&\nn+2 H\left(0,1,1,1; \frac{1}{a}\right)-H\left(1,-1,0,1; \frac{1}{a}\right)-H\left(1,0,-1,1; \frac{1}{a}\right)\\
&\nn-3 H\left(1,0,0,-1; \frac{1}{a}\right)+3 H\left(1,0,0,1; \frac{1}{a}\right)-H\left(1,0,1,-1; \frac{1}{a}\right)\\
&\nn+H\left(1,0,1,1; \frac{1}{a}\right)-2 H\left(1,1,0,-1; \frac{1}{a}\right)+2 H\left(1,1,0,1; \frac{1}{a}\right)\\
G&\left(a^2,-a,a^2,1; 1\right) = 
H\left(-1,-1,-1,-1; \frac{1}{a}\right)-H\left(-1,-1,-1,1; \frac{1}{a}\right)\\
&\nn-H\left(-1,-1,0,-1; \frac{1}{a}\right)+H\left(-1,-1,0,1; \frac{1}{a}\right)-H\left(-1,-1,1,-1; \frac{1}{a}\right)\\
&\nn-H\left(-1,1,-1,-1; \frac{1}{a}\right)-H\left(-1,1,0,-1; \frac{1}{a}\right)+H\left(-1,1,0,1; \frac{1}{a}\right)\\
&\nn-2 H\left(0,-1,-1,-1; \frac{1}{a}\right)+2 H\left(0,-1,-1,1; \frac{1}{a}\right)+2 H\left(0,-1,0,-1; \frac{1}{a}\right)\\
&\nn-2 H\left(0,-1,0,1; \frac{1}{a}\right)+2 H\left(0,-1,1,-1; \frac{1}{a}\right)+2 H\left(0,1,-1,-1; \frac{1}{a}\right)\\
&\nn+2 H\left(0,1,0,-1; \frac{1}{a}\right)-2 H\left(0,1,0,1; \frac{1}{a}\right)-H\left(1,-1,-1,-1; \frac{1}{a}\right)
\end{align}

\begin{align}
&\nn-H\left(1,-1,0,-1; \frac{1}{a}\right)+H\left(1,-1,0,1; \frac{1}{a}\right)-H\left(1,0,-1,-1; \frac{1}{a}\right)\\
&\nn+3 H\left(1,0,-1,1; \frac{1}{a}\right)+7 H\left(1,0,0,-1; \frac{1}{a}\right)-7 H\left(1,0,0,1; \frac{1}{a}\right)\\
&\nn+3 H\left(1,0,1,-1; \frac{1}{a}\right)-2 H\left(1,0,1,1; \frac{1}{a}\right)+4 H\left(1,1,0,-1; \frac{1}{a}\right)-2 H\left(1,1,0,1; \frac{1}{a}\right)\\
G&\left(a^2,a,a^2,1; 1\right) = 
2 H\left(-1,-1,0,-1; \frac{1}{a}\right)-4 H\left(-1,-1,0,1; \frac{1}{a}\right)\\
&\nn+2 H\left(-1,0,-1,-1; \frac{1}{a}\right)-3 H\left(-1,0,-1,1; \frac{1}{a}\right)-7 H\left(-1,0,0,-1; \frac{1}{a}\right)\\
&\nn+7 H\left(-1,0,0,1; \frac{1}{a}\right)-3 H\left(-1,0,1,-1; \frac{1}{a}\right)+H\left(-1,0,1,1; \frac{1}{a}\right)\\
&\nn-H\left(-1,1,0,-1; \frac{1}{a}\right)+H\left(-1,1,0,1; \frac{1}{a}\right)-H\left(-1,1,1,1; \frac{1}{a}\right)\\
&\nn-2 H\left(0,-1,0,-1; \frac{1}{a}\right)+2 H\left(0,-1,0,1; \frac{1}{a}\right)-2 H\left(0,-1,1,1; \frac{1}{a}\right)\\
&\nn-2 H\left(0,1,-1,1; \frac{1}{a}\right)-2 H\left(0,1,0,-1; \frac{1}{a}\right)+2 H\left(0,1,0,1; \frac{1}{a}\right)\\
&\nn-2 H\left(0,1,1,-1; \frac{1}{a}\right)+2 H\left(0,1,1,1; \frac{1}{a}\right)-H\left(1,-1,0,-1; \frac{1}{a}\right)\\
&\nn+H\left(1,-1,0,1; \frac{1}{a}\right)-H\left(1,-1,1,1; \frac{1}{a}\right)-H\left(1,1,-1,1; \frac{1}{a}\right)\\
&\nn-H\left(1,1,0,-1; \frac{1}{a}\right)+H\left(1,1,0,1; \frac{1}{a}\right)-H\left(1,1,1,-1; \frac{1}{a}\right)+H\left(1,1,1,1; \frac{1}{a}\right)\\
G&\left(a^2,-a,a^2,a^2; 1\right) = 
H\left(-1,-1,-1,-1; \frac{1}{a}\right)-H\left(-1,-1,-1,1; \frac{1}{a}\right)\\
&\nn-H\left(-1,-1,0,-1; \frac{1}{a}\right)+H\left(-1,-1,0,1; \frac{1}{a}\right)-H\left(-1,-1,1,-1; \frac{1}{a}\right)\\
&\nn-H\left(-1,1,-1,-1; \frac{1}{a}\right)-H\left(-1,1,0,-1; \frac{1}{a}\right)+H\left(-1,1,0,1; \frac{1}{a}\right)\\
&\nn-H\left(1,-1,-1,-1; \frac{1}{a}\right)-H\left(1,-1,0,-1; \frac{1}{a}\right)+H\left(1,-1,0,1; \frac{1}{a}\right)\\
&\nn-H\left(1,0,-1,-1; \frac{1}{a}\right)+H\left(1,0,-1,1; \frac{1}{a}\right)+H\left(1,0,1,-1; \frac{1}{a}\right)-H\left(1,0,1,1; \frac{1}{a}\right)\\
G&\left(a^2,a,a^2,a^2; 1\right) = 
H\left(-1,0,-1,-1; \frac{1}{a}\right)-H\left(-1,0,-1,1; \frac{1}{a}\right)\\
&\nn-H\left(-1,0,1,-1; \frac{1}{a}\right)+H\left(-1,0,1,1; \frac{1}{a}\right)-H\left(-1,1,0,-1; \frac{1}{a}\right)\\
&\nn+H\left(-1,1,0,1; \frac{1}{a}\right)-H\left(-1,1,1,1; \frac{1}{a}\right)-H\left(1,-1,0,-1; \frac{1}{a}\right)\\
&\nn+H\left(1,-1,0,1; \frac{1}{a}\right)-H\left(1,-1,1,1; \frac{1}{a}\right)-H\left(1,1,-1,1; \frac{1}{a}\right)\\
&\nn-H\left(1,1,0,-1; \frac{1}{a}\right)+H\left(1,1,0,1; \frac{1}{a}\right)-H\left(1,1,1,-1; \frac{1}{a}\right)+H\left(1,1,1,1; \frac{1}{a}\right)
\end{align}
\begin{align}
G&(a,a,0,0; 1) = 
H\left(0,0,1,1; \frac{1}{a}\right)+H\left(0,1,0,1; \frac{1}{a}\right)+H\left(1,0,0,1; \frac{1}{a}\right)\\
G&(a,a,0,1; 1) = 
-\frac{1}{6} \pi ^2 H\left(1,1; \frac{1}{a}\right)+H\left(0,0,1,1; \frac{1}{a}\right)+H\left(0,1,1,1; \frac{1}{a}\right)\\
&\nn+H\left(1,0,0,1; \frac{1}{a}\right)+H\left(1,0,1,1; \frac{1}{a}\right)\\
G&(-a,-a,0,a; 1) = 
-H\left(-1,-1,0,1; \frac{1}{a}\right)\\
G&(-a,a,0,-a; 1) = 
-H\left(-1,1,0,-1; \frac{1}{a}\right)\\
G&(-a,a,0,a; 1) = 
H\left(-1,1,0,1; \frac{1}{a}\right)\\
G&(a,-a,0,-a; 1) = 
-H\left(1,-1,0,-1; \frac{1}{a}\right)\\
G&(a,-a,0,a; 1) = 
H\left(1,-1,0,1; \frac{1}{a}\right)\\
G&(a,a,0,-a; 1) = 
H\left(1,1,0,-1; \frac{1}{a}\right)\\
G&(a,a,0,a; 1) = 
-H\left(1,1,0,1; \frac{1}{a}\right)\\
G&(a,a,1,0; 1) = 
H\left(0,0,1,1; \frac{1}{a}\right)+H\left(0,1,0,1; \frac{1}{a}\right)+H\left(1,0,1,1; \frac{1}{a}\right)\\
&\nn+H\left(1,1,0,1; \frac{1}{a}\right)\\
G&(a,a,1,1; 1) = 
H\left(0,0,1,1; \frac{1}{a}\right)+H\left(0,1,1,1; \frac{1}{a}\right)+H\left(1,0,1,1; \frac{1}{a}\right)\\
&\nn+H\left(1,1,1,1; \frac{1}{a}\right)\\
G&\left(-a,a^2,0,1; 1\right) = 
-\frac{1}{6} \pi ^2 H\left(-1,-1; \frac{1}{a}\right)-\frac{1}{6} \pi ^2 H\left(0,-1; \frac{1}{a}\right)\\
&\nn+\frac{1}{6} \pi ^2 H\left(0,1; \frac{1}{a}\right)-H\left(-1,0,-1,-1; \frac{1}{a}\right)+H\left(-1,0,0,-1; \frac{1}{a}\right)\\
&\nn-2 H\left(0,-1,-1,-1; \frac{1}{a}\right)-2 H\left(0,-1,0,-1; \frac{1}{a}\right)+2 H\left(0,-1,0,1; \frac{1}{a}\right)\\
&\nn+4 H\left(0,0,-1,1; \frac{1}{a}\right)+12 H\left(0,0,0,-1; \frac{1}{a}\right)-12 H\left(0,0,0,1; \frac{1}{a}\right)\\
&\nn+4 H\left(0,0,1,-1; \frac{1}{a}\right)-4 H\left(0,0,1,1; \frac{1}{a}\right)+4 H\left(0,1,0,-1; \frac{1}{a}\right)-2 H\left(0,1,0,1; \frac{1}{a}\right)\\
G&\left(a,a^2,0,1; 1\right) = 
-\frac{1}{6} \pi ^2 H\left(0,-1; \frac{1}{a}\right)+\frac{1}{6} \pi ^2 H\left(0,1; \frac{1}{a}\right)-\frac{1}{6} \pi ^2 H\left(1,1; \frac{1}{a}\right)\\
&\nn-2 H\left(0,-1,0,-1; \frac{1}{a}\right)+4 H\left(0,-1,0,1; \frac{1}{a}\right)-4 H\left(0,0,-1,-1; \frac{1}{a}\right)\\
&\nn+4 H\left(0,0,-1,1; \frac{1}{a}\right)+12 H\left(0,0,0,-1; \frac{1}{a}\right)-12 H\left(0,0,0,1; \frac{1}{a}\right)
\end{align}

\begin{align}
&\nn+4 H\left(0,0,1,-1; \frac{1}{a}\right)+2 H\left(0,1,0,-1; \frac{1}{a}\right)-2 H\left(0,1,0,1; \frac{1}{a}\right)+2 H\left(0,1,1,1; \frac{1}{a}\right)\\
&\nn+H\left(1,0,0,1; \frac{1}{a}\right)+H\left(1,0,1,1; \frac{1}{a}\right)\\
G&\left(-a,a^2,1,0; 1\right) = 
-H\left(-1,-1,0,-1; \frac{1}{a}\right)-2 H\left(-1,0,-1,-1; \frac{1}{a}\right)\\
&\nn-4 H\left(-1,0,0,-1; \frac{1}{a}\right)+4 H\left(-1,0,0,1; \frac{1}{a}\right)+2 H\left(0,-1,0,1; \frac{1}{a}\right)\\
&\nn+4 H\left(0,0,-1,-1; \frac{1}{a}\right)+12 H\left(0,0,0,-1; \frac{1}{a}\right)-12 H\left(0,0,0,1; \frac{1}{a}\right)\\
&\nn+2 H\left(0,1,0,-1; \frac{1}{a}\right)-2 H\left(0,1,0,1; \frac{1}{a}\right)+2 H\left(1,0,-1,-1; \frac{1}{a}\right)\\
&\nn+5 H\left(1,0,0,-1; \frac{1}{a}\right)-4 H\left(1,0,0,1; \frac{1}{a}\right)\\
G&\left(a,a^2,1,0; 1\right) = 
-4 H\left(-1,0,0,-1; \frac{1}{a}\right)+5 H\left(-1,0,0,1; \frac{1}{a}\right)\\
&\nn-2 H\left(-1,0,1,1; \frac{1}{a}\right)-2 H\left(0,-1,0,-1; \frac{1}{a}\right)+2 H\left(0,-1,0,1; \frac{1}{a}\right)\\
&\nn+12 H\left(0,0,0,-1; \frac{1}{a}\right)-12 H\left(0,0,0,1; \frac{1}{a}\right)+4 H\left(0,0,1,1; \frac{1}{a}\right)+2 H\left(0,1,0,-1; \frac{1}{a}\right)\\
&\nn+4 H\left(1,0,0,-1; \frac{1}{a}\right)-4 H\left(1,0,0,1; \frac{1}{a}\right)+2 H\left(1,0,1,1; \frac{1}{a}\right)+H\left(1,1,0,1; \frac{1}{a}\right)\\
G&\left(-a,a^2,1,1; 1\right) = 
H\left(-1,-1,-1,-1; \frac{1}{a}\right)+H\left(-1,-1,0,-1; \frac{1}{a}\right)\\
&\nn-H\left(-1,-1,0,1; \frac{1}{a}\right)-H\left(-1,0,-1,-1; \frac{1}{a}\right)-H\left(-1,0,-1,1; \frac{1}{a}\right)\\
&\nn-4 H\left(-1,0,0,-1; \frac{1}{a}\right)+4 H\left(-1,0,0,1; \frac{1}{a}\right)-H\left(-1,0,1,-1; \frac{1}{a}\right)\\
&\nn+H\left(-1,0,1,1; \frac{1}{a}\right)-2 H\left(-1,1,0,-1; \frac{1}{a}\right)+H\left(-1,1,0,1; \frac{1}{a}\right)\\
&\nn-H\left(0,-1,-1,-1; \frac{1}{a}\right)-H\left(0,-1,-1,1; \frac{1}{a}\right)-4 H\left(0,-1,0,-1; \frac{1}{a}\right)\\
&\nn+4 H\left(0,-1,0,1; \frac{1}{a}\right)-H\left(0,-1,1,-1; \frac{1}{a}\right)+H\left(0,-1,1,1; \frac{1}{a}\right)\\
&\nn+4 H\left(0,0,-1,1; \frac{1}{a}\right)+12 H\left(0,0,0,-1; \frac{1}{a}\right)-12 H\left(0,0,0,1; \frac{1}{a}\right)\\
&\nn+4 H\left(0,0,1,-1; \frac{1}{a}\right)-4 H\left(0,0,1,1; \frac{1}{a}\right)-H\left(0,1,-1,-1; \frac{1}{a}\right)\\
&\nn+H\left(0,1,-1,1; \frac{1}{a}\right)+6 H\left(0,1,0,-1; \frac{1}{a}\right)-4 H\left(0,1,0,1; \frac{1}{a}\right)+H\left(0,1,1,-1; \frac{1}{a}\right)\\
&\nn-H\left(0,1,1,1; \frac{1}{a}\right)-2 H\left(1,-1,0,-1; \frac{1}{a}\right)+H\left(1,-1,0,1; \frac{1}{a}\right)+H\left(1,0,-1,1; \frac{1}{a}\right)
\end{align}

\begin{align}
&\nn+5 H\left(1,0,0,-1; \frac{1}{a}\right)-4 H\left(1,0,0,1; \frac{1}{a}\right)+H\left(1,0,1,-1; \frac{1}{a}\right)-H\left(1,0,1,1; \frac{1}{a}\right)\\
&\nn+2 H\left(1,1,0,-1; \frac{1}{a}\right)-H\left(1,1,0,1; \frac{1}{a}\right)\\
G&\left(a,a^2,1,1; 1\right) = 
H\left(-1,-1,0,-1; \frac{1}{a}\right)-2 H\left(-1,-1,0,1; \frac{1}{a}\right)\\
&\nn+H\left(-1,0,-1,-1; \frac{1}{a}\right)-H\left(-1,0,-1,1; \frac{1}{a}\right)-4 H\left(-1,0,0,-1; \frac{1}{a}\right)\\
&\nn+5 H\left(-1,0,0,1; \frac{1}{a}\right)-H\left(-1,0,1,-1; \frac{1}{a}\right)-H\left(-1,1,0,-1; \frac{1}{a}\right)\\
&\nn+2 H\left(-1,1,0,1; \frac{1}{a}\right)+H\left(0,-1,-1,-1; \frac{1}{a}\right)-H\left(0,-1,-1,1; \frac{1}{a}\right)\\
&\nn-4 H\left(0,-1,0,-1; \frac{1}{a}\right)+6 H\left(0,-1,0,1; \frac{1}{a}\right)-H\left(0,-1,1,-1; \frac{1}{a}\right)\\
&\nn+H\left(0,-1,1,1; \frac{1}{a}\right)-4 H\left(0,0,-1,-1; \frac{1}{a}\right)+4 H\left(0,0,-1,1; \frac{1}{a}\right)\\
&\nn+12 H\left(0,0,0,-1; \frac{1}{a}\right)-12 H\left(0,0,0,1; \frac{1}{a}\right)+4 H\left(0,0,1,-1; \frac{1}{a}\right)\\
&\nn-H\left(0,1,-1,-1; \frac{1}{a}\right)+H\left(0,1,-1,1; \frac{1}{a}\right)+4 H\left(0,1,0,-1; \frac{1}{a}\right)\\
&\nn-4 H\left(0,1,0,1; \frac{1}{a}\right)+H\left(0,1,1,-1; \frac{1}{a}\right)+H\left(0,1,1,1; \frac{1}{a}\right)\\
&\nn-H\left(1,-1,0,-1; \frac{1}{a}\right)+2 H\left(1,-1,0,1; \frac{1}{a}\right)-H\left(1,0,-1,-1; \frac{1}{a}\right)\\
&\nn+H\left(1,0,-1,1; \frac{1}{a}\right)+4 H\left(1,0,0,-1; \frac{1}{a}\right)-4 H\left(1,0,0,1; \frac{1}{a}\right)\\
&\nn+H\left(1,0,1,-1; \frac{1}{a}\right)+H\left(1,0,1,1; \frac{1}{a}\right)+H\left(1,1,0,-1; \frac{1}{a}\right)\\
&\nn-H\left(1,1,0,1; \frac{1}{a}\right)+H\left(1,1,1,1; \frac{1}{a}\right)\\
G&\left(-a,a^2,1,a^2; 1\right) = 
H\left(-1,-1,-1,-1; \frac{1}{a}\right)+H\left(-1,-1,0,-1; \frac{1}{a}\right)\\
&\nn-H\left(-1,-1,0,1; \frac{1}{a}\right)-H\left(-1,0,-1,-1; \frac{1}{a}\right)-H\left(-1,0,-1,1; \frac{1}{a}\right)\\
&\nn-4 H\left(-1,0,0,-1; \frac{1}{a}\right)+4 H\left(-1,0,0,1; \frac{1}{a}\right)-H\left(-1,0,1,-1; \frac{1}{a}\right)\\
&\nn+H\left(-1,0,1,1; \frac{1}{a}\right)-2 H\left(-1,1,0,-1; \frac{1}{a}\right)+H\left(-1,1,0,1; \frac{1}{a}\right)\\
&\nn+H\left(0,-1,-1,-1; \frac{1}{a}\right)-H\left(0,-1,-1,1; \frac{1}{a}\right)-2 H\left(0,-1,0,-1; \frac{1}{a}\right)\\
&\nn+2 H\left(0,-1,0,1; \frac{1}{a}\right)-H\left(0,-1,1,-1; \frac{1}{a}\right)+H\left(0,-1,1,1; \frac{1}{a}\right)
\end{align}

\begin{align}
&\nn-H\left(0,1,-1,-1; \frac{1}{a}\right)+H\left(0,1,-1,1; \frac{1}{a}\right)+2 H\left(0,1,0,-1; \frac{1}{a}\right)\\
&\nn-2 H\left(0,1,0,1; \frac{1}{a}\right)+H\left(0,1,1,-1; \frac{1}{a}\right)-H\left(0,1,1,1; \frac{1}{a}\right)\\
&\nn-2 H\left(1,-1,0,-1; \frac{1}{a}\right)+H\left(1,-1,0,1; \frac{1}{a}\right)+2 H\left(1,0,-1,-1; \frac{1}{a}\right)\\
&\nn+H\left(1,0,-1,1; \frac{1}{a}\right)+5 H\left(1,0,0,-1; \frac{1}{a}\right)-5 H\left(1,0,0,1; \frac{1}{a}\right)\\
&\nn+H\left(1,0,1,-1; \frac{1}{a}\right)-H\left(1,0,1,1; \frac{1}{a}\right)+4 H\left(1,1,0,-1; \frac{1}{a}\right)-2 H\left(1,1,0,1; \frac{1}{a}\right)\\
G&\left(a,a^2,1,a^2; 1\right) = 
2 H\left(-1,-1,0,-1; \frac{1}{a}\right)-4 H\left(-1,-1,0,1; \frac{1}{a}\right)\\
&\nn+H\left(-1,0,-1,-1; \frac{1}{a}\right)-H\left(-1,0,-1,1; \frac{1}{a}\right)-5 H\left(-1,0,0,-1; \frac{1}{a}\right)\\
&\nn+5 H\left(-1,0,0,1; \frac{1}{a}\right)-H\left(-1,0,1,-1; \frac{1}{a}\right)-2 H\left(-1,0,1,1; \frac{1}{a}\right)\\
&\nn-H\left(-1,1,0,-1; \frac{1}{a}\right)+2 H\left(-1,1,0,1; \frac{1}{a}\right)+H\left(0,-1,-1,-1; \frac{1}{a}\right)\\
&\nn-H\left(0,-1,-1,1; \frac{1}{a}\right)-2 H\left(0,-1,0,-1; \frac{1}{a}\right)+2 H\left(0,-1,0,1; \frac{1}{a}\right)\\
&\nn-H\left(0,-1,1,-1; \frac{1}{a}\right)+H\left(0,-1,1,1; \frac{1}{a}\right)-H\left(0,1,-1,-1; \frac{1}{a}\right)\\
&\nn+H\left(0,1,-1,1; \frac{1}{a}\right)+2 H\left(0,1,0,-1; \frac{1}{a}\right)-2 H\left(0,1,0,1; \frac{1}{a}\right)\\
&\nn+H\left(0,1,1,-1; \frac{1}{a}\right)-H\left(0,1,1,1; \frac{1}{a}\right)-H\left(1,-1,0,-1; \frac{1}{a}\right)\\
&\nn+2 H\left(1,-1,0,1; \frac{1}{a}\right)-H\left(1,0,-1,-1; \frac{1}{a}\right)+H\left(1,0,-1,1; \frac{1}{a}\right)\\
&\nn+4 H\left(1,0,0,-1; \frac{1}{a}\right)-4 H\left(1,0,0,1; \frac{1}{a}\right)+H\left(1,0,1,-1; \frac{1}{a}\right)\\
&\nn+H\left(1,0,1,1; \frac{1}{a}\right)+H\left(1,1,0,-1; \frac{1}{a}\right)-H\left(1,1,0,1; \frac{1}{a}\right)+H\left(1,1,1,1; \frac{1}{a}\right)\\
G&(a,a,a,0; 1) = 
H\left(0,1,1,1; \frac{1}{a}\right)+H\left(1,0,1,1; \frac{1}{a}\right)+H\left(1,1,0,1; \frac{1}{a}\right)\\
G&(a,a,a,1; 1) = 
H\left(0,1,1,1; \frac{1}{a}\right)+H\left(1,1,1,1; \frac{1}{a}\right)\\
G&(a,a,a,a; 1) = 
H\left(1,1,1,1; \frac{1}{a}\right)
\end{align}

\begin{align}
G&\left(-a,a^2,a^2,1; 1\right) = 
H\left(-1,-1,-1,-1; \frac{1}{a}\right)+H\left(-1,-1,0,-1; \frac{1}{a}\right)\\
&\nn-H\left(-1,-1,0,1; \frac{1}{a}\right)+H\left(-1,0,-1,-1; \frac{1}{a}\right)-H\left(-1,0,-1,1; \frac{1}{a}\right)\\
&\nn-H\left(-1,0,1,-1; \frac{1}{a}\right)+H\left(-1,0,1,1; \frac{1}{a}\right)-H\left(0,-1,-1,-1; \frac{1}{a}\right)\\
&\nn-H\left(0,-1,-1,1; \frac{1}{a}\right)-2 H\left(0,-1,0,-1; \frac{1}{a}\right)+2 H\left(0,-1,0,1; \frac{1}{a}\right)\\
&\nn-H\left(0,-1,1,-1; \frac{1}{a}\right)+H\left(0,-1,1,1; \frac{1}{a}\right)-4 H\left(0,0,-1,-1; \frac{1}{a}\right)\\
&\nn+4 H\left(0,0,-1,1; \frac{1}{a}\right)+4 H\left(0,0,1,-1; \frac{1}{a}\right)-4 H\left(0,0,1,1; \frac{1}{a}\right)\\
&\nn-H\left(0,1,-1,-1; \frac{1}{a}\right)+H\left(0,1,-1,1; \frac{1}{a}\right)+H\left(0,1,1,-1; \frac{1}{a}\right)\\
&\nn-H\left(0,1,1,1; \frac{1}{a}\right)-2 H\left(1,0,-1,-1; \frac{1}{a}\right)-4 H\left(1,0,0,-1; \frac{1}{a}\right)\\
&\nn+4 H\left(1,0,0,1; \frac{1}{a}\right)-2 H\left(1,1,0,-1; \frac{1}{a}\right)+H\left(1,1,0,1; \frac{1}{a}\right)\\
G&\left(a,a^2,a^2,1; 1\right) = 
-H\left(-1,-1,0,-1; \frac{1}{a}\right)+2 H\left(-1,-1,0,1; \frac{1}{a}\right)\\
&\nn+4 H\left(-1,0,0,-1; \frac{1}{a}\right)-4 H\left(-1,0,0,1; \frac{1}{a}\right)+2 H\left(-1,0,1,1; \frac{1}{a}\right)\\
&\nn+H\left(0,-1,-1,-1; \frac{1}{a}\right)-H\left(0,-1,-1,1; \frac{1}{a}\right)-H\left(0,-1,1,-1; \frac{1}{a}\right)\\
&\nn+H\left(0,-1,1,1; \frac{1}{a}\right)-4 H\left(0,0,-1,-1; \frac{1}{a}\right)+4 H\left(0,0,-1,1; \frac{1}{a}\right)\\
&\nn+4 H\left(0,0,1,-1; \frac{1}{a}\right)-4 H\left(0,0,1,1; \frac{1}{a}\right)-H\left(0,1,-1,-1; \frac{1}{a}\right)\\
&\nn+H\left(0,1,-1,1; \frac{1}{a}\right)+2 H\left(0,1,0,-1; \frac{1}{a}\right)-2 H\left(0,1,0,1; \frac{1}{a}\right)\\
&\nn+H\left(0,1,1,-1; \frac{1}{a}\right)+H\left(0,1,1,1; \frac{1}{a}\right)-H\left(1,0,-1,-1; \frac{1}{a}\right)\\
&\nn+H\left(1,0,-1,1; \frac{1}{a}\right)+H\left(1,0,1,-1; \frac{1}{a}\right)-H\left(1,0,1,1; \frac{1}{a}\right)\\
&\nn+H\left(1,1,0,-1; \frac{1}{a}\right)-H\left(1,1,0,1; \frac{1}{a}\right)+H\left(1,1,1,1; \frac{1}{a}\right)
\end{align}


\section{The analytic expression of the remainder function}
\label{app:R62}

In this appendix we present the full analytic expression of the remainder function. The result is also available in electronic form from {\tt www.arXiv.org}. Using the notation introduced in Eqs.~(\ref{eq:gcaldef}) and (\ref{eq:hcaldef}), the full expression reads,

 \btxtsloppy
\parbox{130mm}{\raggedright\(\displaystyle
R_{6,WL}^{(2)}(u_1,u_2,u_3) =  \)}
\raggedleft\refstepcounter{equation}(\theequation)\label{eq:R62_format}\\
\raggedright\(\displaystyle
\frac{1}{24} \pi ^2 G\left(\frac{1}{1-u_1},\frac{u_2-1}{u_1+u_2-1}; 1\right)+\frac{1}{24} \pi ^2 G\left(\frac{1}{u_1},\frac{1}{u_1+u_2}; 1\right)+\frac{1}{24} \pi ^2 G\left(\frac{1}{u_1},\frac{1}{u_1+u_3}; 1\right)+\frac{1}{24} \pi ^2 G\left(\frac{1}{1-u_2},\frac{u_3-1}{u_2+u_3-1}; 1\right)+\frac{1}{24} \pi ^2 G\left(\frac{1}{u_2},\frac{1}{u_1+u_2}; 1\right)+\frac{1}{24} \pi ^2 G\left(\frac{1}{u_2},\frac{1}{u_2+u_3}; 1\right)+\frac{1}{24} \pi ^2 G\left(\frac{1}{1-u_3},\frac{u_1-1}{u_1+u_3-1}; 1\right)+\frac{1}{24} \pi ^2 G\left(\frac{1}{u_3},\frac{1}{u_1+u_3}; 1\right)+\frac{1}{24} \pi ^2 G\left(\frac{1}{u_3},\frac{1}{u_2+u_3}; 1\right)+\frac{3}{2} G\left(0,0,\frac{1}{u_1},\frac{1}{u_1+u_2}; 1\right)+\frac{3}{2} G\left(0,0,\frac{1}{u_1},\frac{1}{u_1+u_3}; 1\right)+\frac{3}{2} G\left(0,0,\frac{1}{u_2},\frac{1}{u_1+u_2}; 1\right)+\frac{3}{2} G\left(0,0,\frac{1}{u_2},\frac{1}{u_2+u_3}; 1\right)+\frac{3}{2} G\left(0,0,\frac{1}{u_3},\frac{1}{u_1+u_3}; 1\right)+\frac{3}{2} G\left(0,0,\frac{1}{u_3},\frac{1}{u_2+u_3}; 1\right)-\frac{1}{2} G\left(0,\frac{1}{u_1},0,\frac{1}{u_2}; 1\right)+G\left(0,\frac{1}{u_1},0,\frac{1}{u_1+u_2}; 1\right)-\frac{1}{2} G\left(0,\frac{1}{u_1},0,\frac{1}{u_3}; 1\right)+G\left(0,\frac{1}{u_1},0,\frac{1}{u_1+u_3}; 1\right)-\frac{1}{2} G\left(0,\frac{1}{u_1},\frac{1}{u_1},\frac{1}{u_1+u_2}; 1\right)-\frac{1}{2} G\left(0,\frac{1}{u_1},\frac{1}{u_1},\frac{1}{u_1+u_3}; 1\right)-\frac{1}{2} G\left(0,\frac{1}{u_1},\frac{1}{u_2},\frac{1}{u_1+u_2}; 1\right)-\frac{1}{2} G\left(0,\frac{1}{u_1},\frac{1}{u_3},\frac{1}{u_1+u_3}; 1\right)-\frac{1}{2} G\left(0,\frac{1}{u_2},0,\frac{1}{u_1}; 1\right)+G\left(0,\frac{1}{u_2},0,\frac{1}{u_1+u_2}; 1\right)-\frac{1}{2} G\left(0,\frac{1}{u_2},0,\frac{1}{u_3}; 1\right)+G\left(0,\frac{1}{u_2},0,\frac{1}{u_2+u_3}; 1\right)-\frac{1}{2} G\left(0,\frac{1}{u_2},\frac{1}{u_1},\frac{1}{u_1+u_2}; 1\right)-\frac{1}{2} G\left(0,\frac{1}{u_2},\frac{1}{u_2},\frac{1}{u_1+u_2}; 1\right)-\frac{1}{2} G\left(0,\frac{1}{u_2},\frac{1}{u_2},\frac{1}{u_2+u_3}; 1\right)-\frac{1}{2} G\left(0,\frac{1}{u_2},\frac{1}{u_3},\frac{1}{u_2+u_3}; 1\right)+\frac{1}{4} G\left(0,\frac{u_2-1}{u_1+u_2-1},0,\frac{1}{1-u_1}; 1\right)+\frac{1}{4} G\left(0,\frac{u_2-1}{u_1+u_2-1},\frac{1}{1-u_1},0; 1\right)-\frac{1}{4} G\left(0,\frac{u_2-1}{u_1+u_2-1},\frac{1}{1-u_1},1; 1\right)+\frac{1}{4} G\left(0,\frac{u_2-1}{u_1+u_2-1},\frac{1}{1-u_1},\frac{1}{1-u_1}; 1\right)-\frac{1}{4} G\left(0,\frac{u_2-1}{u_1+u_2-1},\frac{u_2-1}{u_1+u_2-1},\frac{1}{1-u_1}; 1\right)-\frac{1}{2} G\left(0,\frac{1}{u_3},0,\frac{1}{u_1}; 1\right)-\frac{1}{2} G\left(0,\frac{1}{u_3},0,\frac{1}{u_2}; 1\right)+G\left(0,\frac{1}{u_3},0,\frac{1}{u_1+u_3}; 1\right)+G\left(0,\frac{1}{u_3},0,\frac{1}{u_2+u_3}; 1\right)-\frac{1}{2} G\left(0,\frac{1}{u_3},\frac{1}{u_1},\frac{1}{u_1+u_3}; 1\right)-\frac{1}{2} G\left(0,\frac{1}{u_3},\frac{1}{u_2},\frac{1}{u_2+u_3}; 1\right)-\frac{1}{2} G\left(0,\frac{1}{u_3},\frac{1}{u_3},\frac{1}{u_1+u_3}; 1\right)-\frac{1}{2} G\left(0,\frac{1}{u_3},\frac{1}{u_3},\frac{1}{u_2+u_3}; 1\right)+\frac{1}{4} G\left(0,\frac{u_1-1}{u_1+u_3-1},0,\frac{1}{1-u_3}; 1\right)+\frac{1}{4} G\left(0,\frac{u_1-1}{u_1+u_3-1},\frac{1}{1-u_3},0; 1\right)-\frac{1}{4} G\left(0,\frac{u_1-1}{u_1+u_3-1},\frac{1}{1-u_3},1; 1\right)+\frac{1}{4} G\left(0,\frac{u_1-1}{u_1+u_3-1},\frac{1}{1-u_3},\frac{1}{1-u_3}; 1\right)-\frac{1}{4} G\left(0,\frac{u_1-1}{u_1+u_3-1},\frac{u_1-1}{u_1+u_3-1},\frac{1}{1-u_3}; 1\right)+\frac{1}{4} G\left(0,\frac{u_3-1}{u_2+u_3-1},0,\frac{1}{1-u_2}; 1\right)+\frac{1}{4} G\left(0,\frac{u_3-1}{u_2+u_3-1},\frac{1}{1-u_2},0; 1\right)-\frac{1}{4} G\left(0,\frac{u_3-1}{u_2+u_3-1},\frac{1}{1-u_2},1; 1\right)+\frac{1}{4} G\left(0,\frac{u_3-1}{u_2+u_3-1},\frac{1}{1-u_2},\frac{1}{1-u_2}; 1\right)-\frac{1}{4} G\left(0,\frac{u_3-1}{u_2+u_3-1},\frac{u_3-1}{u_2+u_3-1},\frac{1}{1-u_2}; 1\right)-\frac{1}{4} G\left(\frac{1}{1-u_1},1,\frac{1}{u_3},0; 1\right)+\frac{1}{2} G\left(\frac{1}{1-u_1},\frac{1}{1-u_1},1,\frac{1}{1-u_1}; 1\right)+\frac{1}{4} G\left(\frac{1}{1-u_1},\frac{u_2-1}{u_1+u_2-1},0,1; 1\right)-\frac{1}{4} G\left(\frac{1}{1-u_1},\frac{u_2-1}{u_1+u_2-1},0,\frac{1}{1-u_1}; 1\right)+\frac{1}{4} G\left(\frac{1}{1-u_1},\frac{u_2-1}{u_1+u_2-1},1,0; 1\right)-\frac{1}{4} G\left(\frac{1}{1-u_1},\frac{u_2-1}{u_1+u_2-1},\frac{1}{1-u_1},0; 1\right)+\frac{1}{4} G\left(\frac{1}{1-u_1},\frac{u_2-1}{u_1+u_2-1},\frac{1}{1-u_1},1; 1\right)-\frac{1}{4} G\left(\frac{1}{1-u_1},\frac{u_2-1}{u_1+u_2-1},\frac{1}{1-u_1},\frac{1}{1-u_1}; 1\right)-\frac{1}{4} G\left(\frac{1}{1-u_1},\frac{u_2-1}{u_1+u_2-1},\frac{u_2-1}{u_1+u_2-1},1; 1\right)+\frac{1}{4} G\left(\frac{1}{1-u_1},\frac{u_2-1}{u_1+u_2-1},\frac{u_2-1}{u_1+u_2-1},\frac{1}{1-u_1}; 1\right)-G\left(\frac{1}{u_1},0,0,\frac{1}{u_2}; 1\right)+\frac{1}{2} G\left(\frac{1}{u_1},0,0,\frac{1}{u_1+u_2}; 1\right)-G\left(\frac{1}{u_1},0,0,\frac{1}{u_3}; 1\right)+\frac{1}{2} G\left(\frac{1}{u_1},0,0,\frac{1}{u_1+u_3}; 1\right)-\frac{1}{4} G\left(\frac{1}{u_1},0,\frac{1}{u_1},\frac{1}{u_1+u_2}; 1\right)-\frac{1}{4} G\left(\frac{1}{u_1},0,\frac{1}{u_1},\frac{1}{u_1+u_3}; 1\right)-\frac{1}{4} G\left(\frac{1}{u_1},0,\frac{1}{u_2},\frac{1}{u_1+u_2}; 1\right)-\frac{1}{4} G\left(\frac{1}{u_1},0,\frac{1}{u_3},\frac{1}{u_1+u_3}; 1\right)-\frac{1}{4} G\left(\frac{1}{1-u_2},1,\frac{1}{u_1},0; 1\right)+\frac{1}{2} G\left(\frac{1}{1-u_2},\frac{1}{1-u_2},1,\frac{1}{1-u_2}; 1\right)+\frac{1}{4} G\left(\frac{1}{1-u_2},\frac{u_3-1}{u_2+u_3-1},0,1; 1\right)-\frac{1}{4} G\left(\frac{1}{1-u_2},\frac{u_3-1}{u_2+u_3-1},0,\frac{1}{1-u_2}; 1\right)+\frac{1}{4} G\left(\frac{1}{1-u_2},\frac{u_3-1}{u_2+u_3-1},1,0; 1\right)-\frac{1}{4} G\left(\frac{1}{1-u_2},\frac{u_3-1}{u_2+u_3-1},\frac{1}{1-u_2},0; 1\right)+\frac{1}{4} G\left(\frac{1}{1-u_2},\frac{u_3-1}{u_2+u_3-1},\frac{1}{1-u_2},1; 1\right)-\frac{1}{4} G\left(\frac{1}{1-u_2},\frac{u_3-1}{u_2+u_3-1},\frac{1}{1-u_2},\frac{1}{1-u_2}; 1\right)-\frac{1}{4} G\left(\frac{1}{1-u_2},\frac{u_3-1}{u_2+u_3-1},\frac{u_3-1}{u_2+u_3-1},1; 1\right)+\frac{1}{4} G\left(\frac{1}{1-u_2},\frac{u_3-1}{u_2+u_3-1},\frac{u_3-1}{u_2+u_3-1},\frac{1}{1-u_2}; 1\right)-G\left(\frac{1}{u_2},0,0,\frac{1}{u_1}; 1\right)+\frac{1}{2} G\left(\frac{1}{u_2},0,0,\frac{1}{u_1+u_2}; 1\right)-G\left(\frac{1}{u_2},0,0,\frac{1}{u_3}; 1\right)+\frac{1}{2} G\left(\frac{1}{u_2},0,0,\frac{1}{u_2+u_3}; 1\right)-\frac{1}{4} G\left(\frac{1}{u_2},0,\frac{1}{u_1},\frac{1}{u_1+u_2}; 1\right)-\frac{1}{4} G\left(\frac{1}{u_2},0,\frac{1}{u_2},\frac{1}{u_1+u_2}; 1\right)-\frac{1}{4} G\left(\frac{1}{u_2},0,\frac{1}{u_2},\frac{1}{u_2+u_3}; 1\right)-\frac{1}{4} G\left(\frac{1}{u_2},0,\frac{1}{u_3},\frac{1}{u_2+u_3}; 1\right)-\frac{1}{4} G\left(\frac{1}{1-u_3},1,\frac{1}{u_2},0; 1\right)+\frac{1}{2} G\left(\frac{1}{1-u_3},\frac{1}{1-u_3},1,\frac{1}{1-u_3}; 1\right)+\frac{1}{4} G\left(\frac{1}{1-u_3},\frac{u_1-1}{u_1+u_3-1},0,1; 1\right)-\frac{1}{4} G\left(\frac{1}{1-u_3},\frac{u_1-1}{u_1+u_3-1},0,\frac{1}{1-u_3}; 1\right)+\frac{1}{4} G\left(\frac{1}{1-u_3},\frac{u_1-1}{u_1+u_3-1},1,0; 1\right)-\frac{1}{4} G\left(\frac{1}{1-u_3},\frac{u_1-1}{u_1+u_3-1},\frac{1}{1-u_3},0; 1\right)+\frac{1}{4} G\left(\frac{1}{1-u_3},\frac{u_1-1}{u_1+u_3-1},\frac{1}{1-u_3},1; 1\right)-\frac{1}{4} G\left(\frac{1}{1-u_3},\frac{u_1-1}{u_1+u_3-1},\frac{1}{1-u_3},\frac{1}{1-u_3}; 1\right)-\frac{1}{4} G\left(\frac{1}{1-u_3},\frac{u_1-1}{u_1+u_3-1},\frac{u_1-1}{u_1+u_3-1},1; 1\right)-\frac{79 \pi ^4}{360}
+\frac{1}{4} G\left(\frac{1}{1-u_3},\frac{u_1-1}{u_1+u_3-1},\frac{u_1-1}{u_1+u_3-1},\frac{1}{1-u_3}; 1\right)-G\left(\frac{1}{u_3},0,0,\frac{1}{u_1}; 1\right)-G\left(\frac{1}{u_3},0,0,\frac{1}{u_2}; 1\right)+\frac{1}{2} G\left(\frac{1}{u_3},0,0,\frac{1}{u_1+u_3}; 1\right)+\frac{1}{2} G\left(\frac{1}{u_3},0,0,\frac{1}{u_2+u_3}; 1\right)-\frac{1}{4} G\left(\frac{1}{u_3},0,\frac{1}{u_1},\frac{1}{u_1+u_3}; 1\right)-\frac{1}{4} G\left(\frac{1}{u_3},0,\frac{1}{u_2},\frac{1}{u_2+u_3}; 1\right)-\frac{1}{4} G\left(\frac{1}{u_3},0,\frac{1}{u_3},\frac{1}{u_1+u_3}; 1\right)-\frac{1}{4} G\left(\frac{1}{u_3},0,\frac{1}{u_3},\frac{1}{u_2+u_3}; 1\right)-\frac{1}{24} \pi ^2 \gcal\left(\frac{1}{1-u_1},u_{123}; 1\right)+\frac{1}{8} \pi ^2 \gcal\left(\frac{1}{1-u_1},v_{123}; 1\right)+\frac{1}{8} \pi ^2 \gcal\left(\frac{1}{1-u_1},v_{132}; 1\right)-\frac{1}{24} \pi ^2 \gcal\left(\frac{1}{1-u_2},u_{231}; 1\right)+\frac{1}{8} \pi ^2 \gcal\left(\frac{1}{1-u_2},v_{213}; 1\right)+\frac{1}{8} \pi ^2 \gcal\left(\frac{1}{1-u_2},v_{231}; 1\right)-\frac{1}{24} \pi ^2 \gcal\left(\frac{1}{1-u_3},u_{312}; 1\right)+\frac{1}{8} \pi ^2 \gcal\left(\frac{1}{1-u_3},v_{312}; 1\right)+\frac{1}{8} \pi ^2 \gcal\left(\frac{1}{1-u_3},v_{321}; 1\right)-\frac{1}{4} \gcal\left(0,0,\frac{1}{1-u_1},v_{123}; 1\right)-\frac{1}{4} \gcal\left(0,0,\frac{1}{1-u_1},v_{132}; 1\right)-\frac{1}{4} \gcal\left(0,0,\frac{1}{1-u_2},v_{213}; 1\right)-\frac{1}{4} \gcal\left(0,0,\frac{1}{1-u_2},v_{231}; 1\right)-\frac{1}{4} \gcal\left(0,0,\frac{1}{1-u_3},v_{312}; 1\right)-\frac{1}{4} \gcal\left(0,0,\frac{1}{1-u_3},v_{321}; 1\right)-\frac{1}{4} \gcal\left(0,0,v_{123},\frac{1}{1-u_1}; 1\right)+\gcal\left(0,0,v_{132},0; 1\right)-\frac{1}{4} \gcal\left(0,0,v_{132},\frac{1}{1-u_1}; 1\right)+\gcal\left(0,0,v_{213},0; 1\right)-\frac{1}{4} \gcal\left(0,0,v_{213},\frac{1}{1-u_2}; 1\right)-\frac{1}{4} \gcal\left(0,0,v_{231},\frac{1}{1-u_2}; 1\right)-\frac{1}{4} \gcal\left(0,0,v_{312},\frac{1}{1-u_3}; 1\right)+\gcal\left(0,0,v_{321},0; 1\right)-\frac{1}{4} \gcal\left(0,0,v_{321},\frac{1}{1-u_3}; 1\right)-\frac{1}{4} \gcal\left(0,\frac{1}{1-u_1},0,v_{123}; 1\right)-\frac{1}{4} \gcal\left(0,\frac{1}{1-u_1},0,v_{132}; 1\right)-\frac{1}{2} \gcal\left(0,\frac{1}{1-u_1},\frac{1}{1-u_1},v_{123}; 1\right)-\frac{1}{2} \gcal\left(0,\frac{1}{1-u_1},\frac{1}{1-u_1},v_{132}; 1\right)-\frac{1}{4} \gcal\left(0,\frac{1}{1-u_1},v_{123},1; 1\right)-\frac{1}{4} \gcal\left(0,\frac{1}{1-u_1},v_{123},\frac{1}{1-u_1}; 1\right)-\frac{1}{4} \gcal\left(0,\frac{1}{1-u_1},v_{132},1; 1\right)-\frac{1}{4} \gcal\left(0,\frac{1}{1-u_1},v_{132},\frac{1}{1-u_1}; 1\right)-\frac{1}{4} \gcal\left(0,\frac{1}{1-u_2},0,v_{213}; 1\right)-\frac{1}{4} \gcal\left(0,\frac{1}{1-u_2},0,v_{231}; 1\right)-\frac{1}{2} \gcal\left(0,\frac{1}{1-u_2},\frac{1}{1-u_2},v_{213}; 1\right)-\frac{1}{2} \gcal\left(0,\frac{1}{1-u_2},\frac{1}{1-u_2},v_{231}; 1\right)-\frac{1}{4} \gcal\left(0,\frac{1}{1-u_2},v_{213},1; 1\right)-\frac{1}{4} \gcal\left(0,\frac{1}{1-u_2},v_{213},\frac{1}{1-u_2}; 1\right)-\frac{1}{4} \gcal\left(0,\frac{1}{1-u_2},v_{231},1; 1\right)-\frac{1}{4} \gcal\left(0,\frac{1}{1-u_2},v_{231},\frac{1}{1-u_2}; 1\right)-\frac{1}{4} \gcal\left(0,\frac{1}{1-u_3},0,v_{312}; 1\right)-\frac{1}{4} \gcal\left(0,\frac{1}{1-u_3},0,v_{321}; 1\right)-\frac{1}{2} \gcal\left(0,\frac{1}{1-u_3},\frac{1}{1-u_3},v_{312}; 1\right)-\frac{1}{2} \gcal\left(0,\frac{1}{1-u_3},\frac{1}{1-u_3},v_{321}; 1\right)-\frac{1}{4} \gcal\left(0,\frac{1}{1-u_3},v_{312},1; 1\right)-\frac{1}{4} \gcal\left(0,\frac{1}{1-u_3},v_{312},\frac{1}{1-u_3}; 1\right)-\frac{1}{4} \gcal\left(0,\frac{1}{1-u_3},v_{321},1; 1\right)-\frac{1}{4} \gcal\left(0,\frac{1}{1-u_3},v_{321},\frac{1}{1-u_3}; 1\right)-\frac{1}{4} \gcal\left(0,u_{123},0,\frac{1}{1-u_1}; 1\right)-\frac{1}{4} \gcal\left(0,u_{123},\frac{1}{1-u_1},0; 1\right)+\frac{1}{4} \gcal\left(0,u_{123},\frac{1}{1-u_1},1; 1\right)-\frac{1}{4} \gcal\left(0,u_{123},\frac{1}{1-u_1},\frac{1}{1-u_1}; 1\right)-\frac{1}{4} \gcal\left(0,u_{123},\frac{u_2-1}{u_1+u_2-1},1; 1\right)+\frac{1}{4} \gcal\left(0,u_{123},\frac{u_2-1}{u_1+u_2-1},\frac{1}{1-u_1}; 1\right)-\frac{1}{4} \gcal\left(0,u_{123},\frac{1}{u_3},0; 1\right)-\frac{1}{4} \gcal\left(0,u_{231},0,\frac{1}{1-u_2}; 1\right)-\frac{1}{4} \gcal\left(0,u_{231},\frac{1}{u_1},0; 1\right)-\frac{1}{4} \gcal\left(0,u_{231},\frac{1}{1-u_2},0; 1\right)+\frac{1}{4} \gcal\left(0,u_{231},\frac{1}{1-u_2},1; 1\right)-\frac{1}{4} \gcal\left(0,u_{231},\frac{1}{1-u_2},\frac{1}{1-u_2}; 1\right)-\frac{1}{4} \gcal\left(0,u_{231},\frac{u_3-1}{u_2+u_3-1},1; 1\right)+\frac{1}{4} \gcal\left(0,u_{231},\frac{u_3-1}{u_2+u_3-1},\frac{1}{1-u_2}; 1\right)-\frac{1}{4} \gcal\left(0,u_{312},0,\frac{1}{1-u_3}; 1\right)-\frac{1}{4} \gcal\left(0,u_{312},\frac{1}{u_2},0; 1\right)-\frac{1}{4} \gcal\left(0,u_{312},\frac{1}{1-u_3},0; 1\right)+\frac{1}{4} \gcal\left(0,u_{312},\frac{1}{1-u_3},1; 1\right)-\frac{1}{4} \gcal\left(0,u_{312},\frac{1}{1-u_3},\frac{1}{1-u_3}; 1\right)-\frac{1}{4} \gcal\left(0,u_{312},\frac{u_1-1}{u_1+u_3-1},1; 1\right)+\frac{1}{4} \gcal\left(0,u_{312},\frac{u_1-1}{u_1+u_3-1},\frac{1}{1-u_3}; 1\right)+\frac{1}{4} \gcal\left(0,v_{123},0,\frac{1}{1-u_1}; 1\right)-\frac{1}{2} \gcal\left(0,v_{123},1,\frac{1}{1-u_1}; 1\right)+\frac{1}{4} \gcal\left(0,v_{123},\frac{1}{1-u_1},0; 1\right)-\frac{1}{2} \gcal\left(0,v_{123},\frac{1}{1-u_1},1; 1\right)+\frac{1}{4} \gcal\left(0,v_{123},\frac{1}{1-u_1},\frac{1}{1-u_1}; 1\right)-\frac{1}{4} \gcal\left(0,v_{132},0,\frac{1}{1-u_1}; 1\right)-\frac{1}{4} \gcal\left(0,v_{132},\frac{1}{1-u_1},0; 1\right)-\frac{1}{4} \gcal\left(0,v_{132},\frac{1}{1-u_1},\frac{1}{1-u_1}; 1\right)-\frac{1}{4} \gcal\left(0,v_{213},0,\frac{1}{1-u_2}; 1\right)-\frac{1}{4} \gcal\left(0,v_{213},\frac{1}{1-u_2},0; 1\right)-\frac{1}{4} \gcal\left(0,v_{213},\frac{1}{1-u_2},\frac{1}{1-u_2}; 1\right)+\frac{1}{4} \gcal\left(0,v_{231},0,\frac{1}{1-u_2}; 1\right)-\frac{1}{2} \gcal\left(0,v_{231},1,\frac{1}{1-u_2}; 1\right)+\frac{1}{4} \gcal\left(0,v_{231},\frac{1}{1-u_2},0; 1\right)-\frac{1}{2} \gcal\left(0,v_{231},\frac{1}{1-u_2},1; 1\right)+\frac{1}{4} \gcal\left(0,v_{231},\frac{1}{1-u_2},\frac{1}{1-u_2}; 1\right)+\frac{1}{4} \gcal\left(0,v_{312},0,\frac{1}{1-u_3}; 1\right)-\frac{1}{2} \gcal\left(0,v_{312},1,\frac{1}{1-u_3}; 1\right)+\frac{1}{4} \gcal\left(0,v_{312},\frac{1}{1-u_3},0; 1\right)-\frac{1}{2} \gcal\left(0,v_{312},\frac{1}{1-u_3},1; 1\right)+\frac{1}{4} \gcal\left(0,v_{312},\frac{1}{1-u_3},\frac{1}{1-u_3}; 1\right)-\frac{1}{4} \gcal\left(0,v_{321},0,\frac{1}{1-u_3}; 1\right)
-\frac{1}{4} \gcal\left(0,v_{321},\frac{1}{1-u_3},0; 1\right)-\frac{1}{4} \gcal\left(0,v_{321},\frac{1}{1-u_3},\frac{1}{1-u_3}; 1\right)-\frac{1}{4} \gcal\left(\frac{1}{1-u_1},0,0,v_{123}; 1\right)-\frac{1}{4} \gcal\left(\frac{1}{1-u_1},0,0,v_{132}; 1\right)-\frac{1}{2} \gcal\left(\frac{1}{1-u_1},0,\frac{1}{1-u_1},v_{123}; 1\right)-\frac{1}{2} \gcal\left(\frac{1}{1-u_1},0,\frac{1}{1-u_1},v_{132}; 1\right)-\frac{1}{4} \gcal\left(\frac{1}{1-u_1},0,v_{123},1; 1\right)-\frac{1}{4} \gcal\left(\frac{1}{1-u_1},0,v_{123},\frac{1}{1-u_1}; 1\right)-\frac{1}{4} \gcal\left(\frac{1}{1-u_1},0,v_{132},1; 1\right)-\frac{1}{4} \gcal\left(\frac{1}{1-u_1},0,v_{132},\frac{1}{1-u_1}; 1\right)-\frac{1}{2} \gcal\left(\frac{1}{1-u_1},\frac{1}{1-u_1},0,v_{123}; 1\right)-\frac{1}{2} \gcal\left(\frac{1}{1-u_1},\frac{1}{1-u_1},0,v_{132}; 1\right)-\frac{3}{4} \gcal\left(\frac{1}{1-u_1},\frac{1}{1-u_1},\frac{1}{1-u_1},v_{123}; 1\right)-\frac{3}{4} \gcal\left(\frac{1}{1-u_1},\frac{1}{1-u_1},\frac{1}{1-u_1},v_{132}; 1\right)-\frac{1}{2} \gcal\left(\frac{1}{1-u_1},\frac{1}{1-u_1},v_{123},1; 1\right)-\frac{1}{4} \gcal\left(\frac{1}{1-u_1},\frac{1}{1-u_1},v_{123},\frac{1}{1-u_1}; 1\right)-\frac{1}{2} \gcal\left(\frac{1}{1-u_1},\frac{1}{1-u_1},v_{132},1; 1\right)-\frac{1}{4} \gcal\left(\frac{1}{1-u_1},\frac{1}{1-u_1},v_{132},\frac{1}{1-u_1}; 1\right)-\frac{1}{4} \gcal\left(\frac{1}{1-u_1},u_{123},0,1; 1\right)+\frac{1}{4} \gcal\left(\frac{1}{1-u_1},u_{123},0,\frac{1}{1-u_1}; 1\right)-\frac{1}{4} \gcal\left(\frac{1}{1-u_1},u_{123},1,0; 1\right)+\frac{1}{4} \gcal\left(\frac{1}{1-u_1},u_{123},\frac{1}{1-u_1},0; 1\right)-\frac{1}{4} \gcal\left(\frac{1}{1-u_1},u_{123},\frac{1}{1-u_1},1; 1\right)+\frac{1}{4} \gcal\left(\frac{1}{1-u_1},u_{123},\frac{1}{1-u_1},\frac{1}{1-u_1}; 1\right)+\frac{1}{4} \gcal\left(\frac{1}{1-u_1},u_{123},\frac{u_2-1}{u_1+u_2-1},1; 1\right)-\frac{1}{4} \gcal\left(\frac{1}{1-u_1},u_{123},\frac{u_2-1}{u_1+u_2-1},\frac{1}{1-u_1}; 1\right)+\frac{1}{4} \gcal\left(\frac{1}{1-u_1},u_{123},\frac{1}{u_3},0; 1\right)+\frac{1}{4} \gcal\left(\frac{1}{1-u_1},v_{123},0,0; 1\right)-\frac{1}{4} \gcal\left(\frac{1}{1-u_1},v_{123},0,1; 1\right)+\frac{1}{4} \gcal\left(\frac{1}{1-u_1},v_{123},0,\frac{1}{1-u_1}; 1\right)-\frac{1}{4} \gcal\left(\frac{1}{1-u_1},v_{123},1,0; 1\right)-\frac{1}{2} \gcal\left(\frac{1}{1-u_1},v_{123},1,\frac{1}{1-u_1}; 1\right)+\frac{1}{4} \gcal\left(\frac{1}{1-u_1},v_{123},\frac{1}{1-u_1},0; 1\right)-\frac{1}{2} \gcal\left(\frac{1}{1-u_1},v_{123},\frac{1}{1-u_1},1; 1\right)+\frac{1}{4} \gcal\left(\frac{1}{1-u_1},v_{123},\frac{1}{1-u_1},\frac{1}{1-u_1}; 1\right)+\frac{1}{4} \gcal\left(\frac{1}{1-u_1},v_{132},0,0; 1\right)-\frac{1}{4} \gcal\left(\frac{1}{1-u_1},v_{132},0,1; 1\right)+\frac{1}{4} \gcal\left(\frac{1}{1-u_1},v_{132},0,\frac{1}{1-u_1}; 1\right)-\frac{1}{4} \gcal\left(\frac{1}{1-u_1},v_{132},1,0; 1\right)-\frac{1}{2} \gcal\left(\frac{1}{1-u_1},v_{132},1,\frac{1}{1-u_1}; 1\right)+\frac{1}{4} \gcal\left(\frac{1}{1-u_1},v_{132},\frac{1}{1-u_1},0; 1\right)-\frac{1}{2} \gcal\left(\frac{1}{1-u_1},v_{132},\frac{1}{1-u_1},1; 1\right)+\frac{1}{4} \gcal\left(\frac{1}{1-u_1},v_{132},\frac{1}{1-u_1},\frac{1}{1-u_1}; 1\right)-\frac{1}{4} \gcal\left(\frac{1}{1-u_2},0,0,v_{213}; 1\right)-\frac{1}{4} \gcal\left(\frac{1}{1-u_2},0,0,v_{231}; 1\right)-\frac{1}{2} \gcal\left(\frac{1}{1-u_2},0,\frac{1}{1-u_2},v_{213}; 1\right)-\frac{1}{2} \gcal\left(\frac{1}{1-u_2},0,\frac{1}{1-u_2},v_{231}; 1\right)-\frac{1}{4} \gcal\left(\frac{1}{1-u_2},0,v_{213},1; 1\right)-\frac{1}{4} \gcal\left(\frac{1}{1-u_2},0,v_{213},\frac{1}{1-u_2}; 1\right)-\frac{1}{4} \gcal\left(\frac{1}{1-u_2},0,v_{231},1; 1\right)-\frac{1}{4} \gcal\left(\frac{1}{1-u_2},0,v_{231},\frac{1}{1-u_2}; 1\right)-\frac{1}{2} \gcal\left(\frac{1}{1-u_2},\frac{1}{1-u_2},0,v_{213}; 1\right)-\frac{1}{2} \gcal\left(\frac{1}{1-u_2},\frac{1}{1-u_2},0,v_{231}; 1\right)-\frac{3}{4} \gcal\left(\frac{1}{1-u_2},\frac{1}{1-u_2},\frac{1}{1-u_2},v_{213}; 1\right)-\frac{3}{4} \gcal\left(\frac{1}{1-u_2},\frac{1}{1-u_2},\frac{1}{1-u_2},v_{231}; 1\right)-\frac{1}{2} \gcal\left(\frac{1}{1-u_2},\frac{1}{1-u_2},v_{213},1; 1\right)-\frac{1}{4} \gcal\left(\frac{1}{1-u_2},\frac{1}{1-u_2},v_{213},\frac{1}{1-u_2}; 1\right)-\frac{1}{2} \gcal\left(\frac{1}{1-u_2},\frac{1}{1-u_2},v_{231},1; 1\right)-\frac{1}{4} \gcal\left(\frac{1}{1-u_2},\frac{1}{1-u_2},v_{231},\frac{1}{1-u_2}; 1\right)-\frac{1}{4} \gcal\left(\frac{1}{1-u_2},u_{231},0,1; 1\right)+\frac{1}{4} \gcal\left(\frac{1}{1-u_2},u_{231},0,\frac{1}{1-u_2}; 1\right)-\frac{1}{4} \gcal\left(\frac{1}{1-u_2},u_{231},1,0; 1\right)+\frac{1}{4} \gcal\left(\frac{1}{1-u_2},u_{231},\frac{1}{u_1},0; 1\right)+\frac{1}{4} \gcal\left(\frac{1}{1-u_2},u_{231},\frac{1}{1-u_2},0; 1\right)-\frac{1}{4} \gcal\left(\frac{1}{1-u_2},u_{231},\frac{1}{1-u_2},1; 1\right)+\frac{1}{4} \gcal\left(\frac{1}{1-u_2},u_{231},\frac{1}{1-u_2},\frac{1}{1-u_2}; 1\right)+\frac{1}{4} \gcal\left(\frac{1}{1-u_2},u_{231},\frac{u_3-1}{u_2+u_3-1},1; 1\right)-\frac{1}{4} \gcal\left(\frac{1}{1-u_2},u_{231},\frac{u_3-1}{u_2+u_3-1},\frac{1}{1-u_2}; 1\right)+\frac{1}{4} \gcal\left(\frac{1}{1-u_2},v_{213},0,0; 1\right)-\frac{1}{4} \gcal\left(\frac{1}{1-u_2},v_{213},0,1; 1\right)+\frac{1}{4} \gcal\left(\frac{1}{1-u_2},v_{213},0,\frac{1}{1-u_2}; 1\right)-\frac{1}{4} \gcal\left(\frac{1}{1-u_2},v_{213},1,0; 1\right)-\frac{1}{2} \gcal\left(\frac{1}{1-u_2},v_{213},1,\frac{1}{1-u_2}; 1\right)+\frac{1}{4} \gcal\left(\frac{1}{1-u_2},v_{213},\frac{1}{1-u_2},0; 1\right)-\frac{1}{2} \gcal\left(\frac{1}{1-u_2},v_{213},\frac{1}{1-u_2},1; 1\right)+\frac{1}{4} \gcal\left(\frac{1}{1-u_2},v_{213},\frac{1}{1-u_2},\frac{1}{1-u_2}; 1\right)+\frac{1}{4} \gcal\left(\frac{1}{1-u_2},v_{231},0,0; 1\right)-\frac{1}{4} \gcal\left(\frac{1}{1-u_2},v_{231},0,1; 1\right)+\frac{1}{4} \gcal\left(\frac{1}{1-u_2},v_{231},0,\frac{1}{1-u_2}; 1\right)-\frac{1}{4} \gcal\left(\frac{1}{1-u_2},v_{231},1,0; 1\right)-\frac{1}{2} \gcal\left(\frac{1}{1-u_2},v_{231},1,\frac{1}{1-u_2}; 1\right)+\frac{1}{4} \gcal\left(\frac{1}{1-u_2},v_{231},\frac{1}{1-u_2},0; 1\right)-\frac{1}{2} \gcal\left(\frac{1}{1-u_2},v_{231},\frac{1}{1-u_2},1; 1\right)+\frac{1}{4} \gcal\left(\frac{1}{1-u_2},v_{231},\frac{1}{1-u_2},\frac{1}{1-u_2}; 1\right)-\frac{1}{4} \gcal\left(\frac{1}{1-u_3},0,0,v_{312}; 1\right)-\frac{1}{4} \gcal\left(\frac{1}{1-u_3},0,0,v_{321}; 1\right)-\frac{1}{2} \gcal\left(\frac{1}{1-u_3},0,\frac{1}{1-u_3},v_{312}; 1\right)-\frac{1}{2} \gcal\left(\frac{1}{1-u_3},0,\frac{1}{1-u_3},v_{321}; 1\right)-\frac{1}{4} \gcal\left(\frac{1}{1-u_3},0,v_{312},1; 1\right)-\frac{1}{4} \gcal\left(\frac{1}{1-u_3},0,v_{312},\frac{1}{1-u_3}; 1\right)-\frac{1}{4} \gcal\left(\frac{1}{1-u_3},0,v_{321},1; 1\right)-\frac{1}{4} \gcal\left(\frac{1}{1-u_3},0,v_{321},\frac{1}{1-u_3}; 1\right)-\frac{1}{2} \gcal\left(\frac{1}{1-u_3},\frac{1}{1-u_3},0,v_{312}; 1\right)-\frac{1}{2} \gcal\left(\frac{1}{1-u_3},\frac{1}{1-u_3},0,v_{321}; 1\right)-\frac{3}{4} \gcal\left(\frac{1}{1-u_3},\frac{1}{1-u_3},\frac{1}{1-u_3},v_{312}; 1\right)-\frac{3}{4} \gcal\left(\frac{1}{1-u_3},\frac{1}{1-u_3},\frac{1}{1-u_3},v_{321}; 1\right)-\frac{1}{2} \gcal\left(\frac{1}{1-u_3},\frac{1}{1-u_3},v_{312},1; 1\right)-\frac{1}{4} \gcal\left(\frac{1}{1-u_3},\frac{1}{1-u_3},v_{312},\frac{1}{1-u_3}; 1\right)-\frac{1}{2} \gcal\left(\frac{1}{1-u_3},\frac{1}{1-u_3},v_{321},1; 1\right)-\frac{1}{4} \gcal\left(\frac{1}{1-u_3},\frac{1}{1-u_3},v_{321},\frac{1}{1-u_3}; 1\right)
-\frac{1}{4} \gcal\left(\frac{1}{1-u_3},u_{312},0,1; 1\right)+\frac{1}{4} \gcal\left(\frac{1}{1-u_3},u_{312},0,\frac{1}{1-u_3}; 1\right)-\frac{1}{4} \gcal\left(\frac{1}{1-u_3},u_{312},1,0; 1\right)+\frac{1}{4} \gcal\left(\frac{1}{1-u_3},u_{312},\frac{1}{u_2},0; 1\right)+\frac{1}{4} \gcal\left(\frac{1}{1-u_3},u_{312},\frac{1}{1-u_3},0; 1\right)-\frac{1}{4} \gcal\left(\frac{1}{1-u_3},u_{312},\frac{1}{1-u_3},1; 1\right)+\frac{1}{4} \gcal\left(\frac{1}{1-u_3},u_{312},\frac{1}{1-u_3},\frac{1}{1-u_3}; 1\right)+\frac{1}{4} \gcal\left(\frac{1}{1-u_3},u_{312},\frac{u_1-1}{u_1+u_3-1},1; 1\right)-\frac{1}{4} \gcal\left(\frac{1}{1-u_3},u_{312},\frac{u_1-1}{u_1+u_3-1},\frac{1}{1-u_3}; 1\right)+\frac{1}{4} \gcal\left(\frac{1}{1-u_3},v_{312},0,0; 1\right)-\frac{1}{4} \gcal\left(\frac{1}{1-u_3},v_{312},0,1; 1\right)+\frac{1}{4} \gcal\left(\frac{1}{1-u_3},v_{312},0,\frac{1}{1-u_3}; 1\right)-\frac{1}{4} \gcal\left(\frac{1}{1-u_3},v_{312},1,0; 1\right)-\frac{1}{2} \gcal\left(\frac{1}{1-u_3},v_{312},1,\frac{1}{1-u_3}; 1\right)+\frac{1}{4} \gcal\left(\frac{1}{1-u_3},v_{312},\frac{1}{1-u_3},0; 1\right)-\frac{1}{2} \gcal\left(\frac{1}{1-u_3},v_{312},\frac{1}{1-u_3},1; 1\right)+\frac{1}{4} \gcal\left(\frac{1}{1-u_3},v_{312},\frac{1}{1-u_3},\frac{1}{1-u_3}; 1\right)+\frac{1}{4} \gcal\left(\frac{1}{1-u_3},v_{321},0,0; 1\right)-\frac{1}{4} \gcal\left(\frac{1}{1-u_3},v_{321},0,1; 1\right)+\frac{1}{4} \gcal\left(\frac{1}{1-u_3},v_{321},0,\frac{1}{1-u_3}; 1\right)-\frac{1}{4} \gcal\left(\frac{1}{1-u_3},v_{321},1,0; 1\right)-\frac{1}{2} \gcal\left(\frac{1}{1-u_3},v_{321},1,\frac{1}{1-u_3}; 1\right)+\frac{1}{4} \gcal\left(\frac{1}{1-u_3},v_{321},\frac{1}{1-u_3},0; 1\right)-\frac{1}{2} \gcal\left(\frac{1}{1-u_3},v_{321},\frac{1}{1-u_3},1; 1\right)+\frac{1}{4} \gcal\left(\frac{1}{1-u_3},v_{321},\frac{1}{1-u_3},\frac{1}{1-u_3}; 1\right)+\frac{1}{2} \gcal\left(v_{123},0,1,\frac{1}{1-u_1}; 1\right)+\frac{1}{2} \gcal\left(v_{123},0,\frac{1}{1-u_1},1; 1\right)+\frac{1}{2} \gcal\left(v_{123},1,0,\frac{1}{1-u_1}; 1\right)-\frac{5}{4} \gcal\left(v_{123},1,1,\frac{1}{1-u_1}; 1\right)+\frac{1}{2} \gcal\left(v_{123},1,\frac{1}{1-u_1},0; 1\right)-\frac{5}{4} \gcal\left(v_{123},1,\frac{1}{1-u_1},1; 1\right)+\frac{1}{2} \gcal\left(v_{123},1,\frac{1}{1-u_1},\frac{1}{1-u_1}; 1\right)+\frac{1}{2} \gcal\left(v_{123},\frac{1}{1-u_1},0,1; 1\right)+\frac{1}{2} \gcal\left(v_{123},\frac{1}{1-u_1},1,0; 1\right)-\frac{5}{4} \gcal\left(v_{123},\frac{1}{1-u_1},1,1; 1\right)+\frac{1}{2} \gcal\left(v_{123},\frac{1}{1-u_1},1,\frac{1}{1-u_1}; 1\right)+\frac{1}{2} \gcal\left(v_{123},\frac{1}{1-u_1},\frac{1}{1-u_1},1; 1\right)-\frac{1}{4} \gcal\left(v_{132},1,1,\frac{1}{1-u_1}; 1\right)-\frac{1}{4} \gcal\left(v_{132},1,\frac{1}{1-u_1},1; 1\right)-\frac{1}{4} \gcal\left(v_{132},\frac{1}{1-u_1},1,1; 1\right)-\frac{1}{4} \gcal\left(v_{213},1,1,\frac{1}{1-u_2}; 1\right)-\frac{1}{4} \gcal\left(v_{213},1,\frac{1}{1-u_2},1; 1\right)-\frac{1}{4} \gcal\left(v_{213},\frac{1}{1-u_2},1,1; 1\right)+\frac{1}{2} \gcal\left(v_{231},0,1,\frac{1}{1-u_2}; 1\right)+\frac{1}{2} \gcal\left(v_{231},0,\frac{1}{1-u_2},1; 1\right)+\frac{1}{2} \gcal\left(v_{231},1,0,\frac{1}{1-u_2}; 1\right)-\frac{5}{4} \gcal\left(v_{231},1,1,\frac{1}{1-u_2}; 1\right)+\frac{1}{2} \gcal\left(v_{231},1,\frac{1}{1-u_2},0; 1\right)-\frac{5}{4} \gcal\left(v_{231},1,\frac{1}{1-u_2},1; 1\right)+\frac{1}{2} \gcal\left(v_{231},1,\frac{1}{1-u_2},\frac{1}{1-u_2}; 1\right)+\frac{1}{2} \gcal\left(v_{231},\frac{1}{1-u_2},0,1; 1\right)+\frac{1}{2} \gcal\left(v_{231},\frac{1}{1-u_2},1,0; 1\right)-\frac{5}{4} \gcal\left(v_{231},\frac{1}{1-u_2},1,1; 1\right)+\frac{1}{2} \gcal\left(v_{231},\frac{1}{1-u_2},1,\frac{1}{1-u_2}; 1\right)+\frac{1}{2} \gcal\left(v_{231},\frac{1}{1-u_2},\frac{1}{1-u_2},1; 1\right)+\frac{1}{2} \gcal\left(v_{312},0,1,\frac{1}{1-u_3}; 1\right)+\frac{1}{2} \gcal\left(v_{312},0,\frac{1}{1-u_3},1; 1\right)+\frac{1}{2} \gcal\left(v_{312},1,0,\frac{1}{1-u_3}; 1\right)-\frac{5}{4} \gcal\left(v_{312},1,1,\frac{1}{1-u_3}; 1\right)+\frac{1}{2} \gcal\left(v_{312},1,\frac{1}{1-u_3},0; 1\right)-\frac{5}{4} \gcal\left(v_{312},1,\frac{1}{1-u_3},1; 1\right)+\frac{1}{2} \gcal\left(v_{312},1,\frac{1}{1-u_3},\frac{1}{1-u_3}; 1\right)+\frac{1}{2} \gcal\left(v_{312},\frac{1}{1-u_3},0,1; 1\right)+\frac{1}{2} \gcal\left(v_{312},\frac{1}{1-u_3},1,0; 1\right)-\frac{5}{4} \gcal\left(v_{312},\frac{1}{1-u_3},1,1; 1\right)+\frac{1}{2} \gcal\left(v_{312},\frac{1}{1-u_3},1,\frac{1}{1-u_3}; 1\right)+\frac{1}{2} \gcal\left(v_{312},\frac{1}{1-u_3},\frac{1}{1-u_3},1; 1\right)-\frac{1}{4} \gcal\left(v_{321},1,1,\frac{1}{1-u_3}; 1\right)-\frac{1}{4} \gcal\left(v_{321},1,\frac{1}{1-u_3},1; 1\right)-\frac{1}{4} \gcal\left(v_{321},\frac{1}{1-u_3},1,1; 1\right)-\frac{3}{4} G\left(0,\frac{1}{u_1},\frac{1}{u_1+u_2}; 1\right) H\left(0; u_1\right)-\frac{3}{4} G\left(0,\frac{1}{u_1},\frac{1}{u_1+u_3}; 1\right) H\left(0; u_1\right)-\frac{1}{4} G\left(0,\frac{1}{u_2},\frac{1}{u_1+u_2}; 1\right) H\left(0; u_1\right)-\frac{1}{4} G\left(0,\frac{1}{u_3},\frac{1}{u_1+u_3}; 1\right) H\left(0; u_1\right)-\frac{1}{4} G\left(0,\frac{u_1-1}{u_1+u_3-1},\frac{1}{1-u_3}; 1\right) H\left(0; u_1\right)+\frac{1}{4} G\left(0,\frac{u_3-1}{u_2+u_3-1},\frac{1}{1-u_2}; 1\right) H\left(0; u_1\right)-\frac{3}{4} G\left(\frac{1}{u_1},0,\frac{1}{u_1+u_2}; 1\right) H\left(0; u_1\right)-\frac{3}{4} G\left(\frac{1}{u_1},0,\frac{1}{u_1+u_3}; 1\right) H\left(0; u_1\right)+\frac{1}{2} G\left(\frac{1}{u_1},\frac{1}{u_1},\frac{1}{u_1+u_2}; 1\right) H\left(0; u_1\right)+\frac{1}{2} G\left(\frac{1}{u_1},\frac{1}{u_1},\frac{1}{u_1+u_3}; 1\right) H\left(0; u_1\right)+\frac{1}{4} G\left(\frac{1}{u_1},\frac{1}{u_2},\frac{1}{u_1+u_2}; 1\right) H\left(0; u_1\right)+\frac{1}{4} G\left(\frac{1}{u_1},\frac{1}{u_3},\frac{1}{u_1+u_3}; 1\right) H\left(0; u_1\right)-\frac{1}{4} G\left(\frac{1}{1-u_2},1,\frac{1}{u_1}; 1\right) H\left(0; u_1\right)+\frac{1}{4} G\left(\frac{1}{1-u_2},\frac{u_3-1}{u_2+u_3-1},1; 1\right) H\left(0; u_1\right)-\frac{1}{4} G\left(\frac{1}{1-u_2},\frac{u_3-1}{u_2+u_3-1},\frac{1}{1-u_2}; 1\right) H\left(0; u_1\right)+\frac{1}{2} G\left(\frac{1}{u_2},0,\frac{1}{u_1}; 1\right) H\left(0; u_1\right)-\frac{1}{4} G\left(\frac{1}{u_2},0,\frac{1}{u_1+u_2}; 1\right) H\left(0; u_1\right)+\frac{1}{4} G\left(\frac{1}{u_2},\frac{1}{u_1},\frac{1}{u_1+u_2}; 1\right) H\left(0; u_1\right)+\frac{1}{4} G\left(\frac{1}{1-u_3},\frac{u_1-1}{u_1+u_3-1},0; 1\right) H\left(0; u_1\right)+\frac{1}{4} G\left(\frac{1}{1-u_3},\frac{u_1-1}{u_1+u_3-1},\frac{1}{1-u_3}; 1\right) H\left(0; u_1\right)-\frac{1}{4} G\left(\frac{1}{1-u_3},\frac{u_1-1}{u_1+u_3-1},\frac{u_1-1}{u_1+u_3-1}; 1\right) H\left(0; u_1\right)+\frac{1}{2} G\left(\frac{1}{u_3},0,\frac{1}{u_1}; 1\right) H\left(0; u_1\right)-\frac{1}{4} G\left(\frac{1}{u_3},0,\frac{1}{u_1+u_3}; 1\right) H\left(0; u_1\right)+\frac{1}{4} G\left(\frac{1}{u_3},\frac{1}{u_1},\frac{1}{u_1+u_3}; 1\right) H\left(0; u_1\right)+\frac{1}{4} \gcal\left(0,\frac{1}{1-u_1},v_{123}; 1\right) H\left(0; u_1\right)+\frac{1}{4} \gcal\left(0,\frac{1}{1-u_1},v_{132}; 1\right) H\left(0; u_1\right)+\frac{1}{4} \gcal\left(0,\frac{1}{1-u_2},v_{213}; 1\right) H\left(0; u_1\right)-\frac{1}{4} \gcal\left(0,\frac{1}{1-u_2},v_{231}; 1\right) H\left(0; u_1\right)+\frac{1}{4} \gcal\left(0,\frac{1}{1-u_3},v_{312}; 1\right) H\left(0; u_1\right)-\frac{1}{4} \gcal\left(0,\frac{1}{1-u_3},v_{321}; 1\right) H\left(0; u_1\right)
-\frac{1}{4} \gcal\left(0,u_{231},\frac{1}{u_1}; 1\right) H\left(0; u_1\right)-\frac{1}{4} \gcal\left(0,u_{231},\frac{1}{1-u_2}; 1\right) H\left(0; u_1\right)+\frac{1}{4} \gcal\left(0,u_{312},\frac{1}{1-u_3}; 1\right) H\left(0; u_1\right)-\frac{1}{4} \gcal\left(0,u_{312},\frac{u_1-1}{u_1+u_3-1}; 1\right) H\left(0; u_1\right)+\frac{1}{4} \gcal\left(0,v_{123},\frac{1}{1-u_1}; 1\right) H\left(0; u_1\right)+\frac{1}{4} \gcal\left(0,v_{132},\frac{1}{1-u_1}; 1\right) H\left(0; u_1\right)-\frac{1}{2} \gcal\left(0,v_{231},\frac{1}{1-u_2}; 1\right) H\left(0; u_1\right)+\frac{1}{2} \gcal\left(0,v_{312},\frac{1}{1-u_3}; 1\right) H\left(0; u_1\right)+\frac{1}{4} \gcal\left(\frac{1}{1-u_1},0,v_{123}; 1\right) H\left(0; u_1\right)+\frac{1}{4} \gcal\left(\frac{1}{1-u_1},0,v_{132}; 1\right) H\left(0; u_1\right)+\frac{1}{2} \gcal\left(\frac{1}{1-u_1},\frac{1}{1-u_1},v_{123}; 1\right) H\left(0; u_1\right)+\frac{1}{2} \gcal\left(\frac{1}{1-u_1},\frac{1}{1-u_1},v_{132}; 1\right) H\left(0; u_1\right)+\frac{1}{4} \gcal\left(\frac{1}{1-u_1},v_{123},1; 1\right) H\left(0; u_1\right)+\frac{1}{4} \gcal\left(\frac{1}{1-u_1},v_{123},\frac{1}{1-u_1}; 1\right) H\left(0; u_1\right)+\frac{1}{4} \gcal\left(\frac{1}{1-u_1},v_{132},1; 1\right) H\left(0; u_1\right)+\frac{1}{4} \gcal\left(\frac{1}{1-u_1},v_{132},\frac{1}{1-u_1}; 1\right) H\left(0; u_1\right)+\frac{1}{4} \gcal\left(\frac{1}{1-u_2},0,v_{213}; 1\right) H\left(0; u_1\right)-\frac{1}{4} \gcal\left(\frac{1}{1-u_2},0,v_{231}; 1\right) H\left(0; u_1\right)+\frac{1}{2} \gcal\left(\frac{1}{1-u_2},\frac{1}{1-u_2},v_{213}; 1\right) H\left(0; u_1\right)-\frac{1}{2} \gcal\left(\frac{1}{1-u_2},\frac{1}{1-u_2},v_{231}; 1\right) H\left(0; u_1\right)-\frac{1}{4} \gcal\left(\frac{1}{1-u_2},u_{231},1; 1\right) H\left(0; u_1\right)+\frac{1}{4} \gcal\left(\frac{1}{1-u_2},u_{231},\frac{1}{u_1}; 1\right) H\left(0; u_1\right)+\frac{1}{4} \gcal\left(\frac{1}{1-u_2},u_{231},\frac{1}{1-u_2}; 1\right) H\left(0; u_1\right)+\frac{1}{4} \gcal\left(\frac{1}{1-u_2},v_{213},0; 1\right) H\left(0; u_1\right)+\frac{1}{2} \gcal\left(\frac{1}{1-u_2},v_{213},\frac{1}{1-u_2}; 1\right) H\left(0; u_1\right)-\frac{1}{4} \gcal\left(\frac{1}{1-u_2},v_{231},0; 1\right) H\left(0; u_1\right)-\frac{1}{2} \gcal\left(\frac{1}{1-u_2},v_{231},\frac{1}{1-u_2}; 1\right) H\left(0; u_1\right)+\frac{1}{4} \gcal\left(\frac{1}{1-u_3},0,v_{312}; 1\right) H\left(0; u_1\right)-\frac{1}{4} \gcal\left(\frac{1}{1-u_3},0,v_{321}; 1\right) H\left(0; u_1\right)+\frac{1}{2} \gcal\left(\frac{1}{1-u_3},\frac{1}{1-u_3},v_{312}; 1\right) H\left(0; u_1\right)-\frac{1}{2} \gcal\left(\frac{1}{1-u_3},\frac{1}{1-u_3},v_{321}; 1\right) H\left(0; u_1\right)-\frac{1}{4} \gcal\left(\frac{1}{1-u_3},u_{312},0; 1\right) H\left(0; u_1\right)-\frac{1}{4} \gcal\left(\frac{1}{1-u_3},u_{312},\frac{1}{1-u_3}; 1\right) H\left(0; u_1\right)+\frac{1}{4} \gcal\left(\frac{1}{1-u_3},u_{312},\frac{u_1-1}{u_1+u_3-1}; 1\right) H\left(0; u_1\right)+\frac{1}{4} \gcal\left(\frac{1}{1-u_3},v_{312},0; 1\right) H\left(0; u_1\right)+\frac{1}{2} \gcal\left(\frac{1}{1-u_3},v_{312},\frac{1}{1-u_3}; 1\right) H\left(0; u_1\right)-\frac{1}{4} \gcal\left(\frac{1}{1-u_3},v_{321},0; 1\right) H\left(0; u_1\right)-\frac{1}{2} \gcal\left(\frac{1}{1-u_3},v_{321},\frac{1}{1-u_3}; 1\right) H\left(0; u_1\right)+\frac{1}{4} \gcal\left(v_{123},1,\frac{1}{1-u_1}; 1\right) H\left(0; u_1\right)+\frac{1}{4} \gcal\left(v_{123},\frac{1}{1-u_1},1; 1\right) H\left(0; u_1\right)+\frac{1}{4} \gcal\left(v_{132},1,\frac{1}{1-u_1}; 1\right) H\left(0; u_1\right)+\frac{1}{4} \gcal\left(v_{132},\frac{1}{1-u_1},1; 1\right) H\left(0; u_1\right)+\frac{1}{4} \gcal\left(v_{213},1,\frac{1}{1-u_2}; 1\right) H\left(0; u_1\right)+\frac{1}{4} \gcal\left(v_{213},\frac{1}{1-u_2},1; 1\right) H\left(0; u_1\right)-\frac{3}{4} \gcal\left(v_{231},1,\frac{1}{1-u_2}; 1\right) H\left(0; u_1\right)-\frac{3}{4} \gcal\left(v_{231},\frac{1}{1-u_2},1; 1\right) H\left(0; u_1\right)+\frac{3}{4} \gcal\left(v_{312},1,\frac{1}{1-u_3}; 1\right) H\left(0; u_1\right)+\frac{3}{4} \gcal\left(v_{312},\frac{1}{1-u_3},1; 1\right) H\left(0; u_1\right)-\frac{1}{4} \gcal\left(v_{321},1,\frac{1}{1-u_3}; 1\right) H\left(0; u_1\right)-\frac{1}{4} \gcal\left(v_{321},\frac{1}{1-u_3},1; 1\right) H\left(0; u_1\right)-\frac{1}{4} G\left(0,\frac{1}{u_1},\frac{1}{u_1+u_2}; 1\right) H\left(0; u_2\right)-\frac{3}{4} G\left(0,\frac{1}{u_2},\frac{1}{u_1+u_2}; 1\right) H\left(0; u_2\right)-\frac{3}{4} G\left(0,\frac{1}{u_2},\frac{1}{u_2+u_3}; 1\right) H\left(0; u_2\right)-\frac{1}{4} G\left(0,\frac{u_2-1}{u_1+u_2-1},\frac{1}{1-u_1}; 1\right) H\left(0; u_2\right)-\frac{1}{4} G\left(0,\frac{1}{u_3},\frac{1}{u_2+u_3}; 1\right) H\left(0; u_2\right)+\frac{1}{4} G\left(0,\frac{u_1-1}{u_1+u_3-1},\frac{1}{1-u_3}; 1\right) H\left(0; u_2\right)+\frac{1}{4} G\left(\frac{1}{1-u_1},\frac{u_2-1}{u_1+u_2-1},0; 1\right) H\left(0; u_2\right)+\frac{1}{4} G\left(\frac{1}{1-u_1},\frac{u_2-1}{u_1+u_2-1},\frac{1}{1-u_1}; 1\right) H\left(0; u_2\right)-\frac{1}{4} G\left(\frac{1}{1-u_1},\frac{u_2-1}{u_1+u_2-1},\frac{u_2-1}{u_1+u_2-1}; 1\right) H\left(0; u_2\right)+\frac{1}{2} G\left(\frac{1}{u_1},0,\frac{1}{u_2}; 1\right) H\left(0; u_2\right)-\frac{1}{4} G\left(\frac{1}{u_1},0,\frac{1}{u_1+u_2}; 1\right) H\left(0; u_2\right)+\frac{1}{4} G\left(\frac{1}{u_1},\frac{1}{u_2},\frac{1}{u_1+u_2}; 1\right) H\left(0; u_2\right)-\frac{3}{4} G\left(\frac{1}{u_2},0,\frac{1}{u_1+u_2}; 1\right) H\left(0; u_2\right)-\frac{3}{4} G\left(\frac{1}{u_2},0,\frac{1}{u_2+u_3}; 1\right) H\left(0; u_2\right)+\frac{1}{4} G\left(\frac{1}{u_2},\frac{1}{u_1},\frac{1}{u_1+u_2}; 1\right) H\left(0; u_2\right)+\frac{1}{2} G\left(\frac{1}{u_2},\frac{1}{u_2},\frac{1}{u_1+u_2}; 1\right) H\left(0; u_2\right)+\frac{1}{2} G\left(\frac{1}{u_2},\frac{1}{u_2},\frac{1}{u_2+u_3}; 1\right) H\left(0; u_2\right)+\frac{1}{4} G\left(\frac{1}{u_2},\frac{1}{u_3},\frac{1}{u_2+u_3}; 1\right) H\left(0; u_2\right)-\frac{1}{4} G\left(\frac{1}{1-u_3},1,\frac{1}{u_2}; 1\right) H\left(0; u_2\right)+\frac{1}{4} G\left(\frac{1}{1-u_3},\frac{u_1-1}{u_1+u_3-1},1; 1\right) H\left(0; u_2\right)-\frac{1}{4} G\left(\frac{1}{1-u_3},\frac{u_1-1}{u_1+u_3-1},\frac{1}{1-u_3}; 1\right) H\left(0; u_2\right)+\frac{1}{2} G\left(\frac{1}{u_3},0,\frac{1}{u_2}; 1\right) H\left(0; u_2\right)-\frac{1}{4} G\left(\frac{1}{u_3},0,\frac{1}{u_2+u_3}; 1\right) H\left(0; u_2\right)+\frac{1}{4} G\left(\frac{1}{u_3},\frac{1}{u_2},\frac{1}{u_2+u_3}; 1\right) H\left(0; u_2\right)+\frac{1}{4} \gcal\left(0,\frac{1}{1-u_1},v_{123}; 1\right) H\left(0; u_2\right)-\frac{1}{4} \gcal\left(0,\frac{1}{1-u_1},v_{132}; 1\right) H\left(0; u_2\right)+\frac{1}{4} \gcal\left(0,\frac{1}{1-u_2},v_{213}; 1\right) H\left(0; u_2\right)+\frac{1}{4} \gcal\left(0,\frac{1}{1-u_2},v_{231}; 1\right) H\left(0; u_2\right)-\frac{1}{4} \gcal\left(0,\frac{1}{1-u_3},v_{312}; 1\right) H\left(0; u_2\right)+\frac{1}{4} \gcal\left(0,\frac{1}{1-u_3},v_{321}; 1\right) H\left(0; u_2\right)+\frac{1}{4} \gcal\left(0,u_{123},\frac{1}{1-u_1}; 1\right) H\left(0; u_2\right)-\frac{1}{4} \gcal\left(0,u_{123},\frac{u_2-1}{u_1+u_2-1}; 1\right) H\left(0; u_2\right)-\frac{1}{4} \gcal\left(0,u_{312},\frac{1}{u_2}; 1\right) H\left(0; u_2\right)-\frac{1}{4} \gcal\left(0,u_{312},\frac{1}{1-u_3}; 1\right) H\left(0; u_2\right)+\frac{1}{2} \gcal\left(0,v_{123},\frac{1}{1-u_1}; 1\right) H\left(0; u_2\right)+\frac{1}{4} \gcal\left(0,v_{213},\frac{1}{1-u_2}; 1\right) H\left(0; u_2\right)+\frac{1}{4} \gcal\left(0,v_{231},\frac{1}{1-u_2}; 1\right) H\left(0; u_2\right)-\frac{1}{2} \gcal\left(0,v_{312},\frac{1}{1-u_3}; 1\right) H\left(0; u_2\right)+\frac{1}{4} \gcal\left(\frac{1}{1-u_1},0,v_{123}; 1\right) H\left(0; u_2\right)-\frac{1}{4} \gcal\left(\frac{1}{1-u_1},0,v_{132}; 1\right) H\left(0; u_2\right)+\frac{1}{2} \gcal\left(\frac{1}{1-u_1},\frac{1}{1-u_1},v_{123}; 1\right) H\left(0; u_2\right)-\frac{1}{2} \gcal\left(\frac{1}{1-u_1},\frac{1}{1-u_1},v_{132}; 1\right) H\left(0; u_2\right)-\frac{1}{4} \gcal\left(\frac{1}{1-u_1},u_{123},0; 1\right) H\left(0; u_2\right)-\frac{1}{4} \gcal\left(\frac{1}{1-u_1},u_{123},\frac{1}{1-u_1}; 1\right) H\left(0; u_2\right)+\frac{1}{4} \gcal\left(\frac{1}{1-u_1},u_{123},\frac{u_2-1}{u_1+u_2-1}; 1\right) H\left(0; u_2\right)+\frac{1}{4} \gcal\left(\frac{1}{1-u_1},v_{123},0; 1\right) H\left(0; u_2\right)+\frac{1}{2} \gcal\left(\frac{1}{1-u_1},v_{123},\frac{1}{1-u_1}; 1\right) H\left(0; u_2\right)-\frac{1}{4} \gcal\left(\frac{1}{1-u_1},v_{132},0; 1\right) H\left(0; u_2\right)-\frac{1}{2} \gcal\left(\frac{1}{1-u_1},v_{132},\frac{1}{1-u_1}; 1\right) H\left(0; u_2\right)+\frac{1}{4} \gcal\left(\frac{1}{1-u_2},0,v_{213}; 1\right) H\left(0; u_2\right)
+\frac{1}{4} \gcal\left(\frac{1}{1-u_2},0,v_{231}; 1\right) H\left(0; u_2\right)+\frac{1}{2} \gcal\left(\frac{1}{1-u_2},\frac{1}{1-u_2},v_{213}; 1\right) H\left(0; u_2\right)+\frac{1}{2} \gcal\left(\frac{1}{1-u_2},\frac{1}{1-u_2},v_{231}; 1\right) H\left(0; u_2\right)+\frac{1}{4} \gcal\left(\frac{1}{1-u_2},v_{213},1; 1\right) H\left(0; u_2\right)+\frac{1}{4} \gcal\left(\frac{1}{1-u_2},v_{213},\frac{1}{1-u_2}; 1\right) H\left(0; u_2\right)+\frac{1}{4} \gcal\left(\frac{1}{1-u_2},v_{231},1; 1\right) H\left(0; u_2\right)+\frac{1}{4} \gcal\left(\frac{1}{1-u_2},v_{231},\frac{1}{1-u_2}; 1\right) H\left(0; u_2\right)-\frac{1}{4} \gcal\left(\frac{1}{1-u_3},0,v_{312}; 1\right) H\left(0; u_2\right)+\frac{1}{4} \gcal\left(\frac{1}{1-u_3},0,v_{321}; 1\right) H\left(0; u_2\right)-\frac{1}{2} \gcal\left(\frac{1}{1-u_3},\frac{1}{1-u_3},v_{312}; 1\right) H\left(0; u_2\right)+\frac{1}{2} \gcal\left(\frac{1}{1-u_3},\frac{1}{1-u_3},v_{321}; 1\right) H\left(0; u_2\right)-\frac{1}{4} \gcal\left(\frac{1}{1-u_3},u_{312},1; 1\right) H\left(0; u_2\right)+\frac{1}{4} \gcal\left(\frac{1}{1-u_3},u_{312},\frac{1}{u_2}; 1\right) H\left(0; u_2\right)+\frac{1}{4} \gcal\left(\frac{1}{1-u_3},u_{312},\frac{1}{1-u_3}; 1\right) H\left(0; u_2\right)-\frac{1}{4} \gcal\left(\frac{1}{1-u_3},v_{312},0; 1\right) H\left(0; u_2\right)-\frac{1}{2} \gcal\left(\frac{1}{1-u_3},v_{312},\frac{1}{1-u_3}; 1\right) H\left(0; u_2\right)+\frac{1}{4} \gcal\left(\frac{1}{1-u_3},v_{321},0; 1\right) H\left(0; u_2\right)+\frac{1}{2} \gcal\left(\frac{1}{1-u_3},v_{321},\frac{1}{1-u_3}; 1\right) H\left(0; u_2\right)+\frac{3}{4} \gcal\left(v_{123},1,\frac{1}{1-u_1}; 1\right) H\left(0; u_2\right)+\frac{3}{4} \gcal\left(v_{123},\frac{1}{1-u_1},1; 1\right) H\left(0; u_2\right)-\frac{1}{4} \gcal\left(v_{132},1,\frac{1}{1-u_1}; 1\right) H\left(0; u_2\right)-\frac{1}{4} \gcal\left(v_{132},\frac{1}{1-u_1},1; 1\right) H\left(0; u_2\right)+\frac{1}{4} \gcal\left(v_{213},1,\frac{1}{1-u_2}; 1\right) H\left(0; u_2\right)+\frac{1}{4} \gcal\left(v_{213},\frac{1}{1-u_2},1; 1\right) H\left(0; u_2\right)+\frac{1}{4} \gcal\left(v_{231},1,\frac{1}{1-u_2}; 1\right) H\left(0; u_2\right)+\frac{1}{4} \gcal\left(v_{231},\frac{1}{1-u_2},1; 1\right) H\left(0; u_2\right)-\frac{3}{4} \gcal\left(v_{312},1,\frac{1}{1-u_3}; 1\right) H\left(0; u_2\right)-\frac{3}{4} \gcal\left(v_{312},\frac{1}{1-u_3},1; 1\right) H\left(0; u_2\right)+\frac{1}{4} \gcal\left(v_{321},1,\frac{1}{1-u_3}; 1\right) H\left(0; u_2\right)+\frac{1}{4} \gcal\left(v_{321},\frac{1}{1-u_3},1; 1\right) H\left(0; u_2\right)+\frac{1}{4} G\left(\frac{1}{u_1},\frac{1}{u_1+u_2}; 1\right) H\left(0; u_1\right) H\left(0; u_2\right)+\frac{1}{4} G\left(\frac{1}{u_2},\frac{1}{u_1+u_2}; 1\right) H\left(0; u_1\right) H\left(0; u_2\right)+\frac{1}{4} G\left(\frac{1}{1-u_3},\frac{u_1-1}{u_1+u_3-1}; 1\right) H\left(0; u_1\right) H\left(0; u_2\right)-\frac{1}{4} \gcal\left(\frac{1}{1-u_3},u_{312}; 1\right) H\left(0; u_1\right) H\left(0; u_2\right)-\frac{1}{4} \gcal\left(\frac{1}{1-u_3},v_{312}; 1\right) H\left(0; u_1\right) H\left(0; u_2\right)-\frac{1}{4} \gcal\left(\frac{1}{1-u_3},v_{321}; 1\right) H\left(0; u_1\right) H\left(0; u_2\right)+\frac{5}{24} \pi ^2 H\left(0; u_1\right) H\left(0; u_2\right)-\frac{1}{4} G\left(0,\frac{1}{u_1},\frac{1}{u_1+u_3}; 1\right) H\left(0; u_3\right)-\frac{1}{4} G\left(0,\frac{1}{u_2},\frac{1}{u_2+u_3}; 1\right) H\left(0; u_3\right)+\frac{1}{4} G\left(0,\frac{u_2-1}{u_1+u_2-1},\frac{1}{1-u_1}; 1\right) H\left(0; u_3\right)-\frac{3}{4} G\left(0,\frac{1}{u_3},\frac{1}{u_1+u_3}; 1\right) H\left(0; u_3\right)-\frac{3}{4} G\left(0,\frac{1}{u_3},\frac{1}{u_2+u_3}; 1\right) H\left(0; u_3\right)-\frac{1}{4} G\left(0,\frac{u_3-1}{u_2+u_3-1},\frac{1}{1-u_2}; 1\right) H\left(0; u_3\right)-\frac{1}{4} G\left(\frac{1}{1-u_1},1,\frac{1}{u_3}; 1\right) H\left(0; u_3\right)+\frac{1}{4} G\left(\frac{1}{1-u_1},\frac{u_2-1}{u_1+u_2-1},1; 1\right) H\left(0; u_3\right)-\frac{1}{4} G\left(\frac{1}{1-u_1},\frac{u_2-1}{u_1+u_2-1},\frac{1}{1-u_1}; 1\right) H\left(0; u_3\right)+\frac{1}{2} G\left(\frac{1}{u_1},0,\frac{1}{u_3}; 1\right) H\left(0; u_3\right)-\frac{1}{4} G\left(\frac{1}{u_1},0,\frac{1}{u_1+u_3}; 1\right) H\left(0; u_3\right)+\frac{1}{4} G\left(\frac{1}{u_1},\frac{1}{u_3},\frac{1}{u_1+u_3}; 1\right) H\left(0; u_3\right)+\frac{1}{4} G\left(\frac{1}{1-u_2},\frac{u_3-1}{u_2+u_3-1},0; 1\right) H\left(0; u_3\right)+\frac{1}{4} G\left(\frac{1}{1-u_2},\frac{u_3-1}{u_2+u_3-1},\frac{1}{1-u_2}; 1\right) H\left(0; u_3\right)-\frac{1}{4} G\left(\frac{1}{1-u_2},\frac{u_3-1}{u_2+u_3-1},\frac{u_3-1}{u_2+u_3-1}; 1\right) H\left(0; u_3\right)+\frac{1}{2} G\left(\frac{1}{u_2},0,\frac{1}{u_3}; 1\right) H\left(0; u_3\right)-\frac{1}{4} G\left(\frac{1}{u_2},0,\frac{1}{u_2+u_3}; 1\right) H\left(0; u_3\right)+\frac{1}{4} G\left(\frac{1}{u_2},\frac{1}{u_3},\frac{1}{u_2+u_3}; 1\right) H\left(0; u_3\right)-\frac{3}{4} G\left(\frac{1}{u_3},0,\frac{1}{u_1+u_3}; 1\right) H\left(0; u_3\right)-\frac{3}{4} G\left(\frac{1}{u_3},0,\frac{1}{u_2+u_3}; 1\right) H\left(0; u_3\right)+\frac{1}{4} G\left(\frac{1}{u_3},\frac{1}{u_1},\frac{1}{u_1+u_3}; 1\right) H\left(0; u_3\right)+\frac{1}{4} G\left(\frac{1}{u_3},\frac{1}{u_2},\frac{1}{u_2+u_3}; 1\right) H\left(0; u_3\right)+\frac{1}{2} G\left(\frac{1}{u_3},\frac{1}{u_3},\frac{1}{u_1+u_3}; 1\right) H\left(0; u_3\right)+\frac{1}{2} G\left(\frac{1}{u_3},\frac{1}{u_3},\frac{1}{u_2+u_3}; 1\right) H\left(0; u_3\right)-\frac{1}{4} \gcal\left(0,\frac{1}{1-u_1},v_{123}; 1\right) H\left(0; u_3\right)+\frac{1}{4} \gcal\left(0,\frac{1}{1-u_1},v_{132}; 1\right) H\left(0; u_3\right)-\frac{1}{4} \gcal\left(0,\frac{1}{1-u_2},v_{213}; 1\right) H\left(0; u_3\right)+\frac{1}{4} \gcal\left(0,\frac{1}{1-u_2},v_{231}; 1\right) H\left(0; u_3\right)+\frac{1}{4} \gcal\left(0,\frac{1}{1-u_3},v_{312}; 1\right) H\left(0; u_3\right)+\frac{1}{4} \gcal\left(0,\frac{1}{1-u_3},v_{321}; 1\right) H\left(0; u_3\right)-\frac{1}{4} \gcal\left(0,u_{123},\frac{1}{1-u_1}; 1\right) H\left(0; u_3\right)-\frac{1}{4} \gcal\left(0,u_{123},\frac{1}{u_3}; 1\right) H\left(0; u_3\right)+\frac{1}{4} \gcal\left(0,u_{231},\frac{1}{1-u_2}; 1\right) H\left(0; u_3\right)-\frac{1}{4} \gcal\left(0,u_{231},\frac{u_3-1}{u_2+u_3-1}; 1\right) H\left(0; u_3\right)-\frac{1}{2} \gcal\left(0,v_{123},\frac{1}{1-u_1}; 1\right) H\left(0; u_3\right)+\frac{1}{2} \gcal\left(0,v_{231},\frac{1}{1-u_2}; 1\right) H\left(0; u_3\right)+\frac{1}{4} \gcal\left(0,v_{312},\frac{1}{1-u_3}; 1\right) H\left(0; u_3\right)+\frac{1}{4} \gcal\left(0,v_{321},\frac{1}{1-u_3}; 1\right) H\left(0; u_3\right)-\frac{1}{4} \gcal\left(\frac{1}{1-u_1},0,v_{123}; 1\right) H\left(0; u_3\right)+\frac{1}{4} \gcal\left(\frac{1}{1-u_1},0,v_{132}; 1\right) H\left(0; u_3\right)-\frac{1}{2} \gcal\left(\frac{1}{1-u_1},\frac{1}{1-u_1},v_{123}; 1\right) H\left(0; u_3\right)+\frac{1}{2} \gcal\left(\frac{1}{1-u_1},\frac{1}{1-u_1},v_{132}; 1\right) H\left(0; u_3\right)-\frac{1}{4} \gcal\left(\frac{1}{1-u_1},u_{123},1; 1\right) H\left(0; u_3\right)+\frac{1}{4} \gcal\left(\frac{1}{1-u_1},u_{123},\frac{1}{1-u_1}; 1\right) H\left(0; u_3\right)+\frac{1}{4} \gcal\left(\frac{1}{1-u_1},u_{123},\frac{1}{u_3}; 1\right) H\left(0; u_3\right)-\frac{1}{4} \gcal\left(\frac{1}{1-u_1},v_{123},0; 1\right) H\left(0; u_3\right)-\frac{1}{2} \gcal\left(\frac{1}{1-u_1},v_{123},\frac{1}{1-u_1}; 1\right) H\left(0; u_3\right)+\frac{1}{4} \gcal\left(\frac{1}{1-u_1},v_{132},0; 1\right) H\left(0; u_3\right)+\frac{1}{2} \gcal\left(\frac{1}{1-u_1},v_{132},\frac{1}{1-u_1}; 1\right) H\left(0; u_3\right)-\frac{1}{4} \gcal\left(\frac{1}{1-u_2},0,v_{213}; 1\right) H\left(0; u_3\right)+\frac{1}{4} \gcal\left(\frac{1}{1-u_2},0,v_{231}; 1\right) H\left(0; u_3\right)-\frac{1}{2} \gcal\left(\frac{1}{1-u_2},\frac{1}{1-u_2},v_{213}; 1\right) H\left(0; u_3\right)+\frac{1}{2} \gcal\left(\frac{1}{1-u_2},\frac{1}{1-u_2},v_{231}; 1\right) H\left(0; u_3\right)-\frac{1}{4} \gcal\left(\frac{1}{1-u_2},u_{231},0; 1\right) H\left(0; u_3\right)-\frac{1}{4} \gcal\left(\frac{1}{1-u_2},u_{231},\frac{1}{1-u_2}; 1\right) H\left(0; u_3\right)+\frac{1}{4} \gcal\left(\frac{1}{1-u_2},u_{231},\frac{u_3-1}{u_2+u_3-1}; 1\right) H\left(0; u_3\right)-\frac{1}{4} \gcal\left(\frac{1}{1-u_2},v_{213},0; 1\right) H\left(0; u_3\right)-\frac{1}{2} \gcal\left(\frac{1}{1-u_2},v_{213},\frac{1}{1-u_2}; 1\right) H\left(0; u_3\right)+\frac{1}{4} \gcal\left(\frac{1}{1-u_2},v_{231},0; 1\right) H\left(0; u_3\right)+\frac{1}{2} \gcal\left(\frac{1}{1-u_2},v_{231},\frac{1}{1-u_2}; 1\right) H\left(0; u_3\right)+\frac{1}{4} \gcal\left(\frac{1}{1-u_3},0,v_{312}; 1\right) H\left(0; u_3\right)+\frac{1}{4} \gcal\left(\frac{1}{1-u_3},0,v_{321}; 1\right) H\left(0; u_3\right)+\frac{1}{2} \gcal\left(\frac{1}{1-u_3},\frac{1}{1-u_3},v_{312}; 1\right) H\left(0; u_3\right)
+\frac{1}{2} \gcal\left(\frac{1}{1-u_3},\frac{1}{1-u_3},v_{321}; 1\right) H\left(0; u_3\right)+\frac{1}{4} \gcal\left(\frac{1}{1-u_3},v_{312},1; 1\right) H\left(0; u_3\right)+\frac{1}{4} \gcal\left(\frac{1}{1-u_3},v_{312},\frac{1}{1-u_3}; 1\right) H\left(0; u_3\right)+\frac{1}{4} \gcal\left(\frac{1}{1-u_3},v_{321},1; 1\right) H\left(0; u_3\right)+\frac{1}{4} \gcal\left(\frac{1}{1-u_3},v_{321},\frac{1}{1-u_3}; 1\right) H\left(0; u_3\right)-\frac{3}{4} \gcal\left(v_{123},1,\frac{1}{1-u_1}; 1\right) H\left(0; u_3\right)-\frac{3}{4} \gcal\left(v_{123},\frac{1}{1-u_1},1; 1\right) H\left(0; u_3\right)+\frac{1}{4} \gcal\left(v_{132},1,\frac{1}{1-u_1}; 1\right) H\left(0; u_3\right)+\frac{1}{4} \gcal\left(v_{132},\frac{1}{1-u_1},1; 1\right) H\left(0; u_3\right)-\frac{1}{4} \gcal\left(v_{213},1,\frac{1}{1-u_2}; 1\right) H\left(0; u_3\right)-\frac{1}{4} \gcal\left(v_{213},\frac{1}{1-u_2},1; 1\right) H\left(0; u_3\right)+\frac{3}{4} \gcal\left(v_{231},1,\frac{1}{1-u_2}; 1\right) H\left(0; u_3\right)+\frac{3}{4} \gcal\left(v_{231},\frac{1}{1-u_2},1; 1\right) H\left(0; u_3\right)+\frac{1}{4} \gcal\left(v_{312},1,\frac{1}{1-u_3}; 1\right) H\left(0; u_3\right)+\frac{1}{4} \gcal\left(v_{312},\frac{1}{1-u_3},1; 1\right) H\left(0; u_3\right)+\frac{1}{4} \gcal\left(v_{321},1,\frac{1}{1-u_3}; 1\right) H\left(0; u_3\right)+\frac{1}{4} \gcal\left(v_{321},\frac{1}{1-u_3},1; 1\right) H\left(0; u_3\right)+\frac{1}{4} G\left(\frac{1}{u_1},\frac{1}{u_1+u_3}; 1\right) H\left(0; u_1\right) H\left(0; u_3\right)+\frac{1}{4} G\left(\frac{1}{1-u_2},\frac{u_3-1}{u_2+u_3-1}; 1\right) H\left(0; u_1\right) H\left(0; u_3\right)+\frac{1}{4} G\left(\frac{1}{u_3},\frac{1}{u_1+u_3}; 1\right) H\left(0; u_1\right) H\left(0; u_3\right)-\frac{1}{4} \gcal\left(\frac{1}{1-u_2},u_{231}; 1\right) H\left(0; u_1\right) H\left(0; u_3\right)-\frac{1}{4} \gcal\left(\frac{1}{1-u_2},v_{213}; 1\right) H\left(0; u_1\right) H\left(0; u_3\right)-\frac{1}{4} \gcal\left(\frac{1}{1-u_2},v_{231}; 1\right) H\left(0; u_1\right) H\left(0; u_3\right)+\frac{5}{24} \pi ^2 H\left(0; u_1\right) H\left(0; u_3\right)+\frac{1}{4} G\left(\frac{1}{1-u_1},\frac{u_2-1}{u_1+u_2-1}; 1\right) H\left(0; u_2\right) H\left(0; u_3\right)+\frac{1}{4} G\left(\frac{1}{u_2},\frac{1}{u_2+u_3}; 1\right) H\left(0; u_2\right) H\left(0; u_3\right)+\frac{1}{4} G\left(\frac{1}{u_3},\frac{1}{u_2+u_3}; 1\right) H\left(0; u_2\right) H\left(0; u_3\right)-\frac{1}{4} \gcal\left(\frac{1}{1-u_1},u_{123}; 1\right) H\left(0; u_2\right) H\left(0; u_3\right)-\frac{1}{4} \gcal\left(\frac{1}{1-u_1},v_{123}; 1\right) H\left(0; u_2\right) H\left(0; u_3\right)-\frac{1}{4} \gcal\left(\frac{1}{1-u_1},v_{132}; 1\right) H\left(0; u_2\right) H\left(0; u_3\right)+\frac{5}{24} \pi ^2 H\left(0; u_2\right) H\left(0; u_3\right)+3 H\left(0; u_2\right) H\left(0,0; u_1\right) H\left(0; u_3\right)+3 H\left(0; u_1\right) H\left(0,0; u_2\right) H\left(0; u_3\right)+\frac{1}{4} H\left(0; u_2\right) H\left(0,1; \frac{u_1+u_2-1}{u_2-1}\right) H\left(0; u_3\right)+\frac{1}{2} H\left(0; u_1\right) H\left(0,1; \left(u_1+u_3\right)\right) H\left(0; u_3\right)+\frac{1}{4} H\left(0; u_1\right) H\left(0,1; \frac{u_2+u_3-1}{u_3-1}\right) H\left(0; u_3\right)+\frac{1}{2} H\left(0; u_2\right) H\left(0,1; \left(u_2+u_3\right)\right) H\left(0; u_3\right)+\frac{3}{4} H\left(0; u_2\right) H\left(1,0; u_1\right) H\left(0; u_3\right)+\frac{3}{4} H\left(0; u_1\right) H\left(1,0; u_2\right) H\left(0; u_3\right)+\frac{1}{4} \gcal\left(\frac{1}{1-u_2},v_{213}; 1\right) H\left(0,0; u_1\right)+\frac{1}{4} \gcal\left(\frac{1}{1-u_2},v_{231}; 1\right) H\left(0,0; u_1\right)+\frac{1}{4} \gcal\left(\frac{1}{1-u_3},v_{312}; 1\right) H\left(0,0; u_1\right)+\frac{1}{4} \gcal\left(\frac{1}{1-u_3},v_{321}; 1\right) H\left(0,0; u_1\right)-\frac{23}{24} \pi ^2 H\left(0,0; u_1\right)+\frac{1}{4} \gcal\left(\frac{1}{1-u_1},v_{123}; 1\right) H\left(0,0; u_2\right)+\frac{1}{4} \gcal\left(\frac{1}{1-u_1},v_{132}; 1\right) H\left(0,0; u_2\right)+\frac{1}{4} \gcal\left(\frac{1}{1-u_3},v_{312}; 1\right) H\left(0,0; u_2\right)+\frac{1}{4} \gcal\left(\frac{1}{1-u_3},v_{321}; 1\right) H\left(0,0; u_2\right)-\frac{25}{4} H\left(0,0; u_1\right) H\left(0,0; u_2\right)-\frac{23}{24} \pi ^2 H\left(0,0; u_2\right)+\frac{1}{4} \gcal\left(\frac{1}{1-u_1},v_{123}; 1\right) H\left(0,0; u_3\right)+\frac{1}{4} \gcal\left(\frac{1}{1-u_1},v_{132}; 1\right) H\left(0,0; u_3\right)+\frac{1}{4} \gcal\left(\frac{1}{1-u_2},v_{213}; 1\right) H\left(0,0; u_3\right)+\frac{1}{4} \gcal\left(\frac{1}{1-u_2},v_{231}; 1\right) H\left(0,0; u_3\right)+3 H\left(0; u_1\right) H\left(0; u_2\right) H\left(0,0; u_3\right)-\frac{25}{4} H\left(0,0; u_1\right) H\left(0,0; u_3\right)-\frac{25}{4} H\left(0,0; u_2\right) H\left(0,0; u_3\right)-\frac{23}{24} \pi ^2 H\left(0,0; u_3\right)+\frac{1}{12} \pi ^2 H\left(0,1; u_1\right)+\frac{1}{12} \pi ^2 H\left(0,1; u_2\right)-\frac{1}{24} \pi ^2 H\left(0,1; \frac{u_1+u_2-1}{u_2-1}\right)+\frac{1}{2} H\left(0; u_1\right) H\left(0; u_2\right) H\left(0,1; \left(u_1+u_2\right)\right)+\frac{1}{12} \pi ^2 H\left(0,1; \left(u_1+u_2\right)\right)+\frac{1}{12} \pi ^2 H\left(0,1; u_3\right)+\frac{1}{4} H\left(0; u_1\right) H\left(0; u_2\right) H\left(0,1; \frac{u_1+u_3-1}{u_1-1}\right)-\frac{1}{24} \pi ^2 H\left(0,1; \frac{u_1+u_3-1}{u_1-1}\right)+\frac{1}{12} \pi ^2 H\left(0,1; \left(u_1+u_3\right)\right)-\frac{1}{24} \pi ^2 H\left(0,1; \frac{u_2+u_3-1}{u_3-1}\right)+\frac{1}{12} \pi ^2 H\left(0,1; \left(u_2+u_3\right)\right)-\frac{1}{2} G\left(0,\frac{1}{u_1+u_2}; 1\right) H\left(1,0; u_1\right)-\frac{1}{2} G\left(0,\frac{1}{u_1+u_3}; 1\right) H\left(1,0; u_1\right)+\frac{1}{4} G\left(\frac{1}{u_1},\frac{1}{u_1+u_2}; 1\right) H\left(1,0; u_1\right)+\frac{1}{4} G\left(\frac{1}{u_1},\frac{1}{u_1+u_3}; 1\right) H\left(1,0; u_1\right)+\frac{1}{4} G\left(\frac{1}{u_2},\frac{1}{u_1+u_2}; 1\right) H\left(1,0; u_1\right)+\frac{1}{4} G\left(\frac{1}{1-u_3},\frac{u_1-1}{u_1+u_3-1}; 1\right) H\left(1,0; u_1\right)+\frac{1}{4} G\left(\frac{1}{u_3},\frac{1}{u_1+u_3}; 1\right) H\left(1,0; u_1\right)-\frac{1}{4} \gcal\left(\frac{1}{1-u_3},u_{312}; 1\right) H\left(1,0; u_1\right)-\frac{3}{4} H\left(0,0; u_2\right) H\left(1,0; u_1\right)-\frac{3}{4} H\left(0,0; u_3\right) H\left(1,0; u_1\right)+\frac{1}{4} H\left(0,1; \frac{u_1+u_3-1}{u_1-1}\right) H\left(1,0; u_1\right)-\frac{1}{3} \pi ^2 H\left(1,0; u_1\right)-\frac{1}{2} G\left(0,\frac{1}{u_1+u_2}; 1\right) H\left(1,0; u_2\right)-\frac{1}{2} G\left(0,\frac{1}{u_2+u_3}; 1\right) H\left(1,0; u_2\right)+\frac{1}{4} G\left(\frac{1}{1-u_1},\frac{u_2-1}{u_1+u_2-1}; 1\right) H\left(1,0; u_2\right)+\frac{1}{4} G\left(\frac{1}{u_1},\frac{1}{u_1+u_2}; 1\right) H\left(1,0; u_2\right)+\frac{1}{4} G\left(\frac{1}{u_2},\frac{1}{u_1+u_2}; 1\right) H\left(1,0; u_2\right)+\frac{1}{4} G\left(\frac{1}{u_2},\frac{1}{u_2+u_3}; 1\right) H\left(1,0; u_2\right)+\frac{1}{4} G\left(\frac{1}{u_3},\frac{1}{u_2+u_3}; 1\right) H\left(1,0; u_2\right)-\frac{1}{4} \gcal\left(\frac{1}{1-u_1},u_{123}; 1\right) H\left(1,0; u_2\right)-\frac{3}{4} H\left(0,0; u_1\right) H\left(1,0; u_2\right)-\frac{3}{4} H\left(0,0; u_3\right) H\left(1,0; u_2\right)+\frac{1}{4} H\left(0,1; \frac{u_1+u_2-1}{u_2-1}\right) H\left(1,0; u_2\right)-\frac{1}{4} H\left(1,0; u_1\right) H\left(1,0; u_2\right)-\frac{1}{3} \pi ^2 H\left(1,0; u_2\right)-\frac{1}{2} G\left(0,\frac{1}{u_1+u_3}; 1\right) H\left(1,0; u_3\right)-\frac{1}{2} G\left(0,\frac{1}{u_2+u_3}; 1\right) H\left(1,0; u_3\right)+\frac{1}{4} G\left(\frac{1}{u_1},\frac{1}{u_1+u_3}; 1\right) H\left(1,0; u_3\right)+\frac{1}{4} G\left(\frac{1}{1-u_2},\frac{u_3-1}{u_2+u_3-1}; 1\right) H\left(1,0; u_3\right)+\frac{1}{4} G\left(\frac{1}{u_2},\frac{1}{u_2+u_3}; 1\right) H\left(1,0; u_3\right)-\frac{1}{3} \pi ^2 H\left(1,0; u_3\right)
+\frac{1}{4} G\left(\frac{1}{u_3},\frac{1}{u_1+u_3}; 1\right) H\left(1,0; u_3\right)+\frac{1}{4} G\left(\frac{1}{u_3},\frac{1}{u_2+u_3}; 1\right) H\left(1,0; u_3\right)-\frac{1}{4} \gcal\left(\frac{1}{1-u_2},u_{231}; 1\right) H\left(1,0; u_3\right)+\frac{3}{4} H\left(0; u_1\right) H\left(0; u_2\right) H\left(1,0; u_3\right)-\frac{3}{4} H\left(0,0; u_1\right) H\left(1,0; u_3\right)-\frac{3}{4} H\left(0,0; u_2\right) H\left(1,0; u_3\right)+\frac{1}{4} H\left(0,1; \frac{u_2+u_3-1}{u_3-1}\right) H\left(1,0; u_3\right)-\frac{1}{4} H\left(1,0; u_1\right) H\left(1,0; u_3\right)-\frac{1}{4} H\left(1,0; u_2\right) H\left(1,0; u_3\right)+\frac{1}{24} \pi ^2 H\left(1,1; u_1\right)+\frac{1}{24} \pi ^2 H\left(1,1; u_2\right)+\frac{1}{24} \pi ^2 H\left(1,1; u_3\right)+\frac{1}{2} H\left(0; u_2\right) H\left(0,0,0; u_1\right)+\frac{1}{2} H\left(0; u_3\right) H\left(0,0,0; u_2\right)+\frac{1}{2} H\left(0; u_1\right) H\left(0,0,0; u_3\right)-\frac{1}{2} H\left(0; u_2\right) H\left(0,0,1; \frac{u_1+u_2-1}{u_2-1}\right)-\frac{1}{2} H\left(0; u_3\right) H\left(0,0,1; \frac{u_1+u_2-1}{u_2-1}\right)-H\left(0; u_1\right) H\left(0,0,1; \left(u_1+u_2\right)\right)-H\left(0; u_2\right) H\left(0,0,1; \left(u_1+u_2\right)\right)-\frac{1}{2} H\left(0; u_1\right) H\left(0,0,1; \frac{u_1+u_3-1}{u_1-1}\right)-\frac{1}{2} H\left(0; u_2\right) H\left(0,0,1; \frac{u_1+u_3-1}{u_1-1}\right)-H\left(0; u_1\right) H\left(0,0,1; \left(u_1+u_3\right)\right)-H\left(0; u_3\right) H\left(0,0,1; \left(u_1+u_3\right)\right)-\frac{1}{2} H\left(0; u_1\right) H\left(0,0,1; \frac{u_2+u_3-1}{u_3-1}\right)-\frac{1}{2} H\left(0; u_3\right) H\left(0,0,1; \frac{u_2+u_3-1}{u_3-1}\right)-H\left(0; u_2\right) H\left(0,0,1; \left(u_2+u_3\right)\right)-H\left(0; u_3\right) H\left(0,0,1; \left(u_2+u_3\right)\right)-\frac{1}{2} H\left(0; u_2\right) H\left(0,1,0; u_1\right)-\frac{1}{2} H\left(0; u_3\right) H\left(0,1,0; u_2\right)-\frac{1}{2} H\left(0; u_1\right) H\left(0,1,0; u_3\right)+\frac{1}{4} H\left(0; u_2\right) H\left(0,1,1; \frac{u_1+u_2-1}{u_2-1}\right)-\frac{1}{4} H\left(0; u_3\right) H\left(0,1,1; \frac{u_1+u_2-1}{u_2-1}\right)+\frac{1}{4} H\left(0; u_1\right) H\left(0,1,1; \frac{u_1+u_3-1}{u_1-1}\right)-\frac{1}{4} H\left(0; u_2\right) H\left(0,1,1; \frac{u_1+u_3-1}{u_1-1}\right)-\frac{1}{4} H\left(0; u_1\right) H\left(0,1,1; \frac{u_2+u_3-1}{u_3-1}\right)+\frac{1}{4} H\left(0; u_3\right) H\left(0,1,1; \frac{u_2+u_3-1}{u_3-1}\right)+\frac{1}{2} H\left(0; u_2\right) H\left(1,0,0; u_1\right)-\frac{1}{2} H\left(0; u_3\right) H\left(1,0,0; u_1\right)-\frac{1}{2} H\left(0; u_1\right) H\left(1,0,0; u_2\right)+\frac{1}{2} H\left(0; u_3\right) H\left(1,0,0; u_2\right)+\frac{1}{2} H\left(0; u_1\right) H\left(1,0,0; u_3\right)-\frac{1}{2} H\left(0; u_2\right) H\left(1,0,0; u_3\right)-\frac{1}{4} H\left(0; u_3\right) H\left(1,0,1; \frac{u_1+u_2-1}{u_2-1}\right)-\frac{1}{4} H\left(0; u_2\right) H\left(1,0,1; \frac{u_1+u_3-1}{u_1-1}\right)-\frac{1}{4} H\left(0; u_1\right) H\left(1,0,1; \frac{u_2+u_3-1}{u_3-1}\right)-7 H\left(0,0,0,0; u_1\right)-7 H\left(0,0,0,0; u_2\right)-7 H\left(0,0,0,0; u_3\right)+\frac{3}{2} H\left(0,0,0,1; \frac{u_1+u_2-1}{u_2-1}\right)+3 H\left(0,0,0,1; \left(u_1+u_2\right)\right)+\frac{3}{2} H\left(0,0,0,1; \frac{u_1+u_3-1}{u_1-1}\right)+3 H\left(0,0,0,1; \left(u_1+u_3\right)\right)+\frac{3}{2} H\left(0,0,0,1; \frac{u_2+u_3-1}{u_3-1}\right)+3 H\left(0,0,0,1; \left(u_2+u_3\right)\right)+\frac{9}{4} H\left(0,0,1,0; u_1\right)+\frac{9}{4} H\left(0,0,1,0; u_2\right)+\frac{9}{4} H\left(0,0,1,0; u_3\right)-\frac{1}{2} H\left(0,1,0,0; u_1\right)-\frac{1}{2} H\left(0,1,0,0; u_2\right)-\frac{1}{2} H\left(0,1,0,0; u_3\right)+\frac{1}{2} H\left(0,1,0,1; \frac{u_1+u_2-1}{u_2-1}\right)+\frac{1}{2} H\left(0,1,0,1; \frac{u_1+u_3-1}{u_1-1}\right)+\frac{1}{2} H\left(0,1,0,1; \frac{u_2+u_3-1}{u_3-1}\right)+H\left(0,1,1,0; u_1\right)+H\left(0,1,1,0; u_2\right)+H\left(0,1,1,0; u_3\right)-\frac{1}{4} H\left(0,1,1,1; \frac{u_1+u_2-1}{u_2-1}\right)-\frac{1}{4} H\left(0,1,1,1; \frac{u_1+u_3-1}{u_1-1}\right)-\frac{1}{4} H\left(0,1,1,1; \frac{u_2+u_3-1}{u_3-1}\right)+H\left(1,0,0,1; \frac{u_1+u_2-1}{u_2-1}\right)+H\left(1,0,0,1; \frac{u_1+u_3-1}{u_1-1}\right)+H\left(1,0,0,1; \frac{u_2+u_3-1}{u_3-1}\right)+2 H\left(1,0,1,0; u_1\right)+2 H\left(1,0,1,0; u_2\right)+2 H\left(1,0,1,0; u_3\right)+\frac{1}{4} H\left(1,1,0,1; \frac{u_1+u_2-1}{u_2-1}\right)+\frac{1}{4} H\left(1,1,0,1; \frac{u_1+u_3-1}{u_1-1}\right)+\frac{1}{4} H\left(1,1,0,1; \frac{u_2+u_3-1}{u_3-1}\right)+\frac{1}{2} H\left(1,1,1,0; u_1\right)+\frac{1}{2} H\left(1,1,1,0; u_2\right)+\frac{1}{2} H\left(1,1,1,0; u_3\right)-\frac{1}{24} \pi ^2 H\left(0; u_3\right) \hcal\left(1; \frac{1}{u_{123}}\right)-\frac{1}{24} \pi ^2 H\left(0; u_1\right) \hcal\left(1; \frac{1}{u_{231}}\right)-\frac{1}{24} \pi ^2 H\left(0; u_2\right) \hcal\left(1; \frac{1}{u_{312}}\right)+\frac{1}{8} \pi ^2 H\left(0; u_2\right) \hcal\left(1; \frac{1}{v_{123}}\right)-\frac{1}{8} \pi ^2 H\left(0; u_3\right) \hcal\left(1; \frac{1}{v_{123}}\right)+\frac{1}{24} \pi ^2 H\left(0; u_2\right) \hcal\left(1; \frac{1}{v_{132}}\right)-\frac{1}{24} \pi ^2 H\left(0; u_3\right) \hcal\left(1; \frac{1}{v_{132}}\right)-\frac{1}{24} \pi ^2 H\left(0; u_1\right) \hcal\left(1; \frac{1}{v_{213}}\right)+\frac{1}{24} \pi ^2 H\left(0; u_3\right) \hcal\left(1; \frac{1}{v_{213}}\right)-\frac{1}{8} \pi ^2 H\left(0; u_1\right) \hcal\left(1; \frac{1}{v_{231}}\right)+\frac{1}{8} \pi ^2 H\left(0; u_3\right) \hcal\left(1; \frac{1}{v_{231}}\right)+\frac{1}{8} \pi ^2 H\left(0; u_1\right) \hcal\left(1; \frac{1}{v_{312}}\right)-\frac{1}{8} \pi ^2 H\left(0; u_2\right) \hcal\left(1; \frac{1}{v_{312}}\right)+\frac{1}{24} \pi ^2 H\left(0; u_1\right) \hcal\left(1; \frac{1}{v_{321}}\right)-\frac{1}{24} \pi ^2 H\left(0; u_2\right) \hcal\left(1; \frac{1}{v_{321}}\right)-\frac{1}{4} H\left(0; u_2\right) H\left(0; u_3\right) \hcal\left(0,1; \frac{1}{u_{123}}\right)-\frac{1}{4} H\left(1,0; u_2\right) \hcal\left(0,1; \frac{1}{u_{123}}\right)+\frac{1}{24} \pi ^2 \hcal\left(0,1; \frac{1}{u_{123}}\right)+\frac{1}{24} \pi ^2 \hcal\left(0,1; \frac{1}{u_{231}}\right)
-\frac{1}{4} H\left(0; u_1\right) H\left(0; u_3\right) \hcal\left(0,1; \frac{1}{u_{231}}\right)-\frac{1}{4} H\left(1,0; u_3\right) \hcal\left(0,1; \frac{1}{u_{231}}\right)-\frac{1}{4} H\left(0; u_1\right) H\left(0; u_2\right) \hcal\left(0,1; \frac{1}{u_{312}}\right)-\frac{1}{4} H\left(1,0; u_1\right) \hcal\left(0,1; \frac{1}{u_{312}}\right)+\frac{1}{24} \pi ^2 \hcal\left(0,1; \frac{1}{u_{312}}\right)-\frac{1}{4} H\left(0; u_2\right) H\left(0; u_3\right) \hcal\left(0,1; \frac{1}{v_{123}}\right)+\frac{1}{4} H\left(0,0; u_2\right) \hcal\left(0,1; \frac{1}{v_{123}}\right)+\frac{1}{4} H\left(0,0; u_3\right) \hcal\left(0,1; \frac{1}{v_{123}}\right)+\frac{1}{6} \pi ^2 \hcal\left(0,1; \frac{1}{v_{123}}\right)-\frac{1}{4} H\left(0; u_2\right) H\left(0; u_3\right) \hcal\left(0,1; \frac{1}{v_{132}}\right)+\frac{1}{4} H\left(0,0; u_2\right) \hcal\left(0,1; \frac{1}{v_{132}}\right)+\frac{1}{4} H\left(0,0; u_3\right) \hcal\left(0,1; \frac{1}{v_{132}}\right)+\frac{1}{6} \pi ^2 \hcal\left(0,1; \frac{1}{v_{132}}\right)-\frac{1}{4} H\left(0; u_1\right) H\left(0; u_3\right) \hcal\left(0,1; \frac{1}{v_{213}}\right)+\frac{1}{4} H\left(0,0; u_1\right) \hcal\left(0,1; \frac{1}{v_{213}}\right)+\frac{1}{4} H\left(0,0; u_3\right) \hcal\left(0,1; \frac{1}{v_{213}}\right)+\frac{1}{6} \pi ^2 \hcal\left(0,1; \frac{1}{v_{213}}\right)-\frac{1}{4} H\left(0; u_1\right) H\left(0; u_3\right) \hcal\left(0,1; \frac{1}{v_{231}}\right)+\frac{1}{4} H\left(0,0; u_1\right) \hcal\left(0,1; \frac{1}{v_{231}}\right)+\frac{1}{4} H\left(0,0; u_3\right) \hcal\left(0,1; \frac{1}{v_{231}}\right)+\frac{1}{6} \pi ^2 \hcal\left(0,1; \frac{1}{v_{231}}\right)-\frac{1}{4} H\left(0; u_1\right) H\left(0; u_2\right) \hcal\left(0,1; \frac{1}{v_{312}}\right)+\frac{1}{4} H\left(0,0; u_1\right) \hcal\left(0,1; \frac{1}{v_{312}}\right)+\frac{1}{4} H\left(0,0; u_2\right) \hcal\left(0,1; \frac{1}{v_{312}}\right)+\frac{1}{6} \pi ^2 \hcal\left(0,1; \frac{1}{v_{312}}\right)-\frac{1}{4} H\left(0; u_1\right) H\left(0; u_2\right) \hcal\left(0,1; \frac{1}{v_{321}}\right)+\frac{1}{4} H\left(0,0; u_1\right) \hcal\left(0,1; \frac{1}{v_{321}}\right)+\frac{1}{4} H\left(0,0; u_2\right) \hcal\left(0,1; \frac{1}{v_{321}}\right)+\frac{1}{6} \pi ^2 \hcal\left(0,1; \frac{1}{v_{321}}\right)-\frac{1}{2} H\left(0; u_2\right) H\left(0; u_3\right) \hcal\left(1,1; \frac{1}{v_{123}}\right)+\frac{1}{2} H\left(0,0; u_2\right) \hcal\left(1,1; \frac{1}{v_{123}}\right)+\frac{1}{2} H\left(0,0; u_3\right) \hcal\left(1,1; \frac{1}{v_{123}}\right)+\frac{11}{24} \pi ^2 \hcal\left(1,1; \frac{1}{v_{123}}\right)-\frac{1}{24} \pi ^2 \hcal\left(1,1; \frac{1}{v_{132}}\right)-\frac{1}{24} \pi ^2 \hcal\left(1,1; \frac{1}{v_{213}}\right)-\frac{1}{2} H\left(0; u_1\right) H\left(0; u_3\right) \hcal\left(1,1; \frac{1}{v_{231}}\right)+\frac{1}{2} H\left(0,0; u_1\right) \hcal\left(1,1; \frac{1}{v_{231}}\right)+\frac{1}{2} H\left(0,0; u_3\right) \hcal\left(1,1; \frac{1}{v_{231}}\right)+\frac{11}{24} \pi ^2 \hcal\left(1,1; \frac{1}{v_{231}}\right)-\frac{1}{2} H\left(0; u_1\right) H\left(0; u_2\right) \hcal\left(1,1; \frac{1}{v_{312}}\right)+\frac{1}{2} H\left(0,0; u_1\right) \hcal\left(1,1; \frac{1}{v_{312}}\right)+\frac{1}{2} H\left(0,0; u_2\right) \hcal\left(1,1; \frac{1}{v_{312}}\right)+\frac{11}{24} \pi ^2 \hcal\left(1,1; \frac{1}{v_{312}}\right)-\frac{1}{24} \pi ^2 \hcal\left(1,1; \frac{1}{v_{321}}\right)+\frac{1}{2} H\left(0; u_2\right) \hcal\left(0,0,1; \frac{1}{u_{123}}\right)+\frac{1}{2} H\left(0; u_3\right) \hcal\left(0,0,1; \frac{1}{u_{123}}\right)+\frac{1}{2} H\left(0; u_1\right) \hcal\left(0,0,1; \frac{1}{u_{231}}\right)+\frac{1}{2} H\left(0; u_3\right) \hcal\left(0,0,1; \frac{1}{u_{231}}\right)+\frac{1}{2} H\left(0; u_1\right) \hcal\left(0,0,1; \frac{1}{u_{312}}\right)+\frac{1}{2} H\left(0; u_2\right) \hcal\left(0,0,1; \frac{1}{u_{312}}\right)+\frac{1}{4} H\left(0; u_3\right) \hcal\left(0,1,1; \frac{1}{u_{123}}\right)+\frac{1}{4} H\left(0; u_1\right) \hcal\left(0,1,1; \frac{1}{u_{231}}\right)+\frac{1}{4} H\left(0; u_2\right) \hcal\left(0,1,1; \frac{1}{u_{312}}\right)+\frac{1}{4} H\left(0; u_2\right) \hcal\left(0,1,1; \frac{1}{v_{123}}\right)-\frac{1}{4} H\left(0; u_3\right) \hcal\left(0,1,1; \frac{1}{v_{123}}\right)-\frac{1}{4} H\left(0; u_2\right) \hcal\left(0,1,1; \frac{1}{v_{132}}\right)+\frac{1}{4} H\left(0; u_3\right) \hcal\left(0,1,1; \frac{1}{v_{132}}\right)+\frac{1}{4} H\left(0; u_1\right) \hcal\left(0,1,1; \frac{1}{v_{213}}\right)-\frac{1}{4} H\left(0; u_3\right) \hcal\left(0,1,1; \frac{1}{v_{213}}\right)-\frac{1}{4} H\left(0; u_1\right) \hcal\left(0,1,1; \frac{1}{v_{231}}\right)+\frac{1}{4} H\left(0; u_3\right) \hcal\left(0,1,1; \frac{1}{v_{231}}\right)+\frac{1}{4} H\left(0; u_1\right) \hcal\left(0,1,1; \frac{1}{v_{312}}\right)-\frac{1}{4} H\left(0; u_2\right) \hcal\left(0,1,1; \frac{1}{v_{312}}\right)-\frac{1}{4} H\left(0; u_1\right) \hcal\left(0,1,1; \frac{1}{v_{321}}\right)+\frac{1}{4} H\left(0; u_2\right) \hcal\left(0,1,1; \frac{1}{v_{321}}\right)+\frac{1}{4} H\left(0; u_3\right) \hcal\left(1,0,1; \frac{1}{u_{123}}\right)+\frac{1}{4} H\left(0; u_1\right) \hcal\left(1,0,1; \frac{1}{u_{231}}\right)+\frac{1}{4} H\left(0; u_2\right) \hcal\left(1,0,1; \frac{1}{u_{312}}\right)+\frac{1}{4} H\left(0; u_2\right) \hcal\left(1,0,1; \frac{1}{v_{123}}\right)-\frac{1}{4} H\left(0; u_3\right) \hcal\left(1,0,1; \frac{1}{v_{123}}\right)-\frac{1}{4} H\left(0; u_2\right) \hcal\left(1,0,1; \frac{1}{v_{132}}\right)+\frac{1}{4} H\left(0; u_3\right) \hcal\left(1,0,1; \frac{1}{v_{132}}\right)+\frac{1}{4} H\left(0; u_1\right) \hcal\left(1,0,1; \frac{1}{v_{213}}\right)-\frac{1}{4} H\left(0; u_3\right) \hcal\left(1,0,1; \frac{1}{v_{213}}\right)-\frac{1}{4} H\left(0; u_1\right) \hcal\left(1,0,1; \frac{1}{v_{231}}\right)+\frac{1}{4} H\left(0; u_3\right) \hcal\left(1,0,1; \frac{1}{v_{231}}\right)+\frac{1}{4} H\left(0; u_1\right) \hcal\left(1,0,1; \frac{1}{v_{312}}\right)-\frac{1}{4} H\left(0; u_2\right) \hcal\left(1,0,1; \frac{1}{v_{312}}\right)-\frac{1}{4} H\left(0; u_1\right) \hcal\left(1,0,1; \frac{1}{v_{321}}\right)+\frac{1}{4} H\left(0; u_2\right) \hcal\left(1,0,1; \frac{1}{v_{321}}\right)+H\left(0; u_2\right) \hcal\left(1,1,1; \frac{1}{v_{123}}\right)-H\left(0; u_3\right) \hcal\left(1,1,1; \frac{1}{v_{123}}\right)-H\left(0; u_1\right) \hcal\left(1,1,1; \frac{1}{v_{231}}\right)+H\left(0; u_3\right) \hcal\left(1,1,1; \frac{1}{v_{231}}\right)+H\left(0; u_1\right) \hcal\left(1,1,1; \frac{1}{v_{312}}\right)-H\left(0; u_2\right) \hcal\left(1,1,1; \frac{1}{v_{312}}\right)-\frac{3}{2} \hcal\left(0,0,0,1; \frac{1}{u_{123}}\right)-\frac{3}{2} \hcal\left(0,0,0,1; \frac{1}{u_{231}}\right)-\frac{3}{2} \hcal\left(0,0,0,1; \frac{1}{u_{312}}\right)-3 \hcal\left(0,0,0,1; \frac{1}{v_{132}}\right)-3 \hcal\left(0,0,0,1; \frac{1}{v_{213}}\right)-3 \hcal\left(0,0,0,1; \frac{1}{v_{321}}\right)-\frac{1}{2} \hcal\left(0,0,1,1; \frac{1}{u_{123}}\right)-\frac{1}{2} \hcal\left(0,0,1,1; \frac{1}{u_{231}}\right)-\frac{1}{2} \hcal\left(0,0,1,1; \frac{1}{u_{312}}\right)-\frac{1}{2} \hcal\left(0,1,0,1; \frac{1}{u_{123}}\right)-\frac{1}{2} \hcal\left(0,1,0,1; \frac{1}{u_{231}}\right)-\frac{1}{2} \hcal\left(0,1,0,1; \frac{1}{u_{312}}\right)+\frac{1}{4} \hcal\left(0,1,1,1; \frac{1}{v_{123}}\right)+\frac{1}{4} \hcal\left(0,1,1,1; \frac{1}{v_{132}}\right)
+\zeta_3 H\left(0; u_1\right)+\zeta_3 H\left(0; u_2\right)+\zeta_3 H\left(0; u_3\right)+\frac{5}{2} \zeta_3 H\left(1; u_1\right)+\frac{5}{2} \zeta_3 H\left(1; u_2\right)+\frac{5}{2} \zeta_3 H\left(1; u_3\right)+\frac{1}{2} \zeta_3 \hcal\left(1; \frac{1}{u_{123}}\right)+\frac{1}{2} \zeta_3 \hcal\left(1; \frac{1}{u_{231}}\right)+\frac{1}{2} \zeta_3 \hcal\left(1; \frac{1}{u_{312}}\right)-\frac{1}{2} \hcal\left(1,0,0,1; \frac{1}{u_{123}}\right)-\frac{1}{2} \hcal\left(1,0,0,1; \frac{1}{u_{231}}\right)-\frac{1}{2} \hcal\left(1,0,0,1; \frac{1}{u_{312}}\right)+\frac{1}{4} \zeta_3 \hcal\left(1; \frac{1}{v_{123}}\right)+\frac{1}{4} \zeta_3 \hcal\left(1; \frac{1}{v_{132}}\right)+\frac{1}{4} \zeta_3 \hcal\left(1; \frac{1}{v_{213}}\right)+\frac{1}{4} \zeta_3 \hcal\left(1; \frac{1}{v_{231}}\right)+\frac{1}{4} \zeta_3 \hcal\left(1; \frac{1}{v_{312}}\right)+\frac{1}{4} \zeta_3 \hcal\left(1; \frac{1}{v_{321}}\right)+\frac{1}{4} \hcal\left(0,1,1,1; \frac{1}{v_{213}}\right)+\frac{1}{4} \hcal\left(0,1,1,1; \frac{1}{v_{231}}\right)+\frac{1}{4} \hcal\left(0,1,1,1; \frac{1}{v_{312}}\right)+\frac{1}{4} \hcal\left(0,1,1,1; \frac{1}{v_{321}}\right)+\frac{1}{4} \hcal\left(1,0,1,1; \frac{1}{v_{123}}\right)+\frac{1}{4} \hcal\left(1,0,1,1; \frac{1}{v_{132}}\right)+\frac{1}{4} \hcal\left(1,0,1,1; \frac{1}{v_{213}}\right)+\frac{1}{4} \hcal\left(1,0,1,1; \frac{1}{v_{231}}\right)+\frac{1}{4} \hcal\left(1,0,1,1; \frac{1}{v_{312}}\right)+\frac{1}{4} \hcal\left(1,0,1,1; \frac{1}{v_{321}}\right)+\frac{1}{4} \hcal\left(1,1,0,1; \frac{1}{v_{123}}\right)+\frac{1}{4} \hcal\left(1,1,0,1; \frac{1}{v_{132}}\right)+\frac{1}{4} \hcal\left(1,1,0,1; \frac{1}{v_{213}}\right)+\frac{1}{4} \hcal\left(1,1,0,1; \frac{1}{v_{231}}\right)+\frac{1}{4} \hcal\left(1,1,0,1; \frac{1}{v_{312}}\right)+\frac{1}{4} \hcal\left(1,1,0,1; \frac{1}{v_{321}}\right)+\frac{3}{2} \hcal\left(1,1,1,1; \frac{1}{v_{123}}\right)+\frac{3}{2} \hcal\left(1,1,1,1; \frac{1}{v_{231}}\right)+\frac{3}{2} \hcal\left(1,1,1,1; \frac{1}{v_{312}}\right)

\)
\etxtsloppy

\end{document}